\documentclass[english,12pt,reqno]{article}

\usepackage{latexsym, amsmath, amssymb,amsthm, natbib,verbatim, xy}
\xyoption{all}

\usepackage[dvipsnames]{xcolor}
\usepackage{tgpagella}
\usepackage{mathpazo}
\usepackage{float}
\usepackage[left=1in,top=1in,right=1in,bottom=1in]{geometry}
\usepackage{setspace,color}
\usepackage{graphicx}
\usepackage{enumerate}
\usepackage[multiple]{footmisc}
\usepackage{hyperref}
\usepackage{fancyhdr}
\usepackage{rotating}
\usepackage{soul}
\usepackage{caption}
\usepackage{subcaption}
\usepackage[section]{placeins} 

\usepackage{appendix}

\usepackage{epigraph}
\usepackage{etoolbox}
\patchcmd{\section}{\scshape}{\bfseries}{}{}

\usepackage{siunitx}

\newtheorem{theorem}{Theorem}
\newtheorem{corollary}{Corollary}
\newtheorem{definition}{Definition}

\newtheorem{proposition}{Proposition}

\newtheorem{example}{Example}
\theoremstyle{definition}

\newcommand{\adsoplus}{\textsuperscript{+}}

\renewenvironment{quote}{
   \list{}{
     \leftmargin1.5cm  
     \rightmargin\leftmargin}
   \item\relax}
{\endlist}

\newcounter{subexample}[example]
\renewcommand{\thesubexample}{\theexample\alph{subexample}}

\newenvironment{subex}
  {\refstepcounter{subexample}
   \par\noindent\textbf{Example \thesubexample.}\ }
  {\par}

\usepackage{scalerel}
\DeclareRobustCommand{\bigempty}{
  \scalerel*{\ensuremath{\varnothing}}{\ensuremath{X(1)}}}
\DeclareRobustCommand{\biggerempty}{
  \ensuremath{\scaleobj{0.85}{\bigempty}}}

\usepackage{booktabs,array,ragged2e,tabularx,makecell}
\newcolumntype{Y}[1]{>{\RaggedRight\arraybackslash\hsize=#1\hsize}X}
\newcolumntype{L}[1]{>{\RaggedRight\arraybackslash}p{#1}}

\renewcommand{\arraystretch}{1.15}

\usepackage{amssymb}  
\usepackage{booktabs,threeparttable}
\usepackage{array}     
\usepackage{pifont}    
\newcommand{\cmark}{\text{\large\checkmark}}
\newcommand{\xmark}{\text{\sffamily X}}

\begin{document}

\title{Minimalist Market Design: A Framework for Economists with Policy Aspirations}

\author{Tayfun S\"{o}nmez\thanks{Department of Economics, Boston College, email: sonmezt@bc.edu. 
This is the fourth draft of a monograph on the philosophy, evolution, and impact of my minimalist approach to market design.
Neither this approach nor the monograph would exist without the sacrifices of my beloved wife, Banu Bedestenci S\"{o}nmez, whom we lost to cancer in August 2016.
Equally critical was the guidance of my Ph.D. advisor, William Thomson.
Banu left our homeland, Turkey---along with her family, a flourishing career, and a rich community---so that I could pursue our shared vision of building knowledge that could touch lives in the United States.
William, on the other hand, is the only reason I was able to accomplish any academic feat in the first place.
This monograph, along with my broader contributions to research and policy, has significantly benefited from the constant emotional, motivational,
and research-funding support of my best friend, Dalınç Arıburnu, over the years. He was always there for me when I needed him most.
I had the privilege of collaborating with a large number of brilliant researchers.
Naturally, my approach to market design benefited from these experiences.
Above all, my long-term collaborations with Parag Pathak and Utku \"{U}nver deeply
shaped how I think about market design at its most fundamental level. These two amazing people also stuck with me through thick and thin.
I am deeply grateful to them for never leaving me alone in my personal and professional quest.  
Other researchers who made notable contributions to various aspects of the minimalist approach or this monograph include 
Tommy Andersson, Philippe van Basshuysen, Eric Budish, Sylvain Chassang, Yeon-Koo Che, Yan Chen, Laura Doval,  
Federico Echenique, Zo\"{e} Hitzig, Matthew Jackson, Fuhito Kojima, Scott Duke Kominers, Rohit Lamba, David Levine, Shengwu Li, Paul Milgrom, Herv\'{e} Moulin, Marek Pycia, 
Alex Rees-Jones, James Schummer, Joel Sobel, and Alex Teytelboym. 
I am deeply grateful to these colleagues. I also thank Dhiraj Agrawal, Anna Alexandrova, Vahit Atakan, P\'{e}ter Bir\'{o}, In\'{a}cio B\'{o}, 
Nicolas Brisset, O\u{g}uzhan \c{C}elebi, 
Andrew Copland, \'{A}gnes Cseh, Pietro Dall'Ara, Piotr Dworczak, Lars Ehlers, Aytek Erdil, Guillaume Haeringer, Jacob Leshno, 
Arthur Lewbel, David Manlove, Steven Postrel, Joe Quinn, Dani Rodrik, Ran Shorrer, Alex Tordjman,   
Can Urgun, Rakesh Vohra, and Glen Weyl for their detailed comments, which greatly improved both the content and exposition of the monograph. 
Finally, I am especially grateful to Editor Alessandro Pavan and the three anonymous reviewers, whose guidance further enhanced the clarity and quality of the monograph.}}

\date{\normalsize First Version: February 2023\\ This Version: October 2025}

\maketitle

\begin{center}
Forthcoming in the \textit{Econometric Society Monographs Series}
\end{center}

\newpage

\mbox{}\\ \bigskip \bigskip \bigskip \bigskip \bigskip \bigskip

\begin{center}
\mbox{} \textit{This monograph is dedicated to}\\ \medskip
\mbox{} \textit{the loving memory of my soulmate Banu Bedestenci S\"{o}nmez, who left her homeland behind\\ 
so that I could pursue my policy aspirations as an economic designer},\\ 
\mbox{} \textit{and}  \\
\mbox{} \textit{William Thomson, who gave me the gift of rigor} \\
\end{center}

\newpage

\begin{abstract}
Minimalist market design is an economic design framework developed from the perspective of an outsider---one seeking to improve real institutions without a commission or official mandate.
It offers a structured, ``minimally invasive'' method for reforming institutions from within: identify their mission as understood by stakeholders, diagnose the root causes of failure, and refine only those elements that compromise that mission.
By fixing what is broken and leaving the rest intact, the framework respects the tacit knowledge embedded in long-standing institutions, minimizes unintended consequences, and secures legitimacy that facilitates adoption.

Such targeted interventions often call for novel, use-inspired theory tailored to the institutional context.
In this way, minimalist market design advances both theory and practice through a reciprocal process fostering collaboration across disciplines and between academic research and real-world practice.

Tracing the framework’s evolution over twenty-five years of intertwined progress in theory and real-world implementation across a range of matching market applications---including housing allocation, 
school choice, living-donor organ exchange for kidney and liver, military branch assignment in the U.S. Army, the allocation of vaccines and therapies during the COVID-19 pandemic, 
and the allocation of public jobs and college seats under India’s reservation system---this monograph reveals a consistent ``less is more'' ethos, 
showing how restrained, precisely targeted reforms can yield substantial policy improvements while advancing fundamental knowledge.
\end{abstract}

\newpage

\tableofcontents

\newpage

\section{Minimalist Market Design: A Proven Framework}  \label{sec:MMD}

Beginning with the first spectrum auction held by the Federal Communications Commission (FCC) in 1994 \citep{milgrom:2000, Milgrom:2004, Klemperer:2004},\footnote{Key contributors to the design of the 
FCC spectrum auction included John McMillan (consultant to the FCC), Paul Milgrom and Robert Wilson (consultants to Pacific Telesis), and Preston McAfee (consultant to AirTouch Communications) \citep{mcmillan:94, mcafeemcmillan:96}.}
the field of \textit{market design} has emerged over the past three decades, with researchers in auction theory and matching theory playing increasingly active roles in designing and reforming economic and social institutions.

In the decade following the FCC spectrum auction, other prominent applications of this emerging field included reforms of the entry-level labor market for medical doctors in the U.S. \citep{roth/peranson:99}, 
and of school choice mechanisms in Boston and New York City \citep{abdulkadiroglu/sonmez:03, apr:05, aprs:05}, as well as the design of a kidney exchange clearinghouse in New England \citep{roth/sonmez/unver:04, kidneyaea, roth/sonmez/unver:05}.

The increasing prominence of market design in mainstream economics has been recognized through several notable awards. 
In 2012, Alvin Roth and Lloyd Shapley shared the \textit{Nobel Memorial Prize in Economic Sciences} for their contributions to ``Stable Allocations and the Practice of Market Design.'' 
Parag Pathak received the 2018 \textit{John Bates Clark Medal} for his work in market design and the economics of education. 
In 2020, Paul Milgrom and Robert Wilson shared the \textit{Nobel Memorial Prize in Economic Sciences} for their contributions to ``Auction Theory and Inventions of New Auction Formats.'' 
As \cite{jackson:13} notes, ``this is an area where microeconomic theory has had its largest direct impact.''

How did market design come to influence policy across such a broad range of areas? Which factors and research paradigms have contributed to its success? 
And how can its policy relevance and effectiveness be further enhanced? In this monograph, I aim to shed light on these questions by addressing a call made by Roth in his 1999 \textit{Fisher-Schultz Lecture}:

\begin{quote} 
``Whether economists will often be in a position to give highly practical advice on design depends in part on whether we report what we learn, 
and what we do, in sufficient detail to allow scientific knowledge about design to accumulate.''\\
\mbox{} \hfill \citealp{Roth:2002}, p. 1342
\end{quote}

Drawing on over 25 years of experience, I present an institutional design framework---\textit{minimalist market design}---to add to the arsenal of policy-oriented researchers.

\subsection{What is Minimalist Market Design?}

\begin{figure}[!tp]
    \begin{center}
       \includegraphics[scale=1.1]{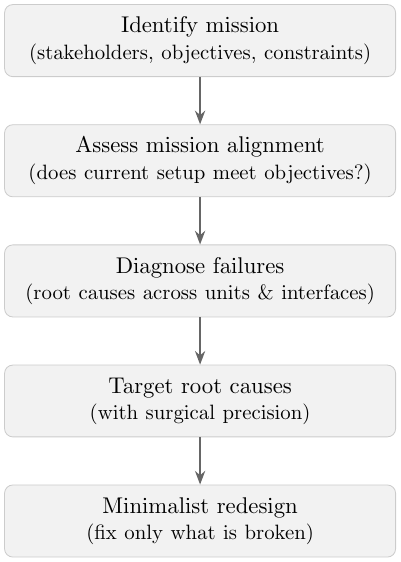}
           \end{center}
\caption{Minimalist market design workflow: identify the mission, assess mission alignment, diagnose failures, target root causes, and implement a minimalist redesign---\emph{a scalpel, not a chainsaw}.}
\label{fig:workflow-MMD}
\end{figure}

At its core, minimalist market design introduces solutions to failures in economic, political, and social institutions---whether through mechanisms, 
governing rules, or even less structured ad hoc systems---by surgically removing design flaws while adhering as closely as possible to current policy goals and the existing setup; 
Figure~\ref{fig:workflow-MMD} summarizes the workflow.

To illustrate this framework, consider an institution established to achieve multiple objectives, composed of several units, each with a specific role and interacting in complex ways. 
When such an institution fails to fulfill its mission, the shortcomings may stem from individual units or the interfaces between them. How can an economic designer effectively address these failures?

Before tackling that question, let's examine how experts in other fields approach similar challenges. A surgeon treating a patient or a mechanic repairing a car begins with a diagnosis---identifying what is wrong. 
They then determine the root causes of the problem. Once identified, they address these underlying causes directly: a surgeon might perform surgery or a transplant, while a mechanic repairs or replaces faulty components.

This is exactly how an economic designer organizes their efforts and operates within the minimalist market design framework.\footnote{Throughout this monograph, the term 
\textit{economic designer} is used interchangeably with \textit{market designer} to refer to researchers in the \textit{market design} community who use formal methods to study or design institutions. 
They are not necessarily trained as economists; many, in fact, come from fields such as computer science, operations research, and electrical engineering.} 

They start by identifying the mission of the institution---the primary objectives of decision-makers, 
system operators, or other stakeholders involved in the institution's design. Understanding the institution's historical evolution is often crucial for recognizing these objectives.

Next, the economist assesses whether the current setup meets these objectives and fulfills the institution's mission. If not, this misalignment presents an opportunity for policy improvement through better design.

To seize this opportunity, the economist identifies the root causes of the failures, pinpointing problematic units or interactions.
The economist then targets these root causes for correction---and nothing else---much like a surgeon performing a minimally invasive procedure.\footnote{This operational principle 
of minimalist market design resonates with Keynes’s famous remark: ``If economists could manage to get themselves thought of as humble, competent people, on a level with dentists, that would be splendid!'' \citep{Keynes:1930}.}
This focused approach bestows the ``minimalist'' characteristic upon the paradigm, distinguishing it from more mainstream approaches in market design.

While minimalism is the hallmark of this framework, it is not its only distinctive feature. Another is its conception through the lens of a researcher aiming to 
influence policy as a ``critical outsider''---a perspective that mirrors my own early career experiences.

In the late 1990s, as a junior economic designer, I observed that emerging market design approaches---for example, the \textit{engineering approach to market design} \citep{roth/peranson:99, Roth:2002}---predominantly, 
if not exclusively, reflected the experiences of leading economists commissioned to design or reform institutions.
The commissioning of such projects by decision-makers indicates their readiness for design or reform, often driven by urgent needs or crises, which makes them particularly receptive to proposed changes.

Recognizing that similar opportunities were unlikely to come my way soon, I realized that these experiences---shaped by more receptive decision-makers---might not be as informative for economic designers like myself, 
who aim to influence policy as critical outsiders. Therefore, I adopted an alternative approach that incorporated a robust persuasion strategy, 
one that could be effective even when decision-makers have vested interests in maintaining the status quo---think of it as integrating political economy constraints into market design.

This pursuit led to another key feature of minimalist market design, further distinguishing it from mainstream approaches: 
a more direct role for theory, often custom-made for specific applications. My goal was to embed an effective persuasion strategy within minimalist market design by leveraging the full power of formal methods to influence policy. 
This approach required a stronger-than-usual link between real-world applications and their formal models, giving rise to "use-inspired" theories.\footnote{As discussed later in Section \ref{sec:Pasteur},
the term ``use-inspired'' was famously coined by \cite{Stokes:1997} when he introduced the concept of \textit{use-inspired basic research}, 
which consists of research where the rationale, conceptualization, and directions are driven by the potential use of the knowledge.}

With this mindset, I charted my own path, striving to develop theories directly relevant to policy. I aimed to make these theories so compelling that, even as an outsider critic, my efforts toward implementation would bear fruit.
Over the past 25 years, I have collaborated with a wide range of co-authors and partners---most regularly with Parag Pathak and Utku \"{U}nver---to develop 
and apply this framework to various resource allocation problems of social value. Beyond my contributions to market design, many of these use-inspired theories have become my primary contributions to matching theory.

Reflecting on this journey, I have seen how time has affirmed the merits of this approach, both in informing policy on problems of social value and in advancing matching theory. 
A central theme of this monograph is the evolution of my holistic approach to theory and practice, culminating in minimalist market design. 
It illustrates how these use-inspired theories and their practical impact have grown together, dynamically shaping and influencing each other over the years.

\subsection{Policy Impact vs. External Validity} \label{sec:policyimpact-externalvalidity}

In his thought-provoking article ``Is the Scientific Paper a Fraud?'', 
Peter Medawar, recipient of the 1960 \textit{Nobel Prize in Physiology or Medicine}, 
criticized the conventional format of scientific writing \citep{medawar:63}. 
He contended that ``the scientific paper may be a fraud because it misrepresents the processes of thought 
that accompany or give rise to the work that is described in the paper.'' 
This critique was later formalized by \cite{Kerr:1998} as \textit{HARKing}---``Hypothesizing After the Results are Known.''\footnote{The replication-crisis literature 
subsequently grouped HARKing with other questionable research practices, including \textit{p}-hacking, selective reporting, 
and undisclosed flexibility in data collection \citep{simmons/nelson/simonsohn:2011, john/loewenstein/prelec:2012}.} 
In a similar vein, \citet{howitt/wilson:2014} argue that even in hypothesis-driven research, ``the trend toward more explicit framing of a hypothesis is often misleading, 
as hypotheses may be framed to explain a set of observations \textit{post hoc}, suggesting a linear process that does not describe the actual discovery.''

Although HARKing---as formalized by \citet{Kerr:1998} and anticipated by \cite{medawar:63}---was originally discussed in the context of 
discovery process in scholarly inquiry, similar challenges 
arise when attributing policy successes to specific scholarly research in applied economics, particularly in market design. 
Often, and I would argue misleadingly, the policy impact attributed to market design scholarship reflects \textit{post hoc} scholarly progress that follows institutional changes, 
especially when the design team is commissioned for the project.\footnote{The point is not to deny credit for the substantive contributions of commissioned designers, 
but to caution against attributing policy impact to the \textit{post hoc} research they may later publish.}

In this monograph, I aim to avoid such distortions in attributing policy impact to specific research by making explicit the underlying thought processes of minimalist market design.
To this end, analogous to the use of \textit{pre-registration} of research hypotheses in many fields,
I focus on applications where policy impact both follows and aligns with the initial research findings, presenting research and policy as a holistic interplay that captures their interaction and mutual evolution.

Accordingly, I discuss two types of policy relevance in this monograph: policy impact and external validity.

\paragraph{Policy Impact.}
If a team of economic designers successfully informs policy---and potentially changes an institution---directly through their research, 
often with their involvement but not necessarily always, I refer to this as \textit{policy impact}.
Crucially, in this definition, research must precede real-world impact. 
If the research is conducted instead after the practical impact has already occurred, 
akin to concerns about HARKing in empirical and experimental studies,
I find it problematic to attribute the impact to the follow-up research.\footnote{Impact may legitimately stem from a researcher’s 
earlier scholarship or practical involvement, but it should not be attributed to their subsequent \textit{post hoc} work.}

\paragraph{External Validity.}
Sometimes, practical relevance comes after the research
but occurs independently of it, when policymakers or other stakeholders develop and adopt solutions akin to those identified in earlier research.
Strictly speaking, I do not classify such outcomes as policy impact; I refer to them instead as \textit{external validity}. In this sense, whenever I speak of external validity, ``life imitates science.''

In the minimalist market design framework---where a key motivation is to uncover the institution’s intended design and to recover it in the least disruptive way by 
addressing the root causes of failure---this scenario is particularly likely. 
If the minimalist exercise proves sound, decision-makers may independently reach similar conclusions and amend their institutions in ways that align with the underlying minimalist design.

While achieving direct policy impact may be considered by many as the more ``glorifying'' outcome, I personally believe that external validity is at least as important and serves a unique role. 
Achieving policy impact---though much less likely for an outsider critic than for a commissioned insider---can be influenced by factors unrelated to the merits of the proposed design, such as personal charisma, 
strong connections, an urgent need to avert a crisis, or sheer luck. In contrast, external validity serves as direct proof that the minimalist framework is working exactly as intended. 
It strongly confirms that the analysis has addressed the root causes of an institution’s failures and formulated a desirable improvement.

\subsection{Evolution of Minimalist Market Design}

Minimalist market design has evolved through three overlapping phases, with each subsequent phase building upon the lessons and experiences of the earlier ones.

Phase 1 began in the late 1990s with my initial efforts in crafting use-inspired theories for two applications: on-campus housing allocation \citep{abdulkadiroglu/sonmez:98, abdulkadiroglu/sonmez:99} 
and Turkish college admissions \citep{balinski/sonmez:99}. Although my first policy initiative in Turkish college admissions was unsuccessful, these experiences were invaluable in shaping my future work.

Phase 2 emerged from that initial setback and spanned from the late 1990s to the late 2000s. During this period, minimalist market design evolved through a series of challenges and successes. 
Notable breakthroughs included work in school choice \citep{abdulkadiroglu/sonmez:03} and kidney exchange \citep{roth/sonmez/unver:04, roth/sonmez/unver:05, rsuds:06, roth/sonmez/unver:07}. 
These achievements greatly benefited from the use-inspired theories developed in Phase 1, even when the original target applications differed. I began deploying customized use-inspired theories 
as a means of persuasion, which proved increasingly effective.

Phase 3---which began in the early 2010s and is currently ongoing---marks my deliberate and systematic application of minimalist market design in shaping both my research and policy initiatives. 
In this phase, working with diverse teams, I have applied---often successfully, though not always---the framework across various areas, including:

\begin{enumerate}
\item Analysis and reform of the U.S. Army's branching process for assigning cadets to military specialties \citep{sonmez/switzer:13, sonmez:13, greenberg/pathak/sonmez:24}.
\item Design, analysis, and implementation of a single-center living-donor liver exchange system in Turkey, which became a global leader within two years of its launch \citep{ergin/sonmez/unver:20, Malatya:2023, Malatya:2024}.
\item Development and implementation of medical resource allocation policies and procedures during the COVID-19 pandemic \citep{Sonmezetal-categorized:2021, rubin-et-al:21, 
white-et-al:22, McCreary:23, pathak/sonmez/unver/yenmez:24}.
\item Critical analysis of key judgments by the Supreme Court of India on affirmative action, some of which were later reflected in subsequent reversals of these decisions, 
providing external validity for minimalist market design \citep{sonyen22, sonmez/yenmez:24, sonmez-unver2022}.
\end{enumerate}

Reflecting on these phases, I see how each has contributed to a progressively richer and more mature paradigm of minimalist market design, 
continuously evolving through new applications and refined strategies. 
This journey has not only shaped my professional growth but has also advanced the field, demonstrating the power of a focused, rigorous, and adaptable framework.

\subsection{Aims, Theses and Outline of the Monograph} \label{sec:aims}

In this monograph, I have three interrelated aims, each with a corresponding thesis.

First, I aim to outline a coherent economic design framework---the \textit{minimalist market design}---which adopts a holistic approach to research and policy. 
Intended for policy-oriented scholars who aspire to influence policy from outside the system, this framework enables even outsider critics to build persuasive cases for change grounded in practical, use-inspired basic research. 
The thesis associated with this aim is that by adopting minimalist market design, these scholars can increase the likelihood that their proposals will be adopted by decision-makers and implemented in practice.

Second, I aim to demonstrate that basic theory can be advanced through policy-oriented research, even in complex settings. I propose that minimalist market design serves as a natural framework for 
conducting use-inspired basic research within the type of scholarship that \cite{Stokes:1997} refers to as \textit{Pasteur's Quadrant}. Exemplified by Louis Pasteur's 19th-century research in microbiology, 
this approach challenges the traditional division between basic and applied science, embracing a holistic path to both scientific discovery and practical invention.

Third, I aim to provide a detailed intellectual history of key innovations and policy breakthroughs in the theory and practice of matching markets over the past 25 years. 
By connecting these innovations and breakthroughs through a series of \textit{Discovery--Invention Cycles}  \`{a} la \cite{Narayanamurti/Odumosu:2016}, 
I also highlight the role of minimalist market design in these achievements, as illustrated in Figures \ref{fig:DIC-SC-arm} and \ref{fig:DIC-KE-arm}.

As a roadmap and preview, Table~\ref{tab:minimalist-policy-relevance} compiles a broad set of minimalist designs discussed in this monograph that achieved policy relevance---both contributions from 
my research collaborations and notable designs by others---indicating the year of adoption, the form of policy relevance (whether direct policy impact, external validation, or commissioned design), and the triggers for adoption.


\begin{table}[!htbp]
\footnotesize
\centering
\begin{threeparttable}
\caption{Minimalist designs in the monograph that influenced real-world practice, with year of adoption, form of policy relevance of the underlying research, and adoption triggers.}
\label{tab:minimalist-policy-relevance}
\setlength{\tabcolsep}{4pt}
\renewcommand{\arraystretch}{1.06}
\begin{tabular}{@{}p{0.22\textwidth} p{0.10\textwidth} p{0.31\textwidth} p{0.27\textwidth} p{0.10\textwidth}@{}}
\toprule
\addlinespace[4pt] 
\textbf{Application} & \textbf{Year} & \textbf{Form of Policy Relevance} & \textbf{Trigger of Adoption} & \textbf{Section} \\ 
\addlinespace[4pt] 
\midrule
\addlinespace[2pt] 

Second-price Auction
&  
&  
&  \\

\addlinespace[2pt] 
\cline{2-5}
\addlinespace[3pt] 

\hspace{6pt}Stamp auctions 
& 1893
& Practitioner-driven \mbox{(External Validation)}    
& \mbox{Sealed bids}; \mbox{Parity with English auction}
& \ref{sec:vickrey} \\

\addlinespace[2pt] 
\cline{2-5}
\addlinespace[3pt] 
\hspace{6pt}\mbox{New Zealand} \mbox{\hspace{4pt} Spectrum Auction}  
&1989--90
& Research-driven \mbox{(Policy Impact)}   
& Transparency  
& \ref{sec:vickrey} \\

\midrule

\mbox{FCC Spectrum Auctions} \mbox{(Collusion mitigation)}
& 1997; \mbox{2008}
& Policy-maker-driven \mbox{(External Validation)}   
& \mbox{Tacit collusion}; \mbox{Transparency}  
& \ref{sec:FCC} \\

\midrule

School Choice
&
&  
&  
&  \\

\addlinespace[2pt] 
\cline{2-5}
\addlinespace[3pt] 
\hspace{6pt}Boston
& 2005 
& Research-driven \mbox{(Policy Impact)} 
& Incentive compatibility  
& \ref{sec:Boston}  \\

\addlinespace[2pt] 
\cline{2-5}
\addlinespace[3pt] 
\hspace{6pt}England
& 2007
& Policy-maker-driven \mbox{(External Validation)} 
& Incentive compatibility  
& \ref{sec:England} \\

\addlinespace[2pt] 
\cline{2-5}
\addlinespace[3pt] 
\hspace{6pt}Chicago
& 2009  
& Policy-maker-driven \mbox{(External Validation)}   
& \mbox{No justified envy}; \mbox{Incentive compatibility}  
& \ref{sec:Chicago} \\

\midrule

Kidney Exchange
&
&  
&  
&  \\

\addlinespace[2pt] 
\cline{2-5}
\addlinespace[3pt] 
\hspace{6pt}NEPKE
& 2004 
& Research-driven \mbox{(Policy Impact)} 
& Efficiency; Ethics  
&  \ref{sec:NEPKE} \\

\addlinespace[2pt] 
\cline{2-5}
\addlinespace[3pt] 
\hspace{6pt}3-way cycles
& 2006
& Research-driven \mbox{(Policy Impact)}  
& Efficiency  
& \ref{sec:3-way-ndd-chain} \\

\addlinespace[2pt] 
\cline{2-5}
\addlinespace[3pt] 
\hspace{6pt}NDD chains
& 2007 
& Research-driven \mbox{(Policy Impact)}   
& Efficiency 
&  \ref{sec:3-way-ndd-chain}\\

\midrule

\mbox{AEA Job market} \mbox{(Signaling)}  
& 2006
& Commissioned  
& Coordination failures
& \ref{sec:AEA} \\

\midrule

\mbox{Reputation Systems} (Taobao) 
& 2012
& Research-driven \mbox{(Policy Impact)}   
& \mbox{Low response rate}; Inflated ratings; Cold-start barrier  
& \ref{sec:Taobao} \\

\midrule

Platform Economy 
&  
&  
&  \\

\addlinespace[2pt] 
\cline{2-5}
\addlinespace[3pt] 

\hspace{6pt}Airbnb 
& 2016; \mbox{2018}
& Research-driven \mbox{(Policy Impact)}      
& Racial bias
& \ref{sec:Airbnb-Uber} \\

\addlinespace[2pt] 
\cline{2-5}
\addlinespace[3pt] 
\hspace{6pt}Uber
& 2024
& Research-driven \mbox{(Policy Impact)}   
& Racial bias  
& \ref{sec:Airbnb-Uber} \\

\midrule

\mbox{Army Branching} \mbox{(USMA, ROTC)} 
& 2020
& Research-driven  \mbox{(Policy Impact)}   
& No justified envy; Incentive compatibility; Transparency
& \ref{sec:USMA-ROTC}\\

\midrule

\mbox{Affirmative Action} \mbox{(India)}  
& 2020
& \mbox{Supreme Court mandate} \mbox{(External Validation)}   
& No justified envy 
& \ref{sec:India-EV} \\

\midrule

COVID-19 \mbox{Resource Allocation} 
&  2020 
& Research-driven \mbox{(Policy Impact)}   
& Ethics; Transparency
& \ref{sec:vaccine}--\ref{sec:Oregon} \\

\midrule

\mbox{Liver Exchange} (Turkey)  
& 2022
& Research-driven  \mbox{(Policy Impact)} 
& Efficiency; Ethics  
& \ref{sec:BBS--LPE} \\

\bottomrule
\end{tabular}
\end{threeparttable}
\end{table}


The remainder of this monograph is organized as follows:

Section \ref{sec:foundations} introduces the foundations and key features of minimalist market design.
Section \ref{sec:theory} examines the interaction between theory and policy within the idealized framework of minimalist market design presented in this monograph. 
It also discusses the \textit{linear model} of innovation outlined by \cite{Bush:1945}, which shaped U.S. innovation policy for five decades following World War II, 
alongside two alternative paradigms: \textit{Pasteur's Quadrant}, proposed by \cite{Stokes:1997} and influential since the late 1990s, and \textit{Discovery--Invention Cycles}, introduced by \cite{Narayanamurti/Odumosu:2016}.
Section \ref{sec:guide} offers advice on conducting research and engaging with policy using minimalist market design and includes a practical guide.
Sections \ref{sec:HA} through \ref{sec:LE} present a series of matching market applications, including: \textit{housing allocation} (Section \ref{sec:HA}), 
\textit{school choice} (Section \ref{sec:schoolchoice}), \textit{kidney exchange} (Section \ref{sec:KE}), \textit{cadet--branch matching} (Section \ref{sec:Army}), 
\textit{reserve systems} (Section \ref{sec:reserve-systems}), \textit{affirmative action in India} (Section \ref{sec:India}), \textit{emergency rationing of scarce medical resources} 
(Section \ref{sec:pandemic}), and \textit{liver exchange} (Section \ref{sec:LE}). 
Section \ref{sec:behavioral} presents insights, contributions, and future directions in behavioral market design. 
Section \ref{sec:additionalapplications} highlights additional applications advanced by other researchers and practitioners across a wide variety of contexts, framed through the lens of minimalist market design.
Finally, Section \ref{sec:epilogue} concludes the monograph.

Throughout this monograph, I have integrated ideas and results from several of my earlier publications. 
The purpose here is not to simply re-present prior work, but to synthesize, extend, and reinterpret it under the overarching theme of minimalist market design. 
Where exposition or results closely parallel earlier work, the relevant articles are cited in the corresponding sections. 
In particular, several parts of the monograph draw directly on prior papers, including Section \ref{sec:HA} on \cite{abdulkadiroglu/sonmez:99}; 
Section \ref{sec:schoolchoice} on \cite{balinski/sonmez:99, abdulkadiroglu/sonmez:03, pathak/sonmez:13}; Section \ref{sec:KE} on \cite{roth/sonmez/unver:04, sonmez/unver:23}; 
Section \ref{sec:Army} on \cite{sonmez/switzer:13, greenberg/pathak/sonmez:24}; Section \ref{sec:reserve-systems} on \cite{dur/kominers/pathak/sonmez:18}; 
Section \ref{sec:India} on \cite{sonyen22, sonmez/yenmez:24}; and Section \ref{sec:pandemic} on \cite{pathak/sonmez/unver/yenmez:24}.

\section{Foundations and Features of Minimalist Market Design} \label{sec:foundations}

I have been fortunate to actively participate in this exciting field since the late 1990s. Thanks to minimalist market design, 
my efforts have resonated not only with the academic community but also with decision-makers and laypeople across various areas. 

In this section, I explore the foundations that have shaped my journey, as well as the key characteristics and additional features of my approach.

\subsection{Navigating the Challenges of Policy Influence as an Outsider Critic} \label{sec:commissioned-vs-aspired}

Let’s begin by exploring the motivation behind minimalist market design. Why did I, as an aspiring economic designer, venture into uncharted territory and experiment with an untested approach, 
rather than follow the methods established by the pioneers of this emerging field?

At first glance, this path might seem unconventional. However, my decision was driven by a clear priority: to produce research that drives meaningful policy change rather than work aimed solely at an academic audience. 
This focus naturally steered me toward a different approach, one that diverged from the methods favored by the field's pioneers.

At that time, most market design literature and methodologies focused on commissioned design or reform projects, where the need for change was already recognized 
and expert input was actively sought---often in response to an urgent crisis. As an aspiring economic designer, I realized that such opportunities might not come my way for a long time if ever, 
posing a significant challenge to my ambition of making a positive impact through economic design. 
Moreover, I believed these experiences were less informative for situations where decision-makers had yet to acknowledge the 
need for change and were likely to have vested interests in preserving the status quo.

This is not to say that commissioned market design projects do not yield significant theoretical contributions, which were central to my focus. 
They certainly do, much like the valuable scholarly work produced from these projects using other methods. However, these contributions often come after the project's completion, 
reflecting lessons learned during or after its execution. As a result, many of these contributions---finalized post-project---are directed more toward an academic audience than practitioners. 
More importantly, it was not this research that enabled the original commission.

Therefore, I questioned whether lessons from commissioned market design projects---typically written for an academic audience---would be equally effective for critics like myself seeking to 
inform policy from outside the system. While they might be, I had my doubts. Consequently, I developed a framework tailored to the needs of an aspiring economic designer offering unsolicited policy advice as an outsider, 
often challenging established institutions.

\subsection{Three Pillars of Minimalist Market Design} \label{sec:pillars}

To address these challenges, minimalist market design is built on three foundational pillars, illustrated in Figure \ref{fig:pillars}:

\begin{figure}[!tp]
    \begin{center}
       \includegraphics[scale=1.0]{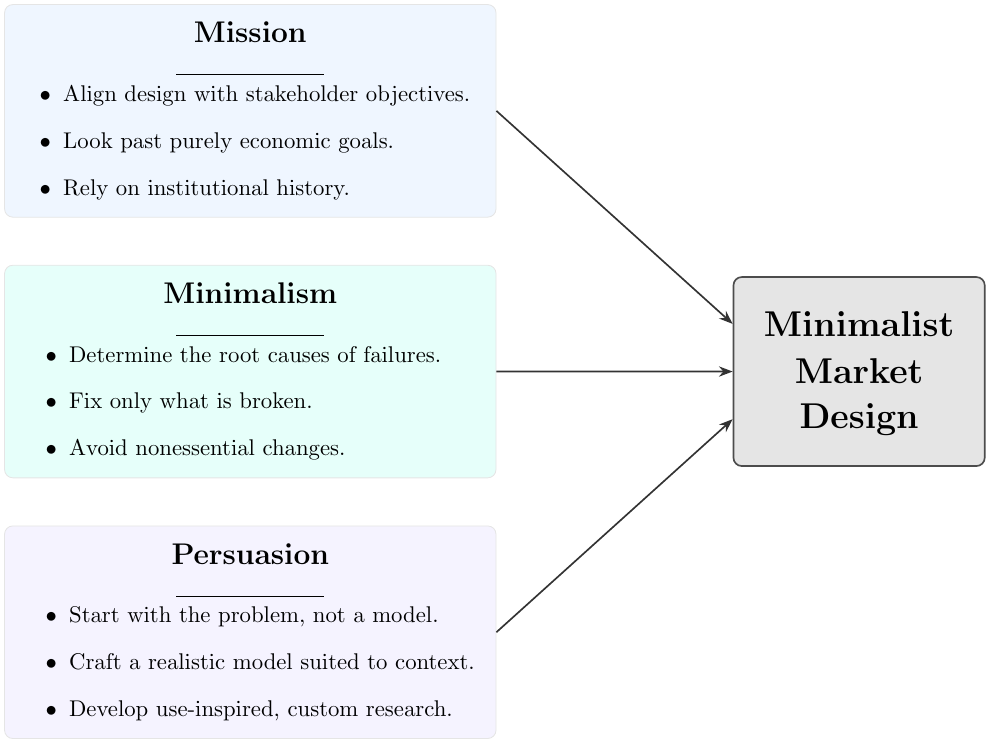}
           \end{center}
\caption{Pillars of Minimalist Market Design.}
\label{fig:pillars}
\end{figure}

\begin{enumerate}
\item \textbf{Identify the Mission.} Decision-makers and other stakeholders across economic, social, and political environments may be flexible about the 
technical aspects of an institution---such as its allocation mechanisms or operational rules---but they often hold firm views about the underlying policy objectives. 
Thoroughly understanding these objectives is crucial for effective design.

\item \textbf{Avoid Nonessential Changes.} As an outside critic, you may not fully understand the role of every aspect of an institution. 
Therefore, avoid altering elements that do not affect the primary objectives; doing so minimizes the risk of unintended consequences.

\item \textbf{Incorporate a Persuasion Strategy.} As an outsider, your ideas are likely to face resistance from decision-makers. 
To overcome this, it is essential to build a compelling case to persuade them to undertake potentially costly reforms. 
\end{enumerate}

Based on these pillars, minimalist economic designers confronting a flawed institution first identify the key policy objectives, 
then determine the root causes of failure, and finally address them by targeting those causes directly. To maximize practical relevance and to persuade stakeholders, 
they typically develop custom, application-specific models and advance new theory to help identify and resolve those root causes.

Let's explore each of these pillars in more detail.

\subsubsection{Establishing the Institutional Mission}

An institution's mission may not always align with the policy objectives typically pursued by academic economists. 
As part of the first pillar of minimalist market design, studying an institution's history can provide valuable insights into its underlying policy goals. 

At this point, a clarification is useful. 
In this monograph, ``policy objectives within an institution's mission'' refer to goals that can be publicly discussed.
I call such objectives \textit{legitimate}. Thus, if an objective cannot be openly stated, it does not qualify as legitimate under my definition. 
While the legitimacy of an objective may not always be clear-cut, to be a viable objective within the context of minimalist market design, 
it should at least be open to public discussion.\footnote{To be clear, I do not mean to imply that an objective that cannot be publicly stated is necessarily illegitimate, although it can be.}

With this clarification on terminology, minimalist market design is conceived as a framework that serves legitimate objectives,\footnote{In principle, the existing institution may embed various illegitimate objectives. 
By maintaining as much of the existing institution as possible, a reform guided by minimalist market design may also inherit some of these aspects. 
In applications where the legitimate objectives imply a single design, this cannot happen. However, in applications where multiple designs satisfy the legitimate objectives, it can. 
Therefore, the best practice in those cases is to provide a complete description of these institutions, including institutions where the illegitimate objectives are weeded out. 
This exercise is referred to as a ``full characterization'' in axiomatic methodology, an analytical framework that often plays a key role in minimalist market design.} 
and is largely intended for institutions governed by honest and democratically elected officials who genuinely aim to enhance these institutions.\footnote{Although the framework 
could be misused to serve the selfish objectives of a politically motivated or malicious decision-maker or interest group, 
I would not consider such misuse as an exercise in minimalist market design, since the framework was developed for public service.}

While \textit{preference utilitarianism}---a \textit{consequentialist} principle focused solely on welfare derived from outcomes---is the default normative position in mainstream economics, 
this perspective is not universally shared across other disciplines, among policymakers, or within the broader population. In practical institution design, particularly where issues of social justice are paramount, 
institutions often pursue normative objectives that diverge from mainstream economic principles, incorporating non-consequentialist considerations that account for factors beyond outcome desirability \citep{Li2017, hitzig2020}. 
Therefore, as an interdisciplinary field, market design must accommodate a range of perspectives extending beyond the traditional focus on outcomes.

One important non-consequentialist principle is \textit{transparency} \citep{hitzig2018, vanBasshuysen:2022}. 
A related non-consequentialist principle, central to many applications discussed in this monograph, is \textit{strategy-proofness}. This principle ensures that participants are never penalized for honestly revealing their private information. 
Even when some participants may not fully understand that truthful revelation is in their best interests, it is often the responsibility of the system operator to create conditions that encourage honest behavior.

By implementing a \textit{strategy-proof} mechanism, policymakers can reduce the risk of misinformation and potential liability, while fostering transparency and trust. 
Though participants may not always follow this guidance for various reasons, the ability to provide clear, straightforward advice remains one of the most compelling features of a \textit{strategy-proof} mechanism.

An illustrative example of this can be found in Section \ref{sec:England}. In 2007, a \textit{strategy-proof} mechanism was mandated across England for the allocation of 
state school seats through a reform of the Education Code. According to Alan Johnson, the Education Secretary at the time, the motivation was that the previous mechanism 
compelled many parents to engage in strategic behavior in order to secure school places for their children. 
Thus, \textit{strategy-proofness} was an objective of the policymakers not solely for its consequences on the outcomes  but rather because of the values it promoted.

Minimalist market design can accommodate a wide range of normative principles, including \textit{non-consequentialist} ones.\footnote{As such, this framework aligns with \cite{Sen:1987}, 
who advocates for reconnecting the two roots of economics in ``ethics'' (normative origin) and ``engineering'' (positive origin), 
which diverged with the rise of neoclassical economics. \cite{Jackson2019} credits \cite{Keynes:1904} for distinguishing positive economics from normative economics.}
The axiomatic methodology, which involves defining desirable properties for solutions and identifying solutions that satisfy these properties, 
is particularly useful in this context \citep{moulin:88, moulin_fair2004, thomson:01, thomson:2011}.\footnote{According to \cite{thomson:01}, 
an axiomatic study begins by specifying problems and listing desirable properties for their solutions, then describes solutions that meet these properties. 
For a discussion of the role of axiomatic methodology in market design, see  \cite{Schummer2019}.} 

Assisting policymakers in formulating and accommodating their normative goals has been a productive avenue for minimalist economic designers.
As noted by \cite{Li2017}, while policymakers may possess deep knowledge of their specific environments, they often struggle to articulate their ethical requirements with precision. 
Economists, with their comparative advantage in defining the normatively relevant properties of complex systems, can help translate these ethical considerations into formal concepts or mechanisms.

In line with this, recent research in minimalist market design has effectively supported officials in both formulating and implementing their normative objectives. 
Notable examples include the 2020 reform of the U.S. Army's branching system (see  Section \ref{sec:Army}) 
and the equitable allocation of scarce medical resources during the COVID-19 pandemic across several U.S. jurisdictions (see  Section  \ref{sec:pandemic}). 
In addition, minimalist market design has demonstrated external validity in recent years, particularly in the context of Indian Supreme Court judgments on affirmative action (see  Section \ref{sec:India}). 

Understanding the mission is crucial, but it is equally important to ensure that changes do not inadvertently disrupt essential components of the existing system.

\subsubsection{Avoiding Nonessential Changes}

Although an economic designer may not uncover every policy goal of stakeholders, existing institutions often contain critical information about the system that is difficult, if not impossible, to obtain otherwise. 
Even when an institution is deeply flawed, it may still reflect important aspects of the problem that an aspiring market designer might overlook. 
In this context, the second pillar of minimalist market design aligns closely with \textit{Chesterton's Fence}, a well-known principle articulated by G.K. Chesterton in the 
opening paragraph of the chapter ``The Drift from Domesticity'' in his book \textit{The Thing} \citep{Chesterton:1929}:

\begin{quote} 
``In the matter of reforming things, as distinct from deforming them, there is one plain and simple principle; a principle which will probably be called a paradox. 
There exists in such a case a certain institution or law; let us say, for the sake of simplicity, a fence or gate erected across a road. 
The more modern type of reformer goes gaily up to it and says, `I don't see the use of this; let us clear it away.' To which the more intelligent type of reformer will do well to answer: 
`If you don't see the use of it, I certainly won't let you clear it away. Go away and think. Then, when you can come back and tell me that you do see the use of it, I may allow you to destroy it.' '' 
\end{quote}

This principle underscores the second pillar of minimalist market design: avoiding nonessential changes to prevent unintended consequences. 
Many things that seem superfluous can serve important roles not immediately apparent.

In minimalist market design, Chesterton's Fence is reflected in the following operational principle: when proposing a reform, only alter elements that are ``clearly broken'' with respect to stakeholders' objectives---namely, 
the root causes of failure, which often stem from a lack of formal expertise. Beyond these essential changes, refrain from making any others. Consistent with Chesterton's main message, 
this approach helps prevent unintended consequences by avoiding the disruption of essential institutional elements. Moreover, it presents stakeholders with designs that closely resemble the original institution, 
minus its shortcomings. Ideally, the central planner should feel, ``this is the institution we intended to design in the first place,'' thereby supporting the third pillar of minimalist market design as well.

This caution has deep antecedents in political philosophy and policy analysis. 
Edmund Burke’s defense of gradual reform stressed that long-standing institutions embody tacit knowledge accumulated over generations; 
well-meaning attempts at a full-scale overhaul risk destroying functions not yet understood \citep{Burke:1790}. 
For Burke, legitimate improvement is piecemeal and evolutionary: it involves correcting faults while preserving the inherited structure, 
and placing the burden of proof on those who would overturn established arrangements. Together with Chesterton’s Fence, 
Burkean conservatism provides a normative rationale for changing only what is demonstrably defective---fix what is broken, and respect the rest \citep{Burke:1790}.

Charles Lindblom’s classic account of policymaking as ``muddling through'' provides a pragmatic complement to this ethos. 
According to Lindblom, rather than comprehensive, blueprint-style redesign, effective policy change typically proceeds by successive limited modifications---small moves from the status quo that target identifiable problems, 
invite feedback, and minimize unforeseen spillovers \citep{Lindblom:1959}. Incrementalism is not timidity but a method for learning under complexity and political constraint. 
In practice, it directs analytic attention toward feasible amendments while preserving institutional components that are functioning tolerably well. 
That is precisely the minimalist approach: target the root cause, avoid unnecessary interference.

In public finance, Martin Feldstein's theory of tax reform exemplifies how targeted adjustments can deliver durable change without uprooting foundational structures. 
Feldstein draws a sharp distinction between designing a system \textit{de novo} and reforming an existing one, urging policymakers to 
treat the status quo as the baseline and focus on correcting clear deficiencies rather than pursuing an idealized design \citep{Feldstein:1976}.
His approach---identifying concrete failures in the status quo, devising analytically well-grounded and administratively feasible corrections, 
and implementing these targeted fixes while leaving the rest of the system intact---embodies the minimalist ethos of repairing what is demonstrably broken and nothing more.

Taken together, these perspectives reinforce the prudence of avoiding nonessential changes. Burke frames the normative foundation of inherited wisdom; 
Lindblom offers an operational logic for incremental correction under complexity; and Feldstein clarifies the distinction between reform and unrestricted design from scratch.
Each points to the same rule of thumb that guides minimalist market design: alter only those components that compromise the mission, and preserve the rest, thereby reducing unintended consequences and easing persuasion.

This rule of thumb is not entirely new to the scholarship and practice of market design. A prevailing view holds that all else equal---such as the expected value generated by the intervention---the closer a proposed reform 
aligns with the existing system, the more likely it is to gain stakeholder support. What distinguishes minimalist market design is the degree to which it operationalizes this insight---elevating the avoidance of 
nonessential changes from a rule of thumb to a design constraint. 
No other economic design framework places as much emphasis on avoiding changes not directly related to institutional failures, and this steadfast discipline is precisely what gives the minimalist approach its name.

With a clear understanding of the mission and a focus on avoiding nonessential changes, the next step involves persuading stakeholders to embrace the proposed reforms.

\subsubsection{Forging a Design Framework with a Built-in Persuasion Strategy}

Building on the first two pillars---identifying the mission and avoiding nonessential changes---the third pillar highlights the crucial role of persuasion in driving reforms. 
Influencing real-world policies or prompting costly institutional changes from the outside can be challenging, even with exceptional research, 
as decision-makers often resist external critiques due to vested interests in maintaining the status quo.

One of the primary motivations for minimalist market design is to address these challenges when an economic designer functions as an external critic rather than a commissioned guide. 
In such contexts, the need for change is rarely acknowledged, and a critical perspective from an eager researcher is often met with resistance.

To overcome this resistance, the research model must be highly realistic, capturing all essential aspects of the problem---especially when addressing decision-makers who are reluctant to change. 
In my experience, this often necessitates developing a custom-made theory rather than relying on simplified, off-the-shelf research. 
Unlike experts who are commissioned to guide reforms---often brought in to address urgent crises---an aspiring economic designer cannot depend on simplified analyses to succeed, 
particularly when attempting to dismantle an existing institution. Therefore, for such ambitions to bear fruit, the model must be meticulously realistic, and the arguments for reform must be compelling and airtight.

Consistent with the first two pillars of minimalist market design, once a realistic model is crafted, the proposed solutions must closely align with the aims of policymakers and other stakeholders, 
ideally without inadvertently compromising any unidentified objectives. Essentially, it must be demonstrated that the proposed design better meets the stakeholders' goals than the current system. 
Such reforms provide genuine value to decision-makers and are easier to justify to superiors or constituents.

From its inception, a research program guided by minimalist market design aims to inform policy directly. The underlying policy aspirations shape the research questions, 
making it a holistic approach to both research and policy, with an emphasis on targeted interventions to the status quo. Here, ``targeted interventions'' do not imply minor tweaks 
but rather adjustments that directly address issues undermining the institution's mission. Instead of creating a brand-new institution, the primary goal is to improve the existing one. 
This focus is crucial in determining which elements to retain and which to amend, thereby helping to establish the primitives of the analytical model.

Relatedly, except in very rare cases, minimalist market design almost always begins with a practical problem to solve; modeling and analysis come next. 
You cannot realistically introduce and solve an analytical model and then hope it happens to address a significant practical problem. The motivating practical problem must be taken very seriously. 
In this sense, underlying policy aspirations shape the research questions in minimalist market design. This approach stands in sharp contrast to most theory papers, 
where the application is treated merely as an ex-post reflection or an excuse to write down a model the authors wanted to solve.

Within this framework, the market designer's initial role is that of a supportive partner. The only interference with the existing institution involves correcting or enhancing flawed elements that require formal expertise. 
This approach helps establish trust and allows stakeholders to ``own'' the proposed design. By adhering to this framework, 
economic designers avoid adopting any normative positions---intentional or unintentional---beyond aligning with the stakeholders' own objectives during initial interactions with likely skeptical parties. 
This strategy helps convince stakeholders that the proposer has no ideological agenda or ulterior motive. 
Providing pro bono support can also be highly valuable---perhaps even necessary---in building trust.

\subsection{Characteristics and Features of Minimalist Market Design}

In light of these pillars, minimalist market design embodies several characteristics that operationalize its foundational principles:

\begin{itemize} 
\item \textbf{Holistic Approach:} Utilizes an integrated methodology that connects research and policy, ensuring that policy objectives shape research questions, 
thereby increasing the likelihood that research will directly inform practical applications.

\item \textbf{Mission-Driven Focus:} Centers on the objectives of decision-makers, system operators, and other stakeholders, extending beyond the typical focus of mainstream economists. 
Frequently derives these objectives from historical contexts.

\item \textbf{Minimally Invasive Design:} Aims to achieve policy objectives through institutional designs that require minimal changes to existing structures, often guided by rigorous axioms or welfare functions.

\item \textbf{Built-in Political Economy Constraints:} Embeds a strong persuasion strategy to win over skeptical stakeholders with vested interests in maintaining the status quo.

\item \textbf{Economist as Detective:} Highlights the economist's role in uncovering the intended institutional design in settings where it is evident but requires 
formalism and technical expertise to design.\footnote{In many such settings, the gap between the intended institution and the one in place arises from a lack of formalism and technical expertise. 
For a prominent illustration, see Section Section \ref{sec:SCIAKG-failures}, which discusses a highly consequential flawed ruling of the Supreme Court of India.}

\item \textbf{Custom-Tailored Theory:} Utilizes formal theory as the primary analytical tool, developing custom-made, use-inspired theories tailored to specific applications.
\end{itemize}

Taken together, these characteristics operationalize the three foundational pillars of minimalist market design, providing a practical framework for effective institutional reform.

While not a universal element of the framework, many of the practical applications presented in this monograph utilize an additional methodological feature to capture the institutional mission:

\begin{itemize} 
\item \textbf{Axiomatic Methodology:}
Often (though not always) employs the axiomatic methodology as a central technique for formal analysis. 
\end{itemize}

\section{From Theory to Practice: The Role of Innovation Models} \label{sec:theory}

Embedding a persuasion strategy within the design and enhancing practical relevance often calls for custom-made theory tailored to specific applications. 
This use-inspired theory underpins the holistic approach to research and policy within minimalist market design.

In this section, I examine how the role of theory in minimalist market design aligns with and diverges from more mainstream approaches in market design. 
A particularly notable distinction lies with the \textit{engineering approach to market design} \citep{roth/peranson:99, Roth:2002}, which reflects the unique demands and perspectives shaped by commissioned market design projects.

According to common wisdom, market design is considered a field that builds on mechanism design \citep{Hurwicz:1960, hurwicz:72, hurwicz:73} as its main theoretical foundation, 
but it places more emphasis on institutional details and aims to provide policy-relevant insights. Throughout the monograph, 
I refer to the spectrum of research that encompasses both foundational research in mechanism design and application-oriented market design 
as \textit{economic design}.\footnote{This spectrum also includes research in \textit{applied mechanism design}, 
which aims not to inform policy within a specific institutional setting but to understand properties that hold across a variety of settings, provided they meet certain technical assumptions.}

Assuming \textit{self-interested} and \textit{rational} behavior from the participants, mechanism design  \`{a} la Hurwicz  studies  how 
design of institutions affects behavior of participants along with the outcomes chosen by these institutions.\footnote{Formally, a mechanism consists of a 
message space for each participant and a  function which selects an outcome for each message profile. 
Using various solution concepts from noncooperative game theory, there are two main approaches in mechanism design. 
Under the first approach, the focus is on the design of an \textit{optimal} mechanism, 
where optimality typically corresponds to the maximization of a given objective function subject to various constraints.
Under the second approach, the objective is attaining ``desirable'' outcomes (represented as a \textit{social choice correspondence}) at equilibrium. 
This segment of mechanism design is known as \textit{implementation theory}. 
See \cite{jackson:14} and  \cite{Jackson:2001} respectively for excellent surveys on mechanism design and implementation theory.} 

Consistent with the common wisdom in the field, 
in a mutual interview, \cite{rothwilson:19} identify non-cooperative game theory, cooperative game theory and mechanism design  
as antecedents of market design:
\begin{quote}
``Designs of auctions and matching markets evolved from the disparate branches
of noncooperative and cooperative game theory. Agents’ private information is
the main consideration in auctions, and designs focus on procedures that elicit
demands and yield good outcomes. The goal is to implement Walras and Hayek using Hurwicz’s scheme.'' (p. 125)
\end{quote} 

Although \cite{rothwilson:19} identify game theory and mechanism design as antecedents of market design, they emphasize that formal analysis plays a less central role in 
their vision than in these highly rigorous fields. They state:
\begin{quote}
``Mathematical models themselves play a less heroic, stand-alone role in market
design than in the theoretical mechanism design literature. A lot of other kinds of
investigation, communication, and persuasion play a role in crafting a workable
design and in helping it to be adopted and implemented, and then maintained and adapted.'' (p. 119)
\end{quote}

This perspective largely reflects an engineering approach to market design, first discussed and advocated by \cite{roth/peranson:99} and \cite{Roth:2002}. 
Under this approach, theory---typically ``off-the-shelf'' and developed without specific consideration for practical use---plays a less prominent role than in minimalist market design, 
primarily offering intuition and guiding practical designs in the right direction. Roth and Wilson describe this role as ``less heroic,'' a perspective frequently echoed by other field leaders involved in major commissioned market design projects:

\begin{quote}
``A lesson from this experience of theorists in policymaking is that the real value of the theory is in developing intuition. [$\ldots$] 
What the theorists found to be the most useful in designing the auction and advising the
bidders was not complicated models that try to capture a lot of reality at the
cost of relying on special functional forms.''\\
\mbox{} \hfill \citealp{mcafeemcmillan:96}, p. 172 \medskip

``Two important lessons that I learned
from working on high-stakes auctions are that they operate in an almost infinite variety
of contexts, and that this variety is the reason for the paradoxical importance
of including unrealistic assumptions in models built to understand and illuminate
reality. No single set of assumptions is adequate to describe all the various settings
in which auctions are used, and too much specificity in models can blind the analyst
to important general insights.''\\
\mbox{} \hfill \citealp{Milgrom:2021}, p. 1383-84
\end{quote}

This perspective---favoring simple models over those tailored specifically to individual projects---is particularly compelling for commissioned projects under tight deadlines, 
where developing novel theoretical contributions is often impractical. As \cite{Narayanamurti/Odumosu:2016} highlight:

\begin{quote}
``Research, which we understand as an \textit{unscheduled} quest for new knowledge and the creation of new inventions, whose outcome cannot be predicted in advance [$\cdots$]
Development, which is a \textit{scheduled} activity with a well-defined outcome in a specified time frame, aimed at the marketplace.'' (p. 11)   
\end{quote}

The engineering approach to market design is primarily associated with such commissioned projects. 
For example, \cite{Roth:2002}, in his seminal work on the foundations and early successes of market design, emphasizes:

\begin{quote}
``Aside from technical issues, these design efforts also shared some features that seem to be characteristic of the context in which markets are designed.
First, design is often required to be \textit{fast}. In each of these three cases, only about a year elapsed between the commissioning of a new market design and its delivery.'' (p. 1345)
\end{quote}

The indirect relationship between theory and practice in the engineering approach to market design is largely shaped by the demands of commissioned projects. 
Economic designers in such settings lack the flexibility to choose their problems or timelines; they must deliver practical solutions under strict deadlines. As \cite{Narayanamurti/Odumosu:2016} note, 
the pressures of completing a ``scheduled'' activity often leave little room for advancing theoretical boundaries. This dynamic, however, is not unique to engineering approach to market design---it reflects a broader paradigm articulated by \cite{Bush:1945}, which envisions a linear, unidirectional relationship between scientific discovery and technological innovation. This model has profoundly influenced U.S. science and technology policy since World War II.

In contrast to the pioneers of market design who were engaged in commissioned projects, I---as an aspiring economic designer---had two significant flexibilities unavailable to them in the late 1990s 
when I ventured into the world of policy design. I could actively seek out real-world applications with a stronger role for novel theory, and I had ample time to pursue these projects as an 
``unscheduled quest for new knowledge and the creation of new inventions.''

Freed from these constraints, minimalist market design emerged as an exception to the broader trend shaped by Bush’s linear model. 
It embraces a holistic approach that integrates research and practice, aligning instead with alternative innovation frameworks advocated by \cite{Stokes:1997} and \cite{Narayanamurti/Odumosu:2016}. 
To clarify these distinctions, we will first review Bush’s post-war science and technology framework in Section \ref{sec:Linear}, then examine Stokes’s \textit{Pasteur’s Quadrant} model in Section \ref{sec:Pasteur}, 
and finally discuss the \textit{Discovery--Invention Cycles} model introduced by Narayanamurti and Odumosu in Section \ref{sec:DIC}.
 
\subsection{Science, the Endless Frontier: Vannevar Bush's Linear Model of Innovation} \label{sec:Linear}

In late 1944, President Franklin Roosevelt commissioned Vannevar Bush to prepare a report on U.S. peacetime scientific policy.
Bush---who led the \textit{Office of Scientific Research and Development} during World War II and played a pivotal role in the \textit{Manhattan Project}---delivered a report titled
\textit{Science, the Endless Frontier} \citep{Bush:1945}.
One of the key themes of this influential document was the relationship between curiosity-driven ``basic'' research, practice-oriented ``applied'' research, technological innovation, 
and practical implementation. It urged the government to institutionalize and prioritize basic research during peacetime.

The success of the Manhattan Project and the resulting widespread support for scientific research in the U.S. during that era helped garner broad approval for Bush’s ideas. 
This led to the establishment of the \textit{National Science Foundation (NSF)} in 1950 and significantly influenced post-war policies on knowledge creation and technological innovation.

In his report, Bush put forward two theses that would become profoundly influential, with important implications for science and technology policy:

\paragraph{Dichotomy Between Basic and Applied Research.} Bush argued that ``basic research is performed without thought of practical ends.'' 
He suggested that basic research, aiming to broaden knowledge without concern for practical applications, and applied research, seeking to solve specific practical problems, should remain distinct. 
According to Bush, focusing on one detracts from the other, as ``applied research invariably drives out pure research.'' 

\begin{figure}[!htbp]
    \begin{center}
       \includegraphics[scale=1.0]{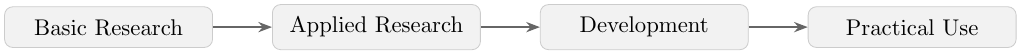}
           \end{center}
\caption{Bush’s linear innovation model.}
\label{fig:bush-linear-paradigm}
\end{figure}

\paragraph{Linearity of the Four Phases of Innovation.}
Bush further asserted that ``basic research is the pacemaker of technological progress.'' Figure \ref{fig:bush-linear-paradigm} depicts the linear sequence: 
basic research expands knowledge; applied research creates practical models and innovations; development converts findings into usable outputs; 
and practical use implements them. Each stage is successive and depends on the previous one.

Many critics of Bush's \textit{linear model}, including \cite{Stokes:1997} and \cite{Narayanamurti/Odumosu:2016}, contend that the division between ``basic'' and ``applied'' research is flawed and influenced, in part, 
by political considerations related to government funding.
Before World War II, funding for curiosity-driven research was scarce in the U.S., 
whereas industry heavily supported research with practical objectives, such as the development of the light bulb, telephone, and automobile since the late 1800s. 
By portraying curiosity-driven research as the driving force behind practical advancements and framing practical research as a threat to curiosity-driven efforts, 
Bush advocated for a strict division between ``basic'' and ``applied'' research.\footnote{The distinction between basic and applied science dates back to ancient Greece, 
where ``epistêmê'' represented the more esteemed theoretical knowledge, while ``technê'' referred to the less prestigious practical knowledge. 
For a detailed exploration of the historical divide between these two forms of science, see \cite{Stokes:1997}.} 
This categorization enabled him to advocate for government funding institutions that would protect curiosity-driven research from being overshadowed by more practical pursuits.

The fifty years following Bush's report witnessed unprecedented research advances, during which the U.S. emerged as a global leader in science. 
Bush's linear model of innovation, as well as its influence on shaping institutions like the NSF and other government funding bodies, went largely unchallenged until the mid-1990s. 
This historical legacy, coupled with the funding structures of U.S. federal agencies like the NSF during that period, contributed to the prominence of Bush's influential theses in many fields, including economic design.

For example, traditional mechanism design often prioritizes expanding theoretical knowledge without immediate practical applications, aligning primarily with basic research. 
Hurwicz, in his 1972 \textit{Richard T. Ely Lecture}, emphasized this perspective:

\begin{quote} 
``The new mechanisms are somewhat like synthetic chemicals: even if not usable for practical purposes, 
they can be studied in a pure form and so contribute to our understanding of the difficulties and potentialities of design.''\\
\mbox{} \hfill  \citealp{hurwicz:73}, p. 27
\end{quote}

In contrast, the engineering approach to market design is driven by practical needs within specific institutional contexts. 
In his seminal work, ``The Economist as Engineer: Game Theory, Experimentation, and Computation as Tools for Design Economics,'' 
Roth envisions the role of economics in market design as analogous to the relationship between engineering and physics:

\begin{quote}
``Consider the design of suspension bridges. The simple theoretical model in
which the only force is gravity, and beams are perfectly rigid, is elegant and general.
But bridge design also concerns metallurgy and soil mechanics, and the sideways
forces of water and wind. Many questions concerning these complications
can’t be answered analytically, but must be explored using physical or computational
models. These complications, and how they interact with the parts of the
physics captured by the simple model, are the domain of the engineering literature.
Engineering is often less elegant than the simple underlying physics, but it
allows bridges designed on the same basic model to be built longer and stronger
over time, as the complexities and how to deal with them become better understood.''\\ 
\mbox{} \hfill \citealp{Roth:2002}, p. 1342
\end{quote}

Roth’s engineering approach encompasses not only applied research and development but also, at times, implementation, following the stages of the linear model of innovation. 
Unlike traditional mechanism design, which focuses on advancing theoretical understanding, Roth’s engineering approach emphasizes the practical application of theories, 
utilizing complementary techniques such as experimental and computational economics. 
As articulated in \cite{rothwilson:19}, this approach positions theory in a supportive---somewhat ``less heroic''---role during the policy development phase.

\subsection{Beyond the Linear Model: Pasteur's Quadrant} \label{sec:Pasteur}

In the early 1990s, I pursued my Economics Ph.D. at the University of Rochester under the guidance of William Thomson. 
My coursework with him included traditional mechanism design, the axiomatic approach to resource allocation, and two-sided matching theory. 
Thanks to Thomson's strong emphasis on equity, I developed a solid foundation in normative economics alongside positive economics. 
In 1995, half a century after Bush's influential 1945 report was published, I defended my thesis---an exercise in pure theory---on ``Strategy-Proofness and Implementation in Matching Markets.''

The 1990s marked the rise of market design, with economists playing pivotal roles in shaping real-world policies. 
This era saw exciting early applications, such as the FCC Spectrum Auction in 1994 and the redesign of the U.S. Medical Residency Match in 1997. 
Inspired by these achievements, I too aspired to apply my newly acquired theoretical skills to improve real-life institutions. Shortly after earning my Ph.D., 
I launched projects focused on the allocation of on-campus housing (see Section \ref{sec:HA}) and nationwide college admissions in Turkey (see Section \ref{sec:schoolchoice})---two practical resource allocation problems 
with which I was intimately familiar.

Little did I know at the time, the approach I adopted in these projects would not only mark the beginning of minimalist market design 
but also closely align with a competing paradigm to Bush's linear model of innovation---one articulated around the same time by Donald Stokes in his posthumously published work \citep{Stokes:1997}.

As the field of market design was emerging and I was completing my Ph.D., Columbia University organized a three-part conference in 1994 and 1995 titled 
``Science, the Endless Frontier 1945-1995: Learning from the Past, Designing for the Future.'' The conference aimed both to celebrate Bush's influential report and to critically assess it, 
seeking ideas for a new science and technology policy model. In his opening speech, Columbia Provost Jonathan Cole remarked \citep{cole:94}:
\begin{quote}
``Science and technology policy would appear to be in a state of crisis. There are many indicators that a crisis does exist in the partnership between the federal government 
and the American research universities, that the terms of the partnership are increasingly being questioned and re-examined. [$\cdots$]

Even if Vannevar Bush is as much a symbol of this period of dramatic progress as its putative architect, it seems fitting that in the 50th anniversary year of the publication of \textit{Science: The Endless Frontier}, 
we celebrate the work and the period: reflect on the origins of the Bush paradigm for scientific excellence, take stock of where we currently are in the relationship between science and government, 
consider whether we are in a period of crisis, and do some serious work on the future shape of the national system of scientific and technological innovation. [$\cdots$]

Now the intent of this conference is more than a celebration of Vannevar Bush or a celebration of the 50th anniversary of the publication of Bush's pioneering report. 
It is our intent to have this series serve as a forum for the thought analysis of the current and historic policy environment and as a forum for the presentation of new concepts regarding 
the design of a new science and technology policy model.''
\end{quote}

Donald Stokes, the first speaker at the conference, later expanded his speech into his celebrated book \textit{Pasteur’s Quadrant: Basic Science and Technological Innovation} \citep{Stokes:1997}. 
In this work, Stokes critically examines Bush's model of innovation, arguing that the relationship between science and technology is interactive and reciprocal rather than linear and unidirectional. 
He highlights a significant flaw in Bush's model: the assumption that technological development flows solely from science. In reality, the influence works both ways, with technology also driving scientific advancements. 
Stokes advocates viewing science and technology as an integrated whole, recognizing the substantial contributions each makes to the other, thus challenging Bush's linear model.

To address the limitations of Bush's unidimensional approach, Stokes introduces a framework that incorporates a second dimension: 
the desire to solve practical problems alongside the quest to expand fundamental understanding. 
As illustrated in Figure \ref{fig:pasteur-quadrant}, his model divides research into three primary quadrants,\footnote{There is also a fourth quadrant where research neither extends fundamental understanding nor serves
a practical purpose.} 
each named after a key historical figure:


\begin{figure}[!tp]
    \begin{center}
       \includegraphics[scale=1.1]{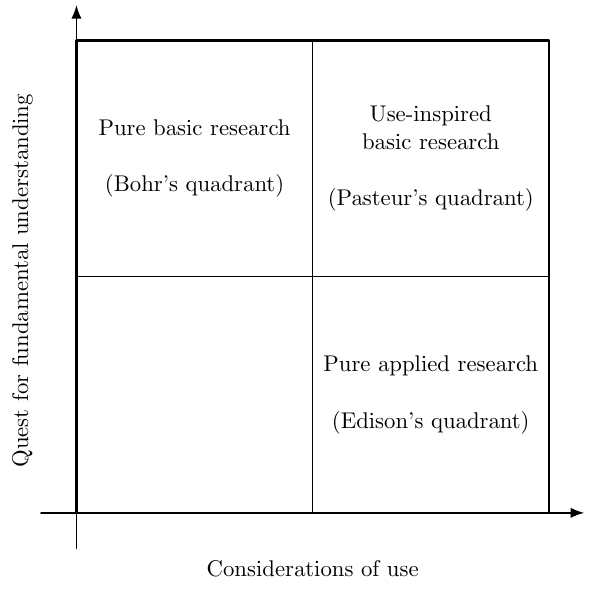}
           \end{center}
\caption{\citet{Stokes:1997}'s classification of research programs: the vertical axis measures relevance for advancing knowledge, and the horizontal axis measures relevance for direct use.}
\label{fig:pasteur-quadrant}
\end{figure}

\begin{itemize} 
\item \textbf{Bohr's Quadrant} (Pure Basic Research): Driven by the quest for fundamental understanding. 
\item \textbf{Edison's Quadrant} (Pure Applied Research): Focused on solving societal practical problems. 
\item \textbf{Pasteur's Quadrant} (Use-Inspired Basic Research): A blend of both, engaging in research that expands knowledge while addressing real-world challenges.
\end{itemize}

Within this classification, much of the research in economic theory, including mechanism design and information design, falls into \textit{Bohr's Quadrant}. 
In contrast, research in the engineering approach to market design and artificial intelligence belongs to \textit{Edison's Quadrant}. 
Bush's linear model of innovation is most closely associated with these two quadrants, though research in \textit{Edison's Quadrant} often includes product development as well.

Stokes, however, emphasizes the importance of conducting scientific research in \textit{Pasteur’s Quadrant}.\footnote{This quadrant is named after Louis Pasteur, who expanded knowledge---such as 
establishing the role of microorganisms in fermentation and developing the germ theory of disease---while also addressing practical challenges, like inventing the pasteurization process and developing 
vaccines for rabies and anthrax.} He describes work in this quadrant as ``use-inspired basic research.'' Minimalist market design naturally fits within \textit{Pasteur’s Quadrant}.

Since its introduction in the late 1990s, Stokes's framework has profoundly influenced innovation policy in the U.S. 
For example, it inspired the NSF to incorporate a ``broader impacts'' criterion into its merit review process for grant applications in the early 2010s.\footnote{Stokes’s framework has also influenced innovation policy beyond the U.S. 
For example, since 2011, researchers submitting proposals to the 
Swiss National Science Foundation have been able to designate their applications as ``use-inspired.'' In the European Union, Horizon 2020 (2014--2020) and now Horizon Europe (2021--2027) 
explicitly evaluate projects on both scientific excellence and their ``impact pathways.'' 
In the United Kingdom, from 2009 to 2020, applicants to the Economic and Social Research Council 
and other Research Councils were required to provide a ``Pathways to Impact'' statement outlining how their work would generate societal benefits. 
Since 2020, this requirement has been integrated directly into the main research proposal to embed impact planning within the project’s core design. 
In Canada, beginning in the late 2000s, the Social Sciences and Humanities Research Council, the Natural Sciences and Engineering Research Council, 
and the Canadian Institutes of Health Research have required applicants to include a ``Knowledge Mobilization Plan.'' 
In Australia, the Australian Research Council introduced ``Impact and Engagement'' criteria in the mid-2010s, which were formalized in the 2018 Engagement and Impact Assessment. 
In Japan, the Japan Science and Technology Agency has promoted ``problem-oriented basic research'' as part of the government’s Science and Technology Basic Plans since the early 2000s.}
Its influence has been so significant that, in 2014, 
in response to those advocating otherwise, the NSF Assistant Director for Mathematical and Physical Sciences emphasized that the NSF should not entirely 
abandon support for pure basic research in \textit{Bohr's Quadrant} by forcing all basic research into \textit{Pasteur's Quadrant} \citep{Crim:14}:

\begin{quote}
``It is alluring to think that all basic research should reside in Pasteur's quadrant. After all, looking for near term payoffs is a human trait. 
However, that view limits the range of our vision and risks missing out on those unforeseen discoveries that propel us, sooner or later, in completely unanticipated directions.''
\end{quote}

To better understand how these quadrants manifest in the field of economic design, let us explore specific examples that illustrate research practices within each quadrant. 
We will focus on seminal works that have not only advanced the field of economic design but have also contributed to their authors receiving the \textit{Nobel Memorial Prize in Economic Sciences}, 
thereby underscoring their significance. We begin with \textit{Bohr's Quadrant}.

\subsubsection{Bohr's Quadrant: Nash Implementation} \label{sec:Bohr-quadrant}

Much of the research in economic theory falls into \textit{Bohr's Quadrant}, and mechanism design is no exception. 
Consider Eric Maskin's seminal 1977 working paper on \textit{Nash implementation}, later published as \cite{Maskin:99}.

Maskin examines the following abstract setting: There is a set of individuals $I$ and a set of outcomes $A$, where each individual $i \in I$ has a weak preference relation $R_i$ over $A$. 
For each individual $i \in I$, let $\mathcal{R}_i$ represent the set of possible preferences. Let $R = (R_i)_{i \in I}$ denote a generic preference profile, 
and let $\mathcal{R}$ denote the set of all possible preference profiles. Importantly, preferences are private information.

A central planner aims to select an outcome based on individuals' true preferences, formalized as a \textit{social choice function} $f: \mathcal{R} \to A$ that maps each preference profile $R \in \mathcal{R}$ to an outcome $\alpha \in A$.\footnote{Maskin's theory is more general, allowing for \textit{social choice correspondences} that map each preference profile to a set of outcomes.}

If preferences were publicly known, the planner could directly select the desired outcome for any given preference profile. However, because preferences are private, this is not feasible. 
A natural approach is to solicit individuals to report their preferences and then select an outcome accordingly. The challenge with this approach is that it creates a game form in which individuals 
may benefit from misrepresenting their preferences. Consequently, even at Nash equilibrium, the outcome may differ from the desired one.

To overcome this, the planner designs a \textit{mechanism} $(S, g)$ à la Hurwicz, where $S_i$ represents the strategy space for each individual $i \in I$, and $g: S \to A$ is the outcome function selecting an outcome 
$\alpha \in A$ for each strategy profile $s = (s_i)_{i \in I} \in S = \prod_{i \in I} S_i$. The objective is to design a mechanism such that, for any preference profile $R$, the Nash equilibrium outcomes of the 
induced game coincide with the desired outcome $f(R)$. This allows the planner to implement $f$ indirectly, despite preferences being private.

Maskin's fundamental question is: Which social choice functions can be implemented in this way?

To address this, he identifies a necessary condition for implementability, now known as \textit{Maskin monotonicity}. This condition states that if an outcome $\alpha$ is selected by $f$ under a preference profile $R$, 
and $\alpha$ remains at least as attractive under another preference profile $R'$ for every individual relative to any other outcome, then $f$ must also select $\alpha$ under $R'$.

It is fairly easy to see that \textit{Maskin monotonicity} is necessary for Nash implementability. Suppose $f$ selects outcome $\alpha$ for preference profile $R$. If $f$ is implementable, 
there exists a mechanism $(S, g)$ such that $\alpha$ is a Nash equilibrium outcome under $R$. This implies the existence of a Nash equilibrium strategy profile $s$ where the outcome function $g$ selects $\alpha$, i.e., $g(s) = \alpha$.

Now, consider a new preference profile $R'$ where, for every individual $i$, any outcome $\beta$ that was less preferred than $\alpha$ under $R_i$ remains less preferred under $R'_i$. 
Observe that the strategy profile $s$ remains a Nash equilibrium of the mechanism $(S, g)$ under $R'$ as well, because $\alpha$ is at least as preferred by each individual under $R'$ as it was under $R$. 
Otherwise, if $s$ were no longer a Nash equilibrium, there would exist a profitable deviation for some individual under $R'$, which would also represent a profitable deviation under $R$---a contradiction.

Therefore, the mechanism $(S, g)$ must still yield $\alpha$ as an equilibrium outcome under $R'$. Since the mechanism implements $f$ at Nash equilibrium by assumption, 
$f$ must also select $\alpha$ under $R'$. Thus, \textit{Maskin monotonicity} is a necessary condition for Nash implementability.

However, \textit{Maskin monotonicity} alone is not sufficient for Nash implementability. Maskin introduces an additional condition, \textit{no veto power}, which, together with \textit{Maskin monotonicity}, 
becomes sufficient when there are at least three individuals. The \textit{no veto power} condition stipulates that if all individuals except possibly one rank an outcome $\alpha$ as their top choice, then $\alpha$ should be selected by $f$.

To prove that any social choice function satisfying \textit{Maskin monotonicity} and \textit{no veto power} is Nash implementable, 
Maskin envisions a constructive proof involving a highly general mechanism $(S, g)$ that achieves this goal regardless of the setting.

In Maskin's mechanism, individuals are indexed as $i_1, i_2, i_3$, and so on.

The strategy space for each individual $i$ is given as  $S_i = \mathcal{R} \times A \times \mathbb{N}$, where $\mathbb{N}$ represents the set of non-negative integers.
In this setup, each individual selects an entire preference profile, an outcome, and an integer.

The outcome function $g$ operates as follows in three cases:

\begin{enumerate}
\item If all individuals select the same strategy $(R, \alpha, n)$ with $f(R) = \alpha$, then $g$ selects $\alpha$.
\item If all but one individual, say individual $i$, select the same strategy $(R, \alpha, n)$ with $f(R) = \alpha$, while $i$ selects a different strategy $(R', \beta, m)$, 
then $g$ selects the less preferred of $\alpha$ and $\beta$ for $i$ under $R_i$---the preference relation assigned to $i$ by the other individuals in their strategies.
\item In all other cases, $g$ selects the outcome chosen by the lowest-indexed individual who picked the highest integer.
\end{enumerate}

Assuming there are at least three individuals, this mechanism always implements the social choice function $f$ in Nash equilibrium, 
provided that $f$ satisfies \textit{Maskin monotonicity} and the \textit{no veto power} condition \citep{Maskin:99}.\footnote{Although the sufficiency of \textit{Maskin monotonicity} along with the \textit{no veto power} 
condition for Nash implementability with three or more individuals was originally proposed in Maskin's 1977 working paper, the proof of this result using his mechanism was incomplete. 
Before the publication of \cite{Maskin:99}, complete proofs were provided using variants of Maskin's mechanism---first by \cite{Williams:86} with additional assumptions, and later by \cite{Repullo:87} 
and \cite{Saijo:88} without them. \cite{Repullo/Moore:90} subsequently introduced a necessary and sufficient condition for Nash implementability in this case, along with a different necessary and sufficient condition for the two-individual case.}

The power and elegance of this result lie in its generality. However, it is evident that this mechanism is not intended for practical use. Beyond the complexity of Maskin's mechanism, 
the broader concept of implementation via a Hurwicz-style mechanism---relying on achieving equilibrium outcomes under various rationality assumptions---renders Maskin's exercise fairly abstract. 
This places the exercise, like many others in the mechanism design literature, firmly within \textit{Bohr's Quadrant}.

While such theoretical contributions are invaluable for expanding fundamental understanding, they do not directly address practical problems. 
This contrasts with research situated in \textit{Edison's Quadrant}, which focuses on solving real-world issues through applied research methods. A representative example of this is the redesign of the U.S. Medical Residency Match.

\subsubsection{Edison's Quadrant: Redesign of the U.S. Medical Residency Match} \label{sec:Edison-quadrant}

In contrast to \textit{mechanism design}, where the underlying research mostly fits within \textit{Bohr's Quadrant}, 
much of the practice-oriented market design research---particularly in its first decade---falls into \textit{Edison's Quadrant}. One of the earliest success stories in market design, 
the redesign of the U.S. Medical Residency Match between 1995 and 1997 directed by Alvin Roth, serves as a prime example \citep{roth/peranson:99}. This reform has an intriguing background spanning several decades.

In the early 1950s, the \textit{National Resident Matching Program (NRMP)} was established to address the chaotic and inefficient decentralized process through which newly graduated doctors 
secured residency positions.\footnote{The National Resident Matching Program is a non-profit organization responsible for placing medical residents into clinical training programs in the U.S. 
It was first established in 1952 under the name \textit{National Interassociation Committee}. In 1953, it incorporated the \textit{National Intern Matching Program} and adopted the name \textit{National Resident Matching Program} in 1978.} 
Prior to its inception, hospitals competed fiercely for promising candidates, leading to the unraveling of appointment dates---offers were made increasingly earlier, sometimes up to two years before graduation. 
This pressure compelled students to commit prematurely, often without fully forming their own preferences or allowing hospitals to properly evaluate their qualifications.

The introduction of the NRMP centralized the matching process, enabling students and hospitals to submit rank-order lists of their preferences. 
The system used an algorithm similar to the \textit{institution-proposing deferred acceptance (DA) algorithm}, independently discovered more than a decade later by \cite{gale/shapley:62}.\footnote{The equivalence 
between the NRMP's matching algorithm and the institution-proposing DA algorithm was identified in \cite{roth:84}, decades after both were invented. 
Since Gale and Shapley formulated their celebrated algorithm without knowledge of the NRMP's algorithm, I regard their equivalence as external validity for their invention.} 
It aimed to produce \textit{stable} outcomes, ensuring that no student and hospital would prefer each other over their assigned matches.\footnote{See Section \ref{sec:two-sided-matching} for an in-depth discussion of \textit{stability} 
and the DA algorithm.} This innovation brought order and efficiency to the residency placement process for several decades.

By the 1970s, demographic shifts within medical schools led to new challenges. The increasing number of women entering the field resulted in more dual-physician couples seeking residencies in the 
same geographic area to live and train together. The original NRMP algorithm treated applicants individually and did not account for joint preferences, often assigning members of couples to distant locations.

Consequently, many dual-physician couples began to circumvent the NRMP, negotiating directly with hospitals to secure placements that accommodated their needs. 
This exodus threatened to destabilize the market, reminiscent of the pre-NRMP era's disorder. Recognizing the severity of the issue, the NRMP began making initial attempts to adjust the matching process starting in 1984.  
Couples were allowed to register as a unit, with one partner designated as the ``leading member,'' but this approach was not sufficiently powerful to fully capture their joint preferences for pairs of positions.
In response to the increasing technical demands, the NRMP contracted the newly established Canadian software company \textit{National Matching Services (NMS)}, 
founded by Elliott Peranson, in 1985 to develop and operate a new matching system.

As NMS modified the institution-proposing DA algorithm over the years to accommodate match variations such as the presence of dual-physician couples that differentiated NRMP's practical problem from the basic setting, 
various groups started to question whether the company was acting in the best interests of the residents. This was an era when the literature on two-sided matching markets had substantially matured, 
culminating in the celebrated monograph by \cite{roth/sotomayor:90}, which broadly exposed the literature. The increased understanding of matching algorithms among medical doctors in the early 1990s 
led some of them to challenge the use of software based on the institution-proposing DA algorithm.

For example, two concerns raised by \cite{Williams:95} were:

\begin{enumerate} 
\item The NRMP's algorithm favored programs over residents, and residents were not informed of this, even though the NRMP had been aware of it since at least 1976. 
\item The NRMP's algorithm contained incentives for students to misrepresent their true preferences, and residents were provided misleading information by the NRMP about this issue. 
\end{enumerate}

In response to these concerns, \cite{Peranson/Randlett:95} argued that while the NRMP algorithm produces the program-optimal stable outcome in the absence of match variations, 
there may be no \textit{stable} outcome when variations are present. They noted that even if such outcomes exist, evidence suggests that the differences between program-optimal 
and individual-optimal stable outcomes are likely to be small. Furthermore, they suggested that additional research was needed to assess the extent of these differences.

Amid mounting dissatisfaction and the threat of market failure, the NRMP commissioned Alvin Roth in the fall of 1995 to redesign the matching process. 
Roth collaborated with Elliott Peranson to develop a new heuristic algorithm. 
These efforts culminated in the \textit{Roth and Peranson design}, a procedure based on Gale and Shapley’s celebrated individual-proposing DA algorithm for the simplified version of the problem, 
but modified to address the complexities specific to the NRMP’s situation.

\cite{Roth:2002} clearly articulated the objective of his commission:

\begin{quote} 
``My status was that of an outside expert hired to design a new algorithm that would be able to handle all the match complications, and to evaluate the scope for favoring one side of the 
market over the other (i.e., applicants and residency programs) while achieving a stable matching.'' (p. 1363) 
\end{quote}

In cases without dual-physician couples, the individual-proposing DA algorithm is an obvious solution to the crisis. However, when dual-physician couples are present, 
there may be no \textit{stable} outcome, and no algorithm guarantees that truth-telling is a dominant strategy for all doctors \citep{roth:84}. 
The goal of the commissioned design was to create a system that met the underlying objectives as fully as possible, addressing the ongoing controversies.

With the new design, addressing the first concern raised by \cite{Williams:95} was straightforward. Simply replacing the institution-proposing DA algorithm with the individual-proposing 
DA algorithm as the starting point of the matching process would resolve this issue. To tackle potential \textit{instability} arising from match variations such as dual-physician couples, 
Roth and Peranson adopted an engineering approach. They leveraged an earlier technique by \cite{roth/vande:90}, which had proven effective in simpler settings. 
The expectation was that this approach would perform reasonably well even in the more complex current setting, given that the fraction of residents in the system who were members of dual-physician couples was relatively low.

The earlier work of \cite{roth/vande:90} established that, in the basic model, a \textit{stable} outcome could always be reached through a carefully constructed sequence of interventions, 
regardless of the initial outcome. At each step, a doctor-program pair that preferred each other over their current matches would be matched together, potentially at the expense of their existing matches.

Although \textit{stable} outcomes might not always exist in the NRMP's more complex setting, computational experiments using several years of NRMP data demonstrated that a 
similar approach was effective in an environment with a relatively low fraction of residents who were members of dual-physician couples \citep{roth/peranson:99}. 
In every instance tested, this method produced a \textit{stable} matching. Roth and Peranson explored various sequencing scenarios for instability-clearing interventions 
and settled on one that yielded the best results. To account for the possibility of cycles---since a \textit{stable} outcome might not exist---they incorporated a stopping rule into the procedure.

Additional computational experiments with simulated data showed that \textit{stable} outcomes would almost certainly exist in their setting, 
their design would identify such outcomes, and almost no doctor would benefit from misrepresenting their preferences.

These findings confirmed the earlier suggestion by \cite{Peranson/Randlett:95} that the criticisms in \cite{Williams:95} were overblown and that the initial starting outcome 
has little impact in any matching process that reaches a \textit{stable} outcome---a near certainty given their findings. Nevertheless, the new \textit{Roth and Peranson design} 
chose the individual-proposing DA algorithm as its starting point to remove any doubt about the fairness of the procedure for the residents. Even though the new design might generate outcomes very 
similar to those of the previous system that raised concerns, by addressing these concerns, the new design would be a success. After all, Roth was called in to help ease the confidence crisis in the process.

Based on these findings and positive feedback from all stakeholders, the \textit{Roth and Peranson design} was adopted starting with the 1998 match (\citealp{Roth:2002}, pp. 1363--64):

\begin{quote} 
``Once my design and evaluation were complete, I conferred at length with the various interested parties (including traveling to present the results to representatives 
of the various organizations of residency directors). Although it was widely anticipated that the results of the study would provoke bitter disagreement, 
the fact that the set of stable matchings proved to be so small was widely understood to mean that making the match as favorable as possible to applicants would not create any 
systematic problems for any segment of residency programs. Consequently, my reports were received without provoking much controversy, and the NRMP board voted in 
May of 1997 to adopt the new algorithm for the match starting in 1998.'' 
\end{quote}

\cite{roth/peranson:99} reported both the technical details of the design and the engineering approach taken during the reform process, 
which was further elaborated in \cite{Roth:2002}. These papers established the foundation of the \textit{engineering approach to market design}, mainly intending to provide guidance on commissioned projects.

Even though \cite{roth/peranson:99} was not primarily motivated by expanding fundamental understanding---as is characteristic of research conducted in \textit{Edison's Quadrant}, 
often performed within the constraints of a schedule---it led to several follow-up studies that contributed to basic knowledge. This illustrates that market design research following the 
engineering approach and situated in \textit{Edison's Quadrant} can nonetheless lead to research in \textit{Bohr's} or \textit{Pasteur's Quadrants} once the timing pressures of the original commission cease to exist.

Indeed, many potential directions for follow-up research on the \textit{Roth and Peranson design} were outlined in \cite{Roth:2002}:

\begin{quote}
``That is, by unanticipated good luck, some of the knotty problems
posed by couples, and other complementarities, could be solved without losing
the most attractive design options that the simple theory suggested.
When I speak of unanticipated good luck, I mean that these computational and
experimental results, while suggested by the theory of simple matching markets,
are not explained by that theory. These results point to a need for new theory, to
explain and further explore the behavior of large labor markets with couples and
linked jobs. We also need new theory to explore the dynamics of market failures
like unraveling, and their cures.'' (p. 1372-73)
\end{quote}

These follow-up studies not only made significant progress in addressing these questions but also provided concrete directions for improving the \textit{Roth and Peranson design} 
once the time constraints and specific needs arising from the NRMP's crisis were no longer restricting the design process. The most notable advances focused on tackling the 
challenges introduced into the system by the inclusion of dual-physician couples.

\cite{Klaus/Klijn:05} examined the preference structures of couples that guarantee the existence of a \textit{stable} outcome. They identified \textit{responsive preferences} 
as one such domain. This structure is particularly plausible in settings where programs are in close proximity to one another. It assumes that each member of a couple has their own preferences over individual programs, 
and their joint preferences are such that, when the assignment of one member is fixed, 
an improvement for the other member based on their individual preferences also improves the couple's 
joint preferences.\footnote{\textit{Responsive preferences} were originally introduced by \cite{Roth:85} in a two-sided matching market to derive institution preferences 
over groups of individuals from their preferences over individual members, in a class of settings without complementarities.}  
In such settings, the individual-proposing DA algorithm, which disregards joint preferences and treats couples as separate individuals, generates a \textit{stable} outcome.

In a surprising and illuminating follow-up, \cite{Klaus/Klijn/Masso:07} showed that the \textit{Roth and Peranson design} may not only fail to find a \textit{stable} outcome in such domains, even though one always exists, 
but also allow couples to profit by presenting themselves as single individuals.

Subsequent literature illuminated the reasons behind the virtually guaranteed existence of \textit{stable} outcomes in the computational experiments with simulated data reported in \cite{roth/peranson:99}. 
These experiments reflected the small proportion of residents in the NRMP pool who are members of dual-physician couples. \cite{Kojima/Pathak/Roth:13} established that, 
with random preferences and a vanishingly small proportion of couples, a \textit{stable} outcome almost always exists. 
However, in the same model, \cite{Ashlagi/Braverman/Hassidim:14} showed that as the proportion of couples increases, the probability of a \textit{stable} outcome existing sharply decreases. 

In the case of the NRMP, dual-physician couples comprise 5--10\% of all residents. Even at this modest rate, however, \cite{Kojima/Pathak/Roth:13}, based on personal communication with Peranson, 
reported that the \textit{Roth and Peranson design} failed to generate a \textit{stable} outcome in multiple years. This could be due either to the non-existence of a \textit{stable} outcome in those years 
or to the design's inability to find one, even though it existed---a possibility earlier established in \cite{Klaus/Klijn/Masso:07}.

Finally, in an exercise of use-inspired theory that is minimalist in spirit, \cite{nguyen/vohra:18} showed that, regardless of the fraction of couples or their preferences in a problem, 
there is always a ``nearby'' problem where a \textit{stable} outcome is guaranteed to exist. This nearby problem is constructed by increasing the total number of positions by no more than four across all programs, 
with the number of positions at any individual program changing by no more than two.

These more recent theoretical advances, inspired by practical challenges, illustrate how applied research can also contribute to fundamental understanding. 
This convergence of basic and applied research---use-inspired basic research---embodies the essence of \textit{Pasteur's Quadrant}. To further illustrate this, we turn to William Vickrey's foundational work in auction theory.

\subsubsection{Pasteur's Quadrant: Vickey Auction and Revenue Equivalence Theorem} \label{sec:vickrey}

William Vickrey's groundbreaking paper \cite{vickrey:61} is widely regarded as the foundational work in auction theory and serves as a prime example of use-inspired basic research, falling into \textit{Pasteur's Quadrant}.

In this paper, Vickrey not only advanced fundamental understanding by establishing the earliest version of the \textit{Revenue Equivalence Theorem},\footnote{\cite{Klemperer:99} provides 
the following definition of the \textit{Revenue Equivalence Theorem}: Assuming a fixed number of risk-neutral potential buyers, each with a privately known signal independently drawn from a common, 
strictly increasing, and atomless distribution, any auction mechanism that satisfies the following conditions---(1) the object is always allocated to the buyer with the highest signal, 
and (2) a bidder with the lowest feasible signal expects zero surplus---yields the same expected revenue under the Bayesian Nash equilibrium of the resulting game form.}
one of the foundational results in auction theory, but also introduced the \textit{second-price sealed-bid auction}, the simplest case of
what are now known as \textit{Vickrey} auctions.\footnote{A \textit{Vickrey auction} is a static auction format for selling any number of identical items, in which the highest bidders each win one item and pay a price equal to the highest losing bid.} 
Interestingly, he devised the second-price sealed-bid auction as a minimalist reform of the first-price sealed-bid auction.

Vickrey considered a setting in which a single, indivisible good is auctioned, with buyers having independently distributed valuations. He carried out a game-theoretic analysis of three auction formats 
commonly used in various markets: the \textit{English (ascending-price) auction}, widely employed across diverse real-world settings; the \textit{Dutch (descending-price) auction}, popular in flower markets in the Netherlands; 
and the \textit{first-price sealed-bid auction}, frequently used for real estate and securities.

In the English auction, after the seller sets the initial price, buyers incrementally raise their bids until no one is willing to bid higher. The final bidder wins the item at their last bid. 
Analyzing bidder behavior under the English auction is straightforward. Since buyers’ valuations are independent, under optimal behavior, the winning bidder will have the highest valuation, 
and the winning price will be within one bidding increment of the second-highest valuation, resulting in a \textit{Pareto efficient} allocation:

\begin{quote}
``The normal result (among rational bidders!) is that the bidding will stop at a level approximately equal to the second highest value among the values that 
the purchasers place on the item, since at that point there will be only one interested bidder left; the object will then be purchased at that price 
by the bidder to whom it has the highest value. (For simplicity, we shall assume that price can vary continuously and that there is no minimum increment between bids.)
This result is obviously Pareto-optimal.''\\
\mbox{} \hfill  \citealp{vickrey:61}, p. 14
\end{quote}

In the Dutch auction, the seller begins with a high initial price and announces successive reductions in fixed unit increments. The first bidder to accept a price wins the item at that price. 
Analyzing bidder behavior in this format is more challenging because, once the price reaches their valuation, a bidder has an incentive to wait for further reductions, provided they can still be the first to bid. 
Consequently, the optimal strategy for each bidder depends not only on their own valuation but also on the anticipated strategies of others, creating a challenging strategic game.

Despite this complexity, Vickrey established that for risk-neutral bidders with valuations independently and uniformly distributed over the unit interval, a unique Bayesian Nash equilibrium exists. 
Furthermore, making significant progress toward the \textit{Revenue Equivalence Theorem}, he demonstrated that this equilibrium produces the same outcome as the equilibrium of the English auction under these assumptions.

However, Vickrey also cautioned that achieving a \textit{Pareto efficient} outcome is generally less likely in the Dutch auction compared to the English auction. 
This is due to factors such as bidders having insufficient information about other bidders’ valuations or the difficulty of formulating an equilibrium strategy.

After examining the two popular dynamic auction formats, Vickrey shifted his focus to a static auction where bidders do not observe each other's bids: the \textit{first-price (sealed-bid) auction}. 
In this format, each bidder submits a bid, and the item is awarded to the highest bidder, who pays the amount they bid. 
Vickrey made a remarkable discovery, noting that the first-price auction is strategically equivalent to the Dutch auction:

\begin{quote}
``Actually, the usual practice of calling for the tender of the bids on the understanding that highest or lowest bid, as the case may be, 
will be accepted and executed in accordance with its own terms is isomorphic to the Dutch auction just discussed. 
The motivations, strategies, and results of such a procedure can be analyzed in exactly the same way as was done above with the Dutch auction.''\\
\mbox{} \hfill  \citealp{vickrey:61}, p. 20
\end{quote}

This isomorphism can be easily seen by interpreting each buyer's bid in the first-price auction as their planned time to accept a price in the Dutch auction.

The strategic equivalence between the first-price auction and the Dutch auction had two important implications for Vickrey. First, in the straightforward case of risk-neutral bidders with 
valuations independently and uniformly distributed over the unit interval, a unique Bayesian Nash equilibrium exists, yielding the same outcome as the English auction. 
Second, similar to the Dutch auction, Vickrey anticipated that the outcome of the first-price auction would generally be less efficient than that of the English auction.

Thus, among the three real-world auction formats, the English auction was Vickrey's favorite. But what if an auction had to be conducted in a static format, like the first-price auction? 
Wouldn't it be great to find a static auction that is strategically equivalent to the English auction rather than the Dutch auction? Addressing this question, Vickrey made another brilliant observation:

\begin{quote}
`` [$\cdots$] it is of interest to inquire whether there is not some sealed-bid procedure that would be logically isomorphic to the progressive [English] auction. 
It is easily shown that the required procedure is to ask for bids on the understanding that the award will be made to the highest bidder, but on the basis of the price
set by the second highest bidder.''\\
\mbox{} \hfill  \citealp{vickrey:61}, p. 20
\end{quote}

Again, this strategic equivalence becomes evident when we interpret each buyer's bid in the second-price auction as the maximum amount they are willing to bid in the English auction. 
With this breakthrough, Vickrey devised the \textit{second-price (sealed-bid) auction} as a minimalist reform of the first-price auction, creating a static and \textit{strategy-proof} auction format.\footnote{Here, 
the \textit{strategy-proofness} of the second-price auction applies to the bidders. However, a significant drawback of the second-price auction is its vulnerability to manipulation by the seller 
\citep{Rothkopf/Teisberg/Kahn:90, Akbarpour/Li:2020}. For instance, the seller could introduce fake bids just below the highest bid to artificially raise the winning price. 
Other limitations of the second-price auction include its vulnerability to collusion by a coalition of bidders \citep{Robinson:1985, klemperer:02, Ausubel/Milgrom:2005}.}

Together, these discoveries also enabled Vickrey to establish the \textit{Revenue Equivalence Theorem} for all four auction formats he analyzed as a byproduct of his work, 
under the simplified case of risk-neutral bidders with valuations independently and uniformly distributed over the unit interval, though this was not his primary focus.

For similar reasons that led Vickrey to favor the English auction over the Dutch auction, he also concluded that the second-price auction is preferable to the first-price auction. 
This aligns with the view that Vickrey designed the second-price auction as a minimalist reform of the first-price auction.

Vickrey’s second-price auction (and close variants) has seen real-world use across several settings, but outcomes have been mixed---in part 
because the format can be vulnerable to collusion and other strategic behavior in multi-item environments \citep{Robinson:1985, klemperer:02, Ausubel/Milgrom:2005}. 
New Zealand’s brief experience with the second-price rule for allocating radio-spectrum ``management rights'' in 1989--1990 is widely cited as the first large-scale 
governmental deployment of a pure Vickrey format \citep{mcmillan:94, Crandall:1998}.\footnote{After several highly publicized anomalies (e.g., a top bid of NZ\$100{,}000 yet payment of only NZ\$6 
because there was no reserve price and the second-highest bid was NZ\$6), and thin bidding in some lots, New Zealand switched to first-price auction in 1991\citep{mcmillan:94, MED:2005}.
Following international practice, New Zealand later moved to simultaneous ascending (multiple-round) auctions in 1996 \citep{MED:2005}.}

This is where our story on Vickrey's brilliant innovations takes a plot twist.

Though Vickrey was the first to introduce the second-price auction to scholarly literature, he was not the first to invent it \citep{Moldovanu/Tietzel:98, Lucking-Reiley:2000}. 
In fact, the second-price auction had been invented and reinvented multiple times before Vickrey, likely because it represents a minimalist reform of both the first-price auction and the 
English auction.\footnote{Since Vickrey formulated the second-price auction independently, without knowledge of its earlier inventions, and was the first to discuss it in academic literature, 
I regard these earlier instances as external validity for Vickrey's invention.}

In 1797, Johann Wolfgang von Goethe---one of Germany's greatest writers---sent the following letter to his publisher, 
effectively inventing the second-price auction (\citealp{Moldovanu/Tietzel:98}, pp. 854--855):\footnote{\cite{Moldovanu/Tietzel:98} attribute the quote to \cite{Mandelkow:1929}.}

\begin{quote}
``I am inclined to offer Mr. Vieweg from Berlin an epic poem, Hermann and Dorothea, which will have approximately 2000 hexameters. [$\cdots$]
Concerning the royalty we will proceed as follows: I will hand over to Mr. Counsel B\"{o}ttiger a sealed note which contains my demand, and I wait for what Mr. Vieweg will suggest to offer for my work. 
If his offer is lower than my demand, then I take my note back, unopened, and the negotiation is broken. If, however, his offer is higher, 
then I will not ask for more than what is written in the note to be opened by Mr. B\"{o}ttiger.''
\end{quote}

One might be tempted to dismiss Goethe's invention of the second-price auction as a special case involving only two buyers, with the seller being one of them, or as an isolated instance. 
However, \cite{Lucking-Reiley:2000} presents a fascinating history of this auction format, which has been regularly used in the U.S. for paper collectibles such as postage stamps and Civil War soldiers' letters since at least 1893. 
Interestingly, Lucking-Reiley uncovered this history through meticulous research into these auctions, revealing a remarkable parallel between the evolution of the second-price auction in these markets 
and the way new institutions are devised under minimalist market design.

After the first postage stamp was introduced in England in 1840 and in the U.S. in 1847, the hobby of stamp collecting gained popularity in the 1850s, 
eventually leading to the emergence of stamp auctions in the U.S. around 1870. The English auction quickly became the standard format for these events. 
However, from the very beginning, special provisions were made for individuals unable to attend in person. These out-of-town collectors would submit their bids---the maximum price they were willing to pay---and 
the auctioneer would essentially act as a proxy bidder for each absentee participant, bidding in single-unit increments up to their submitted maximum.\footnote{\cite{Lucking-Reiley:2000} notes that such 
``absentee'' bidding in English auctions remains widespread today in many auction houses for a range of collectibles, including art and wine.}

Interestingly, this practice can be viewed, in formal terms, as a generalization of both the dynamic English auction and the static second-price auction.

Starting in 1877, stamp shops in small towns began conducting static ``mail-only'' auctions, where they mainly used the first-price auction.

The earliest documented instance of a Vickrey auction employing the second-price format, as identified by Lucking-Reiley, was conducted by Wainwright \& Lewis of Northampton, Massachusetts, in 1893. 
Their sale announcement states (\citealp{Lucking-Reiley:2000}, p. 187):

\begin{quote}
``Catalogue of a Collection of U.S. and Foreign Stamps To be sold WITHOUT RESERVE except where noted. Bids will be received up to 4 P.M., May 15,
1893. Bids are for the LOT, and, contrary to the usual custom in sales of this kind, we shall make this a genuine AUCTION sale; that is to say, each lot will
be sold at an advance of from 1c to 10c above the second highest bidder. Address all bids to Wainwright \& Lewis, Northampton, Mass.''
\end{quote}

Beyond introducing the second-price auction, just as Vickrey did, Wainwright \& Lewis devised this auction format as a minimalist reform of the first-price auction---the usual mail-only auction format of the time---with 
the objective of mimicking the English---the ``genuine''---auction. 

Remarkably, the thought process behind this practical invention closely aligns with Vickrey's. Throughout this monograph, we will encounter several examples where the 
evolution of real-world institutions mirrors designs earlier or independently guided by minimalist market design, demonstrating how ``life imitates science'' in the development of institutions. 
Since one of the roles of minimalist market design is to devise institutions that were intended but failed to materialize for various reasons, 
this kind of external validity is particularly likely in successful applications of market design.

Thus, the history of the second-price auction illustrates not only the fundamental insights of Vickrey's research but also how practical considerations guided through minimalist reforms 
can lead to similar innovations independently across time. This synergy between theory and practice epitomizes the kind of use-inspired basic research that Stokes advocates in \textit{Pasteur's Quadrant}.

\subsection{Rethinking Innovation Frameworks: Discovery--Invention Cycle} \label{sec:DIC}

Minimalist market design integrates use-inspired basic research with policy initiatives, positioning it firmly within \textit{Pasteur's Quadrant}. 
Yet, it aligns even more closely with another framework that also challenges Bush's linear model: the \textit{Discovery--Invention Cycle} (DIC) model by \cite{Narayanamurti/Odumosu:2016}. 
This model shapes the structure of this monograph, tracing the evolution of innovations across diverse applications in Sections \ref{sec:HA} through \ref{sec:LE} and highlighting the 
``discovery--invention cycles'' that connect them, as illustrated in Figures \ref{fig:DIC-SC-arm} and \ref{fig:DIC-KE-arm}.

Narayanamurti and Odumosu argue that while \cite{Stokes:1997} challenged the linear model, he did not go far enough in reshaping how policymakers think about and organize research. 
Like Bush, Stokes focused on the motivations behind research but introduced a dual motivation---aiming to expand understanding and improve technology---through ``use-inspired basic research'' in Pasteur's Quadrant. 
Importantly, however, he still classified research in this quadrant as ``basic'' research. 

Narayanamurti and Odumosu criticize categorizing research solely by its underlying motivation, 
arguing that research is inherently integrated, constantly moving between discovery and invention, leading to new ideas, tools, and devices. 
They contend that focusing solely on researchers' intentions does not fully break down the distinctions between basic and applied research---a dichotomy that hinders innovation. 
According to them, categorizing research strictly as ``basic'' or ``applied,'' and using funding models that support this division, creates conflict by setting basic research against applied research.

Instead, they suggest the division should not be between different types of research, as all research fundamentally involves exploring the unknown. 
The distinction, they argue, should be between ``research,'' which should remain an unscheduled activity, and ``development,'' which focuses on creating a specific product within a set timeframe 
and is therefore subject to market demands.\footnote{This is the only aspect that differs between minimalist market design and Narayanamurti and Odumosu's vision. 
Since I perceive minimalist market design largely as a framework for critical outsiders, even the (policy) development phase is considered an unscheduled activity. 
Key aspects of policy development are integrated within use-inspired theory in minimalist market design. From this perspective, my framework is even more holistic than the DIC model.}

According to this distinction, much of the work in commissioned market design projects, performed as a scheduled activity leading to the adopted design and ``often required to be \textit{fast},'' 
according to \cite{Roth:2002}, is considered ``development'' rather than ``research,'' unlike the work performed after its completion as an unscheduled activity. 
However, in the spirit of Peter Medawar's concerns discussed in Section \ref{sec:policyimpact-externalvalidity}, this research reflects \textit{post hoc} research progress after the practical or policy impact has already occurred. 
Thus, it is less clear how the adapted design in commissioned settings may be received in cases where the economic designer is an outsider advocate, since it is not stress-tested in such situations. 
Indeed, this has been a key reason why I sought an alternative approach to market design as a junior economist.

To overcome the basic-applied dichotomy, Narayanamurti and Odumosu introduce the concepts of  ``discovery''  and ``invention'' to describe the two channels of research. 
``Discovery'' focuses on generating new understanding about the world, while ``invention'' involves creating knowledge that leads to new tools, devices, or processes. 
Their DIC model, a dynamic alternative to Stokes’s Pasteur’s Quadrant, combines these phases with research motivations and institutional settings, 
offering an integrated view of the research process and its impact on innovation. This approach, they argue, enhances our understanding of innovation over time, 
helps identify potential bottlenecks, and supports the development of effective policy interventions.

My experiences in research and policy since the late 1990s strongly resonate with these ideas. As we progress through the applications in Sections  \ref{sec:HA} through \ref{sec:LE}, 
we will observe how these concepts are reflected in the relationships between various market design innovations, both within specific applications and across them. 
This will illustrate the strong alignment between minimalist market design and the DIC model of innovation.

To support their DIC model, \cite{Narayanamurti/Odumosu:2016} trace the evolution of information and communication technologies. 
They highlight Nobel Prize-winning achievements, such as the 1956 discovery of the transistor effect, which demonstrate several discovery--invention cycles over more than four decades. 
These cases show how, in some instances, scientific discoveries lead to inventions, while in others, inventions drive fundamental scientific breakthroughs. They write:

\begin{quote} 
``Building upon early work on the effect of electric fields on metal semiconductor junctions, the interdisciplinary Bell Labs team built a working bipolar-contact transistor and demonstrated the transistor effect. 
This work and successive refinements enabled a class of devices that successfully replaced electromechanical switches, allowing for successive generations of smaller, more efficient, 
and more intricate circuits. While the Nobel was awarded for the discovery of the transistor effect, the team of Shockley, Bardeen, and Brattain had to invent the bipolar-contact transistor to demonstrate it.''\\
 \mbox{} \hfill   \citealp{Narayanamurti/Odumosu:2016}, p. 49
 \end{quote}

This monograph similarly illustrates how minimalist market design can lead to policy breakthroughs driven by use-inspired basic theory in some cases,
while in others, policy interactions stimulate new theoretical advances.
By organizing Sections \ref{sec:HA} through \ref{sec:LE} to showcase a series of discovery--invention cycles in the theory and practice of matching markets, 
as illustrated in Figures \ref{fig:DIC-SC-arm} and \ref{fig:DIC-KE-arm},
I demonstrate how policy innovations can spur further theoretical developments---and vice versa---analogous to breakthroughs in information and communication technologies.\footnote{The idea of a reciprocal 
relationship between theoretical discovery and practical invention in market design is not original to this monograph.
A similar perspective---developed independently of the DIC model---appears in \cite{Pycia:2019}, who emphasizes the two-way relationship between technological innovation and practical market design:
new technologies enable novel designs, while new designs in turn catalyze the emergence of further technologies.
Drawing on examples from medical, energy, automotive, computing, and data technologies, Pycia highlights how these interactions foster innovation in the theory and practice of market design.
The present monograph builds on this broader reciprocal perspective within minimalist market design, linking use-inspired basic theory with policy reform across multiple domains.}

This structure also aligns with several arguments in Duncan Watts' \textit{Nature Human Behaviour} perspective, ``Should social science be more solution-oriented?'' \citep{Watts:2017}. 
Watts argues that, despite numerous theories on human behavior, social science has struggled to reconcile their inconsistencies. Like \cite{Stokes:1997} and \cite{Narayanamurti/Odumosu:2016}, 
he attributes this challenge to an emphasis on theory over practical problem-solving. Watts proposes that one way to address this ``incoherency problem'' is to eliminate the traditional divide between basic and applied science, 
advancing theory with a focus on real-world solutions. He emphasizes that a holistic approach to research and policy could bring significant benefits to the social sciences:

\begin{quote} 
``[$\cdots$] I am also not suggesting that social scientists do not already devote themselves to solving
practical problems: many do, especially in policy-relevant areas like education, health care, poverty and government. 
Rather, what I am suggesting is that social scientists can profitably view the solution of practical problems as a mechanism for improving the coherency of social science itself.'' \\
\mbox{} \hfill \citealp{Watts:2017}, p. 1
\end{quote}

The organization of Sections \ref{sec:HA} through \ref{sec:LE} demonstrates how theory and practice support each other in discovery--invention cycles, directly aligning with Watts' theses. 
Narayanamurti and Odumosu suggest that ``research should be evaluated not only by its initial motivations, but also by its ability to catalyze other research and innovations.'' 
Through this monograph's structure, the role of minimalist market design in catalyzing a series of innovations becomes apparent, 
including those that culminated during the COVID-19 pandemic when these innovations were most urgently needed.

Drawing inspiration from industrial laboratories like Bell Labs, Narayanamurti and Odumosu argue that Nobel Prize-winning industrial research undermines Bush's notions of applied and basic research. 
They highlight that in the industrial research environment, ``the very best discovery--oriented research was carried out.'' Emphasizing the importance of connections between research and practice, 
they assert that the most productive distinction was not between basic and applied research but between various unscheduled research activities and scheduled product development.

Consequently, inspired by the workings of Bell Labs and insights from the DIC model, they propose policy suggestions for building research institutions that excel in innovation. 
Along similar lines, one of the central theses of this monograph argues that minimalist market design can achieve the same goal for teams of economic designers by advancing theory and informing policy.

\section{Advice for Economic Designers with Policy Aspirations} \label{sec:guide}

Having elaborated on the role of use-inspired theory in minimalist market design in the previous section, 
I will now offer guidance for economic designers seeking to adopt this framework as a policy instrument.

In the policy context, the primary role of minimalist market design is to provide practical advice for designing or redesigning institutions within existing constraints by crafting use-inspired theory tailored to each application. 
Additionally, it serves an epistemic role, offering a framework to analyze institutions based on stakeholders' objectives.

\subsection{Pursuit of Promising Applications} \label{sec:pursuit}

Which practical applications are promising candidates for minimalist market design? How can an economic designer identify them? Here, I offer my perspective.

Minimalist market design proves most valuable when stakeholders have well-defined objectives, and the existing system, while imperfect, 
is not so fundamentally broken that it cannot be substantially preserved.\footnote{An example of a system too difficult to salvage with a minimalist approach is the queuing system used by the 
non-profit \textit{Feeding America} before 2005 to allocate excess food across food banks. This system neither disclosed the type of food being allocated nor accounted for the specific needs of individual food banks, 
resulting in severe welfare loss. Recognizing these significant shortcomings, and with guidance from four University of Chicago faculty members, the organization replaced the queuing system with an elaborate 
auction run twice daily. In this new system, each food bank made ``purchases'' using a specialized currency called ``shares,'' which were determined for each auction based on the bank's characteristics and 
purchase history \citep{Prendergast:17, Prendergast:22}.}
To assess its potential, consider the following three questions:

\begin{enumerate} 
\item Are you commissioned by decision-makers or other stakeholders seeking changes to the existing system, or is it merely an aspiration of an outsider? 
\item Does the proposal involve introducing a new system or modifying an existing one?
\item Are all key stakeholder goals within the purview of economics, or do they include essential objectives beyond its scope? 
\end{enumerate}

This framework is especially useful when the economic designer acts as an external critic, intervening in an existing institution that, while flawed, does not require a complete overhaul, 
and addressing its limitations that go beyond the textbook considerations of economics. Its cautious approach is particularly valuable in areas outside the traditional domain of economists, 
where interventions risk being perceived as intrusive.

At times, finding promising applications feels like detective work. Institutions often begin with simple missions that have straightforward solutions. 
However, as these missions evolve, the solutions can become more complex, especially for decision-makers who lack training in analytical methods. 
In response to changing missions, decision-makers may resort to patchwork modifications of simpler solutions, which can introduce various issues into the system. 
By tracing this history, an economic designer can not only identify a candidate institution for improvement but also uncover the root cause of its failures.

Let's illustrate this strategy with a hypothetical scenario.\footnote{Conceptually, this scenario closely parallels the evolution of the U.S. Army's cadet--branch matching mechanism, 
discussed in Section \ref{sec:Army}, as well as a high school admissions policy implemented across China from the early 1990s to the mid-2010s, analyzed in \cite{wang/zhou:24}.}
 
Suppose the seats at an elite school are allocated to applicants based on standardized test results. 
At this level, the solution is straightforward: the highest-scoring applicants are admitted until capacity is reached.

Now, imagine the institution faces a financial crisis, prompting the system operator to introduce a second allocation criterion for a portion of the seats, say 20\%. 
For these seats, they decide to grant a set score advantage, for instance 10 points, to students willing to pay double the tuition if they cannot secure one of the cheaper regular-tuition seats. 
How should the solution change? 

Intuitively, this version of the problem does not seem much more difficult.
For many system operators, the following modification might seem straightforward. Before the allocation, the system operator inquires which applicants 
are willing to pay the increased tuition to receive an extra 10 points for the last 20\% of the seats. Then, they allocate the first 80\% based on the standardized test results. 
Finally, for the last 20\% of the positions, the system operator uses a modified list of scores, giving an extra 10 points to candidates who indicated willingness to pay the increased tuition. 
Candidates who receive the last 20\% of the positions are charged the increased price only if they opted for this option, thus receiving a boost in their scores.

This solution feels right, but it isn't. Consider an applicant who cannot secure one of the regular-tuition seats with their score but is willing to pay the increased tuition for a score boost. 
This applicant receives one of the last 20\% of seats with the boosted score and is automatically charged the increased tuition. 
But who is to say that the applicant wouldn't have secured one of these seats without the extra boost to their score? 
Indeed, suppose another lower-score applicant receives one of these positions at the regular tuition. The applicant who paid the increased tuition would be rightfully upset.

Without formal training, a system operator is unlikely to devise the correct solution. They need support from an expert trained in formal methods. 
Even though the system operator's objective is clear, the solution is surprisingly elusive.\footnote{The exact solution isn't central to our point, but in case you're curious, here it is:
Before allocating the last 20\% of seats with modified scores, they should first tentatively allocate them based on the regular scores. 
If any unassigned applicants opt to pay the increased tuition for these seats, and the score boost is enough to grant them sufficient priority, 
only then should they be temporarily awarded one of these units at the increased price, replacing the lowest-score tentative recipient. 
If the replaced applicants from the original round are also willing to pay the increased tuition, they should then receive one of the seats available at the increased price, 
again replacing the lowest modified-score applicant who was tentatively holding one of these seats. 
This iterative procedure gives the correct solution.}
To make the situation worse, there is a flawed solution that feels right to a layperson. It is highly likely that, without expert guidance, 
they will arrive at this flawed solution. Applications like this are your best bet to make a real impact. Search for them! 

Even if the system operator is highly skilled and finds the correct solution in this scenario---which, in my experience, would be rare---a subsequent mission 
change could make the solution even more challenging. For instance, suppose several elite schools decide to allocate their seats jointly based on these principles while also considering students' preferences. 
If our system operator avoided errors with the initial change, they will surely struggle with this more complex version of the problem. 
By tracing the institution’s history, our economic designer can act like a detective, identifying such market failures and uncovering their root causes.

The strategy I outlined to identify promising applications naturally calls for a minimalist approach. 
As our case studies will illustrate, this is precisely how I developed this method, and even today, I continue to use it to discover new applications.

Earlier in Section  \ref{sec:Pasteur}, we discussed Duncan Watts' advocacy for a solution-oriented social science. Aligned with our hypothetical scenario, in \cite{Watts:2017}, 
he suggests that identifying the most promising problems involves focusing on those with a ``Goldilocks'' property---problems that are neither too easy nor too hard. 
Drawing inspiration from engineering, he recommends placing ``more emphasis on building tangible devices and systems that have specific, well-defined properties.''

Watts also advises selecting problems that are not so vast and complicated as to require an intractably complex theory, yet challenging enough to justify a scientific approach. 
Crucially, he highlights the advantage of targeting ``modular'' problems, which can be tackled in progressively ambitious versions. In many settings, including several case studies presented in this monograph, 
the evolution of an institution’s mission results in a modular progression of the underlying problem. Consequently, Watts' recommendations on problem selection align well with my proposal to 
adopt a ``detective-like'' approach, naturally blending both strategies.

\subsection{A Roadmap for an Aspiring Market Designer} \label{sec:execution}

Next, I will summarize the key steps and prerequisites for effectively implementing this framework to generate realistic policy advice. 
Depending on the application, certain steps may be omitted for various reasons, such as the absence of an existing system or if the objective is solely analysis.

\paragraph{Step 1. Identify a Real Problem.}  This is a substantial undertaking, so following this framework makes sense only if your scholarly efforts have a practical counterpart. 
Can you articulate the problem and your insights to friends who are not academic economists? Can you envision crafting a news story based on these ideas? Naturally, the more significant the problem, the better.
 
\paragraph{Step 2.  Understand the Problem Inside and Out.}  To effectively address a real-life problem, comprehensive understanding is crucial. 
If all goes well, you will eventually engage with authorities and various stakeholders to advocate for your practical ideas. In many cases, your efforts might be perceived as a ``liability'' 
if they expose the limitations of an existing institution. Interactions with authorities who may have designed the ``flawed'' institution are inevitable, 
and they may be inclined to discredit and dismiss your efforts. 
Therefore, as a minimum requirement, you must have an in-depth understanding of the issue to reasonably expect to convince stakeholders that you can be a valuable ``asset.''

Success in this endeavor demands familiarity not only with literature related to the application but also with other relevant materials from the field. 
For instance, before engaging decision-makers and experts in other fields for our various policy-oriented projects, our teams of economic designers:
\begin{itemize}
\item Studied dozens of papers in the transplantation literature before advocating for a centralized kidney exchange clearinghouse in New England (see Section \ref{sec:KE}).
\item Absorbed dozens of court rulings before proposing a reform of a flawed affirmative action procedure mandated by a Supreme Court judgment in India (see Section \ref{sec:India}).
\item Thoroughly examined key considerations in medical ethics, public healthcare, and emergency medicine literatures before advocating for a class of 
emergency rationing procedures during the COVID-19 pandemic (see Section \ref{sec:pandemic}).
\end{itemize}

\paragraph{Step 3. Identify  Key Policy Objectives of Stakeholders.}  The interests of academic economists may not always align with the critical concerns in real-world settings. 
Conducting market design research without thoroughly understanding stakeholders' primary objectives risks producing insights or instruments of limited practical value. 
If your primary goal is to generate research that informs policy, it is paramount to understand the mission of the institution in question.

An institution's history often provides valuable clues about what truly matters to its stakeholders. Institutions evolve by incorporating features that address issues or concerns identified by earlier policymakers. 
The current generation may lack awareness of the intentions behind these concerns or may not explicitly state them as objectives. 
Consequently, these features may contain information about some ``legitimate'' objectives of policymakers, even if they are no longer explicitly articulated.

Do not assume that the objectives of policymakers or stakeholders need to resemble standard assumptions in neoclassical economics. 
At this phase, thinking like a mainstream economist may not always serve you well. Neoclassical analysis largely relies on a \textit{preference utilitarian} framework, 
where the primary objective is the maximization of some aggregate measure of preference satisfaction. Among my successful endeavors in market design, kidney exchange (see Section \ref{sec:KE}) 
and liver exchange (see Section \ref{sec:LE}) stand out as the only ones where preference utilitarianism effectively captures a key stakeholder objective.
In these cases, a naturally fitting and objectively significant welfare measure revolves around the number of lives saved through donor exchanges. 
Consequently, the welfare gains presented in \cite{roth/sonmez/unver:04, roth/sonmez/unver:05, roth/sonmez/unver:07, ergin/sonmez/unver:20} received a favorable response from members of the transplantation community.

In contrast, in my work on school choice, it was predominantly considerations tied to \textit{incentive compatibility} that propelled a series of reforms (see Section  \ref{sec:schoolchoice}). 
It was largely \textit{equity} considerations that were the driving force for bioethicists and emergency healthcare experts endorsing and adopting our policy recommendations 
for addressing the pandemic allocation of scarce medical resources during COVID-19 (see Section  \ref{sec:pandemic}). 
Regarding India's affirmative action policies, recent reforms aligned with those advocated in \cite{sonyen22} were exclusively steered by \textit{equity} considerations (see Section  \ref{sec:India}). 
Lastly, both \textit{equity} and \textit{incentive compatibility} considerations played central roles in the U.S. Army's reform of its branching process at West Point and ROTC (see Section  \ref{sec:Army}).\footnote{\cite{Kominers/Teytelboym/Crawford:2017} 
similarly highlight that equity considerations play a fundamental role across many market design settings.}

A note of caution is warranted: in many contexts, clearly defining or translating stakeholders' objectives presents significant challenges. 
In my view, especially in its primary role as a tool for formulating practical policy advice, minimalist market design may hold limited promise for these applications.

\paragraph{Step 4. Formulate a Realistic and Analytically Tractable Model.} Understanding the problem thoroughly and identifying key policy objectives are just the beginning. 
Now, you need a model that is both realistic and analytically manageable. Determine which elements of the problem are essential and which are superfluous. 
Make your model as lean as possible without omitting any key considerations. Often, the structure of the existing institution and the policy goals you identified will be instrumental in shaping your model.

\paragraph{Step 5. Evaluate the Suitability of the Original Institution for Key Objectives.}
Your best bet for capturing the attention of various stakeholders is to demonstrate that the existing institution falls short in meeting at least one of their key objectives.
Clearly, if the existing institution already aligns well with the key objectives of stakeholders, a reform is not necessary. 
Nevertheless, the overarching framework remains valuable for studying the existing system. After all, researchers in market design are not solely focused on achieving direct policy impact.

\paragraph{Step 6. Identify Root Causes of Any Failures.}
Even when an existing institution falls short of meeting some of the most critical stakeholder objectives, 
convincing them to embrace a costly change is a formidable task. Stakeholders are particularly unlikely to support a reform that entails a complete overhaul of the entire institution. 
Beyond the heightened difficulty and cost associated with more extensive reform, many elements within an institution may serve additional purposes you are unaware of. 
Therefore, it becomes imperative to enhance the institution while preserving its core structure. To achieve this, diagnosing the ``root causes'' of the failures becomes crucial. 
This stage is also valuable for optimizing your model. For instance, you can establish the primitives of the model 
based on which elements of the existing institution can be retained and which ones need to be modified or eliminated.

\paragraph{Step 7. Tackle the Root Causes of Failures to Develop an Alternative System Aligned with Key Objectives.}
This step embodies a defining aspect of ``minimalist'' market design, lending this paradigm its name.

If you can identify the root causes of the failures, assuming that the key objectives are collectively attainable, the next step involves rectifying them. 
By directly addressing the source of any issue, these corrections are likely to pave the way for an alternative design that resolves the failures with minimal interference with the original system. 
Envision yourself as a ``surgeon'' performing a ``minimally invasive" procedure.

This aspect of the minimalist approach aids in building trust among various stakeholders. After all, you're not discarding the entire institution; instead, you're simply repairing elements detrimental to stakeholders' objectives. 
Consequently, you can easily pitch your design as an enhancement of their own institution, making it more straightforward for various stakeholders to embrace your proposed reform.

An additional effort that can enhance your credibility with stakeholders is to identify all possible designs that meet their key objectives, or at least a representative set, rather than focusing solely on a single solution.
In axiomatic methodology, this type of result is known as an \textit{axiomatic characterization} of a class of rules that satisfy the desired axioms.\footnote{See, 
for example, \cite{moulin:88, moulin_fair2004}, \cite{thomson:01, thomson:2011}, and \cite{Schummer2019}.}
Concentrating on a single design that satisfies the key objectives of stakeholders (when there are many) may potentially introduce biases into the system with important distributional consequences. 
By presenting stakeholders with a comprehensive picture of what is feasible, you maintain \textit{informed neutrality} between reasonable normative positions \citep{Li2017}, 
thereby clarifying that your recommendation is entirely impartial.

In rare cases, there will be a unique design that satisfies the desiderata (see, for example, Theorem \ref{thm:MPCO} in Section \ref{sec:thm-MPCO} and Theorem \ref{thm:2SMG}  in Section \ref{sec:India-EV}).
This is considered the gold standard in axiomatic methodology, provided that all axioms correspond to actual desiderata. 
Naturally, the desired reform is most evident in those cases.\footnote{The value of uniqueness is less clear in practical applications than in scholarly inquiry. 
While a uniqueness result is valuable in research, partly because it avoids selection concerns and partly due to its elegance, the added flexibility of multiple solutions can be beneficial in practical applications.} 

\paragraph{Step 8. Solidify the Practical Value and Outreach of Your Proposal.}
If you have come this far, your research is of potential value to stakeholders. Before engaging them to explore their receptiveness to your policy analysis, 
it is advisable to support your conceptual and theoretical ideas with experimental, computational, or empirical methods. 
Additionally, gathering as much field or anecdotal evidence as possible can demonstrate the relevance of your research.

In this regard, media coverage of your research on a reputable outlet can be especially helpful in convincing stakeholders of the potential value of your work. 
For example, a \textit{Boston Globe} story on \cite{abdulkadiroglu/sonmez:03} was instrumental in facilitating the initial interaction with officials at 
Boston Public Schools to discuss the failures of the \textit{Boston school choice mechanism} and its two potential alternatives (see Section \ref{sec:BostonGlobe}). 
This meeting not only resulted in the city adopting one of the alternative designs---the \textit{individual-proposing deferred acceptance mechanism} based on \cite{gale/shapley:62}'s 
celebrated \textit{deferred acceptance algorithm}---but also helped trigger a cascade of subsequent reforms in the U.S. and other countries, 
making the alternative design one of the most widely used worldwide in recent years (see Section \ref{sec:broader-sc}).

One additional point about media coverage: The systematic approach based on stakeholders' policy objectives and the root causes of existing institutional failures, 
which forms the basis of minimalist market design, will also help you articulate your innovations to laypersons in simple terms. This clarity naturally increases the likelihood of attracting attention from the media.\medskip

Now is a good time to approach policymakers or other stakeholders and explore a potential collaboration aimed at transforming your ideas into policy impact.

\paragraph{Step 9. Forge Strategic Partnerships with Insiders.} 
While you may have started this endeavor as a critical ``outsider,'' you cannot complete it while remaining in this role.
To influence policy and institute change, you need to collaborate with insiders. Forming these partnerships will transition you from an outsider to an insider, 
enabling you to transform your innovations into concrete policies and potentially paving the way for more elaborate interventions.

Identify individuals in influential positions relevant to your contributions. If you have genuinely addressed a significant real-world problem and can articulate your ideas effectively, 
there is a good chance that some insiders will be open to collaborating with you, especially if your persuasion strategy is effective. 
Potential collaborators may include authorities responsible for running institutions, experts in other fields who serve as advisors, or other influential stakeholders with a vested interest in resolving the identified issues.

Often, the issues you address may already be apparent to these individuals, and they might be under pressure to resolve these problems independently. 
Approaching stakeholders when your ideas are most useful to them naturally increases your chances of capturing their interest.
During such times, you are more likely to compel them to consider a partnership. 
Building this collaboration is easier if they believe in your competence and trust you. 
To demonstrate your competence, articulate your ideas in non-technical terms. Here too, minimalist market design will make your job easier.

One effective method I've found for building trust is providing pro-bono assistance. 
Attempting to sell your services without this gesture may not reassure stakeholders about your intentions.\footnote{Charging for your services is common, and often expected, 
if insiders recruit you as a consultant for this endeavor. However, as an outsider seeking to change an institution, offering pro-bono support emphasizes your sincerity, 
highlighting a key difference between commissioned and aspired market design.}

It is also possible that the issues you identify and solve, while genuine, may not be apparent to insiders. 
If you articulate your ideas well and support them with evidence, insiders may still express interest in a partnership. This interest could stem from their desire to excel in their tasks or to proactively avert a crisis.

One crucial consideration is this: If you succeed in forming partnerships with insiders to influence policy and change an institution, 
it will be these authorities who answer for these changes to their superiors, constituents, and various other stakeholders. Therefore, for this collaboration to be successful, 
your partners must believe in the value of your ideas as much as you do. They need to truly ``own'' this endeavor as their own. Make their jobs as easy as possible. 
Support at this stage can include writing white papers explaining the need for reform, assisting with their presentations, 
and providing support in software and analysis. And have I mentioned that changing the world for the better with your research is not easy?

Of course, none of this can happen if the issue you bring lacks a real-life counterpart or if it is not of sufficient importance. 

\paragraph{Step 10. Tinker with the Design, Incorporating Additional Input from Stakeholders if Necessary.} 
Once a partnership is established with authorities, you are no longer bound by the constraints that directly influenced both your research questions and proposed design, as outlined in Steps 1--9. 
After all, you are no longer an outsider. Make the most of this added flexibility.

If there are additional paths for further improvements, either due to various elements of the original institution you hesitated to tinker with or due to feedback from your new collaborators, you can now pursue these ideas as an insider.

Note that this final step fundamentally differs from the earlier ones. Steps 1--9 can be viewed as efforts aimed at convincing decision-makers that you are a ``worthy'' partner in this endeavor---the persuasion 
phase of minimalist market design. Consequently, many elements of the framework are deliberately ``conservative.'' This conservatism helps our work to be seen as an interdisciplinary collaboration rather than an intrusive approach, 
significantly enhancing its reception among experts in other fields and decision-makers (see Section \ref{sec:conservatism}).  

However, once a certain level of trust is established between the relevant parties, the approach in this final step need not be as conservative. 
Similarly, the role of theory can be toned down at this stage. In this sense, there is potentially a more robust role in Step 10 for alternative methodologies, including computational, experimental, and empirical economics. 
Thus, the open-ended tinkering step in minimalist market design can be seen as the stage where you transition to a ``commissioned'' role, guiding the design with greater flexibility.\footnote{It is tempting to associate 
Step 10 of my framework with the ``tinkering'' phase of policy design in Esther Duflo's 
influential \textit{Richard T. Ely Lecture} ``The Economist as Plumber'' \citep{Duflo:17}. While this association is natural, there are some important distinctions 
between the roles of the tinkering phase in the two frameworks. In Duflo's framework, the role of the tinkering phase is making tweaks to 
prescribed policy to correct any issues that may arise in field implementation. Such issues can emerge due to the gap between theoretical analysis and the actual problem. 
Thus, the tinkering phase is particularly valuable in settings where the primary role of theory is building intuition. In my framework, 
in contrast, there should not be a mismatch between theory and practice at a level that could lead to any serious issues. 
The role of the tinkering phase is further improving the institution with the additional flexibility gained by the economic designer at this stage.}\medskip

It is crucial to emphasize that, within the minimalist market design framework, the objective is neither to discover nor propose a globally optimal institution to authorities before reaching the tinkering phase. 
Even if such an optimal institution exists, it might often involve a significant departure from the existing one, making it challenging to persuade authorities to pursue a reform.\footnote{An important exception 
here is applications where authorities are aware of the necessity of reform and have commissioned the design or redesign of the institution to an expert in economic design.} 
Various considerations, including delicate issues that are not explicitly discussed, the potential to upset sensitive balances between interest groups, or even the difficulty of  
``saving face,'' may limit the institutions that authorities are comfortable considering.

Given that many flawed institutions in the field are challenging to analyze, comparing them with optimal institutions may be difficult. 
Therefore, as the initial and most crucial step of an envisioned reform process, it is essential to start with a design that closely resembles the original institution. 
If a partnership with authorities can be established and the initial reform can be implemented, it becomes much easier to introduce more substantial deviations 
from the original institution in later stages of the redesign process. With the removal of major conceptual or technical failures in the original institution through the initial reform, 
it becomes much easier at this stage to analyze and demonstrate the potential value of more extensive reforms by relaxing many other aspects of the original institution. 
Assuming that a ``globally optimal'' institution actually exists, one can think of this tinkering phase as an attempt to move closer to this optimum, 
potentially with a more robust role in principles that are more prominent in mainstream economics research.
 
Having outlined a guide to minimalist market design,  in the next part of the monograph, Sections \ref{sec:HA} through \ref{sec:LE}, 
I will present several applications that exemplify the holistic interaction between research and policy within this framework.\footnote{Although this monograph is not intended as a textbook, 
Sections \ref{sec:HA} through \ref{sec:LE} (or a subset thereof) are highly structured and contain sufficient formalism and examples to support an advanced undergraduate or graduate course on matching markets or market design.
These sections can be supplemented with \cite{Sonmez/Unver:25b, Sonmez/Unver:25a} to increase the mathematical and technical depth of the material.}

\section{Housing Allocation} \label{sec:HA}

Some academic experiences leave a lasting impact, shaping our perspectives and fostering personal growth. 
For me, one such event came early in my career: the Stanford Institute for Theoretical Economics (SITE) Summer Conference, which I attended in July 1995.
This was just before I started my first job as a tenure-track Assistant Professor at the University of Michigan--Ann Arbor. 
Having received my Ph.D. from the University of Rochester a few months earlier, I was thrilled to be invited to present my job market paper, 
``Strategy-Proofness and Singleton Cores in Generalized Matching Problems''---an exercise in pure theory---at such a prestigious conference.\footnote{A generalized 
version of my job market paper was later published with a slightly different title as \cite{Sonmez:1999}.}

The quality and depth of interactions at the conference were outstanding. With only two presentations each day over two weeks, we had ample opportunity for in-depth discussions. In one of these interactions, 
I learned from Paul Milgrom himself about the active roles economists had taken in the FCC's first spectrum auction. 
I was so impressed that leading economic theorists were influencing real-life policies at this level.

Then it was my turn to present. To be honest, I don't remember much of it---I was pretty nervous about presenting my work to so many leading economic theorists. 
What I do remember vividly, though, is the second presentation of the day by Matthew Jackson, someone I have always looked up to as a role model.

I knew Jackson from his earlier fundamental work in mechanism design and implementation theory. 
However, that day he was presenting his joint work with Asher Wolinsky on 
``A Strategic Model of Social and Economic Networks,'' later published in \cite{Jackson/Wolinsky:1996}. 
Their paper was a brilliant example of use-inspired theory, studying efficiency and stability in various networks. 
With the winning combination of elegant theory and practical relevance, it was clear that this type of work was the future.\footnote{\cite{Jackson/Wolinsky:1996} indeed proved to be one of the most successful examples of 
use-inspired theory of the last three decades, initiating a very active literature and leading to many successful interdisciplinary contributions.}

Inspired by Milgrom and Jackson, I realized that I needed to up my game. It was time to move beyond abstract theory and produce work that could influence real-life policies. But where would I start?

I came up with a plan. After spending the last few years on abstract matching models, I decided to find their closest real-life counterparts. During the SITE conference, 
I stayed at Stanford University's graduate residences in Escondido Village. This experience sparked my interest in on-campus housing as a practical application. 
Although I couldn't find details on Stanford's allocation process, I discovered information from other institutions, such as Carnegie Mellon, Duke, Harvard, MIT, 
Northwestern, Pennsylvania, and Rochester. I began contemplating this problem during the SITE conference.

All these universities allocate their housing units through a few similar \textit{preference revelation mechanisms} (also called a \textit{direct mechanism}), 
where each student submits a strict preference ranking of housing units to the centralized housing authority. The authority then uses an algorithm to allocate the units based on these preferences.

Before the mid-1990s, two related problems had been studied. The first, the \textit{housing markets} model of \cite{shapley/scarf:74},
is an exchange economy in which each individual is endowed with a single indivisible good---a ``house''---and has preferences over all houses.
Individuals may either keep their houses or trade them if they wish.
The second, the \textit{house allocation} model of \cite{hylland/zeckhauser:79}, is an allocation problem in which a central planner distributes houses among individuals according to their preferences.
The key distinction between the two lies in property rights: in the first, houses are privately endowed, whereas in the second, individuals may have equal claims or be priority-listed according to various criteria.

\subsection{The Beginning: My Job Market Paper}

I was intimately familiar with Shapley and Scarf's housing markets model, as it played a key role in my job market paper. 
My research focused on direct mechanisms that satisfy three fundamental axioms---each corresponding to an essential principle in mainstream economics---across a broad class of domains, including housing markets.

The first two axioms ensures welfare optimality and respect for private property: 

\begin{definition}
Given a domain and a problem within this domain, an outcome $\mu$ \textbf{Pareto dominates} an outcome $\nu$ if each individual weakly prefers their assignment under $\mu$ to that under $\nu$, 
with at least one individual strictly preferring their assignment under $\mu$.

Given a domain and a problem within this domain, an outcome satisfies \textbf{Pareto efficiency} if it is not Pareto dominated by any other outcome. 
\end{definition}

\begin{definition}
Given a domain and a problem within this domain, an outcome satisfies \textbf{individual rationality} if it provides each individual with an assignment they weakly prefer to their initial endowment.
\end{definition}

For axioms such as \textit{Pareto efficiency} and \textit{individual rationality}, which are defined for outcomes, the same axiom is said to hold for a direct mechanism 
in a given domain if its outcome satisfies the axiom for all problems in that domain. This is a convention I will maintain throughout the applications.

The next property, an incentive-compatibility criterion defined for mechanisms rather than specific outcomes in given problems, 
is widely considered the gold standard in both traditional mechanism design and practical market design.

\begin{definition}
Given a domain, a direct mechanism satisfies \textbf{strategy-proofness} if no individual ever receives a strictly better assignment by misrepresenting their true preferences.
\end{definition}

The following solution concept, central in cooperative game theory, also plays a key role in my analysis.

\begin{definition}
Given a domain and a problem within this domain, an outcome is in the \textbf{core} if there exists no coalition of individuals whose members can benefit, 
all weakly and at least one of them strictly, by pooling their initial endowments and reassigning them among themselves.
\end{definition}

As the main contribution of my job market paper, I established the following result:

\begin{theorem}[\citealp{Sonmez:1999}] \label{thm:Sonmez-99} 
Consider a domain where all individuals have strict preferences over their assignments. Suppose a direct mechanism satisfies individual rationality, Pareto efficiency, and strategy-proofness. Then:
\begin{enumerate}
\item For each problem in the domain, there exists no more than one outcome in the core, and
\item For each problem with a non-empty core, the mechanism selects the unique outcome in the core.
\end{enumerate}
\end{theorem}

Since there are multiple outcomes in the core for at least some problems in the vast majority of domains, my result is fairly negative: in such domains, no direct mechanism can satisfy all three desiderata. 
However, the housing markets model is an important exception. When individuals have strict preferences over houses, each housing market has a unique outcome in the core \citep{shapley/scarf:74, roth/postlewaite:77}. 
Consequently, this domain allows for positive results, which is why I knew every nook and cranny of this model.

\subsection{Housing Markets and the Core}

I firmly believe that many fundamental concepts in consumer theory should be introduced to economics undergraduates through Shapley and Scarf's housing market model. 
Key ideas like Pareto efficiency, competitive equilibrium, and the core can be taught without first relying on technical constructs like utility functions or methods such as constrained optimization. 
Honestly, given its simplicity, I can't think of any other model that offers so much bang for your buck!

The results on the core in this setting are especially compelling when individuals have strict preferences over houses. In this case, each housing market has a unique allocation in the core which also coincides with the
unique competitive allocation \citep{roth/postlewaite:77}. 
Moreover, in any housing market domain, the direct mechanism that selects the unique core outcome for each problem 
satisfies \textit{strategy-proofness} \citep{roth:82}. Lastly, and aligned with Theorem \ref{thm:Sonmez-99},
it is the only direct mechanism that satisfies individual rationality, Pareto efficiency, and strategy-proofness \citep{ma:94}.\footnote{Any allocation in the core, 
by definition, satisfies \textit{Pareto efficiency} and \textit{individual rationality}, and therefore does so for housing markets as well.}

But for me, the most amazing thing about this model is the following procedure, which generates the unique core outcome of a housing market. 
This algorithm was famously invented by David Gale and suggested to Herbert Scarf during one of Scarf’s early presentations of the housing markets model \citep{scarf-ttc2009}.

\paragraph{Gale’s Top Trading Cycles (G--TTC) Algorithm.}
The key technical insight of the G--TTC algorithm is this: there is always at least one \textit{cycle} of individuals---possibly consisting of a single person---where each person prefers the endowment 
house of the next person in the cycle as their first choice. Consequently, all individuals in any cycle can receive their first choices by pooling their initial endowments and reassigning them among themselves.
So, why not form such a cycle in the first round of an algorithm, awarding all individuals in each cycle their first choices, 
and keep repeating this process with the individuals remaining in the market and their endowment houses? What a brilliant idea!

\begin{figure}[!tp]
    \centering
       \includegraphics[scale=1.05]{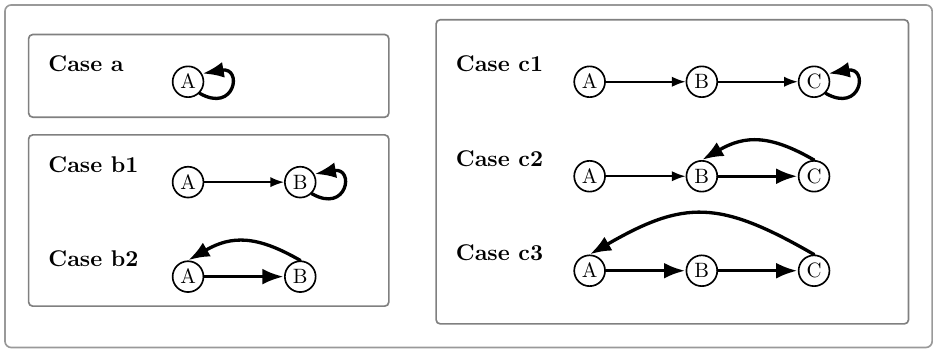}
\caption{Existence of a cycle in Gale’s Top Trading Cycles Algorithm (Example \ref{ex:TTC}). 
Label the first agent as A. If A's top choice is their own house (Case~a), A immediately forms a one-person cycle. 
Otherwise, A's top choice is another house, whose owner we label B (Cases~b and~c). 
If B's top choice belongs to someone already encountered (either A or B), a cycle arises. 
In particular, if B prefers their own house (Case~b1), there is a one-person cycle, 
whereas if B prefers A's house (Case~b2), then B and A form a two-person cycle. 
If instead B's top choice is C's house (Case~c), then C's top choice must belong to someone already encountered, since no one else remains. 
In this case, if C prefers their own house (Case~c1), there is a one-person cycle; 
if C prefers B's house (Case~c2), then C and B form a two-person cycle; 
and if C prefers A's house (Case~c3), then C, B, and A form a three-person cycle. 
Thus, in every scenario, a cycle emerges, denoted with bold arrows.}
\label{fig:GTTC-cycles}
\end{figure}

But why does such a cycle exist in every housing market? We explore this through an example, illustrated in Figure \ref{fig:GTTC-cycles}.\footnote{Throughout the monograph, 
individuals are denoted by the first letter of their names in the figures (e.g., Alp is denoted as A, Banu as B, Cora as C, etc.).}


\begin{example} \label{ex:TTC}
There are three individuals: Alp, Banu, and Cora.

Consider any individual, say Alp. Either his first choice is his own house, forming a one-person cycle, or his first choice is someone else’s house, say Banu’s. In this case, we haven’t found a cycle yet.

Next, consider Banu---the owner of Alp’s first-choice house. Her first choice is either her own house, forming a one-person cycle, or it’s Alp’s house, forming a two-person cycle with Alp. 
If neither, her first choice must be Cora’s house, since no one else remains.

Now, Cora’s top choice must be the house of one of the three. If it’s her own, she forms a one-person cycle. If it’s Banu’s, she forms a two-person cycle with Banu. And if it’s Alp’s, she forms a three-person cycle with Alp and Banu.

Regardless of the situation, a cycle always emerges! 
\end{example}

The observation in Example \ref{ex:TTC} is general. 
No matter how many individuals there are or what their preferences might be, you can always find a cycle as long as the number of people is finite. 
Here’s why: once you start with someone and follow a sequence of individuals based on their first choices, a cycle is formed as soon as any individual is repeated. 
Specifically, everyone between the two appearances of the repeated person---including the repeated person---forms the cycle. This is what drives Gale's brilliant algorithm---simple yet powerful.

Amidst all this beauty, the only downside to this elegant model was that, at the time, it didn’t seem to have any real-world application. Little did I know that my efforts, which began at the SITE conference, 
would help change that years later. 

Through a ``discovery--invention cycle'' driven by minimalist market design---starting with Shapley and Scarf's housing market model and G--TTC, 
and continuing with the key innovations of \cite{abdulkadiroglu/sonmez:99}---the ideas behind Gale’s brilliant algorithm would eventually lead to saving thousands of lives every year through 
living-donor organ exchanges worldwide (see Section \ref{sec:AS99-KE} and Figure \ref{fig:DIC-KE-arm}).

\subsection{House Allocation and Simple/Random Serial Dictatorship} \label{sec:SSD-RSD}

Unlike the abstract housing market model by \cite{shapley/scarf:74}, the house allocation model by \cite{hylland/zeckhauser:79} has concrete real-life applications.

The key difference between these two models lies in their property-rights structures: the first represents a private-ownership economy, whereas in the latter individuals may either be 
priority listed according to various criteria or have equal claims.

A straightforward solution for the house allocation model is to mimic the natural dynamics of a queue. Students are placed in a sequence based on priority criteria such as seniority, 
grade point average (GPA), or, in the case of equal claims, a random lottery. Housing units are then allocated one at a time according to the students’ positions in the queue and their preferences: 
the first student receives their top choice, the next receives their top remaining choice, and so on.

\paragraph{Serial Dictatorship.}
This procedure, when applied with a fixed sequence of students, was first studied by \citet{hylland/zeckhauser:79} and later termed a \textit{serial dictatorship} by \citet{Satterthwaite/Sonnenschein:1981}. 
In subsequent work, it has sometimes been referred to as a \textbf{\textit{simple serial dictatorship}} (SSD), which is the terminology we will adopt throughout this monograph. 
When the sequence of students is drawn uniformly at random, the mechanism is known as \textbf{\textit{random serial dictatorship}} (RSD), a term introduced by \cite{abdulkadiroglu/sonmez:98}.

The SSD and RSD mechanisms perform well with respect to their efficiency and incentive-compatibility properties: 
SSD is \textit{Pareto efficient}, RSD is \textit{ex post Pareto efficient}---its outcome being a probability distribution over deterministic outcomes that assigns positive weight only to Pareto-efficient ones---and both mechanisms are \textit{strategy-proof} \citep{Svensson:1994}.\footnote{\cite{Svensson:1994} focuses on SSD, referring to its outcomes as \textit{queue allocations}.}

Likely because they imitate the natural dynamics of a queue, SSD and RSD are the mechanisms of choice for system operators in a wide variety of real-life settings. 
Simply put, these are the mechanisms laypersons would most likely come up with. Combined with their appealing efficiency and incentive compatibility properties, it’s no wonder they are so popular globally across diverse settings.

\subsection{House Allocation with Existing Tenants} \label{sec:existingtenants}

In many practical applications of on-campus housing, the relevant setting is not fully captured by either Shapley and Scarf's housing markets model or Hylland and Zeckhauser's house allocation model. 
Instead, it represents a hybrid of these two settings.

Graduate housing is typically awarded for several years, and in the annual allocation process, students who already occupy units from the previous year are entitled to keep them if they choose.
Thus, in this version of the problem, there are ``existing tenants'' with their earmarked ``occupied'' units, ``newcomers'' without an earmarked unit, and ``vacant'' units.

Motivated by this more complex setting, \cite{abdulkadiroglu/sonmez:99} model and study this variation, calling it \textit{house allocation with existing tenants}.
This model generalizes both the housing markets model of \cite{shapley/scarf:74}, in which all units are earmarked,
and the house allocation model of \cite{hylland/zeckhauser:79}, in which no units are earmarked and all are available for allocation.

\subsection{Simple/Random Serial Dictatorship with Squatting Rights} \label{sec:SSR-SR}

Even though the house allocation model does not capture all aspects of the practical on-campus housing problem, system operators often adapt solutions from this 
simpler setting to address their needs in more complex scenarios. Indeed, this approach---relying on adaptations from simpler models to solve more challenging problems---is not limited to authorities 
overseeing real-life applications; it is also a common method among economic designers. In this vein, many institutions, including Carnegie Mellon, Duke, Harvard, Northwestern, and Pennsylvania, 
have implemented a straightforward modification of the SSD (or RSD) to accommodate the added complexity of existing tenants.

Before using SSD (or RSD), upperclassmen at these institutions---existing tenants---are given a choice between retaining their units from the previous year, 
a practice called ``squatting,'' or giving them up. If a student gives up their unit, it is added to the pool of available houses, and the student joins the set of applicants. 
Otherwise, the student retains their unit from the previous year. In either case, the complex problem simplifies to a standard house allocation problem, 
allowing the system operator to apply SSD (or RSD). The resulting mechanism is known as \textbf{\textit{SSD}} (or \textbf{\textit{RSD}}) \textbf{\textit{with Squatting Rights}}.

Formally speaking, because of the initial step where existing tenants choose between opting in or out, these mechanisms---unlike their predecessors SSD or RSD---are no longer direct mechanisms. 
The strategy space of existing tenants includes an additional dimension beyond their preferences over houses.

As popular as these mechanisms are across various universities, they unfortunately lack some of the appealing properties of their predecessors.

\subsection{You Request My House--I Get Your Turn: A Minimalist Redesign} \label{sec:minimalist-YRMH--IGYT}

Although SSD and RSD feature attractive efficiency and incentive-compatibility properties in standard house allocation, SSD (or RSD) with Squatting Rights loses these qualities 
when adapted to accommodate existing tenants. Because existing tenants risk receiving a less desirable unit than their current one if they relinquish it to join the applicant pool, 
these mechanisms generally do not incentivize them to give up their earmarked units. This reduces potential gains from trade and ultimately compromises efficiency.

The root cause of these failures is clear: the possibility for existing tenants to receive less desirable houses compared to their earmarked units, indicating a potential failure of \textit{individual rationality}. 
Therefore, a minimalist redesign would address this issue by directly targeting this root cause.

Let's assume there is a natural sequence of individuals that captures a priority order and work through one such redesign for the case of SSD with Squatting Rights. 

For starters, since addressing the root cause of the issues requires assuring \textit{individual rationality}, there is no need to provide existing tenants with an explicit option to exit the market with their earmarked units. 
Thus, we can stick to a direct mechanism for our prospective design. Essentially, we are aiming for a redesign of the full-participation SSD that assures existing tenants assignments at least as desirable as their earmarked units.

For our prospective minimalist redesign, we want to maintain the same underlying sequence of individuals as the original mechanism. 
Consider the dynamics of the SSD, which assigns each individual their most-preferred house that still remains unassigned, one at a time, following their sequence in the queue. 
However, if we insist on the same dynamics, we may encounter an issue with \textit{individual rationality}. Let’s illustrate this possibility, along with a minimalist intervention to proactively avoid such failures, with an example.

\begin{example} \label{ex:YRMH--IGYT}
There are three existing tenants---Alp, Banu, and Cora---along with a newcomer, Diya. Their order in the queue is Alp, Banu, Cora, and then Diya. In addition to the three houses occupied by the existing tenants, there is one vacant house.

Under SSD, Alp receives his first choice. This does not compromise \textit{individual rationality} if Alp's first choice is either his own earmarked unit or the vacant one; 
in these cases, the dynamics of SSD can proceed unchanged. But what if Alp's first choice is another existing tenant's earmarked unit, say Banu's? If Banu's first choice is her own unit, giving it to Alp would violate \textit{individual rationality}.

To prevent this issue, whenever Alp’s first choice is a unit earmarked for another existing tenant--- as in this case with Banu--- we move that tenant to the top of the queue, 
right before Alp, and proceed with the adjusted order. The queue then becomes Banu, Alp, Cora, and Diya.

Now, consider Banu's top choice after her promotion. If Banu's first choice is her own unit, she retains it, and Alp---now second in line---must consider his next choice. 
Alp couldn't have received his first choice anyway, since Banu must receive her earmarked unit to satisfy \textit{individual rationality}. 
On the other hand, if Banu's first choice is the vacant house, that's great news for Alp. Banu gets the vacant house, and Alp receives Banu's unit, which is now available.
If Banu's first choice is Alp's earmarked unit, that's also great news for Alp because 
they can trade their earmarked units in a two-way exchange:
Banu gets Alp's unit, and Alp gets Banu's unit---they exchange units directly and both leave with their preferred houses. 
However, if Banu's first choice is Cora's earmarked unit, we again move Cora to the top of the queue to avoid compromising \textit{individual rationality}, adjusting the order to Cora, Banu, Alp, and Diya.

With the new queue, no matter what Cora's first choice is, awarding her this unit cannot compromise \textit{individual rationality}.  Depending on Cora’s choice, either a ``cycle'' forms or a ``chain'' terminating at the vacant house:
\begin{itemize}
\item If Cora's first choice is her own unit, she forms a single-person cycle and exits the process with her unit.
\item If her first choice is Banu's unit, they form a two-person cycle, exchanging units and leaving the process.
\item If her first choice is Alp's unit, they form a three-person cycle: Cora gets Alp's unit, Alp gets Banu's unit, and Banu gets Cora's unit.
\item Finally, if Cora's first choice is the vacant house, a chain forms where Cora gets the vacant house, Banu gets Cora's unit, and Alp gets Banu's unit---all receiving their first choices.
\end{itemize}
\end{example}

Regardless of the specifics of the problem, this simple adjustment to the dynamics of SSD---where an individual who has not yet received an assignment is 
promoted to the top of the queue once another individual requests their occupied unit---always yields an individually rational outcome through cycles or chains.
Such a chain may terminate either at a vacant unit or at a unit whose occupant has already left the queue for a more preferred one.

This resulting minimalist redesign, first introduced in \cite{abdulkadiroglu/sonmez:99}, is called the \textbf{\textit{You Request My House--I Get Your Turn}} (YRMH--IGYT) algorithm.

To further illustrate how cycles and chains arise in this algorithm, we next present its execution with a more structured example.

\begin{figure}[!tp]
    \begin{center}
       \includegraphics[scale=1.05]{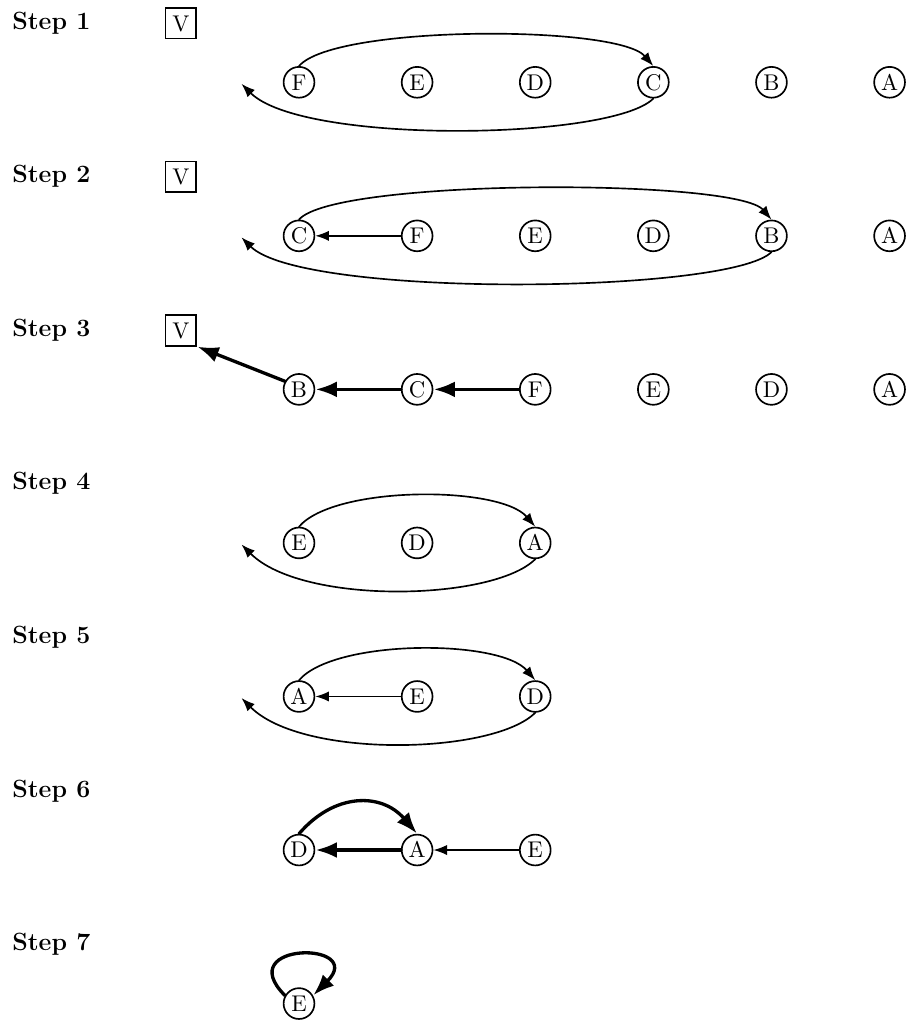}
           \end{center}
\caption{Execution of YRMH--IGYT for Example \ref{ex:YRMI-IGYT-2}.
Arrows to individuals indicate a request for their endowment house,
an arrow to vacant house V indicates a request for it, and arrows to the front of the queue indicate a priority upgrade to the front of the queue
for the next step. Step 3 features a chain, Step 6 a 2-person cycle, and Step 7 a single-person cycle, all depicted with bold arrows.}
\label{fig:YRMH--IGYT}
\end{figure}

\begin{example} \label{ex:YRMI-IGYT-2} 
There are five existing tenants Alp, Banu, Cora, Diya, and Ezra, each endowed with an occupied house $H_A$, $H_B$, $H_C$, $H_D$, and $H_E$, respectively. 
There is also a newcomer, Frank, with no endowment, and a vacant house $V$ that is not tied to any individual. 
All individuals are priority-ordered in a queue in reverse alphabetical order of their names. Preferences of individuals over houses are given as follows:
\[
\begin{array}{llll}
\mbox{Alp}: & H_B - V - H_D - H_C - H_A - H_E \qquad \quad & \mbox{Diya}: & H_C - H_A - H_D - V - H_B - H_E \\
\mbox{Banu}: & V - H_A - H_E - H_C - H_D - H_B & \mbox{Ezra}: & V - H_B - H_A - H_D - H_E - H_C \\
\mbox{Cora}: & H_B - H_E - H_D - V - H_C - H_A & \mbox{Frank}: & H_C - H_A - H_B - H_E - V - H_D
\end{array}
\]

We execute YRMH--IGYT, beginning with Frank in Step 1, the first individual in the queue. His top choice is $H_C$, the endowment of Cora who is still in the line. 
Cora is therefore promoted to the front, and the adjusted queue becomes: Cora, Frank, Ezra, Diya, Banu, Alp. 

In Step 2, Cora’s top choice is $H_B$, the endowment of Banu, who is also still in the line. Banu moves to the front, yielding an adjusted queue: Banu, Cora, Frank, Ezra, Diya, Alp.

In Step 3, Banu’s top choice is the vacant house $V$. This creates a \textit{chain}: Banu receives $V$, Cora takes $H_B$ which has just been vacated by Banu, and Frank takes $H_C$ which has just been vacated by Cora. 
Once this chain exits, the remaining individuals are lined up as follows: Ezra, Diya, Alp. 

In Step 4, Ezra’s top choice among the remaining houses is $H_A$, the endowment of Alp who is still in the line. 
Alp is promoted to the top, yielding the further adjusted queue: Alp, Ezra, Diya. 

In Step 5, Alp’s top choice among the remaining houses is $H_D$, the endowment of Diya who is still in the line. 
Diya is moved ahead, resulting in the adjusted queue: Diya, Alp, Ezra. 

In Step 6, Diya’s top choice among the remaining houses is $H_A$, which generates a 2-person \textit{cycle} with Alp: Diya takes $H_A$ and Alp takes $H_D$. Both then leave the process.  

In Step 7, the only person that remains is Ezra and the only house that remains is his endowment $H_E$. 
His top choice is therefore his own house, forming a 1-person \textit{cycle}, and he leaves with $H_E$. 

See Figure \ref{fig:YRMH--IGYT} for the execution of the YRMH--IGYT algorithm. 
\end{example}


The execution illustrates how chains and cycles naturally emerge under YRMH--IGYT, leading to a structure that helps address the shortcomings of SSD with Squatting Rights. 
The next proposition highlights those properties of YRMH--IGYT that are especially useful in contrasting the mechanism with its real-life counterpart.

\begin{proposition}[\citealp{abdulkadiroglu/sonmez:99}] \label{thm:YRMH--IGYT}
YRMH--IGYT mechanism satisfies individual rationality, Pareto efficiency and strategy-proofness. 
\end{proposition}

Another key feature of the YRMH--IGYT mechanism is its flexibility to accommodate any deterministic or probabilistic priority order of individuals, similar to a serial dictatorship. 
This feature provides system operators with an additional policy lever. For example, in on-campus housing, the system can assign higher priority to upperclassmen, 
students with higher grades, or students from disadvantaged backgrounds.

As we progress through this monograph, we see that the YRMH--IGYT algorithm---one of my early key inventions---plays a crucial role in multiple ``discovery--invention cycles'' discussed throughout the text 
(see Figures  \ref{fig:DIC-SC-arm} and \ref{fig:DIC-KE-arm}). 
To fully appreciate its significance, let’s explore this procedure in greater depth.

In our setting, there are two types of property rights. First, just as in Shapley and Scarf's housing markets model, existing tenants have ``private ownership'' of their earmarked units. 
Additionally, similar to Hylland and Zeckhauser's house allocation model, the sequence of individuals---used as a parameter of the mechanism---effectively creates a ``priority ordering'' 
for uncommitted units.

Under these property rights, the YRMH--IGYT algorithm enhances the basic SSD by integrating potential welfare gains through two types of trades. 
First, similar to the G--TTC algorithm, it facilitates the exchange of earmarked units among existing tenants through ``cycles'' of individuals. 
Second, it enables the trade of a leading spot in the queue---whether held by an existing tenant or a newcomer---for an earmarked unit of an existing tenant. 
This latter trade creates a ``chain'' initiated by the queue leader at various stages, who then receives either a vacant unit or an unassigned unit vacated by an existing tenant who has already exited the procedure with their assignment.

Given a sequence of individuals, the YRMH--IGYT algorithm maintains the exact same dynamics as the SSD when there are no existing tenants, as no one risks 
receiving an inferior assignment to their non-existent earmarked unit. In settings with only existing tenants and no newcomers or vacant houses, 
the YRMH--IGYT algorithm generates the unique core allocation regardless of the sequence of individuals. In this case, the dynamics of the procedure parallel the construction in Example  \ref{ex:TTC}, 
illustrating how Gale's brilliant idea is integrated within the underlying mechanics of YRMH--IGYT.\footnote{Strictly speaking, Gale's invention, which occurred during a presentation by Scarf, 
pertains to the existence of an outcome in the core \citep{scarf-ttc2009}. 
This reflects the abstract focus of Shapley and Scarf in their housing markets model, where the main aim is to invoke Scarf's Theorem \citep{Scarf:1967} to prove the existence of the core. 
Consequently, though it must have been apparent to Gale at the time, a uniqueness result was not provided in \cite{shapley/scarf:74}. 
The uniqueness was later proved by \cite{roth/postlewaite:77} for the case of strict preferences.}

In summary, as a natural generalization of both SSD and G--TTC, YRMH--IGYT overcomes the failures of SSD with Squatting Rights as a minimalist intervention, 
inheriting the compelling properties of its predecessors, as established by Proposition  \ref{thm:YRMH--IGYT}. 

\subsubsection{Top Trading Cycles: A Variant for House Allocation with Existing Tenants} \label{sec:houseTTC}

\cite{abdulkadiroglu/sonmez:99} introduces a second algorithm that yields the same outcome as YRMH--IGYT. 
Employing a technical construct that treats individuals and houses as separate entities, this algorithm transforms each chain of individuals in YRMH--IGYT into a cycle of individuals and houses, 
thereby implementing all trades through cycles.
Accordingly, we denote this algorithm as the \textbf{\textit{House Allocation variant of Top Trading Cycles}} (HA--TTC).\footnote{The term \textit{HA--TTC mechanism} is used interchangeably 
with the \textit{YRMH--IGYT mechanism} in this monograph.}

In this alternative procedure, not only individuals but also houses are represented as nodes in a directed graph. 
At any step, a directed link is formed from each individual to their most preferred house still available. Existing tenants have private ownership of their earmarked units, 
which is reflected by a directed link from each occupied house to its ``owner''---the current occupant. For vacant units and those vacated by previously processed existing tenants, 
the given priority order determines property rights, represented by a directed link from each such house to the highest priority individual remaining in the procedure. 
This mapping enables dynamics analogous to those of the G--TTC algorithm for house allocation with existing tenants.

As we will see in several other cases---and discuss in depth in Section \ref{sec:framing}---different algorithms that induce the same mechanism, 
or even different framings of a mechanism (and, more broadly, of an institution), may offer distinct advantages for practical design.
In this vein, while YRMH--IGYT and HA--TTC generate the same outcome for any given problem and thus induce the same direct mechanism, they offer different benefits.
As presented in Section \ref{sec:minimalist-YRMH--IGYT}, YRMH--IGYT is a minimalist refinement of SSD with Squatting Rights, making it easier to adopt for institutions already using that mechanism.
By contrast, HA--TTC---though more abstract---connects more directly to earlier literature and readily accommodates unit-specific priority queues for vacant units, a feature absent in YRMH--IGYT.
Moreover, as discussed in Section \ref{sec:SC--TTC}, this added flexibility underlies the school choice variant of the Top Trading Cycles algorithm introduced in \cite{abdulkadiroglu/sonmez:03}, 
as part of one of the ``discovery---invention cycles'' highlighted in this monograph (see Figure \ref{fig:DIC-SC-arm}).

\subsubsection{Contrast with a Technocratic Design Based on the Core} \label{sec:G--TTC}

In the late 1990s, before its publication, I presented \cite{abdulkadiroglu/sonmez:99} at several universities. 
Many economic theorists in the audience asked why we felt compelled to design a new mechanism for house allocation with existing tenants when a simpler ``technocratic'' solution based on the G--TTC algorithm could be adopted:

First, convert the problem into a housing market by randomly assigning vacant houses to newcomers as their initial endowments using a uniform distribution. 
Since existing tenants already have their initial endowments, this step yields a housing market as long as the number of newcomers is at least as large as the number of vacant houses.\footnote{If there are more 
newcomers than vacant houses, some newcomers will not be assigned initial endowments and will remain unassigned under the proposed mechanism.} 
Next, select the unique core allocation of the resulting housing market as the outcome. For lack of a better term, 
I refer to this mechanism as the \textbf{\textit{core-based technocratic mechanism}}.\footnote{When the number of houses equals the number of individuals and all individuals are newcomers, 
this mechanism is called the core from random endowments (CRE). Interestingly, for any house allocation problem, 
the CRE generates the same probability distribution over matchings as RSD \citep{Knuth:1996, abdulkadiroglu/sonmez:98}.}

By construction, the core-based technocratic mechanism satisfies \textit{individual rationality}. 
Moreover, by directly invoking results from earlier housing market literature, the mechanism also satisfies \textit{ex-post Pareto efficiency} and \textit{strategy-proofness}. 
What, then, was the point of our scholarly inquiry into the on-campus housing problem? After all, basic theory had already provided a solution to this practical application.

In retrospect, for researchers accustomed to the science and technology ecosystem shaped by Vannevar Bush's linear model (see Section \ref{sec:Linear}), 
this line of criticism feels quite natural. After all, policy development should follow from earlier work in pure and applied research, making the core-based technocratic mechanism an obvious solution. Mission accomplished!

Whenever this question was asked, I provided the following answer: YRMH--IGYT is more general because it also works when there are more houses 
than individuals and includes additional flexibility---an extra policy layer---in the priority order, which is a parameter of the mechanism.

This answer was not bad, and at the time, it seemed to satisfy the audience. However, deep inside, I was not fully satisfied with my response. 
Something was amiss. Nevertheless, the paper was accepted for publication, and my interest shifted to other projects that seemed more important at the time. 
It would take a few years, and embarking on what would become a very fruitful collaboration with another young economist, Utku Ünver, before I returned to this question.

In early 1999, after three and a half years at the University of Michigan in Ann Arbor, I returned to my native Turkey to start as an Assistant Professor at Ko\c{c} University in \.{I}stanbul.
Shortly after, in the fall of 2000, we recruited  \"{U}nver ---a recent Ph.D. from the University of Pittsburgh---for our group. 
Ko\c{c} University was relocating to its new campus in Sar{\i}yer, and  \"{U}nver  would stay in the faculty housing there. During the transition period, he spent a few days at my place.

On his first day as my house guest, I decided to put his talents to good use by bringing up the unresolved issue with the core-based technocratic mechanism. 
Joining forces, we proved the following result within a few months, clarifying what had been bothering me about this mechanism.

\begin{theorem}[\citealp{sonmez/unver:05}] \label{thm:technocratic}
For any house allocation problem with existing tenants, the outcome of the core-based technocratic mechanism is same as the outcome of the 
YRMH--IGYT mechanism with any sequence of individuals that (i) prioritizes each newcomer over each existing tenant, 
and (ii) randomly prioritizes newcomers among themselves with uniform distribution.\footnote{When each newcomer is prioritized over each existing tenant, 
the relative processing sequence of existing tenants among themselves becomes immaterial under the YRMH--IGYT (or HA--TTC) algorithm. }
\end{theorem}

Under the core-based technocratic mechanism, not only is the extra policy lever provided by the flexibility in the underlying priority order of YRMH--IGYT lost, 
but newcomers also systematically gain an advantage at the expense of existing tenants. 
Since on-campus housing policies tend to prioritize upperclassmen, the core-based technocratic mechanism would often result in unintended distributional consequences.

While the mechanism's description does not immediately reveal this bias, its source becomes clear upon reflection.
To transform the problem into a housing market and leverage its well-understood solution---the core---vacant houses are assigned exclusively to newcomers during the artificial construction of an initial endowment.
This technical construct deprives existing tenants of potential trade opportunities, including direct consumption, associated with these units.
Even when existing tenants may have higher or equal claims to vacant houses, as is common for upperclassmen in on-campus housing,
they effectively lose these property rights under the core-based technocratic mechanism, being confined to direct consumption or trades involving only their earmarked units.

Perhaps the most concerning aspect is that this bias is subtly embedded in the system, compromising \textit{transparency} and creating opportunities for exploitation by ill-intentioned 
decision-makers or for unintended consequences by well-meaning ones (see Section  \ref{sec:behavioral} for further discussion on these behavioral considerations). 
This phenomenon is not uncommon in applications where a solution to a complex problem is derived from a well-understood solution to a simpler case. 
With its emphasis on custom-made theory tailored to specific applications, minimalist market design is more resilient against such failures.

 \section{School Choice} \label{sec:schoolchoice}

Although \cite{abdulkadiroglu/sonmez:99} drew its motivation from campus housing at various universities,
I never pursued policy initiatives with their housing offices to reform flawed mechanisms like SSD (or RSD) with Squatting Rights. 
As the project progressed, my interest shifted more towards the theory of these problems, largely due to the richness and elegance of the underlying analytical framework. 
By the time we developed YRMH--IGYT, another application---Turkish college admissions---seemed far more important from a policy perspective.

The evolution of minimalist market design in actual policy can be traced through a series of holistic research and policy initiatives concerning the allocation of 
college seats in Turkey \citep{balinski/sonmez:99} and K--12 public school seats in the U.S. \citep{abdulkadiroglu/sonmez:03, chen/sonmez:06, ergin/sonmez:06} from the late 1990s to mid-2000s. 
Based on these experiences, along with those from another early application---kidney exchange, presented in Section \ref{sec:KE}---I began deliberately 
and rigorously applying minimalist market design in my research and policy initiatives starting in the early 2010s.

\subsection{Student Placement in Turkey} \label{sec:SP-Turkey}

\cite{balinski/sonmez:99} marks a milestone in my career. It was my first project where I engaged with authorities upon completion to explore possible policy impact. 
This represented my first holistic effort in research and policy, utilizing \textit{normative economics} to pursue the reform of a major real-life institution. 
It was also my first failure in policy efforts---the earliest of many to come---but not without teaching me several valuable lessons that later influenced the evolution of my minimalist approach to market design.

The focus of \cite{balinski/sonmez:99} is centralized \textit{student placement} through standardized tests.
The allocation of school seats based on a standardized test (or another form of objective performance measure) is a widespread practice worldwide. 
This has also been the case in my homeland, Turkey, at least since my childhood, for admissions to competitive secondary schools and the centralized allocation of college seats nationwide. 
Assuming there are no distributional objectives, the solution to the student placement problem is straightforward when there is only a single institution with multiple identical seats. 
The institution simply admits the highest-scoring students up to its capacity (or all students in case of insufficient demand).

The student placement problem is not much harder when there are multiple institutions that all prioritize applicants uniformly based on a single standardized test 
and allocate all seats in the system in a coordinated manner. In this case, while the presence of multiple institutions introduces heterogeneity to school seats, 
there is a simple direct mechanism that imitates the natural dynamics of a queue: the simple serial dictatorship (SSD) that we discussed earlier in the context of house allocation in Section \ref{sec:SSD-RSD}.

In the current setting, the pooled seats of all institutions are allocated to students one at a time based on their submitted preferences and their rankings in the standardized test. 
When it is a student's turn in the procedure, they are awarded a position at their highest-ranked school that still has an available seat. 
With a natural instrument, the standardized test, the case for SSD is even stronger for student placement than house allocation.

In Turkey, the student placement problem for college admissions takes a more complex form. Although all college seats are assigned through a centralized mechanism based on a single standardized test, 
the system constructs multiple field-specific rankings by assigning different weights to sections of the test. Each college is exogenously mapped to one of these rankings according to its field. 
For example, the ranking for engineering schools places greater weight on mathematics questions than the ranking for medical schools. 
As a result, SSD is no longer suitable for this problem, since no single priority ranking applies to all schools.

\subsection{No Justified Envy}

Overcoming favoritism and corruption is a key reason why many countries and local authorities allocate school seats centrally using results from objective standardized tests. 
This practice lends legitimacy to the allocation process. Therefore, respecting the results of standardized tests is crucial for institutions that allocate school seats in this way.
The formulation of this principle as the following normative axiom, which is vital not only for the centralized allocation of school seats in 
Turkey but also for the priority-based allocation of various other goods, represents one of the most significant contributions of \cite{balinski/sonmez:99}. 

\begin{definition} \label{def:NJE}
An allocation of school seats to students satisfies \textbf{no justified envy},\footnote{This axiom is called \textit{fairness} in \cite{balinski/sonmez:99}.
In a simpler setting---where a single priority ranking applies at all institutions and the number of items equals the number of individuals---\cite{Svensson:1994} earlier introduced this axiom as \textit{weak fairness}.
A reference to this axiom via the absence of \textit{justified envy} first appears in \cite{abdulkadiroglu/sonmez:03}, where it is called \textit{elimination of justified envy}.}
if there is no school $s$ and two distinct students $i$ and $j$ such that,  
\begin{enumerate}
\item student $j$ is assigned a seat at school $s$,
\item student $i$ strictly prefers school $s$ to their own assignment, and
\item student $i$ has a higher score than student $j$ on the standardized test for school $s$.   
\end{enumerate}
\end{definition}

Clearly, in cases where the axiom fails due to existence of a school $s$ and pair of students $i, j$ which satisfy the above given conditions 1-3,  
student $i$ would have a legitimate concern, and even perhaps channels available for litigation in some countries. 

The axiom of \textit{no justified envy} is closely tied to the common practice of announcing student placement outcomes 
through a vector of \textit{cutoff scores}. These cutoff scores indicate the lowest score that grants admission to each school, effectively defining the ``budget set'' for each student.

\begin{definition}
Given a student placement problem, a vector of cutoff scores supports an allocation of school seats to students if
\begin{enumerate}
\item every student $i$ who is assigned a seat at school $s$ has a score at least as high as the cutoff score for school $s$, and
\item every student $i$ whose score is at least as high as the cutoff score of some school $s$ weakly prefers their assignment to school $s$.  
\end{enumerate}
\end{definition}

The following result provides further justification for the \textit{no justified envy} axiom. 

\begin{theorem}[\citealp{balinski/sonmez:99}] \label{thm:cutoff}
An allocation of school seats to students satisfies \textbf{no justified envy} if and only if there exists a vector of cutoff scores that supports this allocation.
\end{theorem}

Therefore, the practice of announcing outcomes based on cutoff scores---a practice common worldwide---is a hallmark of mechanisms that satisfy \textit{no justified envy}. 

When all schools prioritize applicants uniformly with a single ranking, SSD satisfies \textit{no justified envy}. Indeed, the following result suggests that 
it is the only compelling mechanism in this setting. 

\begin{proposition}[\citealp{balinski/sonmez:99}] \label{prop:SSD}
When the priority ranking of students is identical at all institutions, SSD is the unique direct mechanism that satisfies \textit{no justified envy} and \textit{Pareto efficiency}. 
\end{proposition}

Proposition \ref{prop:SSD} supports the implementation of this mechanism for centralized
admissions to prestigious public middle schools (Anadolu Liseleri) and high schools (Fen Liseleri) in Turkey, 
at the time \cite{balinski/sonmez:99} was published.\footnote{Since then, at least some of these mechanisms are replaced.}

When schools rely on multiple priority rankings of students, on the other hand, enforcing \textit{no justified envy} comes with a cost, as we illustrate 
in the next example. 

\begin{figure}[!tp]
    \begin{center}
       \includegraphics[scale=1.04]{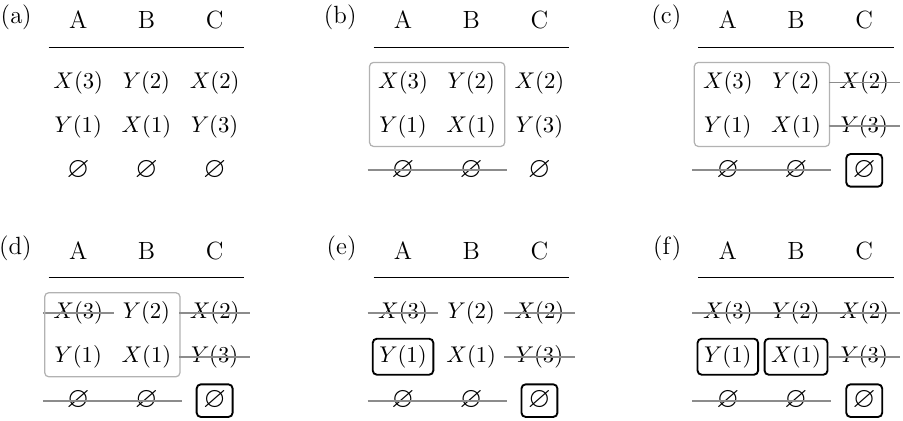}
           \end{center}
\caption{Efficiency cost of \textit{no justified envy} (NJE). Panel (a) presents the environment: individuals A, B, and C rank-order preferences over assignments X, Y, and $\biggerempty$ (unmatched). 
Each school's priority for a student is indicated alongside the student's preferences; for example, under A, $X(3)$ indicates that A has third priority at $X$. 
Because both schools are acceptable to all, NJE requires that no student be assigned an outcome worse than a school that ranks them first. 
Thus, A and B cannot be placed below their second choices; consequently, $X$ and $Y$ must be allocated between them (Panel (b)), leaving C unmatched (Panel (c)). 
Since C prefers $X$ to its assignment $\biggerempty$ and outranks A at $X$, NJE rules out assigning A to $X$ (Panel (d)); therefore, A receives $Y$ (Panel (e)), 
which in turn implies that B receives $X$ as the remaining option (Panel (f)). This allocation gives A and B their second choices and keeps C unmatched; 
it is Pareto dominated by the allocation in which A and B obtain their first choices while C remains unmatched.}
\label{fig:NJE-PE-conflict}
\end{figure}

\begin{example} \label{ex:NJE-PE} There are three students: Alp, Banu, and Cora. There are two colleges, $X$ and $Y$, each with a single seat. 
The preferences of the students and their priority rankings at the two colleges are as follows:
\[
\begin{array}{lll}
\mbox{Alp}:  & X - Y  \qquad \qquad \; \; &  X: \;\; \mbox{Banu, Cora, Alp}  \\
\mbox{Banu}: & Y  - X  \qquad & Y: \;\; \mbox{Alp, Banu, Cora} \\
\mbox{Cora}:  & X - Y \qquad &
\end{array} \]
Since Alp and Banu are each the highest-priority student at their respective second-choice colleges, neither can be assigned a less-preferred school under any allocation that satisfies  \textit{no justified envy}. 
Thus, the two seats at $X$ and $Y$ must be shared between Alp and Banu, leaving Cora unassigned.

However, Cora’s presence is not without consequence for Alp and Banu. Because she is ranked higher than Alp at college $X$, 
Alp cannot be assigned his first-choice college $X$ under any allocation that satisfies  \textit{no justified envy}. Consequently, Alp must receive a seat at his second-choice college $Y$, 
which in turn forces Banu to receive a seat at her second-choice college $X$. 
Yet this allocation is \textit{Pareto dominated} by the one in which Alp and Banu receive their first choices, establishing the potential incompatibility between  \textit{no justified envy} and \textit{Pareto efficiency}.
(See Figure \ref{fig:NJE-PE-conflict} for an illustration.)
\end{example}


Therefore, there is a fundamental incompatibility between \textit{no justified envy} and \textit{Pareto efficiency}. 

\begin{proposition}[\citealp{balinski/sonmez:99}] \label{prop:NJE-PE}
Assuming there are at least two colleges, three students, and that priority rankings of students can differ across two or more schools, 
there exists no mechanism that satisfies both \textit{no justified envy} and \textit{Pareto efficiency}.
\end{proposition}

\subsection{Multi-Category Serial Dictatorship}

What the mechanism SSD represents is so deeply intertwined with the use of priority rankings in unit-demand assignment problems with indivisible goods that policymakers often take SSD 
as the natural starting point for allocation procedures, even in settings where it cannot be directly applied.
Much like many institutions that use a modification of SSD---specifically, SSD with Squatting Rights---for practical applications in house allocation with existing tenants 
(see Section \ref{sec:SSR-SR}), Turkish authorities adopted an iterative procedure that embeds SSD as its core engine to address the more complex problem of college admissions with field-specific student rankings.

At each step of the procedure, SSD is applied to each field-specific ranking of students to ``tentatively'' assign seats in colleges within that field. 
This may result in some students being assigned seats at multiple colleges across different fields. At the end of each step, each student's preference ranking is truncated 
by ``removing''---i.e., deeming unacceptable---all colleges ranked lower than the highest-ranking college where the student holds a seat.

This process is repeated with the modified preferences until no student holds seats at multiple colleges, at which point the tentative assignments are finalized. 
Because the preference list of any student with multiple seats strictly shortens at each step, the procedure terminates after a finite number of iterations. 
The resulting direct mechanism is known as the \textbf{\textit{multi-category serial dictatorship}} (MCSD).

We now illustrate this mechanism with an example.

\begin{figure}[!tp]
    \begin{center}
       \includegraphics[scale=1.0]{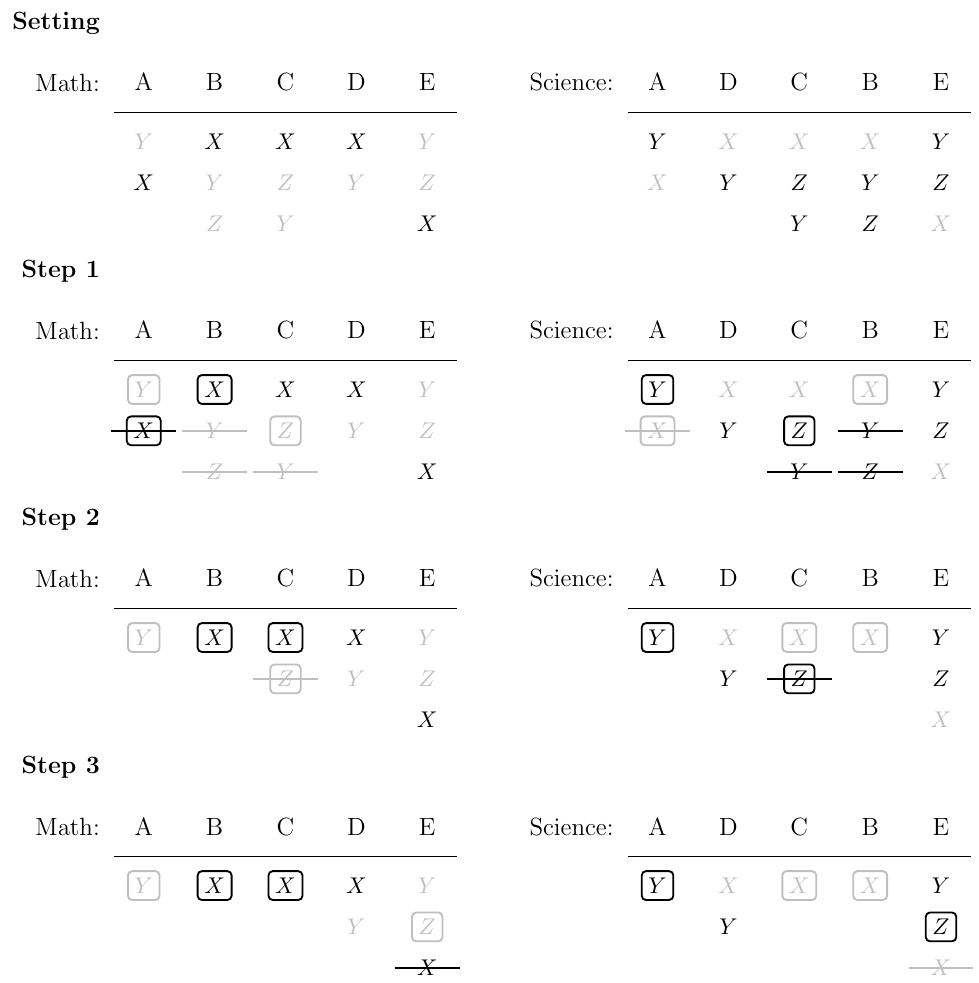}
           \end{center}
\caption{Execution of the MCSD mechanism in Example \ref{ex:MCSD}. The top panel illustrates a setting with five students (A, B, C, D, E) and three schools ($X$, $Y$, $Z$),  
where school $X$ belongs to the Math category and offers two seats, while schools $Y$ and $Z$ belong to the Science category and offer one seat each. 
In each category, students are priority ordered by their field-specific scores, with their school preferences ranked vertically beneath them. 
Within these preference lists, schools from different categories are shown in faint gray text. 
At each step, in both categories, seats are tentatively assigned using simple serial dictatorship (SSD), 
and any school ranked lower than a student’s best tentative assignment is removed from their preferences. 
The procedure terminates in Step 3, when no student holds more than one tentative assignment, finalizing the outcome.}
\label{fig:MCSD}
\end{figure}

\begin{example} \label{ex:MCSD}
There are five students: Alp, Banu, Cora, Diya, and Ezra. There is an engineering college, $X$, with two seats, and two medical colleges, $Y$ and $Z$, each with one seat. 
Students are evaluated in two categories: math and science. They are priority-ranked by their math scores at college $X$ and by their science scores at colleges $Y$ and $Z$. 
The preferences of the students over their acceptable colleges and their priority rankings in math and science are given as follows:
\[
\begin{array}{llllll}
\mbox{Alp}: & Y - X \qquad \quad & \mbox{Diya}: & X - Y \qquad \qquad \qquad & \mbox{Math}: & \; \mbox{Alp, Banu, Cora, Diya, Ezra} \\
\mbox{Banu}: & X - Y - Z & \mbox{Ezra}: & Y - Z - X & \mbox{Science}: & \; \mbox{Alp, Diya, Cora, Banu, Ezra} \\
\mbox{Cora}: & X - Z - Y &&&&
\end{array} \]

We refer to the SSD induced by the math ranking as SSD-math, and the SSD induced by the science ranking as SSD-science. 
Throughout the procedure, the two seats at $X$ are allocated using SSD-math, while the seats at $Y$ and $Z$ are allocated using SSD-science.

In Step 1, SSD-math assigns the two seats at $X$ to Alp and Banu, the students with the two highest math rankings. SSD-science assigns the seat at $Y$ to Alp, 
the highest-ranked in science, and the seat at $Z$ to Cora, the third-ranked in science, since it is not acceptable to Diya, who is ranked second. 
Because Alp holds seats at two colleges, the procedure continues to Step 2. For students holding at least one seat, 
their preferences are truncated after their most preferred tentative assignment, giving the following modified list of preferences:
\[
\begin{array}{llll}
\mbox{Alp}: & Y \qquad \quad & \mbox{Diya}: & X - Y \\
\mbox{Banu}: & X & \mbox{Ezra}: & Y - Z - X \\
\mbox{Cora}: & X - Z &&
\end{array} \]

In Step 2, SSD-math assigns the two seats at $X$ to Banu and Cora, the second- and third-ranked in math, since $X$ is no longer acceptable to Alp. 
As in Step 1, SSD-science assigns the seat at $Y$ to Alp and the seat at $Z$ to Cora, since Diya, ranked second in science, does not find $Z$ acceptable. 
Because Cora now holds seats at two colleges, the procedure continues to Step 3. Again, for students holding at least one seat, preferences are truncated after their most preferred tentative assignment. 
This modification does not further affect Alp and Banu, but reduces Cora’s list, resulting in:
\[
\begin{array}{llll}
\mbox{Alp}: & Y \qquad \quad & \mbox{Diya}: & X - Y \\
\mbox{Banu}: & X & \mbox{Ezra}: & Y - Z - X \\
\mbox{Cora}: & X &&
\end{array} \]

In Step 3, SSD-math assigns the two seats at $X$ to Banu and Cora, just as in Step 2. 
SSD-science assigns the seat at $Y$ to Alp and the seat at $Z$ to Ezra, the fifth-ranked in science, since it is not acceptable to Diya, Cora, or Banu, 
ranked second through fourth. Because no student holds seats at multiple colleges, the procedure terminates, finalizing the tentative assignments. See Figure \ref{fig:MCSD} for the execution of MCSD.
\end{example}

\begin{figure}[!tp]
    \begin{center}
       \includegraphics[scale=1.1]{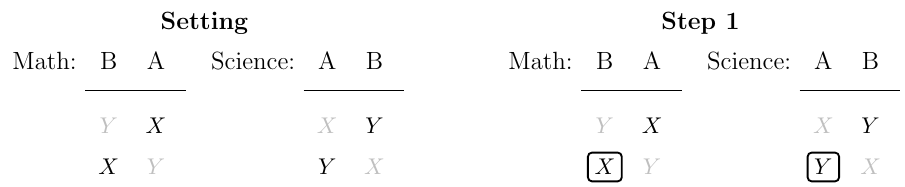}
           \end{center}
\caption{MCSD is not second-best subject to \textit{no justified envy}.
The left panel represents the setting in Example \ref{ex:MCSD-efficiency}, with two students (A and B) and two schools ($X$ and $Y$), each with one seat, where $X$ is in the Math category and $Y$ is in the Science category.
In each category, students are priority-ordered by their field-specific scores, with their school preferences ranked vertically beneath them. Within these preference lists, schools from different categories are shown in faint gray.
The right panel shows Step 1 of the execution of MCSD, in which seats in both categories are tentatively assigned via simple serial dictatorship. Because no student holds more than one tentative assignment, 
the procedure terminates immediately at Step 1, finalizing the tentative assignments.
Under MCSD, both students receive their second choices. Another outcome that satisfies \textit{no justified envy}---one that assigns both students their first choices---\textit{Pareto-dominates} the outcome of MCSD.}
\label{fig:MCSD-not-second-best}
\end{figure}

When SSD is modified into SSD with Squatting Rights for house allocation with existing tenants, as presented in Section \ref{sec:SSR-SR}, 
the adjustment feels somewhat ``forced,'' and the design’s shortcomings become rather evident. In contrast, 
MCSD---though also an SSD-based design---appears quite compelling at first glance. 
For one, it satisfies \textit{no justified envy}, the most critical axiom in the context of Turkish college admissions. 
Yet the following simple example, depicted in Figure \ref{fig:MCSD-not-second-best}, shows that the reliance on SSD in the execution of MCSD also entails an efficiency cost.

\begin{example} \label{ex:MCSD-efficiency}
There are two students: Alp and Banu. 
There is an engineering college, $X$, and a medical college, $Y$, each with one seat. 
Students are evaluated in two categories: math and science. They are priority-ranked by their math scores at college $X$ and by their science scores at college $Y$.
The preferences of the students and their priority rankings for math and science are given as follows:
\[
\begin{array}{llll}
\mbox{Alp}:  & X - Y  \qquad \qquad \; \;  & \mbox{Math}: & \mbox{Banu, Alp}  \\
\mbox{Banu}: & Y  - X   & \mbox{Science}: & \mbox{Alp, Banu} \\
\end{array} \]

Throughout the procedure, the seat at $X$ is allocated using SSD-math and the seat at $Y$ is allocated using SSD-science.

In Step 1, SSD-math assigns the seat at $X$ to Banu as the highest math-ranking student. 
Similarly, SSD-science assigns the seat at $Y$ to Alp as the highest science-ranking student.  
Since no one holds seats at multiple colleges, the procedure terminates right away, assigning both students a seat at their second-choice colleges. 

While this outcome satisfies \textit{no justified envy}, it is \textit{Pareto dominated} by another outcome that assigns both students seats at their first-choice colleges---an outcome that also satisfies \textit{no justified envy}.
\end{example}


In Example \ref{ex:NJE-PE}, we saw that some problems admit no outcome satisfying both \textit{no justified envy} and \textit{Pareto efficiency}. 
By contrast, this is not the case in Example \ref{ex:MCSD-efficiency}. Hence, the potential welfare loss under MCSD cannot be attributed to this fundamental incompatibility between the two axioms. 
Although not immediately apparent, the root cause of MCSD’s shortcomings lies in its excessive reliance on SSD as a procedural subroutine. 
To see why, it is helpful to relate the student placement model to the celebrated \textit{college admissions model} of \cite{gale/shapley:62}.

\subsection{Two-Sided Matching and the Deferred Acceptance Mechanism} \label{sec:two-sided-matching}

\cite{gale/shapley:62} marks the beginning of modern matching theory. In their celebrated paper, 
Gale and Shapley model and study \textit{two-sided matching markets}, making several fundamental contributions.

Their model includes two types of agents: students and colleges. What is distinctive about this framework is that agents and resources are the same entities: 
agents on one side of the market serve as resources for those on the other, and vice versa.

Gale and Shapley present two versions of the model. In the simpler \textit{one-to-one matching} model, each agent has unit demand. 
In the more general \textit{many-to-one matching} model, agents on one side---students---each have unit demand, while agents on the other side---college---can have multi-unit demand. 
In this setting, agents on each side have strict preferences over agents on the other side.

As an exercise in a use-inspired basic theory,\footnote{Gale and Shapley explicitly emphasize that some insights from their analysis can be useful in real life.}
Gale and Shapley focus on the existence of an outcome---called a \textit{matching}---that satisfies the following axiom.\footnote{\cite{gale/shapley:62} consider a setting where every 
individual on one side is acceptable to every individual on the other, whereas we follow \cite{roth/sotomayor:90} in presenting the axiom in a more general context without this assumption.}

\begin{definition} \label{def:stability}
A matching satisfies \textbf{stability} if no agent on either side is assigned an unacceptable partner, and no college-student pair strictly prefer each other over their current assignments, 
including nil assignments where a student is unmatched or a college has unfilled positions.
\end{definition}

One of the most important innovations in matching theory is an iterative procedure that guarantees the existence of a stable matching.

\paragraph{Individual-Proposing Deferred Acceptance Algorithm.}  
The algorithm proceeds in a series of steps, each consisting of two phases.
In the first phase of each step, each student ``proposes'' to their most-preferred acceptable college that has not yet rejected them.  
In the second phase, each college tentatively ``holds'' the most-preferred acceptable proposals it has received in that step, up to its capacity, and rejects the rest.  
The algorithm terminates when no new proposals are made or rejected. At this point, the proposals on hold are finalized as assignments.

\begin{figure}[!tp]
    \begin{center}
       \includegraphics[scale=1.11]{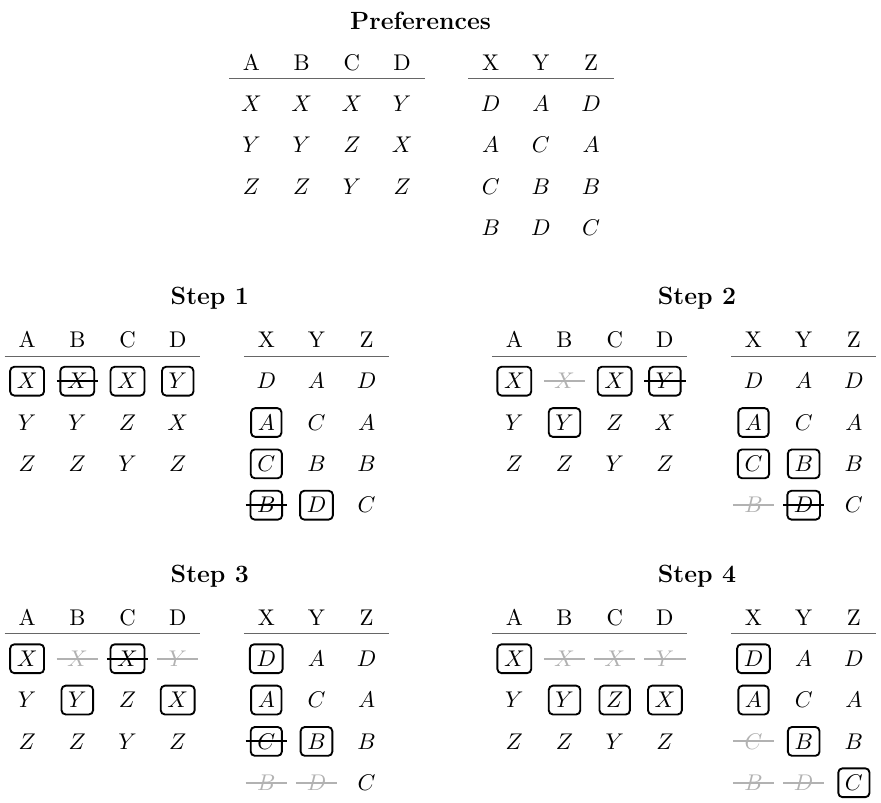}
           \end{center}
\caption{Execution of the individual-proposing deferred acceptance algorithm in Example \ref{ex:IPDA}. The top panel illustrates the setting with student preferences and school rankings.
At each step, every student makes a proposal to their most preferred school that has not rejected them in previous steps, as indicated by a boxed proposal. 
Each school then reviews all proposals received in that step (also shown with boxes) and rejects the lowest-ranked proposals exceeding its capacity (2 for $X$, 1 each for $Y$ and $Z$).
Rejections in the current step are marked by a line crossing the proposal box, while rejections from previous steps appear in faint gray with their crossed lines retained.} \label{fig:IPDA}
\end{figure}

We next illustrate this algorithm with an example, depicted in Figure~\ref{fig:IPDA}.  

\begin{example} \label{ex:IPDA}
There are four students, Alp, Banu, Cora, and Diya, and three schools: $X$, $Y$, and $Z$, where school $X$ has two seats and schools $Y$ and $Z$ each have one.  
The preferences of the students and schools over each other are as follows:
\[
\begin{array}{llll}
\mbox{Alp}:  & X - Y - Z  \qquad \qquad \; \;  & X: & \mbox{Diya -- Alp -- Cora -- Banu}  \\
\mbox{Banu}: & X - Y - Z   & Y: & \mbox{Alp -- Cora -- Banu -- Diya} \\
\mbox{Cora}:  & X - Z - Y  \qquad \qquad \; \;  & Z: & \mbox{Diya -- Alp -- Banu -- Cora} \\
\mbox{Diya}: & Y - X - Z   & & 
\end{array}
\]

\textit{Step 1.} Alp, Banu, and Cora propose to their top-choice school, $X$, while Diya proposes to her top choice, $Y$.  
School $X$ tentatively holds the two best proposals---from Alp and Cora---and rejects Banu.  
School $Y$ tentatively holds its only proposal from Diya.

\textit{Step 2.} Having been rejected by $X$, Banu proposes to her second choice, $Y$.  
Alp, Cora, and Diya maintain their offers from Step 1.  
Because its capacity is not exceeded this step, school $X$ continues to hold the proposals from Alp and Cora.
Of the two proposals received by $Y$, it tentatively holds the better one, from Banu, and rejects Diya.

\textit{Step 3.} Having been rejected by $Y$, Diya proposes to her second choice, $X$.  
Alp, Banu, and Cora maintain their offers from Step 2.  
School $X$ now holds the two best proposals---from Diya and Alp---and rejects Cora.  
School $Y$ tentatively holds its only proposal this step, from Banu.

\textit{Step 4.} Having been rejected by $X$, Cora proposes to her next choice, $Z$.
Alp, Banu, and Diya maintain their offers from Step 3.
Since no school receives more proposals than its capacity, no student is rejected in this step, and all offers are finalized.

The individual-proposing deferred acceptance algorithm concludes with Alp and Diya each receiving a seat at $X$, Banu receiving a seat at $Y$, and Cora receiving a seat at $Z$.
\end{example}


In their celebrated paper, Gale and Shapley proved that the individual-proposing deferred acceptance algorithm always yields a stable matching.
They also showed that each student weakly prefers their assignment under this matching to their assignment under any stable matching, 
thereby called the \textit{individual-optimal stable matching} (or \textit{student-optimal stable matching} in our setting).
They also introduced the analogous \textbf{\textit{college-proposing deferred acceptance algorithm}}, in which the roles of students and colleges are reversed.
This version yields a stable matching that favors the colleges, known as the \textit{college-optimal stable matching}.

For illustration, the reader can easily verify that in Example~\ref{ex:IPDA}, this college-proposing version of the algorithm terminates in two steps, 
assigning seats at $X$ to Alp and Diya---as in the individual-proposing version---but seats at their third-choice schools, $Z$ and $Y$, to Banu and Cora, 
in contrast to their second choices assigned under the individual-proposing version.

Later, in a result he attributed to John Conway, Donald Knuth showed that the optimal stable outcome for one side is the worst stable outcome for agents on the other side \citep{knuth:76}. 
This means that each agent weakly prefers their assignment under any stable outcome to their assignment under the optimal stable outcome for the other side.

The student placement model \citep{balinski/sonmez:99} is inspired by the many-to-one matching model. The key distinction between the two lies in the nature of preferences: 
in the many-to-one matching model, both students and colleges are considered agents with preferences over each other. In contrast, the student placement model treats only students 
as agents with preferences over colleges, while seats at colleges are regarded as goods. Nonetheless, because students are strictly prioritized at each college based on standardized test scores, 
colleges effectively have a strict ranking of students, similar to Gale and Shapley's many-to-one matching model. By interpreting these priority rankings as college preferences over students, 
it becomes straightforward to induce a many-to-one matching problem from any given student placement problem.

Beyond this natural mapping between the two models, there is a deeper connection between the \textit{no justified envy} axiom in student placement and the 
\textit{stability} axiom in many-to-one matching---one that plays a central role in the ``discovery--invention cycles'' linking basic research in two-sided matching with practical advancements in school choice (see Figure \ref{fig:DIC-SC-arm}). 
To illustrate this connection, we first introduce the following mild efficiency axiom for student placement.

\begin{definition} \label{def:non-wastefulness}
An outcome satisfies \textbf{non-wastefulness} if no school $s$ is left with an idle seat unless all students eligible for that seat (including those who remain unassigned) prefer their assignments over a seat at school $s$.
\end{definition}

\begin{proposition}[\citealp{balinski/sonmez:99}] \label{prop:NJE-stability}
An outcome satisfies, \textit{individual rationality, non-wastefulness} and  \textit{no justified envy} for a student placement problem 
if and only if it satisfies \textit{stability} for its induced many-to-one matching problem. 
\end{proposition}

Let the \textit{individual-proposing deferred acceptance mechanism}, henceforth simply referred to as  the \textbf{\textit{deferred acceptance mechanism}} (DA), 
be the direct student placement mechanism that selects the student-optimal stable matching 
of the induced many-to-one matching problem for any given student placement problem.\footnote{\cite{balinski/sonmez:99} and \cite{abdulkadiroglu/sonmez:03} refer to the same mechanism 
as the \textit{Gale-Shapley student-optimal mechanism} and  \textit{Gale-Shapley student-optimal stable mechanism} respectively.} 
Similarly, let the \textbf{\textit{college-proposing deferred acceptance mechanism}} be the direct student placement mechanism that selects the college-optimal stable matching 
of the induced many-to-one matching problem for any given student placement problem.

Having clarified the conceptual links between the student placement and many-to-one matching models,
we can reinterpret the following two fundamental results  from many-to-one matching for the student placement model. 

\begin{theorem}[\citealp{gale/shapley:62}] \label{thm:SP-SOSM-optimality}
For any student placement problem, the student-optimal stable matching \textit{Pareto dominates} any 
other outcome that satisfies \textit{individual rationality, non-wastefulness} and  \textit{no justified envy}. 
\end{theorem}

\begin{theorem}[\citealp{knuth:76}] \label{thm:SP-COSM-pessimality}
For any student placement problem, the college-optimal stable matching is \textit{Pareto dominated} by any 
other outcome that satisfies \textit{individual rationality, non-wastefulness} and  \textit{no justified envy}. 
\end{theorem}

We are finally ready to identify the root cause of the failure of the MCSD mechanism. 

\begin{theorem}[\citealp{balinski/sonmez:99}]  \label{thm:MCSD-COSM}
For any student placement problem, the outcome of the MCSD is the same as the outcome of the college-proposing deferred acceptance mechanism. 
\end{theorem}

On the one hand, Theorem \ref{thm:MCSD-COSM} provides yet another instance of external validity for Gale and Shapley's celebrated invention of the college-proposing deferred acceptance algorithm. 
On the other hand, together with Proposition \ref{prop:NJE-stability}, Theorems \ref{thm:SP-COSM-pessimality} and \ref{thm:MCSD-COSM} immediately imply that, among all outcomes satisfying 
\textit{no justified envy, non-wastefulness}, and \textit{individual rationality}, MCSD produces the worst possible assignment for each student.

In retrospect, reliance on SSD is the root cause of MCSD's shortcomings. To see why, we reassess its underlying iterative procedure.
At each step, within each field-specific ranking, SSD allocates the seats of all colleges in that field. 
When applied, SSD operates as if the entire field makes coordinated offers to students, avoiding unnecessary proposals that students would decline in favor of more-preferred colleges within the same field.
In this sense, the procedure functions as a ``field-proposing'' deferred acceptance algorithm that yields the student-pessimal stable outcome.

\subsection{What Sets Deferred Acceptance Apart} \label{sec:SOSM-superiority}

Beyond clarifying the root cause of the MCSD mechanism's failure, the strong connection between the student placement and many-to-one matching models---especially Theorem \ref{thm:SP-SOSM-optimality}---also positions 
DA as a leading candidate for student placement.
Indeed, as the following result establishes, the appeal of this mechanism on welfare grounds is even stronger than 
Theorem \ref{thm:SP-SOSM-optimality} alone suggests in settings where \textit{no justified envy} is indispensable.

\begin{theorem}[\citealp{balinski/sonmez:99}] \label{thm:SP-SOSM-stronger-optimality}
For any student placement problem, the student-optimal stable matching \textit{Pareto dominates} any other outcome that satisfies \textit{no justified envy}. 
\end{theorem}

In other words, in a system where \textit{no justified envy} is essential, DA not only outperforms MCSD but also any other direct mechanism. 
Moreover, as the next result shows, the advantage of DA extends beyond welfare considerations.


\begin{figure}[!tp]
    \begin{center}
       \includegraphics[scale=1.04]{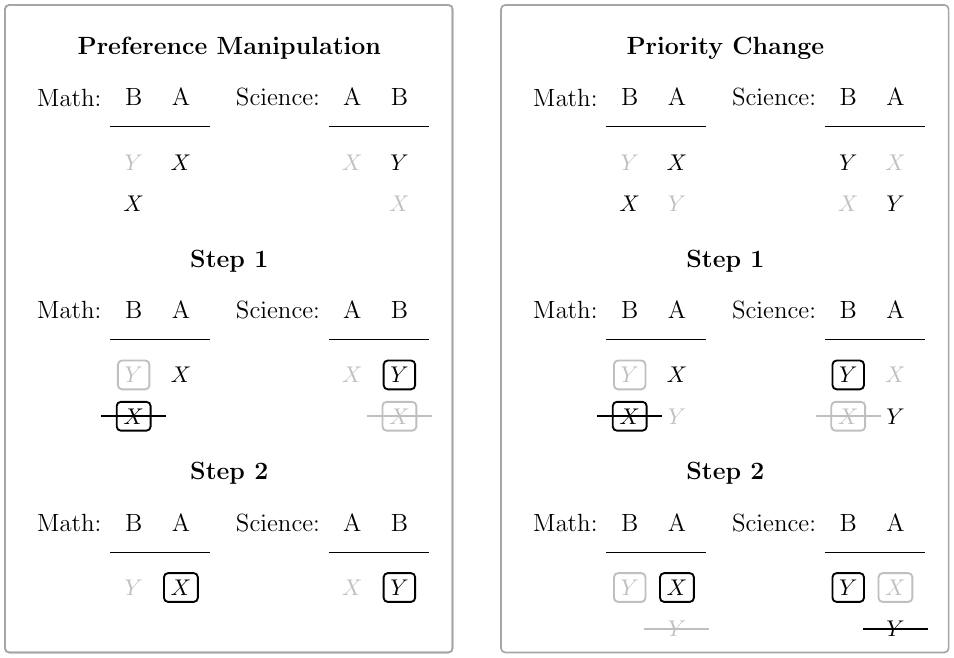}
           \end{center}
\caption{MCSD neither satisfies \textit{strategy-proofness} nor does it \textit{respect priority improvements}.
The left and right panels represent two modifications of Example \ref{ex:MCSD-efficiency}, depicted in Figure \ref{fig:MCSD-not-second-best}. 
The left panel illustrates a case where student A misrepresents their true preferences by removing their second-choice school $Y$ and reporting only their first-choice school $X$. 
The right panel shows a case where student A loses priority in Science, now ranked second in both categories.  
In both scenarios, student B is tentatively assigned seats at both schools in Step 1 and removes their second choice. 
In Step 2, both students receive tentative assignments at their first choices only, and the procedure terminates. 
Since student A receives their second choice under MCSD in Example \ref{ex:MCSD-efficiency}, they benefit either by manipulating their preferences (left panel) or by losing priority (right panel).}
\label{fig:mcsd-incentivefailures}
\end{figure}

\begin{theorem}[\citealp{alcalde/barbera:94, balinski/sonmez:99}] \label{thm:SP-SOSM-strategy-proofness}
DA is the unique mechanism that satisfies \textit{individual rationality, non-wastefulness, no justified envy}, and \textit{strategy-proofness}. 
\end{theorem}

Observe that Theorem \ref{thm:SP-SOSM-strategy-proofness} implies that MCSD does not satisfy \textit{strategy-proofness}.
As illustrated in the left panel of Figure~\ref{fig:mcsd-incentivefailures},
this is also evident in a slight modification of Example~\ref{ex:MCSD-efficiency}, where either student can single-handedly
secure their first choice by reporting it as their only acceptable option, thereby improving upon their outcome in the baseline scenario.

Next, we introduce another criterion that highlights yet another reason why DA stands out as an exceptionally strong candidate for student placement.

\begin{definition} A direct mechanism \textbf{respects priority improvements} if no student ever receives a less preferred assignment due to an increase in one or more of their rankings at a school, 
provided the relative rankings of all other students remain unchanged. 
\end{definition}

Through another slight modification of Example \ref{ex:MCSD-efficiency}, it also becomes evident that MCSD does not \textit{respect priority improvements}.
As illustrated in the right panel of Figure~\ref{fig:mcsd-incentivefailures},
both students would receive their first choices if one of them were ranked last by all colleges in both fields, whereas in the baseline scenario they receive their second choices.
Thus, under MCSD, a student may not only benefit from misreporting preferences but also be penalized for improved performance on standardized tests.

In contrast, DA sets itself apart in this regard too.

\begin{theorem}[\citealp{balinski/sonmez:99}] \label{thm:SP-SOSM-respectingimprovements}
DA is the unique mechanism that satisfies \textit{individual rationality, non-wastefulness, no justified envy}, and \textit{respects priority improvements}. 
\end{theorem}

\subsection{Failed Policy Attempt in Turkey and Lessons Learned} \label{subsec:failed-Turkish-policyeffort}

As a fresh Ph.D., I was exhilarated by these discoveries. My model was an exact fit for the application in Turkey, and, at least in my mind, 
the superiority of my proposed mechanism over the existing one was clear-cut. I had something of value to offer my homeland. Surely, authorities would welcome my discovery and correct their flawed mechanism---or so I thought.

In 1997, even before submitting the paper for scholarly review, I shared my discovery with Turkish officials. 
Following several correspondences by mail and a meeting with the head of the centralized clearinghouse, \"{O}SYM,\footnote{\"{O}SYM is the acronym for \"{O}\u{g}renci Se\c{c}me ve Yerle\c{s}tirme Merkezi, 
which translates to English as the Student Selection and Placement Center.}
I received a formal letter from Ankara politely rejecting my proposal.\footnote{See Figure \ref{OSYM-letter} for a copy of the letter from the head 
of \"{O}SYM rejecting my proposal to replace MCSD with DA in Turkish college admissions, 
and Figure \ref{OSYM-letter-English} for its English translation, both in the Supplemental Materials.}

The letter indicated that the leadership at  \"{O}SYM found academic value in  \cite{balinski/sonmez:99}.
They acknowledged that there can be multiple solutions that respects the cut-off score condition,
MCSD gives one of them, whereas DA gives another, and there can also be other ones. 
They also indicated that, while in theory the two mechanisms could generate very different outcomes, they saw this as a very low probability event. 
They further indicated that, with 1997 data (which includes more than a million students and thousands of colleges) the two
mechanisms generated the same outcome. They concluded that, in light of their findings, there is no reason to reform their mechanism. 

I learned several valuable lessons from this experience. Leadership at \"{O}SYM spent valuable time and resources to explore 
the merits of my policy proposal. While I failed in my ultimate objective of making policy impact with my research, 
my proposal received serious consideration. 
Their diligence gave me the hope that, perhaps, next time I might succeed. 
Given their emphasis on solutions that respect cut-off scores, I was correct in my hypothesis that, at least in the Turkish context 
\textit{no justified envy} was the most important desideratum.  On the other hand, 
the authorities did not even comment about the lack of \textit{strategy-proofness}, \textit{respect for
priority improvements}, or even the potential \textit{Pareto inferiority} of their mechanism. 
They clearly understood that the two mechanisms can generate different outcomes, but since their simulation of DA with
their most recent data for the 1997 placement generated the exact  same outcome as the actual match, they concluded that 
the failures I identified must correspond to low probability events. 

This was an eye-opening experience, leaving me with three inescapable takeaways:
\begin{enumerate}
\item What matters to authorities may differ from what matters to academic economists.
\item How good the mechanism I advocate is unlikely to matter much for authorities with vested interests in maintaining the status quo, 
unless I also show that their current mechanism is really bad.
\item No matter how accurate my model and clean my analysis might be, my theoretical analysis alone will likely not cut it for my policy aspirations. 
I need to present more concrete value to stakeholders in order to influence policy.
\end{enumerate}
A few years later, these takeaways guided my interactions with the authorities at Boston Public Schools.
 
\subsection{School Choice in the U.S.} \label{sec:schoolchoice-US}
 
Shortly after \cite{balinski/sonmez:99} was published, I learned that, in the U.S., students were also being assigned seats at schools annually 
through various centralized clearinghouses---not for college admissions, but for K--12 admissions in several major school districts. 
This centralized assignment of (typically public) school seats to students was promoted as an alternative to neighborhood-based assignment and was commonly referred to as \textit{school choice}. 
 
 At first, the school choice problem appeared completely isomorphic to the student placement problem, except, 
 rather than performance at standardized tests, typically other exogenous criteria were used to determine student priorities at schools. 
 If that was the case, perhaps there was no room for additional formal analysis. After all, the problem was already solved in  \cite{balinski/sonmez:99}
 with DA as its unambiguous winner. 
 
However, after examining the field implementation in several school districts, I came to the conclusion that school choice was somewhat different from Turkey's student placement. 
More specifically, school choice offered more flexibility in its potential solutions, and it wasn’t as clear as it was in Turkey that \textit{no justified envy} was indispensable as a desideratum. I reached this conclusion for two reasons.
 
The first reason was the broad structure of the criteria used to construct priority rankings of students at schools across various school districts. 
Unlike in Turkey, the use of standardized tests to determine the priority ranking of students at schools was rare in the U.S. Instead, factors such as proximity of the home address to the school 
or the presence of a sibling already attending the school were more commonly used to establish priority rankings. When \textit{no justified envy} is indispensable for a mechanism, 
it enforces strict adherence to these priority rankings, ensuring that no student can raise objections to the outcome based on them. 
In Turkey, students would spend as much time preparing for national standardized tests as on their regular school education. 
Thus, it made sense to strictly enforce compliance with the results of these nationwide tests when placing students in competitive schools. 
However, strict enforcement of a priority order based on a home address or even a lucky lottery draw seemed less essential.

The second reason was more direct. Of all the student assignment mechanisms in the U.S. that we documented in \cite{abdulkadiroglu/sonmez:03} 
across more than a dozen school districts, not a single one satisfied \textit{no justified envy}. 
Presumably, if no school district bothered to design a system that abides by this desideratum, it couldn't have been indispensable.

Based on these findings, I decided to explore the formal implications of relaxing the \textit{no justified envy} axiom. In my view, the primary difference between student placement in 
Turkish colleges and school choice was the potential dispensability of \textit{no justified envy} in the latter setting. The flexibility to drop or relax this axiom was valuable to me due to the incompatibility 
between \textit{no justified envy} and \textit{Pareto efficiency}. Naturally, my next objective was to design a \textit{Pareto efficient} mechanism without 
sacrificing any other plausible property of DA, such as \textit{strategy-proofness} or \textit{respect for priority improvements}.\footnote{For school choice, 
\textit{Pareto efficiency} already implies \textit{individual rationality} and \textit{non-wastefulness}.} 
 
\subsection{Top Trading Cycles: A Variant for School Choice} \label{sec:SC--TTC}

As presented earlier in Example \ref{ex:NJE-PE}, there is, in general, an incompatibility between \textit{no justified envy} and \textit{Pareto efficiency}.
There may be a student $i$ whose priority at a school $s$ is not high enough to secure a seat there, yet still sufficient to block a Pareto-improving trade between other students. 
Suppose student $j$ has sufficiently high priority at school $s$, and student $k$ has sufficiently high priority at school $t$. If student $j$ prefers school $t$ over school $s$, 
and student $k$ prefers school $s$ over school $t$, then \textit{no justified envy} prevents them from engaging in such a Pareto-improving trade whenever student $i$ 
has higher priority at school $s$ than student $k$.\footnote{This observation also forms the basis of the \textit{efficiency-adjusted deferred acceptance mechanism} 
(EADAM) in \cite{kesten:10}, along with related ideas in \cite{ehlers/morrill:20} and \cite{reny:22}.}

Thus, enforcing \textit{no justified envy} comes at the cost of forgoing Pareto-improving trades of priorities.
While trading priorities earned via standardized tests may be implausible,
why would authorities not allow such trades in applications where school priorities are determined by factors like proximity to home or even a lottery draw?

This idea naturally connects school choice to the house allocation with existing tenants model (cf. Sections \ref{sec:existingtenants}--\ref{sec:minimalist-YRMH--IGYT}).
In particular, two straightforward modifications to the model---(1) allowing multiple identical positions at each school and 
(2) introducing school-specific priority rankings---lead naturally to an extension of the HA--TTC algorithm (see Section \ref{sec:houseTTC}) for school choice.

In the house allocation with existing tenants model, HA--TTC not only permits trades of privately owned (earmarked) houses by existing tenants,
but also allows individuals to trade their priorities over central-planner-owned houses (vacant or vacated during the algorithm) for these privately owned houses.
Since each central-planner-owned house has the same priority order, there is no need to trade priorities across individuals.

In contrast, in school choice, student priorities differ across schools. We therefore adapted the HA--TTC algorithm by removing private-ownership objects and introducing heterogeneous priority orders across schools with multiple seats.
The resulting algorithm, the \textbf{\textit{School Choice variant of Top Trading Cycles}} (SC--TTC), trades priorities whenever this yields a Pareto improvement.

\begin{figure}[!tp]
    \begin{center}
       \includegraphics[scale=1.04]{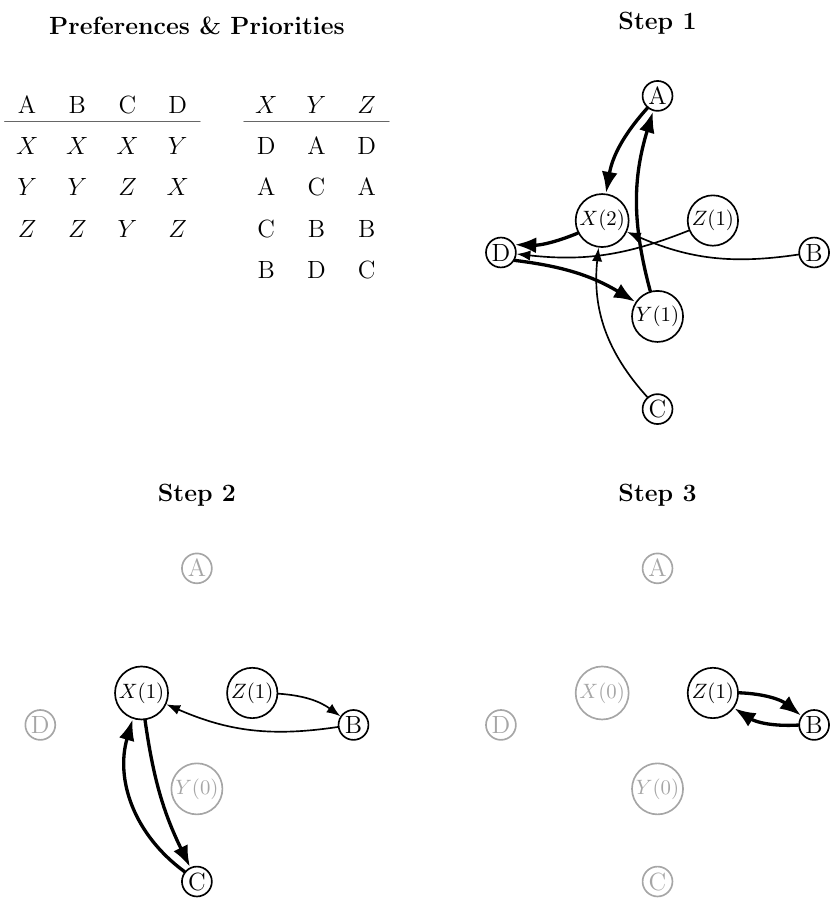}
           \end{center}
\caption{Execution of the SC--TTC algorithm in Example \ref{ex:SC--TTC}. The top-left panel indicates the setting.
In Steps 1–3, the smaller nodes represent students, and the larger nodes represent schools and their unfilled seats.
At each step, students point to their highest-ranked school with unfilled seats, and each school points to its highest-priority student who is still unassigned.
When a cycle forms, indicated by boldface arrows, each participating student leaves the process with a seat at the school they point to.
In Steps 2 and 3, students who have left the process with their assignments, and schools with no remaining seats, are shown as faint gray nodes.} \label{fig:SC--TTC}
\end{figure}

Using the same setting as in Example \ref{ex:IPDA}, we next illustrate the SC--TTC algorithm, depicted in Figure \ref{fig:SC--TTC}. 

\begin{example} \label{ex:SC--TTC}
There are four students, Alp, Banu, Cora, and Diya, and three schools: $X$, $Y$, and $Z$, where school $X$ has two seats and schools $Y$ and $Z$ each have one.  
The preferences of the students and schools over each other are as follows:
\[
\begin{array}{llll}
\mbox{Alp}:  & X - Y - Z  \qquad \qquad \; \;  & X: & \mbox{Diya -- Alp -- Cora -- Banu}  \\
\mbox{Banu}: & X - Y - Z   & Y: & \mbox{Alp -- Cora -- Banu -- Diya} \\
\mbox{Cora}:  & X - Z - Y  \qquad \qquad \; \;  & Z: & \mbox{Diya -- Alp -- Banu -- Cora} \\
\mbox{Diya}: & Y - X - Z   & & 
\end{array}
\]

At each step, each student points to their most-preferred school with available seats, and each school points to its highest-priority student still unassigned.

In Step 1, a cycle forms among Alp, $X$, Diya, and $Y$. Alp and Diya leave the process with seats at their first-choice schools, $X$ and $Y$, respectively.
School $Y$ also leaves the process with all its seats filled, while $X$ remains with one unfilled seat.

In Step 2, a cycle forms between Cora and $X$. Cora leaves the process assigned to the remaining seat at $X$, and $X$ exits the process.

In Step 3, a cycle forms between Banu and $Z$. Banu leaves the process with a seat at $Z$, which then exits the process, terminating the algorithm.

SC--TTC concludes with Alp and Cora each receiving a seat at $X$, Diya receiving a seat at $Y$, and Banu receiving a seat at $Z$.
\end{example}

It is instructive to observe that, under the outcome of SC--TTC in Example \ref{ex:SC--TTC},
not only does Banu prefer Diya’s assignment, $Y$, to her own, $Z$, which is her last-choice school,
but she also has higher priority at school $Y$ than Diya.
That is, Banu has justified envy toward Diya.
This becomes possible under SC--TTC because Alp, who has higher priority than Banu at $Y$, trades his priority with Diya
in exchange for her priority at school $X$, his first choice.

As a mechanism for school choice, SC--TTC is similar to DA in that it satisfies \textit{individual rationality}, \textit{non-wastefulness}, \textit{strategy-proofness},
and \textit{respect for priority improvements}.
However, SC--TTC differs from DA by satisfying \textit{Pareto efficiency}, but at the cost of potentially violating \textit{no justified envy}.
As established by \cite{balinski/sonmez:99}, DA is the clear choice in applications where \textit{no justified envy} is indispensable.
By contrast, \cite{abdulkadiroglu/sonmez:03} proposed SC--TTC as the main alternative for applications where \textit{no justified envy}, though desirable, is secondary and can therefore be relaxed.

At that time, we could not identify a single school district using a mechanism that satisfied \textit{no justified envy}, which led us to consider SC--TTC as a strong contender against DA for school choice in the U.S. 
However, time would show that we might have underestimated the appeal of the \textit{no justified envy} axiom even in the U.S., and that DA would ultimately emerge as the clear winner between our two proposals.

\subsection{Boston Mechanism} \label{sec:Boston-mechanism}

One of the biggest takeaways from my experience with the Turkish authorities was that the availability of a good mechanism---even one that is theoretically perfect---is not, 
by itself, enough to convince policymakers to undertake a costly reform.
One must also demonstrate that the mechanism they currently use is clearly inferior to the proposed alternatives, based on criteria that matter to the stakeholders. 

In my earlier interactions with the Turkish authorities, I failed to establish sufficient value for replacing MCSD with DA. 
At least with the 1997 data, the outcome of MCSD was the same as that of DA. In theory, this equivalence could have been an artifact of potential preference manipulation by students, 
but there was no apparent reason---such as field evidence---to believe this was the case. 
Maybe, I thought, MCSD wasn’t too bad after all... So, I shifted my attention to finding real-life mechanisms that were more visibly flawed according to criteria likely to be important to the stakeholders.

Among the mechanisms we documented in \cite{abdulkadiroglu/sonmez:03} across more than a dozen major U.S. school districts,
one that stood out was the mechanism used to allocate K--12 public school seats at Boston Public Schools (BPS).
It was, in fact, the only mechanism we identified at the time that had been implemented in multiple districts.
After researching its history, we concluded that Boston was likely the first district to adopt it, and we therefore dubbed it the \textbf{\textit{Boston mechanism}}.\footnote{Today, 
the Boston mechanism is also referred to as the \textit{immediate acceptance} mechanism, reflecting its algorithmic dynamics and contrasting with Gale and Shapley’s renowned deferred acceptance algorithm.}
Other major school districts that used this mechanism included Cambridge, Charlotte, Denver, Minneapolis, 
Seattle,\footnote{See Section \ref{sec:Seattle} for Seattle’s unusual experience with the Boston and DA mechanisms.} and St. Petersburg--Tampa.

In terms of its mechanics, the Boston mechanism is one of the simplest direct mechanisms one can imagine for school choice.
Under this procedure, the outcome is determined in several steps: first, seats are assigned at students’ first-choice schools; 
next, students who remain unmatched are considered for their second-choice schools, and so on.
At each step, every school admits the highest-priority students among those applying until all of its seats are filled.
Informally speaking, the mechanism first processes first choices, then second choices, and so forth — literally in the most straightforward way possible.

\begin{figure}[!tp]
    \begin{center}
       \includegraphics[scale=1.1]{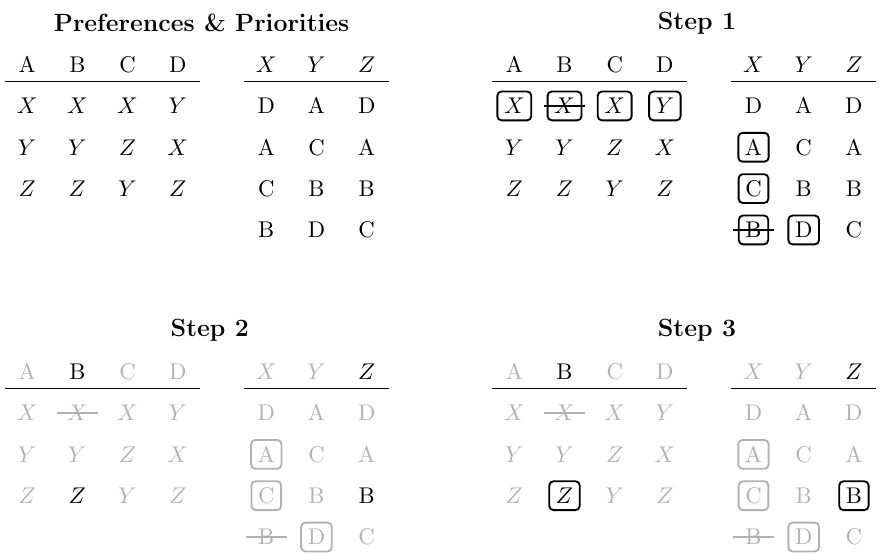}
           \end{center}
\caption{Execution of the Boston mechanism for the setting in Examples \ref{ex:IPDA} and \ref{ex:SC--TTC}. 
The top-left panel shows the setting. In Step 1, all students propose to their first choices, and all schools admit the top-ranked offers up to capacity (two for $X$, one each for $Y$ and $Z$). 
Assignments are final. In Step 2, only student B and school $Z$ are active; the rest are shown in faint gray. Since the second choice of B is not in the process, this round becomes obsolete. 
Student B then proposes to $Z$ as her third choice and is assigned a seat there in Step 3.} \label{fig:Boston}
\end{figure}

Figure \ref{fig:Boston} illustrates the execution of the Boston mechanism for the same setting as in Examples \ref{fig:IPDA} and \ref{fig:SC--TTC}, in which we executed the DA and SC--TTC mechanisms, respectively.

To its credit, if we take the submitted student preferences at face value---that is, assuming they report their preferences truthfully---the outcome of the Boston mechanism is \textit{Pareto efficient}.
After all, this mechanism directly prioritizes accommodating students’ first choices over their second choices, and similarly gives higher priority to any given ranked choice compared to those ranked lower in the submitted preferences.
Indeed, as can be seen in Figure \ref{fig:Boston}, the outcome of the Boston mechanism coincides with that of SC--TTC---a \textit{Pareto-efficient} mechanism---for the setting in Example~\ref{ex:SC--TTC}.

However, whereas SC--TTC achieves \textit{Pareto efficiency} as a \textit{strategy-proof mechanism}, the Boston mechanism creates strong incentives for students to misrepresent their preferences, 
precisely because of the same simple mechanics that make it \textit{Pareto efficient}.
Since it prioritizes assigning seats at any school to students who rank it first in their submitted preferences, it is often unwise for students to ``waste'' their first choice on schools they are unlikely to secure due to low priority at those schools.
Indeed, in Figure \ref{fig:Boston}, Banu can benefit from misrepresenting her preferences by ranking school $Y$ first---thereby avoiding ``wasting'' her first choice by 
ranking her true first-choice $X$ at the top of her preferences under a mechanism that penalizes truthful behavior and ultimately causes her to receive a seat at her last-choice school~$Z$.

We soon discovered that we were not the first to observe that Boston mechanism harbored strong incentives for preference manipulation. 
For example, \cite{Glazerman/Meyer:94} discussed this issue in the education literature:

\begin{quote} 
``It may be optimal for some families to be strategic in listing their school choices. For example, if a parent thinks that their favorite school is oversubscribed 
and they have a close second favorite, they may try to avoid 'wasting' their first choice on a very popular school and instead list their number two school first.'' 
\end{quote}

Several news stories also guided parents on this tricky aspect of the Boston mechanism. For instance, \cite{Tobin:2003} advises:

\begin{quote} 
``Make a realistic, informed selection of the school you list as your first choice. It's the cleanest shot you'll get at a school, but if you aim too high, you might miss. 
Here's why: If the random computer selection rejects your first choice, your chances of getting your second choice are greatly diminished. 
That's because you then fall in line behind everyone who wanted your second choice as their first choice. You can fall even farther back in line as you get bumped down to your third, fourth, and fifth choices.'' 
\end{quote}

Indeed, some school districts using the Boston mechanism explicitly advised parents to strategize when submitting their preferences. As reported in \cite{aprs:06}, the 2004 Boston Public Schools guide states:

\begin{quote} 
``For a better chance of your `first choice' school [...] consider choosing less popular schools. Ask Family Resource Center staff for information on `underchosen' schools.'' 
\end{quote}

As discussed later in Section \ref{sec:leveling-the-playingfield}, there was even an organized parent group in Boston that held meetings 
dedicated to discussing optimal application strategies under the Boston mechanism.

These discoveries convinced me that school districts using the Boston mechanism might be more receptive to our reform ideas than the Turkish authorities had been. 
However, before proceeding, I decided to build a much stronger case against the Boston mechanism than I had been able to make against the MCSD. 
Specifically, I decided to explore how the Boston mechanism's vulnerability to preference manipulation affects its efficiency.

Since the Boston mechanism is not \textit{strategy-proof}, analyzing its efficiency is not a straightforward exercise. In a laboratory experiment \cite{chen/sonmez:06}, 
we examined both the efficiency and the extent to which students engaged in strategic behavior for the Boston mechanism and its competitors DA and SC--TTC. 
Our experimental analysis revealed that, while the Boston mechanism is \textit{Pareto efficient} in the absence of strategic manipulation, 
in our designed environment, it has lower efficiency than both DA and SC--TTC.\footnote{While DA and SC--TTC are both \textit{strategy-proof},  
the subjects were not given this important information in our experiment. As a result, many of them engaged in strategic behavior under these mechanisms and reduced their efficiency, but more so for SC--TTC than DA. 
It is important to emphasize that, in real-life implementation, it is in the best interest of a school district to educate participants on the \textit{strategy-proofness} of these mechanisms. 
Indeed, one of the most important benefits of adopting a \textit{strategy-proof} mechanism is that a school district can encourage students to be truthful without taking any risk.
Hence, by not educating the subjects on the \textit{strategy-proofness of} DA and SC--TTC, in our experimental setup, we analyzed these mechanisms in an environment that is unfavorable for them. 
Despite this treatment, both mechanisms outperformed the Boston mechanism, thus increasing our confidence that they are likely to be more efficient than the Boston mechanism in real-life implementation.}

To further support the experimental evidence in \cite{chen/sonmez:06}, we showed in \cite{ergin/sonmez:06} that DA outperforms the Boston mechanism in efficiency, 
based on a complete-information equilibrium analysis of the latter.
These findings are formalized in Section \ref{sec:leveling-the-playingfield} as Theorem \ref{thm:ergin-sonmez-06} and Corollary \ref{cor:ergin-sonmez-06}, 
where we explore the behavioral aspects of our minimalist market design applications.\footnote{As we elaborate in Footnote \ref{fn:Ergin-Sonmez:2006} in Section \ref{sec:leveling-the-playingfield}, 
\cite{ergin/sonmez:06} shows that the complete-information assumption is key to the unambiguous efficiency advantage of DA over the Boston mechanism. 
Indeed, they also show that, in an incomplete-information environment, a student may prefer a Bayesian Nash equilibrium outcome of the Boston mechanism to the truthful dominant-strategy outcome of DA. 
Along these lines, using tools from Bayesian mechanism design, \cite{abdulkadiroglu/che/yasuda:2011} show that the Boston mechanism can be more efficient than DA in certain environments.}

\subsection{Policy Impact at Boston Public Schools} \label{sec:Boston}

With the main analysis in \cite{balinski/sonmez:99} and \cite{abdulkadiroglu/sonmez:03}, the supporting analysis in \cite{chen/sonmez:06} and \cite{ergin/sonmez:06}, 
and anecdotal evidence of strategic behavior, we were ready to make our move against the Boston mechanism (see Table \ref{tab:mech-comparison} for a summary comparison of 
Boston, DA, SC--TTC, and MCSD with respect to policy-relevant properties). 
However, it was one additional development that provided the impetus to approach the leadership at BPS for a potential reform of their student assignment mechanism.


\begin{table}[!tp]
\centering
\begin{threeparttable}
\caption[Properties of school choice mechanisms]{Properties of school choice mechanisms.}
\label{tab:mech-comparison}
\renewcommand{\arraystretch}{1.3}
\setlength{\tabcolsep}{6pt}
\begin{tabular}{
  l
  >{\centering\arraybackslash}p{2.2cm}
  >{\centering\arraybackslash}p{2.2cm}
  >{\centering\arraybackslash}p{2.2cm}
  >{\centering\arraybackslash}p{2.2cm}
}
\toprule
\textbf{Property}
& \textbf{Boston}
& \textbf{DA}
& \textbf{SC--TTC}
& \textbf{MCSD} \\
\midrule
Strategy-proofness      & \xmark & \cmark & \cmark & \xmark \\
No justified envy       & \xmark & \cmark & \xmark & \cmark \\
Pareto efficiency       & \cmark\textsuperscript{*} & \cmark\textsuperscript{\S} & \cmark & \xmark \\
Respect for priorities  & \xmark & \cmark & \cmark & \xmark \\
\bottomrule
\end{tabular}

\begin{tablenotes}\footnotesize
\item Notes: \cmark\;satisfied; \xmark\;not satisfied.
\item \textsuperscript{*}\;Pareto efficiency holds only under truthful preferences (hence vulnerable to preference manipulation).
\item \textsuperscript{\S}\;Pareto efficiency holds in a restricted, second-best sense within the set of outcomes that satisfy \emph{no justified envy}.
\end{tablenotes}
\end{threeparttable}
\end{table}


\subsubsection{Boston Globe Story on  \cite{abdulkadiroglu/sonmez:03}} \label{sec:BostonGlobe}

A few weeks after \cite{abdulkadiroglu/sonmez:03} was published, a reporter named Gareth Cook interviewed us about our analysis of the Boston mechanism and its alternatives. 
The resulting story in the \textit{Boston Globe} provided us with the perfect opportunity to approach the leadership at BPS \citep{Cook:2003}.

In preparing his story, Cook communicated not only with parents who were frustrated with the Boston mechanism
but also with officials at BPS and the Boston School Committee. The following quotes are from this story:

\begin{quote}
 ``Officials with the Boston public schools and the Boston School Committee readily acknowledge that 
 parents are frustrated with the current system, and officials said at a School Committee meeting 
 this week that they would make changing the system a priority. [...]
 
 `For every parent who feels frustrated about a policy, there is always a parent who will feel frustrated about an alternative,'
 said Christopher M. Horan, chief of staff for the Boston public schools. 
 Horan said he was intrigued by the economists' work and considered their suggestion a serious alternative. [...]
 
 Of course, no new system can create more seats at the most sought-after schools. 
 But all parents interviewed by the Globe said that it would be a huge relief simply to write a truthful answer to the question: 
 What school do you want?

`A lot of the alienation some parents have toward the choice system is solely attributable 
to the alienation of not making your first choice your first choice,' said Neil Sullivan, 
the father of four children who have attended Boston public schools.''
 \end{quote}

These quotes suggested that our critique of the Boston mechanism resonated with the concerns of Boston parents.

\subsubsection{Collaboration with Officials at Boston Public Schools}

Not only did the Boston Globe story give a significant boost to our policy efforts at BPS, but it also reinforced my belief that I was on the right track with my broader strategy to influence policy. 
The vulnerability of the Boston mechanism to strategic manipulation was indeed a significant burden for families. 
Thus, the value of our proposed mechanisms in DA and SC--TTC became more apparent with this story. 

With this important breakthrough, it was clear that the time was right to communicate with officials at BPS. 
In an email message that included the Boston Globe story along with the four papers \cite{balinski/sonmez:99}, \cite{abdulkadiroglu/sonmez:03}, \cite{chen/sonmez:06}, and \cite{ergin/sonmez:06}, 
I expressed to Superintendent Tom Payzant our desire to collaborate with their office for a potential reform of their school choice mechanism.

A few weeks later, I received a reply to my email from Valerie Edwards, then Strategic Planning Manager at BPS, and had a phone conversation with her to explain our ideas and underlying motives. 
BPS officials were initially upset by the mayhem caused by the Boston Globe story and mentioned that they had no funding for our involvement. 
I assured her that our work was motivated by both scholarly interests and a commitment to public service and that all our efforts would be pro bono.

These assurances led to an invitation to present the details of our policy proposal and its potential benefits to the city. In October 2003, I flew from \.{I}stanbul to Boston, 
and at a meeting held at the BPS central office, I presented the merits of our proposed reform. Also in attendance were Atila Abdulkadiro\u{g}lu, Alvin Roth, and Parag Pathak, 
then a first-year graduate student at Harvard University under Roth's supervision. This meeting marked the beginning of my long and fruitful collaboration with Pathak.

During my presentation, it became clear that, above all, the vulnerability of the Boston mechanism to preference manipulation, 
together with the \textit{strategy-proofness} of its proposed alternatives---DA and SC--TTC---was what had secured this meeting.
Given the clearer comparison between DA and the Boston mechanism provided by \cite{ergin/sonmez:06}, I focused in this meeting mainly on DA as a plausible alternative for BPS.
The group at BPS, however, were well prepared for the papers I had sent, and during the meeting they also inquired about our thoughts on SC--TTC.

By the end of this meeting, they were convinced that the Boston mechanism does not serve the city well, 
and it likely  alters the submitted preferences. 
Since the city was using the preference information  also to assess the popularity of its schools, 
officials at BPS were concerned that a disconnect between submitted and true preferences could undermine policies reliant on this data.
They were also convinced that adoption of a  \textit{strategy-proof} mechanism will increase both the \textit{transparency} of 
the system and parental satisfaction. 

At the end of the meeting, officials at BPS decided to form a Student Assignment Task Force to evaluate 
the city's assignment process including our proposed reform. The task force was formed in December 2003, 
and submitted its report and recommendations in September 2004 \citep{BPS-taskforce:2004}.

In order to build a case for reform to the Boston School Committee and other constituents, officials at BPS provided us with student preference data from 
earlier years under the Boston mechanism and requested an empirical analysis of the extent of strategic behavior. 
This analysis corroborated the presence of strategic behavior under the Boston mechanism
and was used by the city to make the case for reform \citep{aprs:06}.\footnote{Several subsequent studies, 
including \cite{pais/pinter:08, He:14, Burgess/Greaves/Vignoles/Deborah:15, 
Agarwal/Somaini:18, Dur/Hammond/Morrill:18, calsamiglia/guell:18, Calsamiglia/Fu/Guell:20, Kapor/Neilson/Zimmerman:20}, 
further validated the prevalence of strategic behavior under the Boston mechanism and its variants.}  
Although the task force recommended adoption of SC--TTC rather than DA, the city ultimately chose to adopt DA.

\subsubsection{2005 School Choice Reform at BPS} \label{sec:BPSreform}

\begin{figure}[!tp]
    \begin{center}
       \includegraphics[scale=1.2]{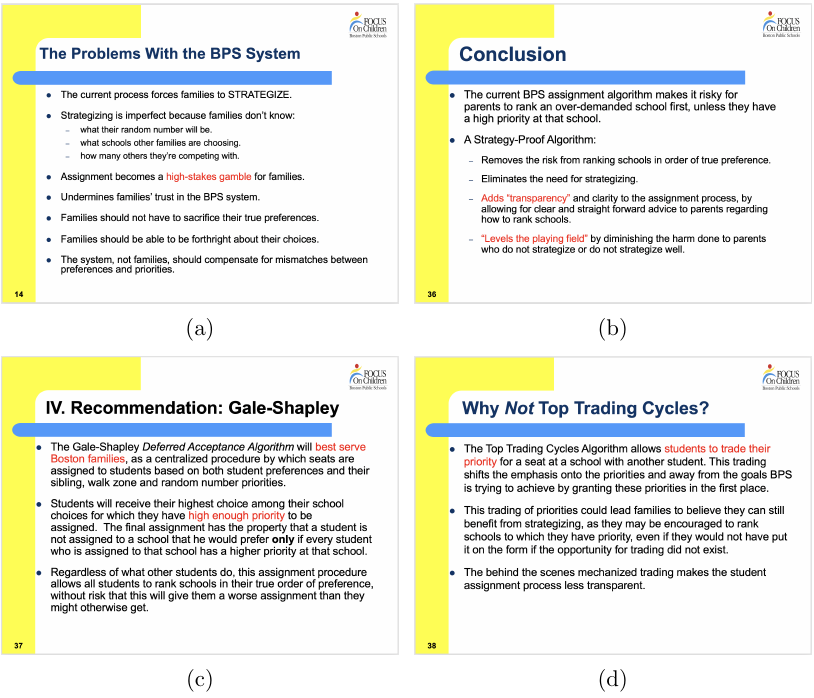}
           \end{center}
  \caption{Selected pages from the Boston Public Schools presentation, 
  \emph{Recommendation to Implement a New BPS Assignment Algorithm}, 
  presented to the Boston School Committee (May 2005). 
  Panels show (a) page 14, (b) page 36, (c) page 37, and (d) page 38.}
  \label{fig:BPS-4panel}
\end{figure}

During the public School Committee meeting in May 2005, Boston Public Schools (BPS) officials presented their rationale for their recommendation to abandon the Boston mechanism. 
This decision was succinctly outlined as follows:\footnote{All bullet-point lists reproduced in this section 
are taken verbatim from the official 2005 Boston Public Schools presentation to the Boston School Committee 
(see Figure \ref{fig:BPS-4panel}, Panels (a)--(d)).}

\begin{quote}
\begin{itemize}
\item The current process forces families to strategize.
\item Strategizing is imperfect because families don’t know:
\begin{itemize}
\item what their random number will be.
\item what schools other families are choosing.
\item how many others they’re competing with.
\end{itemize}
\item Assignment becomes a high-stakes gamble for families.
\item Undermines families’ trust in the BPS system.
\item Families should not have to sacrifice their true preferences.
\item Families should be able to be forthright about their choices.
\item The system, not families, should compensate for mismatches between preferences and priorities.
\end{itemize}
\end{quote}

Consequently, the pivotal factor leading to the abandonment of the Boston mechanism was its susceptibility to strategic manipulation, 
and the far-reaching consequences of this vulnerability played a decisive role in shaping this course of action.
The importance of \textit{strategy-proofness} in their recommendation was further emphasized as follows:

\begin{quote}
\begin{itemize}
\item The current BPS assignment algorithm makes it risky for parents to rank an over-demanded school first, 
unless they have a high priority at that school.
\item A Strategy-Proof Algorithm:
\begin{itemize}
\item Removes the risk from ranking schools in order of true preference.
\item Eliminates the need for strategizing.
\item Adds ``transparency'' and clarity to the assignment process, by allowing for clear and 
straight forward advice to parents regarding how to rank schools.
\item ``Levels the playing field'' by diminishing the harm done to parents who do not strategize or do not strategize well.
\end{itemize}
\end{itemize} 
\end{quote}

Officials at BPS recommended adopting a \textit{strategy-proof} mechanism for several reasons, including the \textit{transparency} it offered by enabling straightforward advice to parents 
and its role in leveling the playing field for underprivileged parents who might struggle to strategize effectively. This latter point was formalized in \cite{pathak/sonmez:08} and is 
discussed further in Section \ref{sec:leveling-the-playingfield} in the context of behavioral market design.

The appeal of adopting a \textit{strategy-proof} mechanism positioned DA and SC--TTC as natural replacements for the Boston mechanism. 
The officials at BPS explained their recommendation of DA as follows:

\begin{quote}
\begin{itemize}
\item The Gale-Shapley Deferred Acceptance Algorithm will best serve Boston families, 
as a centralized procedure by which seats are assigned to students based on both student preferences and their sibling, 
walk zone and random number priorities.
\item Students will receive their highest choice among their school choices for which they have high enough priority to be assigned.  
The final assignment has the property that a student is not assigned to a school that he would prefer 
only if every student who is assigned to that school has a higher priority at that school.
\item Regardless of what other students do, this assignment procedure allows all students to rank schools 
in their true order of preference, without risk that this will give them a worse assignment than they might otherwise get.
\end{itemize}
\end{quote}

Hence, the desiderata of \textit{strategy-proofness}  and \textit{no justified envy} were the primary reasons behind the BPS recommendation for DA.

Finally, officials at BPS explained their reluctance to recommend SC--TTC, despite its \textit{strategy-proofness} and \textit{Pareto efficiency}, as follows:

\begin{quote}
\begin{itemize}
\item The Top Trading Cycles Algorithm allows students to trade their priority for a seat at a school with another student. 
This trading shifts the emphasis onto the priorities and away from the goals BPS is trying to achieve by granting these priorities in the first place.
\item This trading of priorities could lead families to believe they can still benefit from strategizing, 
as they may be encouraged to rank schools to which they have priority, 
even if they would not have put it on the form if the opportunity for trading did not exist.
\item The behind the scenes mechanized trading makes the student assignment process less transparent.
\end{itemize}
\end{quote}

Even when priorities were not determined by standardized exams, officials at BPS were uncomfortable with the idea of trading priorities.\footnote{More specifically, 
BPS officials were concerned about trading ``sibling'' priorities, which were intended to allow siblings to share the schooling experience at the same school. 
Leveraging sibling priorities to secure assignments in other schools was not considered acceptable.
In contrast, they were less concerned about trading ``neighborhood'' priorities. See \cite{Rodriguez-Alvarez/Romero-Medina:2024} 
for an analysis of settings where some priority-awarding student characteristics can be traded while others cannot.}
As a result, the failure to satisfy \textit{no justified envy} was seen as a drawback of this mechanism. Additionally, officials were concerned that, while SC--TTC is \textit{strategy-proof}, 
this feature might not be clear to families, potentially leading them to engage in strategic behavior.

Officials also expressed concern that the mechanics of SC--TTC were less transparent than those of DA, which are effectively demonstrated through cutoff scores, as established in Theorem \ref{thm:cutoff}.

In June 2005, the Boston School Committee voted to replace the Boston mechanism with DA. The city implemented DA starting in the next school year and has used it for the allocation of K--12 public school seats ever since. 
The critical role played by various axioms in both the successful policy efforts in Boston and the earlier unsuccessful ones in Turkey was instrumental in the initial development of minimalist market design.

\subsection{The 2003 NYC High School Reform: Persuasion Takes a Backseat}  

The 2005 reform at BPS was one of two high-profile school choice initiatives led by economic designers in the mid-2000s in the U.S.
The other was the 2003 New York City (NYC) reform of the flawed high school admissions system \citep{apr:05}.
Both systems had suffered from strategic vulnerabilities, and in each case the reform introduced a variant of DA.
Despite these parallels, the processes that produced the two reforms diverged in important ways.  
Boston’s reform was triggered by academic recommendations, whereas the NYC reform originated within the Department of Education. 
 
In 2002, the New York City Department of Education (NYCDOE) faced three major issues in assigning nearly 100,000 students to more than 500 programs \citep{apr:05}:  
\begin{enumerate}  
\item Approximately 30\% of students were assigned to a school not included on their submitted preference lists.  
\item The system was vulnerable to preference manipulation, similar to the Boston mechanism.  
\item Some schools concealed capacity from the central administration to reserve seats for allocation outside the system.\footnote{These phenomena were studied earlier in a two-sided matching setting in \cite{sonmez:97, sonmez:99}.}  
\end{enumerate}  

In response to these failures, the NYCDOE Director of Strategic Planning reached out to Alvin Roth in May 2003 to explore whether the matching process used in the 
U.S. medical residency match could be adapted to high school admissions. Similar to the DA mechanism advocated by \cite{balinski/sonmez:99} 
for student placement and by \cite{abdulkadiroglu/sonmez:03} for school choice, 
the \textit{Roth and Peranson design}---developed for the medical residency match (see Section \ref{sec:Edison-quadrant})---is also based on Gale and Shapley’s celebrated \textit{individual-proposing deferred acceptance algorithm}.  

Given the extent of the crisis in NYC and decision-makers’ initial inclination toward DA, reinforced by Roth’s advocacy, 
the city adopted a version of the mechanism in October 2003 on a fast track---just a few months after \cite{abdulkadiroglu/sonmez:03} appeared in print---for students starting high school in Fall 2004.  
As in Boston, officials emphasized their desire to ``limit the need to game the system,'' as highlighted by Peter Kerr, Director of Communications for the NYCDOE, 
in a \textit{New York Times} op-ed \citep{Kerr:2003}:\footnote{While the \textit{strategy-proofness} of DA was used to justify the reform, 
NYC capped the number of schools students could rank at 12, leaving the system vulnerable to strategic behavior. For experimental analysis of such caps, see \cite{Calsamiglia/Haeringer/Klijn:2010}.}  

\begin{quote} 
``For more than a generation, parents and students have been unhappy with the admissions process to New York City high schools. 
The new process is a vast improvement, as it provides greater choice, equity, and efficiency. For example, for the first time, students will be able to list preferences as true preferences, limiting the need to game the system.  

This means that students will be able to rank schools without the risk that naming a competitive school as their first choice will adversely affect their ability to get into a school they rank lower."  
\end{quote}  

The critical difference between the school choice reforms in Boston and New York City lies in how they were initiated and carried out.
Boston’s reform was driven directly by academic advocacy and followed a thorough due diligence process,
whereas NYC decision-makers, already inclined toward DA, commissioned expert guidance to swiftly resolve an ongoing crisis, bypassing such vetting in the process.
As such, the NYC case represents both a ``commissioned'' reform and an instance of external validity for the theoretical framework in \cite{abdulkadiroglu/sonmez:03}, given the parallels between the NYC reform and this work.
Because swift execution was essential, many stages of the institutional redesign process outlined in Section \ref{sec:execution} were shortened, handled during implementation, or bypassed altogether.

Beyond helping end a major crisis and delivering tangible benefits for students, the NYC high school admissions reform made two key contributions to the market design literature: 
first, it increased the visibility of school choice as a successful application of the field in the early phases of
market design; second, it catalyzed a rich follow-up empirical literature (e.g., \citealp{abdulkadiroglu/angrist/pathak:14, abdulkadiroglu/agarwal/pathak:17}).
Together, the commissioned reform in NYC and the research-driven BPS school choice reform highlight two distinct but complementary pathways through which market design research can help shape policy.

While the NYC case validated earlier market design research in school choice, it did so without the persuasion strategies that had been central in Boston or any minimalism apparent in the reform, 
as much of the abandoned mechanism’s design was not publicly disclosed.
Lacking a clear link to two of the framework’s three pillars, it offered little sense of the broader potential of minimalist market design at that stage.
This potential became much clearer in the policymaker-driven reforms that followed in England and Chicago, where officials independently identified failures similar to those in Boston, 
traced them to the same root causes, and arrived at parallel resolutions---notably without any direct involvement from economic designers.

\subsection{External Validity for Minimalist Market Design: School Choice Reforms in England and Chicago} \label{sec:Chicago-England}

In Section \ref{sec:policyimpact-externalvalidity}, we defined external validity as policy relevance that arises when field implementation aligns with earlier research, 
but importantly, when that alignment occurs independently as policymakers and other stakeholders converge on similar approaches and conclusions.

To illustrate, take the BPS school choice reform discussed in Section \ref{sec:Boston}. A central trigger was vulnerability to preference manipulation: 
Boston officials abandoned the Boston mechanism after recognizing its manipulability and adopted a \textit{strategy-proof} alternative, as recommended in \cite{abdulkadiroglu/sonmez:03}. 
In his memo to the School Committee on May 25, 2005, Superintendent Thomas Payzant wrote \citep{aprs:06}:

\begin{quote} 
``The most compelling argument for moving to a new algorithm is to enable families to list their true choices of schools without jeopardizing their chances of being assigned to any school by doing so. 
[...] A strategy-proof algorithm levels the playing field by diminishing the harm done to parents who do not strategize or do not strategize well.'' 
\end{quote}

To my knowledge, this was the first time \textit{strategy-proofness}---a technical axiom from mechanism design---had been invoked by a policymaker in public deliberations to advocate a position. 
Moreover, Payzant framed \textit{strategy-proofness} not in its usual mechanism-design role as an incentive-compatibility constraint, but as an equity principle.\footnote{In economic theory,
\textit{strategy-proofness} is considered a special case of \textit{dominant-strategy incentive compatibility} (DSIC), where private information typically takes the form of preferences over outcomes.
As far as formal representation through mathematical relations is concerned, the two notions are indeed isomorphic.
However, they have a subtle but very important distinction.

DSIC is a concept in \textit{positive} economics.
In an effort to overcome challenges due to private information, it was formulated by \citet{Hurwicz:1960, hurwicz:72, hurwicz:73} as a \textit{constraint} in his mechanism design framework.
Thus, it mattered only to the extent that incentive considerations affected the outcome of an interaction between self-interested and rational economic agents.
In particular, if the same optimum could be reached through dishonest revelation of private information, this framework would see no use for the constraint.
It is a means to an end, not an end in itself.

In contrast, \textit{strategy-proofness} is both a \textit{positive} and a \textit{normative} concept.
It is usually a key design objective for policymakers, serving as an end in itself.}
This represents direct policy impact, as Payzant’s framing drew explicitly on \cite{abdulkadiroglu/sonmez:03}.

By contrast, when nearly identical reforms unfolded a few years later in England and Chicago, this time without guidance from economic designers, they reflected a different form of policy relevance.
As we wrote in \cite{pathak/sonmez:13}:

\begin{quote}
``The Boston episode challenges a paradigm in traditional mechanism design that
treats incentive compatibility only as a constraint and not as a direct design objective,
at least for the specific context of school choice. Given economists' advocacy
efforts, one might think that this incident is isolated, and the Boston events do not
adequately represent the desirability of nonconsequentialist objectives as design
goals. To demonstrate otherwise, we provide further, and perhaps more striking,
evidence that excessive vulnerability to `gaming' is considered highly undesirable
in the context of school choice. Officials in England and Chicago have taken drastic
measures to attempt to reduce it, and remarkably the Boston mechanism plays a
central role in both incidents.'' (p. 81)
\end{quote}

This notable parallel lends external validity to both the policy guidance of \cite{abdulkadiroglu/sonmez:03} and its interpretation of \textit{strategy-proofness} as a key policy objective. 

More importantly---and central to this monograph---it made clear that something larger was at play.

Beyond correcting flaws in original institutional designs, a central aim of minimalist market design is to recover their ``intended'' form---often by carving away incidental yet detrimental features that have accumulated over time. 
This makes the notion of external validity especially salient for research that follows this framework. 
Consequently, if decision-makers later (1) identify the same types of failures within their own institutions, (2) recognize these failures as stemming from the same underlying causes, 
and (3) independently devise resolutions that parallel those found in earlier research---effectively following the minimalist playbook on their own---this constitutes a particularly clear instance of external validity.

It was in observing precisely such instances---most strikingly in policymaker-driven reforms in England and Chicago---that the conceptual scope of this approach came into sharper focus. 
These reforms served as an intellectual awakening, 
revealing that what I had pursued largely by instinct was in fact part of a broader, coherent research and policy framework.
With this heightened awareness, I began applying the minimalist framework more deliberately and systematically from the early 2010s---a shift catalyzed by the two cases in \cite{pathak/sonmez:13}, discussed next.

\subsubsection{2007 Ban of the Boston Mechanism in England} \label{sec:England}

In England, the nationwide 2003 School Admissions Code mandated that local authorities, an operating body similar to a U.S. school district, coordinate student admissions through a centralized mechanism. 
With the 2003 code, two classes of mechanisms were recommended. 

The first of these two classes was DA and its ``capped'' versions, where the maximum number of schools that can be ranked in student preferences is limited to a fixed number. 
The second was a generalization of the Boston mechanism---called a \textit{first preference first system}---and its capped versions. 
However, four years later, all versions of the first preference first system, including the Boston mechanism, were banned nationwide under the \textit{2007 School Admissions Code}. 
As a result, all local authorities in England were using variants of DA by 2007.

The official justification for banning the first preference first system in England echoed the reasoning provided by BPS officials two years earlier when recommending discarding the Boston mechanism. 
According to officials at England’s Department for Education and Skills, ``the 'first preference first' criterion made the system unnecessarily complex for parents'' (School Code 2007, Foreword, p. 7). 
Additionally, Education Secretary Alan Johnson remarked that the first preference first system ``forces many parents to play an 'admissions game' with their children's future.''

Thus, much like in the 2005 reform at BPS, the Boston mechanism and its first preference first variant were abandoned in England due to their vulnerability to preference manipulation.
Rather than being adopted solely on its own merits, DA emerged as the primary mechanism largely because of the critical weaknesses of its main competitor---failures serious enough to warrant changes to the School Admissions Code.
The parallel is striking: the same flawed mechanism, the same root cause, and the same minimalist resolution.

This policymaker-driven reform provides a clear case of external validity for the research in \cite{abdulkadiroglu/sonmez:03} and 
demonstrates the power of minimalist market design to anticipate institutional evolution and shape enduring, policy-relevant reforms.

\subsubsection{2009 School Choice Reform at Chicago Public Schools} \label{sec:Chicago}

A similar reform to those in BPS and England also took place in Chicago in 2009, albeit in a more abrupt manner. Midway through the allocation process,
the city abandoned a capped version of the Boston mechanism that had been used for allocating seats at selective high schools. 
A few months after preferences were submitted under the capped Boston mechanism---but before placements were announced---officials at Chicago Public Schools (CPS) 
asked students to resubmit their preferences under a capped version of DA. This incident was reported in a \textit{Chicago Sun-Times} story \citep{Rossi:2009}:

\begin{quote}
``Poring over data about eighth-graders who applied to the city’s elite
college preps, Chicago Public Schools officials discovered an alarming
pattern. 

High-scoring kids were being rejected simply because of the order in
which they listed their college prep preferences.

`I couldn`t believe it,' schools CEO Ron Huberman said. `It’s terrible.'

CPS officials said Wednesday they have decided to let any eighth-grader
who applied to a college prep for fall 2010 admission re-rank their preferences
to better conform with a new selection system.''
\end{quote}

In addition to vulnerability to preference manipulation, the mechanism’s failure to satisfy \textit{no justified envy}---evident from Huberman’s outrage---was a major concern in Chicago. 
As in Boston and England, the same failure and the same root cause again led to the same solution: replacement of a capped version of the Boston mechanism with a variant of DA.
These failures were deemed so unacceptable that officials did not wait for the next academic year, instead altering student assignments mid-process.

This episode provides yet another compelling example of external validity for the research in \cite{abdulkadiroglu/sonmez:03} and underscores the broader potential of minimalist market design to generate policy-relevant insights.

\subsection{Broader Impact in Education Market Design} \label{sec:broader-sc}

\cite{balinski/sonmez:99} and \cite{abdulkadiroglu/sonmez:03}, later reinforced by school choice reforms in New York City and Boston in the mid-2000s, 
marked the beginning of one of the most active and successful subfields of economic design: \textit{education market design}.\footnote{See \cite{Pathak:2017} 
and \cite{abdulkadiroglu/andersson:23} for excellent overviews of this literature.} Since then, numerous school districts in the U.S. have adopted the DA mechanism, 
including Chicago Public Schools, Denver Public Schools, Indianapolis Public Schools, Newark Public Schools, the Recovery School District in New Orleans, 
Tulsa Public Schools, and Washington D.C. Public Schools \citep{Pathak:2017, abdulkadiroglu/andersson:23}. 
All these reforms have been shaped by economic designers. 
To my knowledge, no jurisdiction reported using the DA mechanism in the U.S. or elsewhere prior to the publication of this literature.\footnote{See Section \ref{sec:Seattle} for 
Seattle’s unusual experience with the DA mechanism. 
Although the city replaced the Boston mechanism with the DA for its public schools in 1999, after \cite{balinski/sonmez:99} but before \cite{abdulkadiroglu/sonmez:03} was published, 
it did not adequately communicate the reform to the public. It was not until 2008 that it became widely known that Seattle had been using the DA mechanism for nearly a decade. 
Once this information surfaced, the city reverted to the Boston mechanism within a few years.}

The influence of education market design extends well beyond the U.S. By 2024, versions of the DA mechanism have been adopted in at least some jurisdictions for primary education in 17 countries, 
secondary education in 21 countries, and higher education in 32 countries \citep{Neilson:24}. These countries include Chile, England, France, Hungary, India, Spain, and Sweden 
\citep{pathak/sonmez:13, Baswana/Chakrabarti/Chandran/Kanoria/Patange:19, Fack/Grenet/He:19, Andersson/etal:24}.

As apparent from these figures, DA has emerged as the most widely used mechanism for higher education globally over the last two decades, with the Boston mechanism a distant second, being used in 11 countries. 
In contrast, for primary and secondary education, the Boston mechanism remains the most common, with 50 and 47 countries relying on it in at least some of their jurisdictions for primary and secondary education, 
respectively \citep{Neilson:24}.

One possible explanation for why the DA mechanism and its variants have become the clear choice in higher education but not in primary or secondary education lies in 
differing attitudes toward the importance of the \textit{no justified envy} axiom in these settings. As noted earlier in this section, \cite{balinski/sonmez:99} and \cite{abdulkadiroglu/sonmez:03} 
argue that \textit{no justified envy} is more likely to be regarded as indispensable in settings where priorities are determined through centralized exams. If this hypothesis holds, 
it could explain why variants of DA are more prevalent in higher education, where reliance on performance metrics such as centralized exams is more common globally than in primary or secondary education.\footnote{See 
the \cite{Rossi:2009} quote in Section \ref{sec:Chicago} regarding the school choice crisis in Chicago Public Schools for evidence that lends support to this hypothesis.} 

As we have emphasized throughout this section, the vulnerability of the Boston mechanism to preference manipulation has been a key factor in its replacement by variants of the DA mechanism in many jurisdictions. 
Along these lines, but as a partial compromise, numerous jurisdictions have also adopted hybrid mechanisms over the last two decades that combine features of the 
DA and Boston mechanisms to reduce vulnerability to preference manipulation.
Among these, perhaps the most prominent is the \textit{parallel mechanism}, first introduced in China in 2001 in 
Hunan province for military academies, later adopted by Jiangsu in 2005 and Zhejiang in 2007, and eventually implemented by 
28 out of 31 provinces by 2012 \citep{chen/kesten:17}.\footnote{See Section \ref{sec:Taiwan}
for a closely related \textit{Taiwan Deduction} mechanism.}

Versions of the Boston mechanism had been used for college assignments in several Chinese provinces since the 1980s, where baseline priorities depend on 
scores from centralized exams. As with the parental concerns about the vulnerability of the Boston mechanism to preference manipulation in several U.S. school districts, discussed in Section \ref{sec:Boston-mechanism},
Chinese parents also expressed similar worries (\citealp{chen/kesten:17}, p. 101):\footnote{\cite{chen/kesten:17} credit the original source of these parent quotes to \cite{Nie:07}.}

\begin{quote} 
``A good score in the college entrance exam is worth less than a good strategy in the ranking of colleges.'' 
\end{quote}

Another parent lamented:

\begin{quote} 
``My child has been among the best students in his school and school district. He achieved a score of 632 in the college entrance exam last year. 
Unfortunately, he was not accepted by his first choice. After his first choice rejected him, his second and third choices were already full. My child had no choice but to repeat his senior year." 
\end{quote}

To address these concerns, most Chinese provinces adopted the parallel mechanism, introduced by Zhenyi Wu, 
the director of undergraduate admissions at Tsinghua University from 1999 to 2002 \citep{chen/kesten:17}.

Under this mechanism, student preferences are divided into several bands. For example, the first band may contain three schools, the second band five schools, and the third band eight schools. 
Given a preference profile, each college $c$ prioritizes students first based on the band in which they rank college $c$ in their preferences, and then based on their exam scores. 
The outcome of the parallel mechanism is obtained by running the individual-proposing deferred acceptance algorithm with the submitted preferences and adjusted priorities.

When there is a single band, the parallel mechanism reduces to the DA mechanism (or a capped version), and when each band contains only one school, 
it reduces to the Boston mechanism (or a capped version), illustrating how the parallel mechanism functions as a hybrid of these two designs.


\section{Kidney Exchange} \label{sec:KE}

In the mid-2000s, through a series of papers, a team of economic designers consisting of Alvin Roth, Utku \"{U}nver, and me laid the theoretical foundation for organized kidney exchange 
\citep{roth/sonmez/unver:04, roth/sonmez/unver:05, roth/sonmez/unver:07}. Leveraging formal techniques from market design and optimization, kidney exchange rapidly transformed 
living donor kidney donation in numerous countries, saving thousands of lives annually within just a few years \citep{Purtill:2018, Rose:2019}. 
As an unexpected yet transformative application of market design, these developments played a pivotal role in elevating the field's practical relevance and visibility during this period.

Recognition for these efforts reached a milestone in 2012 when our co-author, Alvin Roth, was honored with the \textit{Nobel Memorial Prize in Economic Sciences}, jointly with Lloyd Shapley. 
Notably, kidney exchange was prominently featured in Torsten Persson's award ceremony speech as a committee member \citep{Persson:12}.

But how did a team of economists manage to develop the tools and establish the infrastructure that now consistently touches so many lives? 
This section explores the pivotal role of key ideas in minimalist market design, illustrating their profound influence on the success and widespread adoption of organized kidney exchange clearinghouses worldwide. 
The policy and practical impacts of kidney exchange and school choice in the late 2000s---amplified by the external validity demonstrated by school choice reforms in England and Chicago in the early 2010s---in turn inspired 
my more systematic application of these ideas in subsequent market design efforts, ultimately leading to the conceptualization of minimalist market design as presented in this monograph.

\subsection{Paired Kidney Exchange and List Exchange}

End-Stage Renal Disease (ESRD) is a major global health problem.
As of 2025, more than 800,000 people in the U.S. alone---about 0.2\% of the population---are living with ESRD.
Kidney transplantation is the preferred treatment, but kidneys from deceased donors fall well short of the need, making living donors a crucial source of transplant kidneys.

Unfortunately, immunological barriers can make transplantation between a living donor and their intended recipient infeasible or medically suboptimal. 
One of the most consequential barriers is \textit{ABO blood group} incompatibility.

There are four main human blood types---A blood type, B blood type, AB blood type, and O blood type---which are defined by the presence or absence of two antigens on red blood cells, A and B.
Individuals with type O have neither antigen;\footnote{The term ``O'' comes from the German word \textit{Ohne}, meaning ``without.'' 
In many German-speaking regions, blood type O is still referred to as \textit{Blutgruppe 0} (“blood group zero”).} those with type A have only the A antigen, those with type B have only the B antigen, and those with type AB have both.

\begin{figure}[!tp]
    \begin{center}
       \includegraphics[scale=1.2]{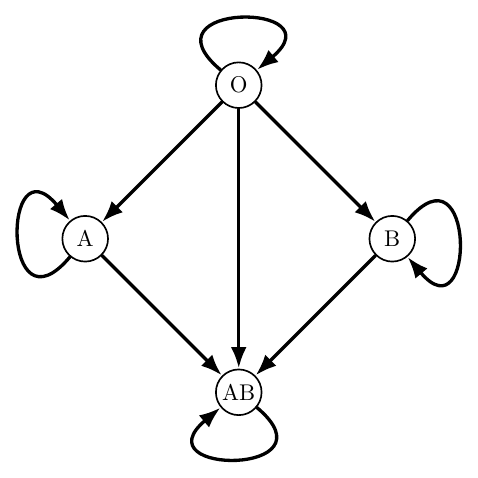}
           \end{center}
\caption{Directed graph representation of blood-type compatibility under the ABO grouping.}
\label{fig:ABO}
\end{figure}

People naturally produce antibodies against the antigens they lack. Those without the A antigen produce anti-A antibodies, and those without the B antigen produce anti-B antibodies. 
As a result, someone lacking the A antigen cannot safely receive blood or a solid organ from someone who has it, and likewise for the B antigen.

This creates a major biological asymmetry that puts type O patients at a clear disadvantage: because they lack both antigens, they can receive kidneys only from type O donors. 
More broadly, in the absence of other complications---most notably tissue-type incompatibility---a kidney from a type O donor can be transplanted into any patient; 
a kidney from a type A donor into patients with type A or AB; a kidney from a type B donor into patients with type B or AB; and a kidney from a type AB donor only into patients with type AB.\footnote{One exception, 
discouraged in much of the world, is \textit{ABO-incompatible kidney transplantation}, which uses intensive immunosuppression to overcome blood-type incompatibility. 
It has been adopted in a few countries such as Japan and South Korea, where cultural or historical factors have severely limited the availability of deceased-donor kidneys, 
but is generally not used in most other countries facing similar or even greater deceased-donor shortages, such as Turkey, where it is not covered by national health insurance. 
Because it often yields inferior outcomes compared to compatible transplants, it is typically avoided due to the extensive preparation it requires and its higher medical risks and costs.}
Figure~\ref{fig:ABO} illustrates these compatibility relations as a directed graph under the ABO grouping.

This structural disadvantage for patients with blood type O has long raised ethical and policy concerns within the transplant community, prompting efforts to develop allocation practices that could mitigate such inequities.

In 2000, representatives from several professional transplantation societies issued the \textit{Consensus Statement on the Live Organ Donor} \citep{PKE-consensus:2000},
aiming to provide standardized ethical, medical, and procedural guidelines for transplant physicians, primary care providers, and healthcare planners.
Central to this discussion, 
the statement discussed two types of donor exchange arrangements to overcome barriers to living donation---\textit{paired kidney exchange} 
and \textit{list exchange}---with strong enthusiasm for the first and caution advised on the second.\smallskip

\begin{figure}[!tp]
    \begin{center}
       \includegraphics[scale=1.0]{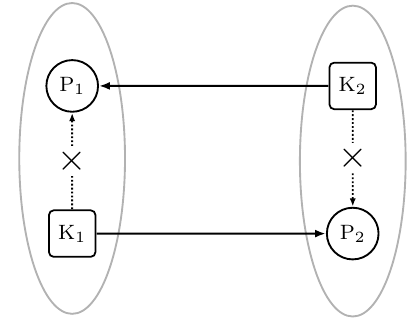}
           \end{center}
\caption{Paired kidney exchange. Patient P$_1$ is incompatible with the kidney K$_1$ from their own living donor but is compatible with the kidney K$_2$ from Patient P$_2$’s living donor. 
Similarly, Patient P$_2$ is incompatible with the kidney K$_2$ from their own living donor but is compatible with the kidney K$_1$ from Patient P$_1$’s living donor. 
By exchanging donors, both patients can receive compatible transplants.}
\label{fig:2-way-ke}
\end{figure}

\paragraph{Paired Kidney Exchange.} This innovation involves an exchange of donors between two kidney patients, each of whom is (typically) incompatible with their own donor but compatible with the other patient’s donor 
(see Figure \ref{fig:2-way-ke} for an illustration). It is essentially an in-kind gift exchange between two patient--donor pairs to overcome biological incompatibility.

Originally proposed by \cite{Rapaport:1986}---a transplant surgeon---this approach has been practiced in South Korea since the early 1990s \citep{park:99}.\footnote{South Korea was the 
first country in the world to implement kidney exchange, launching a program at Hanyang University Hospital in 1991 \citep{park:99}. 
Between 1991--2010, 152 kidney exchange transplants were performed at this center \citep{Kim/Kwon/Kang:2012}. 
Another leading center at Yonsei University reported 129 kidney exchange transplants between 1995--2006 \citep{Huh:2008}. 
However, with the introduction of desensitization protocols and ABO-incompatible living-donor transplantation, kidney exchange declined, 
and by the 2010s only a handful of operations were performed annually in the country.}
It gained traction in the U.S. in the early 2000s, though only a handful of paired kidney exchanges had been conducted before our involvement in 2004. 
The biggest challenge was the absence of an organized system or even a database to facilitate these exchanges. 
At the time, methods for discovering paired kidney exchanges were limited to rudimentary practices, such as discussions between patients and their donors in settings like dialysis rooms.\smallskip

\begin{figure}[!tp]
    \begin{center}
       \includegraphics[scale=1.0]{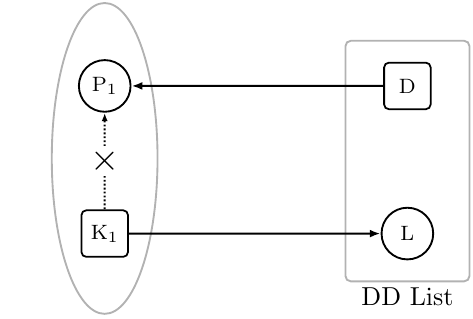}
           \end{center}
\caption{List exchange. Patient P$_1$ is incompatible with kidney K$_1$ from their living donor. 
The donor instead donates their kidney K$_1$ to the deceased-donor list, where it is allocated to the highest-priority patient L. 
In return, Patient P$_1$ receives priority on the deceased-donor list and is allocated a deceased-donor kidney D.}
 \label{fig:listexchange}
\end{figure}

\paragraph{List Exchange.} This innovation, also known as an \textit{indirect exchange}, involves a donor giving a kidney to the deceased-donor (DD) list in exchange 
for their incompatible patient receiving priority on the DD list (see Figure \ref{fig:listexchange} for an illustration).

In their paper introducing this approach, \cite{ross/woodle:2000} also highlighted a major ethical concern: its potential detrimental impact on blood-type O patients without living donors, 
whose options are limited to kidneys from the DD list. Involving only a single patient--donor pair, list exchange is much easier to organize than paired kidney exchange, 
but pairs with a blood-type O donor are far less likely to participate than those with a blood-type O patient.

This disparity is particularly concerning in the U.S., where blood type O is disproportionately common among certain minority groups. 
Therefore, while appealing from a \textit{utilitarian} perspective, list exchange raises significant concerns in terms of \textit{equity}. \medskip

Among these two approaches, paired kidney exchange garnered widespread approval in the \textit{Consensus Statement on the Live Organ Donor} 
\citep{PKE-consensus:2000}, being deemed “ethically acceptable.” In contrast, in line with the equity concerns highlighted in \cite{ross/woodle:2000}, 
the Consensus Statement underscored the ethical challenges associated with list exchange.

\subsection{Relation to  \cite{abdulkadiroglu/sonmez:99}} \label{sec:AS99-KE}

In the early 2000s, \"{U}nver and I were both faculty members at Ko\c{c} University in \.{I}stanbul. 
As discussed in Section \ref{sec:G--TTC}, my first project with him focused on the \textit{house allocation with existing tenants} model (see Sections \ref{sec:existingtenants}--\ref{sec:minimalist-YRMH--IGYT}). 
During the 2002--2003 academic year, while \"{U}nver was visiting Roth at Harvard University, we realized that this model had a direct application in healthcare---kidney exchange. 
Discovering this connection between our work and a healthcare application was exhilarating; never before had economic theory research shown such clear potential to save lives.
Motivated by this promise, the three of us soon began a joint research project on kidney exchange, aiming both to advance the emerging field of market design and to influence policy.

Yet pursuing this line of work was far from straightforward, and our initial enthusiasm was tempered by two immediate concerns.
The first was whether economists could undertake such a project independently, without the involvement of a transplantation expert.
The second, and more pressing, was whether meaningful formal analysis remained to be done in kidney exchange, 
given its close connection to \cite{abdulkadiroglu/sonmez:99} and the HA--TTC (or, equivalently, the YRMH--IGYT) mechanism introduced there as a solution.

After reviewing numerous papers on kidney transplantation, we became confident that economists could pursue this project on their own.
The second concern, however, proved more substantive, especially from a pure research perspective.

The parallel with \cite{abdulkadiroglu/sonmez:99}, 
forming the basis of the ``discovery--invention chain'' that links innovations in house allocation to kidney exchange, rests on the following observations.

As discussed in Section \ref{sec:houseTTC}, the HA--TTC mechanism allocates two types of houses to two types of individuals.
Existing tenants, one type of individual, have the option to retain their current houses if they so choose. The other type, the newcomers, have no claim to any specific house.
The two types of houses are the occupied houses earmarked for their existing tenants and the vacant houses, which are not tied to any specific individual.

This structure has a natural counterpart in kidney exchange, where patients and donor kidneys play roles analogous to tenants and houses:

\begin{itemize}
\item A patient with an incompatible (or less-than-ideal) living donor is analogous to an existing tenant seeking a more-preferred house than the one they currently occupy.
\item The kidney from a patient’s living donor is analogous to an occupied house (i.e., the tenant’s endowment).
\item A patient on the DD list with no living donor is analogous to a newcomer who does not currently occupy any house.
\item A DD kidney or a kidney from a non-directed living donor is analogous to a vacant house.\footnote{In \cite{roth/sonmez/unver:04}, this analogy with vacant houses was drawn only for DD kidneys. 
The same analogy was later extended to kidneys from non-directed donors in \cite{rsuds:06}, which also introduced and advocated non-simultaneous \textit{donor chains} initiated by non-directed (a.k.a.\ Good Samaritan) donors.}
\end{itemize}

Similar to an existing tenant who may retain their current house, a patient with a living donor may keep their own donor’s kidney if they so choose.
Just as a newcomer has no claim to any particular house, a patient on the DD list has no claim to any particular kidney.
Finally, just as a vacant house is not tied to any individual, a DD kidney is not tied to any specific patient.

\begin{figure}[!tp]
    \begin{center}
       \includegraphics[scale=1.2]{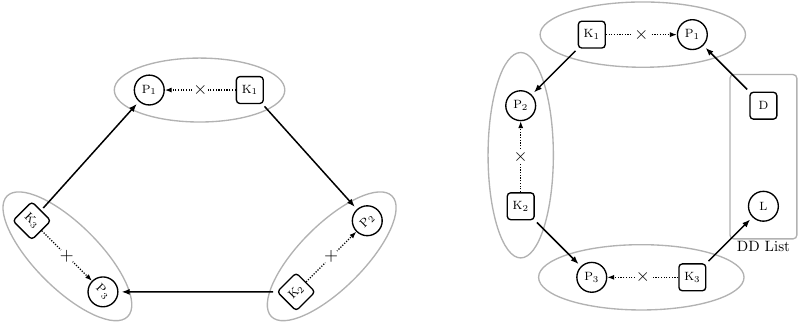}
           \end{center}
\caption{Kidney exchange cycle (left) and kidney exchange chain (right), each involving three patients with living donors. 
In the cycle, the donor of P$_1$ donates kidney K$_1$ to P$_2$, the donor of P$_2$ donates K$_2$ to P$_3$, and the donor of P$_3$ donates K$_3$ to P$_1$. 
In the chain, the donor of P$_1$ donates K$_1$ to P$_2$ in exchange for P$_1$ gaining priority on the deceased-donor list. 
This initiates a sequence: the donor of P$_2$ donates K$_2$ to P$_3$, the donor of P$_3$ donates K$_3$ to L (a patient on the deceased-donor list), 
and with this upgraded priority P$_1$ ultimately receives kidney D from the deceased-donor pool.}
\label{fig:combined-3cycle-listexchange}
\end{figure}

\begin{figure}[!tp]
    \begin{center}
       \includegraphics[scale=1.0]{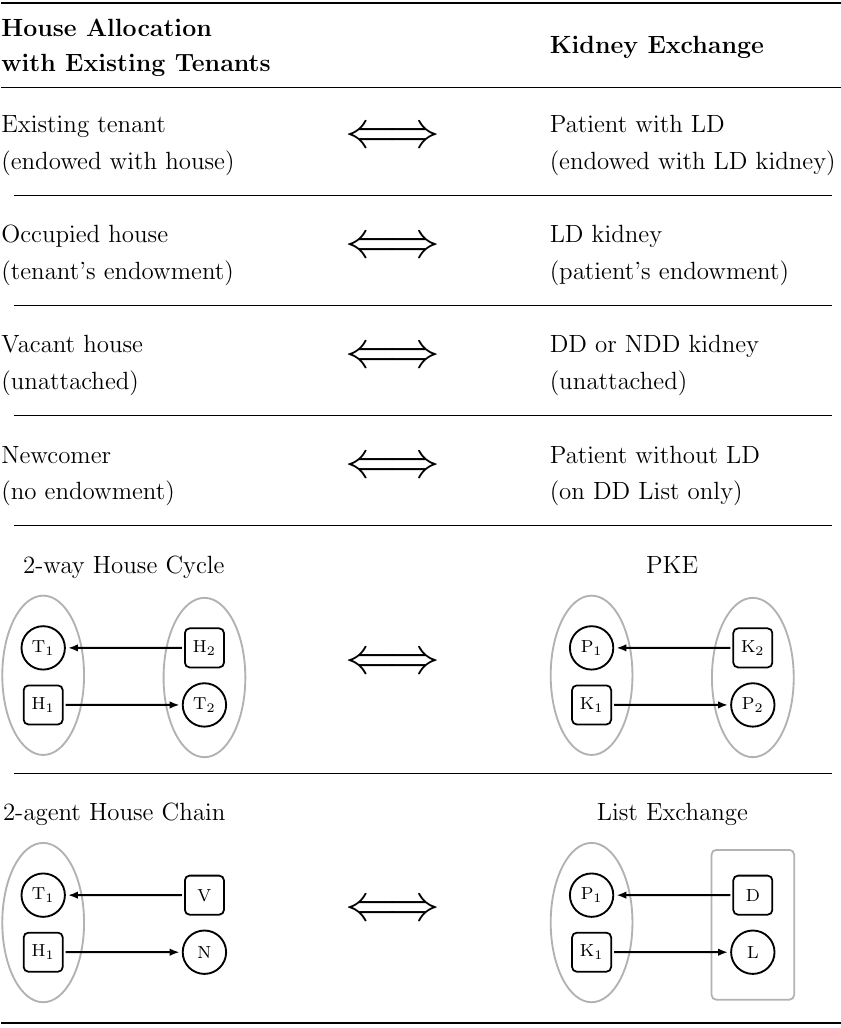}
           \end{center}
             \captionsetup{skip=-4pt}  
\caption{Isomorphism between the house allocation with existing tenants model and kidney exchange. 
\textit{House allocation with existing tenants}: T$_1$ and T$_2$ represent existing tenants endowed with occupied houses H$_1$ and H$_2$, respectively. 
N represents a newcomer with no endowment, and V represents a vacant house unattached to any individual. 
\textit{Kidney exchange}: P$_1$ and P$_2$ represent patients endowed with kidneys K$_1$ and K$_2$, respectively, from their living donors. 
L represents a patient who has no living donor, is on the deceased-donor list only, and has the highest priority for kidney K$_1$ when it is assigned through the deceased-donor list. 
D represents the deceased-donor (or non-directed donor) kidney that is assigned to patient P$_1$ in exchange for kidney K$_1$ from P$_1$’s living donor, which is donated to the deceased-donor list.
Abbreviations: LD: Living donor; DD: Deceased donor; NDD: Non-directed donor; PKE: Paired Kidney Exchange.}
 \label{fig:HA-KE-isomorphism}
\end{figure}

Drawing on these analogies, the HA--TTC mechanism finds a direct application in kidney exchange. 
Figure~\ref{fig:HA-KE-isomorphism} illustrates the isomorphism between the two settings. 
By managing claims to ``unattached'' house---whether vacant or vacated during the procedure---through an exogenous priority list, the HA--TTC mechanism 
organizes two distinct types of transactions: \textbf{\textit{cycles}} and \textbf{\textit{chains}}. 
Figure~\ref{fig:combined-3cycle-listexchange} provides an example of both.\footnote{Strictly speaking, by employing a technical construct that treats individuals and houses as separate entities, 
the HA--TTC algorithm transforms each chain of individuals in YRMH--IGYT into a cycle of individuals and houses. 
To distinguish between the two types of cycles in HA--TTC, we adopt the terminology from its YRMH--IGYT equivalent, using chain nomenclature for the latter.}

In a cycle, individuals with endowments (e.g., existing tenants or patients with living donors) exchange them among themselves. 
In a chain, an individual gives up priority for an unattached object to receive another’s endowment. 
That person then either ends the chain by taking an unattached object or passes their endowment to another individual, allowing the chain to continue. 
The process repeats until the final individual, whose endowment was taken by the previous participant, receives an unattached object. 
Paired kidney exchange corresponds to a cycle involving two patients with living donors, 
while list exchange corresponds to a chain involving one patient with a living donor and the highest-priority patient on the DD list.

\subsection{First Kidney Exchange Model in \cite*{roth/sonmez/unver:04}}

Though kidney exchange closely mirrors house allocation with existing tenants, we saw clear value in pursuing this project. 
Our optimism stemmed both from the prospect of demonstrating substantial welfare gains through computational simulations of how many additional transplants an organized system could enable 
and from the ethical concerns surrounding list exchanges and the possibility of modifying the HA--TTC mechanism to address them. 
Guided by the principles emphasized in the Consensus Statement, we began exploring how such a modification could better align with the ethical priorities of the transplantation community.

\paragraph{Top Trading Cycles and Chains (TTCC) Mechanism.}
While the use of an exogenous priority list, as in HA--TTC, remained a viable way to regulate chains in kidney exchange, we found that alternative chain selection rules could better serve the community. 
We therefore developed a new mechanism, TTCC.\footnote{See Table \ref{tab:ttc-comparison} for a comparison of TTCC with its antecedents---G--TTC, HA--TTC, and SC--TTC---in terms of their property-rights structures, 
the types of trades they generate, and the tools they employ.}


\begin{table}[!tp]
  \centering
  \begin{threeparttable}
  \caption{\normalsize{Comparison of Top Trading Cycles (TTC) variants: 
  G--TTC in housing markets, HA--TTC in house allocation with existing tenants, SC--TTC in school choice, and TTCC in kidney exchange.
  Organized by property-rights structures and the induced trade types.}}
  \label{tab:ttc-comparison}

  \footnotesize
  \setlength{\tabcolsep}{4pt}
  \begin{tabularx}{\textwidth}{L{3cm} Y{0.9} Y{1.1} Y{1} Y{1}}
    \toprule
    & \normalsize{\textbf{G--TTC}}
    & \normalsize{\textbf{HA--TTC}}
    & \normalsize{\textbf{SC--TTC}}
    & \normalsize{\textbf{TTCC}} \\
    \midrule
    \normalsize{Model}
    & Housing markets
    & House allocation w/ existing tenants
    & School choice
    & Kidney exchange \\
    \addlinespace[4pt]\cmidrule(lr){1-5}
    \normalsize{Object types}
    & House
    & OH \& AH
    & School seat
    & LD \& DD \\
    \addlinespace[4pt]\cmidrule(lr){1-5}
    \normalsize{Property rights}
    & Pvt.\ ownership
    & \begin{tabular}[t]{@{}l@{}}Pvt.\ ownership (OH)\\Priority list (AH)\end{tabular}
    & Priority list
    & \begin{tabular}[t]{@{}l@{}}Pvt.\ ownership (LD)\\Priority list (DD)\end{tabular} \\
    \addlinespace[4pt]\cmidrule(lr){1-5}
    \normalsize{Trade types}
    & House $\rightarrow$ house
    & \begin{tabular}[t]{@{}l@{}}OH $\rightarrow$ OH\\OH $\rightarrow$ AH\text{-priority}\end{tabular}
    & Priority $\rightarrow$ priority
    & \begin{tabular}[t]{@{}l@{}}LD $\rightarrow$ LD\\LD $\rightarrow$ DD\text{-priority}\end{tabular} \\
    \bottomrule
  \end{tabularx}

  \begin{tablenotes}\footnotesize
    \item Abbreviations: Pvt.: Private; OH: Occupied House; AH: Available House; LD: Living-Donor; DD: Deceased-Donor.
  \end{tablenotes}

  \end{threeparttable}
\end{table}


Building on HA--TTC,  the TTCC mechanism allows for larger cycles and chains than those in paired kidney exchange or list exchange, 
offering important efficiency gains (see Figure \ref{fig:combined-3cycle-listexchange} for a depiction of a cycle and a chain, each involving three patients with living donors). 
Its flexibility in chain selection provides a valuable policy lever: it allows welfare gains from chains to be incorporated without undermining the balance of 
blood-type O patients on the DD list. In this way, it could mitigate---or even eliminate---the adverse effects of list exchanges on these patients.

Thus, rather than treating kidney exchange as a direct application of the HA--TTC mechanism from \cite{abdulkadiroglu/sonmez:99}, 
TTCC leverages this flexibility to advance both efficiency and equity.
The more we learned from the kidney exchange and living-donor liver transplantation literatures, 
the more confident we became that TTCC offers significant practical and theoretical value.

This theoretical advance also suggested a practical path forward.
At the time, existing kidney exchange programs in the early 2000s suffered from three main root causes: (1) the absence of a comprehensive database containing living-donor information, 
(2) the lack of a systematic mechanism to organize exchanges, and 
(3) the restriction of both paired and list exchanges to their most basic forms, involving only the smallest possible number of patient--donor pairs (two for paired exchange and one for list exchange).
Through TTCC, conceived as a minimalist intervention, 
we showed that addressing these root causes could yield significant improvements in \textit{utilitarianism} and \textit{equity}---the two primary objectives of the transplantation community, 
highlighted in the Consensus Statement.

We circulated our analysis in \cite{roth/sonmez/unver:03} in September 2003 and, a year later, published a more streamlined version in \cite{roth/sonmez/unver:04}.\footnote{In terms of the fit between the 
real-world problem and its formulation in theoretical research, kidney exchange posed a far more challenging application than school choice. 
It involved many more details and considerations from diverse fields (e.g., transplantation, immunology, and medical ethics), and there was often no consensus on key aspects of the problem. 
For example, as thoroughly discussed in \cite{roth/sonmez/unver:04}, the literature was unclear on whether a reduction in HLA mismatch between patient and donor would improve graft survival. 
From the outset, therefore, it was clear that there could be no single best model for kidney exchange.} 
Positioned as the next innovation in a ``discovery---invention cycle''---building on earlier medical innovations such as paired kidney exchange and list exchange, 
and on the foundation laid by \cite{abdulkadiroglu/sonmez:99}---our work provided the transplantation community with both a methodology 
and a mechanism, TTCC, that embodied the principles emphasized in the Consensus Statement.  

As we would soon learn, the timing of our paper was perfect.

\subsection{Initial Phase of Kidney Exchange in New England} \label{sec:InitialPhase-NE-KE}

After receiving approval from the \textit{United Network for Organ Sharing (UNOS)} Board of Trustees in the fall of 2000, a kidney exchange program---the first of its kind in the U.S.---was established in New England 
(UNOS Region 1) in February 2001 \citep{Delmonico:2004}.\footnote{To address the nation’s severe organ donation shortage and improve organ matching, the U.S. Congress 
passed the \textit{National Organ Transplant Act (NOTA)} in 1984. 
The act established the \textit{Organ Procurement and Transplantation Network (OPTN)} to manage a national organ registry, which is operated by a private organization under a federal contract. Since 1986, UNOS 
has held this contract and continues to administer the OPTN.} 
Fourteen transplant centers participated in the program. Despite ethical concerns about list exchanges, this controversial form of kidney exchange was organized alongside the uncontroversial paired kidney exchanges. 
The decision was defended in \cite{Delmonico:2004} by a familiar principle:
\begin{quote}
``This exchange program has a clear utilitarian goal: to have more recipients undergo successful transplantation by expanding the pool of compatible live donors." (p. 1632)
\end{quote}
However, in acknowledgment of the concerns associated with this decision, much of the discussion in \cite{Delmonico:2004} 
revolved around the precautions taken in the region to mitigate the adverse impact of the program on blood type O patients on the DD list.

In its early stages, the program did not establish a unified database of willing living donors for kidney patients that was accessible to all transplant centers involved. 
Consequently, from 2001 to 2004, the program arranged only 5 paired kidney exchanges, providing kidney transplants to a total of 10 patients \citep{roth/sonmez/unver:05}.

Since it involved only a single patient--donor pair, arranging a list exchange might have been expected to be much easier. However, the eligibility criteria for this more controversial 
form of kidney exchange included ensuring that no paired kidney exchange---the preferred option between the two---was feasible with any other patient registered 
across all 14 transplant centers in New England. 
\begin{quote}
``The duration that the RTOC [Renal Transplant Oversight Committee] will wait for a live donor exchange pair to come forward from another center has not been regulated, 
although the general practice has been to ask such pairs to wait a minimum of one month, in order to avoid flooding the system with 'unnecessary' list exchanges. 
If no such pair is identified, the center can proceed with the live donor list exchange process."\\
\mbox{} \hfill  \citealp{Delmonico:2004}, p. 1629
\end{quote}
As a result, arranging list exchanges also involved operational challenges in New England. 
Nonetheless, 17 of these exchanges were carried out in New England between February 2001 and December 2003.

\subsection{Forging a Partnership with New England Medical Professionals} \label{sec:NEPKE}

Shortly after circulating \cite{roth/sonmez/unver:03}, our team secured a meeting with Dr. Francis Delmonico, the Chief Medical Officer at the New England Organ Bank. 
Fortunately for us, this meeting occurred at a time when members of the New England transplantation community were facing challenges with their kidney exchange 
program and were open to ideas for improvement. Naturally, members of the transplantation community had less experience organizing a system of exchanges than with the ethical and medical aspects of the resulting transplants.

During the meeting, Delmonico found our policy proposal intriguing but expressed three reservations about our approach. 
Addressing his first concern was straightforward. Given the substantial potential increase in the number of transplants offered by our proposal, 
Delmonico did not believe it was necessary to include the more controversial list exchanges in the system. However, his second and third reservations were more involved and required additional analysis.

The transplantation community's Consensus Statement recommended that all four operations in a paired kidney exchange---two for patients and two for donors---be conducted simultaneously \citep{PKE-consensus:2000}. 
This recommendation aimed to prevent a scenario in which a donor in a pair donates a kidney but faces a compromised situation if the patient in their pair fails to receive a transplant because the donor from the 
other pair is unable or unwilling to donate later. Due to the logistical challenges posed by this recommendation, Delmonico noted that, at the time, only 2-way exchanges were feasible in New England. 
Furthermore, he opposed the idea of patients expressing strict preferences among compatible kidneys, asserting that the only relevant information about patient preferences should be whether a kidney is compatible.

As far as modeling itself is concerned, 
formulating a new kidney exchange model to address Delmonico's reservations was relatively straightforward. However, the analysis of the resulting model presented a completely different story. 
At that time, none of us were familiar with the formal techniques required to analyze this new model.

This challenge could have posed a serious problem had basic theory not come to our rescue. 
Fortunately, just a few months earlier, economists Anna Bogomolnaia and Herv\'{e}  Moulin had published a paper on a closely related theoretical model, 
albeit in a more restrictive setting that precluded its direct adoption for kidney exchange \citep{bogomolnaia/moulin:2004}.\footnote{In \citet{bogomolnaia/moulin:2004}, agents are partitioned into two groups, 
and each agent can be matched only with an agent on the other side of the economy. In our setting, by contrast, any patient can be matched with any other patient to exchange their donors.} 
If only we could analyze the more general version of the model we needed, we could then address Delmonico's reservations. 
Building on the foundational work of \cite{Gallai:63, Gallai:64} and \cite{edmonds:65, edmonds:71} in discrete optimization, this is exactly what we did in \cite{roth/sonmez/unver:05}.

The progress we made with our second kidney exchange model was well-received by Delmonico and his colleagues in the New England transplantation community. 
This breakthrough laid the groundwork for establishing the \textbf{\textit{New England Program for Kidney Exchange (NEPKE)}}, founded by our team of three economic designers in 
collaboration with doctors Francis Delmonico and Susan Saidman \citep{kidneyaea}. 
Approved by the \textit{Renal Transplant Oversight Committee of New England} in September 2004, NEPKE became the world's first organized kidney exchange system to utilize formal techniques from market design and optimization.

Starting with the \textit{Alliance for Paired Donation} in 2006, many kidney exchange clearinghouses worldwide followed NEPKE's example in the subsequent years, 
transforming the way living-donor kidney transplants are organized in numerous countries.\footnote{See, 
for example, \cite{Manlove:12, APD:2015, Bingaman/Kapturczak/Ashlagi/Murphey:2017, Veale:17, Furian:19, Furian:20, Biro:19, Biro:21, Salman:23, Scandiatransplant:23, Malatya:2023, Malatya:2024}
for living-donor organ exchange programs and techniques where market design experts took active roles.}  
In that sense, the establishment of NEPKE stands out as one of the most consequential policy achievements of my career.

This interaction with Delmonico left me with lasting impressions for two additional reasons. From a methodological perspective, it provided firsthand insight into the importance of 
\textit{custom-made} theory in persuading decision-makers and experts in other fields of the value of our contributions. As elaborated in this monograph, my pursuit of policy impact 
through use-inspired theory, closely aligned with practical realities, has become my primary methodology for numerous applications of market design in the years that followed. 
In my view, this methodology naturally embeds an effective persuasion strategy within minimalist market design, 
going a long way toward satisfying the framework’s third pillar.

The final reason for this lasting impression lies in how this interaction influenced the global evolution of kidney exchange, shaping my future research and policy efforts in living-donor organ exchange. 
Currently, kidney exchange programs worldwide predominantly restrict participation to incompatible patient--donor pairs, resulting in substantial global welfare losses. 
This restriction poses a particular challenge, especially for blood type O patients with donors of blood types A, B, or AB. 
The issue arises because, aside from tissue-type incompatibility, blood type O donors are medically compatible with their patients.

Our initial formulation of kidney exchange in \cite{roth/sonmez/unver:03, roth/sonmez/unver:04} addressed this challenge by defining patient preferences over donors based on 
factors such as blood type compatibility, tissue type compatibility, and donor age as strict preferences. Consequently, the paper incorporated a built-in mechanism that enabled compatible pairs to participate in kidney exchange.

In a notable departure from this approach, \cite{roth/sonmez/unver:05}---developed at the request of Delmonico and marking the beginning of our collaboration---assumes that patients 
are indifferent among all compatible donors. Unfortunately, this modeling choice eliminated the intrinsic mechanism present in \cite{roth/sonmez/unver:03, roth/sonmez/unver:04} that incentivized compatible pairs to participate in kidney exchange. 
Since NEPKE served as a blueprint for kidney exchange programs subsequently launched in the U.S. and many other countries, this led to a substantial global welfare loss relative to the innovation’s true potential. 
In the case of the U.S., \citet{suy:2020} estimates that including blood-type compatible pairs in kidney exchange could increase the number of transplants by as much as 160\%.

To incentivize blood-type compatible pairs to participate in kidney exchange programs, various policies have been proposed in the academic literature or implemented in the field since then. 
Before examining one of these policies in Section \ref{sec:IKE}, let's explore how our partnership with New England medical professionals paved the way for a few other breakthroughs in kidney exchange.

\subsection{Advocating for 3-way Exchanges and  NDD-Initiated Chains} \label{sec:3-way-ndd-chain}

To operationalize NEPKE, it was necessary to establish a web-based data entry system. This system would enable each transplant center to enter their patient and donor information directly. 
The process of setting up this system and addressing bureaucratic requirements for NEPKE spanned about a year. During this period, \"{U}nver developed the software for NEPKE using a priority 
matching algorithm formulated in \cite{roth/sonmez/unver:05}.\footnote{A priority matching works by first arranging patient--donor pairs in a priority order and then considering them one by one. 
At each step, the procedure checks whether the current pair can be part of a feasible outcome while still honoring the transplants promised to higher-priority patients earlier in the process. 
If it is feasible, the patient in the pair is guaranteed a transplant, though the exact exchange is not finalized until the end if there are multiple possible options. In this way, 
the final matching ensures that higher-priority patients receive transplants whenever possible. This method produces a Pareto-efficient outcome and, 
when only two-way exchanges are possible, also maximizes the total number of transplants \citep{roth/sonmez/unver:05}.}
Even though Delmonico initially indicated that only 2-way kidney exchanges were feasible in New England, 
for our research purposes, \"{U}nver also coded priority matching algorithms that allowed for larger size exchanges.

After NEPKE became operational in 2006, kidney exchanges in New England were determined by \"{U}nver's software for a few years. 
Every few months, our team received anonymized patient--donor data from NEPKE, and we ran the software to derive any new
kidney exchanges. This was an exciting period for us, as we were directly impacting and saving lives with our ideas. Hence, upon receiving anonymized data from NEPKE,
we would immediately inspect it to estimate how many transplants the software would determine.

\paragraph{Significance of 3-way Exchanges.}
In one of these instances, the software produced one more transplant than we thought would be feasible when 3-way exchanges were allowed in the system. 
This observation came as a big surprise. At that time, our intuition for the maximal-size kidney exchange was derived from \cite{roth/sonmez/unver:05}, 
admittedly in a model that only permitted 2-way exchanges. To address this limitation, we would then utilize some of the technical details in our proofs to estimate the maximum number of 
transplants if larger-size exchanges were also allowed. This unexpected discovery revealed that our intuition was evidently incomplete.\footnote{The maximum number of transplants can be found in 
polynomial time when only 2-way cycles are allowed, via a reduction to maximum matching \citep{edmonds:65}. Once cycles of length three or longer are allowed, however, 
the problem becomes computationally intractable, as established in earlier work on bounded-length cycle packing \citep{Karp:1972} and highlighted in the kidney exchange setting by \cite{Abraham/Blum/Sandholm:2007}.}

We then conducted a detailed analysis of how allowing larger kidney exchanges affects transplant numbers. 
This analysis revealed that the underlying partial order structure induced by ABO blood type compatibility leads to two key findings:
\begin{enumerate}
\item In sufficiently large patient--donor pools, allowing 3-way and 4-way kidney exchanges achieves the maximum number of transplants, and
\item Most gains over 2-way exchanges come from 3-way exchanges.\footnote{For small pools with 25 patient--donor pairs, 
the gain was approximately 30\% more transplants from kidney exchanges. In large pools with 100 patient--donor pairs, this increase was around 20\% more transplants from kidney exchanges.}
\end{enumerate}

Given the practical significance of these discoveries, and despite Delmonico’s initial reservations, we convinced our New England partners to include 3-way kidney exchanges in the system, 
publishing our findings in \cite{srsud:06} and \cite{roth/sonmez/unver:07}.\footnote{With 3-way exchanges allowed in our system, a priority matching no longer guaranteed the maximization of the number of transplants in the system. 
Consequently, we adopted an alternative algorithm that maximized the number of transplants \`{a} la \cite{roth/sonmez/unver:07}.}

\begin{figure}[!tp]
    \begin{center}
       \includegraphics[scale=1.0]{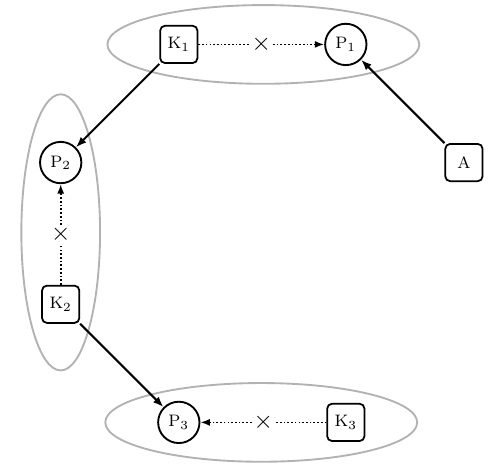}
           \end{center}
\caption{NDD chain with three patient--donor pairs. An altruistic donor donates kidney A to patient P$_1$, initiating the chain. 
The donor of P$_1$ then donates kidney K$_1$ to P$_2$, whose donor donates K$_2$ to P$_3$. 
The kidney K$_3$ from the donor of P$_3$ remains available for subsequent transplantation, either to start another chain in the same role as A 
(effectively extending the current chain over time) or to be allocated to a patient on the deceased-donor list.}
\label{fig:three-pair-NDD}
\end{figure}

\paragraph{Non-Directed Donor (NDD) Donor Chains.}
Our collaboration with New England partners also led to another breakthrough. Before this collaboration, altruistic donors were not a major source of living donor kidneys in the U.S. 
According to the UNOS database, only 8 such transplants were conducted in the U.S. before 2000 \citep{OPTN:25}.
However, this number was increasing in the early 2000s, offering the possibility to initiate donor chains with an altruistic \textit{non-directed donor} (NDD), 
rather than a patient--donor pair volunteering for list exchange (see Figure \ref{fig:three-pair-NDD} for an illustration). 
Moreover, since these NDD-chains do not interfere with the DD lists, they were not subject to the same ethical concerns that hindered the utilization of list exchanges in NEPKE. 

NDD chains presented another possibility. As mentioned earlier, the Consensus Statement on the Live Organ Donor recommended that all procedures in kidney exchange be performed simultaneously, 
primarily to prevent a situation in which the donor of a pair gives a kidney as part of an exchange, but their intended patient might not receive a transplant if the other donor backed out or became unavailable.
However, this concern was significantly reduced with an NDD chain that starts with an altruistic donation. Because all patients receive a transplant before their donor contributes to the next patient in the chain, 
the worst-case outcome is simply a shorter chain than its full potential.\footnote{Over time, even this worst-case scenario proved to be rare and of negligible consequence. 
Using data from 344 non-simultaneous NDD chains implemented by the National Kidney Registry from 2008 through May 2016, which generated 1,568 transplants, 
only 20 of these chains were broken due to reneging \citep{Cowan:17}. Interestingly, the average size of the broken chains was 4.8, compared to 4.5 for unbroken chains.}

Based on these observations, we formulated \textbf{\textit{non-simultaneous NDD chains}} in \cite{rsuds:06}.\footnote{While \cite{rsuds:06} was the first to introduce \textit{non-simultaneous NDD chains} in the literature, 
a subsequent publication by \cite{Lee:2009} revealed that the approach had already been implemented in South Korea. In a multi-center program involving 16 hospitals, 
70 such chains were conducted between February 2001 and July 2007, resulting in 179 transplants.} 
While this idea was not pursued by NEPKE, it was later adopted by two other kidney exchange programs in the U.S.: 
first by the \textit{Alliance for Paired Donation} (APD),\footnote{Starting in 2006, our team also established a partnership with APD. Based on \cite{rsuds:06, roth/sonmez/unver:07}, 
APD included 3-way and 4-way exchanges in their allocation mechanism from its inception and incorporated non-simultaneous NDD chains shortly after. 
Not only did \"{U}nver introduce the concept of non-simultaneous NDD chains to Mike Rees, the head of APD, but the software used by APD from 2006 to 2008--just like the code of NEPKE--was also developed by him. 
See \cite{APD:2015} for details of our subsequent partnership with APD, later renamed \textit{Alliance for Paired Kidney Donation (APKD)}.} 
launched in 2006 in the Midwest, and shortly thereafter by the \textit{National Kidney Registry}, launched in 2008.

Because of its logistical advantages, the role of non-simultaneous NDD chains in U.S. kidney exchange programs grew considerably over time, 
particularly due to the National Kidney Registry's success in enrolling altruistic non-directed donors into its system \citep{Cowan:17, agarwal:2019}.

\subsection{Incentivized Kidney Exchange} \label{sec:IKE} 

While our pioneering role in the evolution and widespread adoption of kidney exchange programs worldwide stands as one of our most consequential policy achievements to date, 
a crucial policy lever is still missing in a vast majority of these programs---one that has the potential to significantly enhance their efficiency.

As we briefly discussed in Section \ref{sec:NEPKE}, partly due to the circumstances surrounding the foundation of NEPKE---a program that became a blueprint for the rest of the world---participation in 
kidney exchange programs remains largely restricted to patients who are biologically incompatible with their donors, leading to a substantial efficiency loss.

A handful of policies have been proposed and promoted by various groups to encourage the participation of compatible pairs in kidney exchange programs. 
Among these, we developed \textit{incentivized kidney exchange} (IKE) in the early 2010s, as detailed in \cite{sonmez/unver:15}  and partially published in  \cite{suy:2020}.  Despite its potential,
however, this policy has yet to be implemented in practice.

IKE is a targeted strategy for encouraging the participation of blood-type compatible pairs in kidney exchange. 
The approach hinges on offering patients a higher deceased-donor (DD) priority should they face another renal failure in the future. Since a kidney from a living donor often lasts just 15 to 20 years, this policy provides a practical and forward-looking incentive.

The ethical aspects of this policy have been discussed favorably by members of the Canadian transplantation community \citep{Gill:17}. 
In principle, IKE could be integrated into the ongoing reform of the UNOS DD allocation system \citep{sonmez/unver:23}. 
Yet because living-donor and DD transplants are managed separately in the U.S., adopting a policy that links the two would require broad national consensus.

Despite receiving favorable evaluations of its ethical principles within the transplantation community, strictly speaking, IKE falls outside the scope of minimalist market design. 
While linking the management of living-donor kidneys and DD kidneys may be a valuable policy for increasing the utilization of willing living donors, 
it is not clear that the absence of this link is the primary cause of the lack of participation by compatible pairs in kidney exchange. In this sense, while IKE offers substantial welfare gains, 
as an innovation that requires the joint management of multiple separate institutions, a reform based on IKE lacks the typical persuasion strategy inherent in minimalist market design. 
As an ambitious policy proposal for an aspiring economic designer, this may be one of the reasons why IKE has fallen short of its practical promise so far, 
despite the significant welfare gains it offers.\footnote{While IKE departs from the idealized version of minimalist market design, 
it still maintains two of the approach's three pillars: it identifies the institutional mission, including its ethical considerations, and employs a compelling persuasion strategy. 
The only pillar IKE does not fully embrace is the avoidance of non-essential changes, as it integrates the management of living-donor and DD kidneys. 
If separating the management of these kidneys compromises the primary objectives of the transplantation community---a plausible yet unexplored hypothesis---IKE may also satisfy the third pillar over time. 
In this sense, it is not far from being a minimalist reform. Other policies that involve linking the DD program with living-donor exchange include the 
\textit{list exchange} by \cite{ross/woodle:2000}, \textit{DD chains} by \cite{roth/sonmez/unver:04}, the \textit{voucher program} by \cite{Veale:17}, and the \textit{unpaired kidney exchange} by \cite{unpaired-ke:2024}.}

\subsection{Kidney Exchange Worldwide: Two Decades On} \label{sec:KE-2decadeslater}

Unlike my research and policy efforts in school choice, discussed in Section \ref{sec:schoolchoice}, which took six years to garner serious attention from academics and policymakers, 
our work on kidney exchange generated widespread enthusiasm within months.

Attending the 1995 Summer Conference of the Stanford Institute for Theoretical Economics (SITE), just months after receiving my Ph.D., 
was a pivotal experience that shifted my research focus from abstract theory toward use-inspired theory. 
Eight years later, in August 2003, I presented our kidney exchange research publicly for the first time at another SITE Summer Conference. 
With its unprecedented potential to save lives through economic design, our work immediately drew attention from economists, computer scientists, and operations researchers alike.

Within a month, we circulated our findings as an NBER Working Paper \citep{roth/sonmez/unver:03} and soon after shared them with Delmonico, 
a prominent leader in transplantation and the Chief Medical Officer at the New England Organ Bank. At his request, we refined our model in preparation for a potential collaboration.

Recognizing the importance of this opportunity, my wife Banu (see Section \ref{sec:Banu}) and I made the life-changing decision to relocate from Turkey to the U.S. in August 2004. 
This transformative step---requiring immense personal sacrifice from Banu---was driven by our shared dream of translating my research in kidney exchange and school choice into public benefit. 
Within a month of our arrival, our collaboration with Delmonico led to a significant milestone: the \textit{Renal Transplant Oversight Committee of New England} approved the establishment of the 
\textit{New England Program for Kidney Exchange (NEPKE)} in September 2004 (see Section \ref{sec:NEPKE}).

\subsubsection{NSF Advocacy and Resolving Legal Complexities in the U.S.} \label{sec:NSF}

This was a period when the National Science Foundation (NSF) was becoming increasingly supportive of use-inspired basic research within Pasteur's Quadrant, 
spurred by the publication of \cite{Stokes:1997}. Consequently, our efforts quickly gained strong support from the NSF. Beyond funding our kidney exchange research, 
the NSF also highlighted our unique collaboration with members of the transplantation community in a feature story published just one year after the NEPKE agreement was reached \citep{NSF:05}:

\begin{quote}
``This year, more than 60,000 people in the United States will need a kidney transplant. Of that number, roughly 15,000 will receive transplants from either cadavers or living donors. 
But for those who aren't so lucky, the prospects have gotten better thanks to a kidney exchange system developed by Harvard economist Alvin Roth, Tayfun Sonmez of Boston College and Utku Unver of University of Pittsburgh. [$\cdots$]

The system, which was based on an idea for dormitory housing allocations, has led to the establishment of a kidney exchange clearinghouse for patients needing transplants in the New England area. 
Applied nationally, the system could shorten the wait time for kidney transplant patients and potentially save thousands of lives.''
\end{quote}

To apply kidney exchange at a national scale, however, one major legal challenge had to be addressed. A subtle aspect of the law’s wording left it uncertain whether kidney exchange was legal.

The 1984 National Organ Transplant Act (NOTA) made it a felony in the U.S. to pay for an organ for transplantation, stating:

\begin{quote}
``it shall be unlawful for any person to knowingly acquire, receive or otherwise transfer any human organ for valuable consideration for use in human transplantation.''
\end{quote}

Since it was unclear whether receiving a kidney in exchange for another constituted ``valuable consideration'' under this act, Delmonico was hesitant to publicize NEPKE widely, 
even though it had already begun performing kidney exchanges. While our partners were not concerned enough to stop performing kidney exchanges, they were cautious enough to avoid advertising them.

This challenge was finally overcome in March 2007, when the Justice Department issued a legal memo stating that kidney exchanges do not constitute ``valuable consideration'' under the provisions of NOTA, 
thereby not violating the act. Later that year, in December 2007, the U.S. Senate passed an amendment to NOTA---the \textit{Charlie W. Norwood Living Organ Donation Act}---explicitly clarifying that kidney exchange is legal in the U.S. 
With this breakthrough, the pathway was opened for national kidney exchange.\footnote{Before the enactment of the \textit{Charlie W. Norwood Living Organ Donation Act} in the U.S., 
kidney exchange had already been deemed legal in the UK under the provisions of the \textit{Human Tissue Acts} of 2004 (covering England, Northern Ireland, and Wales) and 2006 (for Scotland) \citep{Biro:19}. 
Since then, with active support from members of the market design community in many instances, similar laws have been enacted in several other countries, enabling various forms of kidney exchange. 
Examples include the \textit{Belgian Living Donor Exchange Protocol} in Belgium in 2008 \citep{Biro:19}, and the \textit{Bioethics Law} in France---initially in 2011 (allowing only two-way exchanges) 
and later expanded in 2021 (permitting exchanges and NDD chains of up to three-way) \citep{Combe:22, Biro:24}. In addition, as of December 2024, 
a draft law for an amendment to the \textit{Transplantation Law} in Germany allowing for kidney exchange between incompatible pairs as well as 
NDD chains is pending approval in Parliament \citep{Ashlagi/Cseh/Manlove/Ockenfels/Pettersson:24, Ockenfels/Sonmez/Unver:24}.}

Following this legal clarification, the number of kidney exchange transplants in the U.S. surged, surpassing 500 annually by the early 2010s. 
A major driver of this rapid growth was the widespread adoption of \textit{non-simultaneous NDD chains}, initiated by altruistic donors (see Section \ref{sec:3-way-ndd-chain}). 

Although this innovation emerged from our NEPKE partnership \citep{rsuds:06}, it was the \textit{Alliance for Paired Donation (APD)} and the \textit{National Kidney Registry (NKR)} 
that ultimately implemented and scaled it \citep{Rees:09}.
By the early 2010s, as depicted in Figure \ref{fig:DIC-KE-arm}, this kidney exchange modality had become a cornerstone of the \textit{discovery–invention cycle}---a series of life-saving 
innovations that transformed our early theoretical work on housing allocation into tangible, real-world healthcare advancements.

The NKR, in particular, established itself as the highest-volume kidney exchange platform in the U.S., driven largely by its effective use of non-simultaneous NDD chains \citep{agarwal:2019}. 
Between 2008 and May 2016, 89.7\% of the 1,748 kidney exchange transplants facilitated by NKR---1,568 in total---were achieved through these chains \citep{Cowan:17}.

During this period, the National Science Foundation (NSF) continued to support our kidney exchange initiatives, recognizing them as a prime example of how Social, Behavioral, and Economic Sciences (SBE) 
research can deliver immediate public benefits. Our work received significant recognition from NSF officials, who highlighted it on numerous occasions.

In an \textit{NSF Science Nation} article, Nancy Lutz, Program Director in the NSF’s Social and Economic Sciences Division, remarked \citep{NSF:12}:

\begin{quote}
``it's especially rewarding to see such a clear and immediate benefit to the public. This research
moved from abstract, academic theory to real world, direct impact very quickly.''
\end{quote}

Even more notably, in a June 2011 Congressional hearing on SBE research funding, Myron Gutmann, Assistant Director of the NSF-SBE Directorate, 
emphasized the measurable gains of NSF-supported kidney exchange research for American taxpayers (\citealp{US:2011}, p. 16):

\begin{quote}
`` 3.1 SBE research has resulted in measurable gains for the U.S. taxpayer

\textbf{Matching markets and kidney transplants.} Researchers in economics at Harvard
University, the University of Pittsburgh, and Boston College have applied economic
matching theory to develop a system that dramatically improves the ability
of doctors to find compatible kidneys for patients on transplant lists. Organ donation
is an example of an exchange that relies on mutual convergence of need. In this
case, a donor and a recipient. This system allows matches to take place in a string
of exchanges, shortening the waiting time and, in the case of organ transplants, potentially
saving thousands of lives.''
\end{quote}  

Two decades later, the potential highlighted by Gutmann has indeed materialized, establishing kidney exchange as a mainstream modality in kidney transplantation not only in the U.S. but also in many other countries.

\subsubsection{The Global Expansion of Kidney Exchange Since the Mid-2000s} \label{sec:KE-2008}

After the first kidney exchange performed in the U.S. in Rhode Island in 2000, progress was initially slow. 
Between 2000 and 2003---prior to our engagement with the transplantation community---only 31 kidney exchange transplants were performed nationwide. 
In sharp contrast, a decade later the four-year total had risen to 2,078 between 2010 and 2013. 
The number climbed further over the next decade, with 4,387 kidney exchange transplants conducted between 2020 and 2023 \citep{OPTN:25}. 
A record 1,493 kidney exchange transplants were performed in 2024 alone, and by September 2025 the cumulative total in the U.S. exceeded 14,200 patients. 
(For annual transplant numbers in the U.S. from 2000 to 2024, see Figure \ref{fig:KE-US-2000-24}.)

\begin{figure}[!tp]
    \begin{center}
       \label{fig:KE-US-2000-24}
    \includegraphics[scale=0.435]{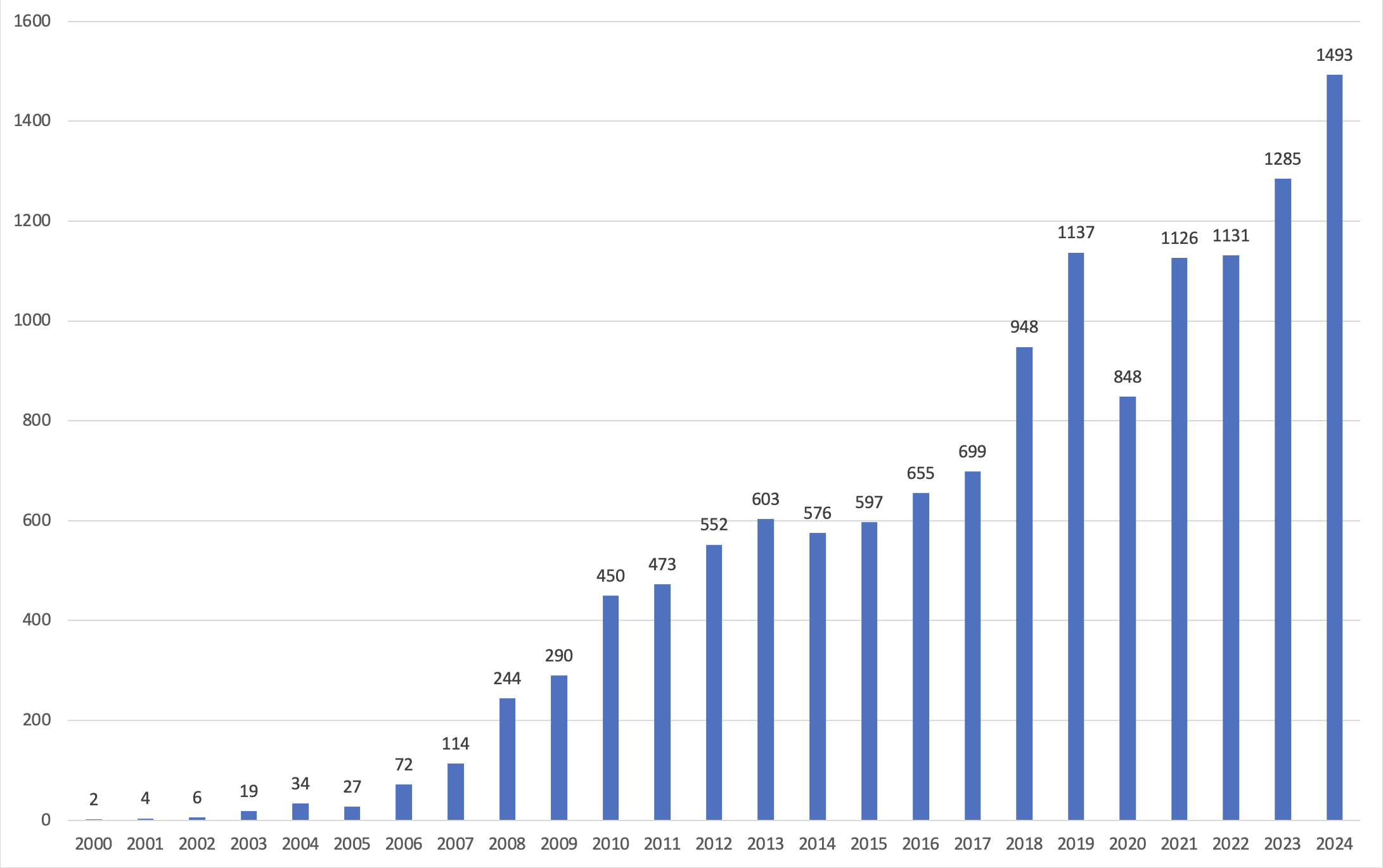}
    \end{center}
    \caption{Annual number of kidney exchange transplants performed in the U.S. from 2000 to 2024, as reported in the UNOS database on August 27, 2025 \citep{OPTN:25}.} \label{fig:KE-US-2000-24} 
\end{figure}

According to estimates by \cite{Teltser:19}, 64\% of these transplants---approximately 9,000---represent new transplants that would not have occurred without kidney exchange. 
Furthermore, \cite{Teltser:19} highlights that kidney exchange has also improved average graft survival rates, despite the fact that kidney exchange recipients often come from more disadvantaged groups. 
These include highly sensitized patients with antibodies against large portions of the population, older patients, and recipients of kidneys from older donors.

Over the past two decades, kidney exchange programs have expanded beyond the U.S., providing life-saving transplants to patients worldwide.
Countries such as the UK, Turkey, Canada, India, Australia, Saudi Arabia, the Netherlands and Spain now perform some of the highest kidney exchange volumes outside the U.S.
Altogether, the number of kidney exchange transplants carried out globally to date exceeds 23,000 (see Table \ref{tab:KE-worldwide} for worldwide kidney exchange activity since early 2000s).


\begin{table}[!tp]
\footnotesize
\centering
\begin{threeparttable}
\caption{Expansion of kidney exchange activity worldwide since the early 2000s.}
\label{tab:KE-worldwide}
\setlength{\tabcolsep}{4pt}
\renewcommand{\arraystretch}{1.06}
\begin{tabular}{@{}p{0.205\textwidth} p{0.44\textwidth} p{0.08\textwidth} p{0.25\textwidth}@{}}
\toprule
\addlinespace[4pt] 
\textbf{Country} & \textbf{Program / Operator} & \textbf{Start} & \textbf{Documented volume} \\ 
\addlinespace[4pt] 
\midrule
\addlinespace[7pt] 

United States
& Three major platforms NKR, APKD, UNOS KPD \& regional/hospital programs
& 2000
& $>$14{,}200 \; (by 08/2025) \\
\midrule
\addlinespace[4pt] 

Turkey
& Multiple single-center programs
& 2006
& $>$2{,}300  \; (by 12/2023) \\
\midrule
\addlinespace[4pt]

United Kingdom
& UK Living Donor Kidney Sharing Scheme (NHS Blood \& Transplant)
& 2007
& $>$2{,}000  \; (by 09/2024) \\
\midrule
\addlinespace[4pt] 

India
& Multiple single-center programs
& 2000
& 1{,}891  \; (09/2000--03/2024) \\
\midrule
\addlinespace[4pt] 

Canada
& Kidney Paired Donation Program
& 2009
& $>$1{,}000  \;  (by 05/2023) \\
& (Canadian Blood Services)
&
& \\
\midrule
\addlinespace[4pt] 

Australia
& Australian Paired Kidney Exchange Program
& 2010
&  \\
\addlinespace[3pt] 
New Zealand
& New Zealand Kidney Exchange
& 2011
& \\
\addlinespace[3pt] 
Bilateral cooperation
& Australian and New Zealand Paired Kidney Exchange (ANZKX)
& 2019
&  $>$600 \; (by 09/2025)   \\
\midrule
\addlinespace[4pt] 

Saudi Arabia
& Kidney Paired Donation Program
& 2011
&  $>$500  \;  (by 10/2024) \\
& (King Faisal Specialist Hospital )
&
& \\
\midrule
\addlinespace[4pt] 

Spain
& Organización Nacional de Trasplantes
& 2009
& 341  \; (by 12/2024) \\
\addlinespace[3pt] 
Italy
& Centro Nazionale Trapianti
& 2006
& 91  \; (2005--2023) \\
\addlinespace[3pt] 
Portugal
& Programa Nacional de Doação Renal Cruzada
& 2010
& $\geq$24  \;  (by 2020) \\
\addlinespace[3pt] 
Trilateral cooperation
& Spain--Italy--Portugal cooperation protocol
& 2017
& $\geq$19  \;  (by 12/2024) \\
\midrule
\addlinespace[4pt] 

Netherlands
& National kidney exchange program
& 2004
& 284  \;  (2004--2016) \\
& Regional / single-center NDD chains
&
& 82  \;  (by 04/2014) \\
\midrule
\addlinespace[4pt] 

Denmark
&
& 
&  \\
\addlinespace[3pt] 
Finland
& Scandiatransplant Kidney Exchange
& 2019
&  95 \; (by Q2 2025) \\
\addlinespace[3pt] 
Norway
& Programme (STEP)
& 
& \\
\addlinespace[3pt] 
Sweden
&
& 
&  \\
\midrule
\addlinespace[4pt] 

Austria
&
& 2015
&  \\
\addlinespace[3pt] 
Czech Republic
& Unified AT--CZ--IL program
& 2015
& 81  \;  (2011--2019) \\
\addlinespace[3pt] 
Israel
&
& 2019
&  \\
\midrule
\addlinespace[4pt] 

Global
& Global Kidney Exchange (GKE)
& 2015
& 52  \;  (2015--02/2022) \\
\addlinespace[2pt] 
\bottomrule
\end{tabular}
\end{threeparttable}
\end{table}

\paragraph{UK.} The UK was one of the first countries to adopt market design tools for kidney exchange---including optimal matching algorithms---and now operates the largest national system. 
Since 2007, the \textit{National Health Service Blood and Transplant} (NHSBT) has coordinated kidney exchanges through the \textit{UK Living Donor Kidney Sharing Scheme}. 
This quarterly program has collaborated with market design researcher David Manlove since 2008 and celebrated its 2,000th kidney exchange transplant in September 2024.

Initially limited to 2-way and 3-way cycles \citep{Biro/Manlove/Rizzi:09}, the UK scheme has also been organizing non-directed donor (NDD) chains of up to three recipients since 2012 \citep{Manlove:14, Ferrari:14}. 
Among the 670 kidney exchange transplants conducted between April 2020 and March 2024, 126 (18.8\%) were carried out in 2-way cycles, 118 (17.6\%) in 3-way cycles, 215 (32.1\%) in 2-way NDD chains, 
and 211 (31.5\%) in 3-way NDD chains \citep{NHS:24}.

Two key factors behind the UK’s success are the substantial participation of altruistic donors and compatible donor-recipient pairs. 
Between April 2015 and March 2024, there were 653 NDDs, with 347 donating directly to the deceased donor list and 306 donating through the kidney sharing scheme \citep{NHS:24}. 
Over this period, the proportion of NDDs participating in kidney exchanges grew from approximately 30\% to 35\% to 55\% to 65\%. Additionally, 119 compatible pairs participated in 
kidney exchanges during the same period.

Taken together, the UK's centrally managed national system---which incorporates up to 3-way cycles and NDD chains, employs an optimal matching algorithm, 
and benefits from significant participation by altruistic donors and compatible pairs---stands out as perhaps the closest large-scale realization of the vision pioneered 
in \cite{roth/sonmez/unver:04, roth/sonmez/unver:05, rsuds:06, roth/sonmez/unver:07}.

\paragraph{Turkey.} Over 80\% of kidney transplants in Turkey are from living donors, and the country led the world in living kidney donation rates in 2023, exceeding 35 donors per million people \citep{Irodat:24}. 
Performing over 3,000 transplants annually, Turkey also ranks third globally in living-donor kidney transplants (LDKTs), behind only India and the U.S. 
These factors make Turkey an ideal candidate to implement a national system. Unfortunately, as discussed later in Section \ref{sec:Banu}, my efforts with \"{U}nver in this direction have not succeeded so far.

Nevertheless, kidney exchange is widely practiced in Turkey through several single-hospital programs, with more than 2,300 transplants performed as of 2023. 
However, NDD chains are not utilized, as altruistic donation is prohibited by law. 

The Turkish Ministry of Health and the Turkish Society of Nephrology have tracked kidney exchange activity nationwide since 2012 \citep{TSN:12-22}. 
According to their registries, 2,067 kidney exchange transplants were performed between 2012 and 2023, accounting for 6.3\% of the country's total LDKTs.\footnote{Kidney exchange activity in Turkey predates 2012, 
with significant activity reported between 2006 and 2011. According to \cite{Yucetin:13}, among the 59 hospitals performing LDKTs nationwide, 
\textit{Antalya Medicalpark Hospital}, \textit{Istanbul G\"{o}ztepe Medicalpark Hospital}, and \textit{Akdeniz University Medical School Transplantation Center} reported the highest kidney exchange volumes during this period. 
Together, they conducted 272 transplants, accounting for 14.4\% of the LDKTs carried out at these centers.} 
This relatively modest rate reflects the absence of a national kidney exchange system and the limited availability of kidney exchange in only some hospitals. 
However, in hospitals with kidney exchange programs, rates are significantly higher, around 15\%. 
For example, \cite{Altun/Yavuz:23} documented 141 kidney exchange transplants at \textit{Istanbul G\"{o}ztepe Medicalpark Hospital} from July 2011 to June 2021, representing 16.1\% of the LDKTs performed at the center during that period.

\paragraph{India.} The most populous country in the world, India leads globally in LDKTs, performing over 10,000 in 2022  \citep{Irodat:24}.
Kidney exchange is widespread and conducted through numerous single-center programs across the country. 
\cite{Kute:2025} reports that 1,891 kidney exchange transplants were performed between September 2000 and March 2024 across 84 transplant centers.

As in Turkey, NDD chains are not practiced in India because altruistic donation is prohibited by law.
Unlike in many other countries, kidney exchange operations in India are often carried out over several days, enabling the execution of large exchange cycles.

At the single-center level, \cite{Kute:24} reports that the \textit{Dr HL Trivedi Institute of Transplantation Sciences} in Ahmedabad---among the largest single-center kidney exchange programs in Asia---performed 
539 kidney exchange transplants between January 2000 and March 2024.
These included one 10-way and two 6-way exchanges conducted over multiple days, representing 10\% of the 5,346 LDKTs performed at the center during this period.
Similarly, \textit{Mahatma Gandhi Medical College} in Jaipur, described by \cite{Mangal:24} as the largest program in Northern India, reported 127 kidney exchange transplants between 2014 and March 2024.

\paragraph{Canada.} The national Kidney Paired Donation program has been coordinated by  \textit{Canadian Blood Services} since 2009. 
The system runs every four months and incorporates cycles as well as NDD chains of up to four pairs. It is regarded as one of the most successful kidney exchange programs worldwide, 
largely because of the strong participation of altruistic donors. In May 2023, the program reached its 1,000th kidney exchange transplant, 68\% of which were facilitated through donations from altruistic donors.

\paragraph{Saudi Arabia.} More than 80\% of kidney transplants in Saudi Arabia are from living donors, a pattern shared with Turkey and many other Asian countries. 
In 2023, the country ranked second worldwide---after Turkey---in the living kidney donation rate, 
with more than 35 donors per million people \citep{Irodat:24}. The \textit{King Faisal Specialist Hospital and Research Centre} celebrated its 500th kidney exchange transplant in October 2024 as part of its single-hospital 
\textit{Kidney Paired Donation Program}, which has been in operation since 2011.

\paragraph{The Netherlands.} The Netherlands is the only European country where transplants from living donors are as prevalent as those from deceased donors, with 400--500 transplants annually from each 
source since the early 2010s. In 2023, the country ranked fourth worldwide---after Turkey, Saudi Arabia, and Israel---in the living kidney donation rate, with more than 28 donors per million people \citep{Irodat:24}. 

With the participation of eight transplant centers and one histocompatibility center, the Netherlands launched its national kidney exchange program in 2004, becoming the first country in 
Europe to establish such a system \citep{deKlerk:05}. The program allows up to 4-way exchanges and facilitated 284 kidney exchange transplants between 2004 and 2016 \citep{Biro:19}. 
Although altruistic donation is legal in the country and NDD chains are performed regionally, they are not managed within the national system. 
\cite{Ferrari:14} reports that 82 transplants had been facilitated through regional NDD chains as of April 2014.

\paragraph{Australia / New Zealand.} National kidney exchange programs were launched in Australia in 2010 and in New Zealand in 2011, 
and the two were merged in 2019 into the \textit{Australian and New Zealand Paired Kidney Exchange} (ANZKX) program. Since their inception, these programs have facilitated more than 600 kidney exchange transplants.

\paragraph{Spain / Italy / Portugal}  As another cross-border initiative, Spain, Italy, and Portugal signed a cooperation protocol in 2017 and achieved their first cross-border kidney exchange in 
2018 \citep{Valentin:19}. Unlike other multi-country collaborations, however, each country continues to operate its own independent national system before engaging in cross-border exchanges.

The \textit{Organización Nacional de Trasplantes} (ONT) launched Spain’s national kidney exchange program in 2009. Allowing 2-way and 3-way exchanges, 
it had facilitated 325 kidney exchange transplants by December 2023, with another 16 in 2024, bringing the total to 341 \citep{GobiernoEspana2024}.

The \textit{Centro Nazionale Trapianti} (CNT) introduced kidney exchange in Italy in 2006, making it one of the earliest adopters in Southern Europe. 
Between 2005 and 2023, 91 transplants were facilitated under the program, including 18 in 2023 alone \citep{CNT2023}.

The \textit{Programa Nacional de Doação Renal Cruzada} (PNDRC) was established in Portugal in 2010. At least 24 transplants were reported through 2020 \citep{IPST2020}, 
and more recent updates show continued activity, including several 2-way and 3-way cycles and two NDD-chain transplants in 2024.  

By July 2023, a total of 16 patients had received transplants through the trilateral collaboration between Spain, Italy, and Portugal \citep{GobiernoEspana2024}, 
with at least three additional international procedures recorded in 2024. This brought the cumulative cross-border total to more than 19 by the end of that year.

\paragraph{Scandinavia.}  One of the most active cross-border collaborations in Europe is the \textit{Scandiatransplant Kidney Exchange Programme} (STEP), with participation from Denmark, Finland, Norway, and Sweden \citep{Weinreich:23}.
Launched in 2019 with the advocacy of economic designer Tommy Andersson and initially limited to 2-way exchanges, STEP has since been organizing 2-way and 3-way exchanges as well as NDD chains since January 2024. 
As of the second quarter of 2025, the program has facilitated a total of 95 transplants \citep{Scandiatransplant:25-Q2}.

\paragraph{Austria / Czech Republic / Israel.} Another cross-border kidney exchange partnership currently operates between Austria, the Czech Republic, and Israel. 
Due to the limited size of their national pools, Austria and the Czech Republic---both of which introduced national kidney exchange systems in 2011---merged their programs in 2015. 
This collaboration led to the first cross-border kidney exchange in 2016 \citep{Bohmig:17} and was further expanded to include Israeli centers in 2019. 
Between 2011 and 2019, these systems facilitated 81 transplants, including 33 after 2015 under the unified program, of which eight were via cross-border exchanges \citep{Viklicky:20}.

\paragraph{Global Kidney Exchange.} Finally, a related cross-border initiative is the \textit{Global Kidney Exchange (GKE)} program, spearheaded by Alvin Roth and Michael Rees \citep{GKE:17,GKE:22}. 
The program funds kidney transplants for recipients from low- and middle-income countries in exchange for living donors who facilitate transplant cycles or NDD chains in high-income countries. 
Since 2015, the \textit{Alliance for Paired Kidney Donation (APKD)} has implemented GKE. Between January 2015 and February 2022, 17 international patients facing financial barriers to 
direct transplantation facilitated kidney transplants for 35 U.S. patients through GKE, resulting in a total of 52 kidney transplants.\footnote{Our earlier efforts on kidney exchange in New England 
and elsewhere aligned with minimalist principles and, most importantly, strictly adhered to the ethical guidelines of the transplantation community as codified in the Consensus Statement \citep{PKE-consensus:2000}. 
Consequently, these initiatives did not encounter major opposition within the transplantation community. In contrast, the GKE program has sparked debate, receiving both endorsements and criticisms. 
Supporters include prominent bioethicists such as Peter Singer \citep{Minerva/Savulescu/Singer:19} and the Italian government, which has recommended GKE to the \textit{World Health Organization} \citep{Pullen:18}. 
However, unlike other kidney exchange initiatives, GKE has also faced strong ethical objections. Opponents include Frank Delmonico, who led criticism of GKE in his role as Executive Director of the \textit{Declaration of Istanbul Custodian Group}, 
as well as the \textit{European Union National Competent Authorities}, which oversee transplant activities across the EU \citep{Delmonico:17, DOICG:18, NCA:18}. For an in-depth discussion of this debate, see \cite{Ambagtsheer:20}.}

\paragraph{European Union.} Since 2008, collaborations among the European market design community, clinicians, and policymakers have driven significant advances in kidney exchange under the guidance of market design experts such as 
Tommy Andersson, Peter Bir\'{o}, David Manlove, Joris van de Klundert, and Ana Viana. A major milestone was the establishment of the \textit{European Network for Collaboration on Kidney Exchange Programs (ENCKEP)}, 
supported by the \textit{European Cooperation in Science and Technology (COST)} from September 2016 to March 2021 \citep{Biro:19, ENCKEP:24}.

ENCKEP led to a follow-on project called \textit{Software for Transnational Kidney Exchange Programmes (KEP-SOFT)}, funded by a COST Innovators Grant from November 2021 to October 2022. 
This effort produced \textit{KEPsoft}, a software solution for national and international kidney exchange \citep{KEP-SOFT:24}.\footnote{\textit{KEPsoft} was jointly developed by the 
University of Glasgow, INESC TEC (Porto), HUN-REN-KRTK (Budapest), and Óbuda University (Budapest) \citep{KEP-SOFT:24}.}

Further affirming the growing enthusiasm for kidney exchange in Europe, the \textit{European Commission} launched a 2023 call for proposals for action grants under its \textit{Health Systems and Healthcare Workforce} initiative, 
specifically targeting ``Facilitating Organ Paired Exchange'' \citep{EU4Health:23}. This initiative aims to harmonize organ donation and transplantation practices across EU member states by developing a 
shared algorithm for matching donor-patient pairs and establishing common protocols, platforms, and governance frameworks. The ultimate goal is to create an EU-wide system that expands kidney exchange 
opportunities and improves patient outcomes.\footnote{The \textit{EURO-KEP} project, led by the \textit{Spanish National Transplant Organization (ONT)}, 
was the sole project funded under this call and is supported by \textit{KEPsoft} software \citep{KEP-SOFT:24}.}

\section{U.S. Army's Branching  Process} \label{sec:Army}

My holistic approach to research and policy in school choice (see Section \ref{sec:schoolchoice}) and kidney exchange (see Section \ref{sec:KE}) 
was primarily driven by instinct. Early in my academic career, I did not consciously follow any specific institutional design framework. In both applications, 
my goal was to address the limitations of existing resource allocation systems in ways that could gain acceptance from decision-makers vested in preserving them.

Clarity emerged as I observed parallels between the research-guided school choice reform in Boston (see Section \ref{sec:Boston}) and the policy-maker-driven reforms in England 
and Chicago (see Section \ref{sec:Chicago-England}). It became evident that the methodology evolving from my holistic research and policy efforts could provide the 
foundation for a promising economic design framework on a broader scale.

I interpreted these developments in England and Chicago as providing external validity for this emerging methodology. 
My approach---uncovering how real-life institutions had diverged from their intended roles and restoring their alignment---resonated with decision-makers. 
This emphasis on aligning institutions with their original intentions enabled me to achieve policy impact, which may explain my success.

Following this realization, I consistently applied minimalist market design in subsequent applications, starting with the U.S. Army's branching process for cadets in the early 2010s.

In this section, I outline my decade-long research and policy efforts on the U.S. Army's branching process, initiated in collaboration with then-U.S. Air Force Colonel Tobias Switzer \citep{sonmez/switzer:13}. 
While these efforts were initially unsuccessful in influencing policy, they ultimately culminated in a major reform of the U.S. Army's branching processes at the United States Military Academy (USMA) 
at West Point and the Reserve Officer Training Corps (ROTC). I led this reform in collaboration with Lieutenant Colonel Kyle Greenberg and Parag Pathak during the fall of 2020 \citep{greenberg/pathak/sonmez:24}. 
The new design took effect with the Class of 2021 branching assignments at both institutions.

\subsection{My Introduction to U.S. Army's Branching Process}

In May 2011, I received an unexpected email from (then) Major Tobias Switzer, a U.S. Air Force officer, 
seeking feedback on his master’s thesis \citep{switzer:11} from the Catholic University of Chile.
Following the basic structure of \cite{balinski/sonmez:99} and adopting the \textit{no justified envy} axiom (see Definition \ref{def:NJE}) for his framework, 
Switzer examined the branching process of cadets into military specialties at the United States Military Academy (USMA).\footnote{Following the terminology introduced in \cite{balinski/sonmez:99}, 
\cite{switzer:11} also refers to the \textit{no justified envy} axiom as \textit{fairness}.}

Upon reading his thesis, I found the problem fascinating and his critique of the mechanism used at West Point---henceforth the \textit{USMA-2006 mechanism}---to be valid.
Little did I know that, building on the use-inspired theory in \cite{balinski/sonmez:99}, his thesis would mark a key step in yet another ``discovery--invention cycle'' (see Figure \ref{fig:DIC-SC-arm}).

Despite my enthusiasm, however, I questioned the practical viability of his proposed alternative mechanism.
At the time, I had just learned about the school choice reforms in England and Chicago and had begun systematically exploring potential applications of institutional redesign with a minimalist mindset.
My skepticism toward Switzer’s proposal stemmed from its departure from the principles of minimalist market design.

The starting point of \cite{switzer:11} aligns well with minimalist market design. 
Switzer begins by identifying several key policy objectives for the Army and demonstrates that the USMA-2006 mechanism falls short in achieving some of them. 
However, when designing an alternative mechanism, Switzer deviates from the minimalist approach by altering a policy parameter unrelated to the shortcomings of the USMA-2006 mechanism, 
as I will discuss further in Section \ref{sec:Switzerproposedreform}. 

From a purely academic standpoint, there is nothing wrong with this methodology. In fact, such policy recommendations are common under other economic design paradigms.
However, in my view, this route posed a significant obstacle to the practical relevance of Switzer’s proposal.

Interpreting the proposed mechanism as interfering with an Army policy choice unrelated to the underlying shortcomings Switzer identified, I made the following point in my email reply:

\begin{quote} 
``If I were to design a mechanism to correct the deficiency, I would have suggested exactly what you did in terms of the cadet preferences, 
but I would have kept the two types of positions separate. For the first type of positions, the priority would be based on cadet merit ranking, and for the second type of position, 
the priority would be adjusted based on whether the cadet is willing to pay the additional price or not. I think this proposal would have been close' to the current algorithm, 
and it would have still handled the deficiencies that you pointed out. In a way, I felt you adjusted the mechanism a little too much.'' 
\end{quote}

Later, Switzer informed me that his proposal had indeed been dismissed by the relevant Army officers. The failures of the USMA-2006 mechanism identified in \cite{switzer:11}, 
combined with my alternative proposal---the \textbf{\textit{Multi-Price Cumulative Offer (MPCO) mechanism}}---led to a collaboration that resulted in \cite{sonmez/switzer:13}.\footnote{MPCO mechanism 
is referred to as \textit{cadet-optimal stable mechanism (COSM)} in \cite{sonmez/switzer:13}.}

\subsection{BRADSO Incentive Program and the USMA-2006 Mechanism} \label{sec:history-BRADSO}

Switzer’s research strategy in his thesis, which followed the basic structure of \cite{balinski/sonmez:99}, was sensible given the many parallels between the U.S. Army’s branching process and the 
Turkish student placement system (cf. Sections \ref{sec:SP-Turkey}--\ref{subsec:failed-Turkish-policyeffort}).
The Army’s branching process assigns USMA cadets to military specialties---known as \textit{branches}---during their senior year.
Branching across the 17 specialties (e.g., Aviation, Infantry, Military Intelligence, etc.) plays a central role in shaping a cadet’s career trajectory.

Each fall, the Army announces the number of positions available to USMA cadets at each branch. Prior to 2006, the allocation of these positions was implemented with a \textit{simple serial dictatorship} (SSD) 
induced by a cadet performance ranking called the \textit{Order of Merit List} (OML). Let's refer to this mechanism as \textbf{\textit{SSD-OML}}.

Under SSD-OML, cadets submit their preference rankings for branches, and the highest OML-ranked cadet receives their first-choice branch.
The next highest-ranked cadet receives their top choice among the remaining positions, and so on. 
Thus, cadet claims over positions at Army branches depended solely on their performance rankings prior to 2006---a practice that makes sense given the importance of hierarchy in the Army.
In addition, because of the importance of trust both among cadets and between cadets and the USMA leadership, another key advantage of this mechanism was its \textit{strategy-proofness}.
Naturally, a system that allows cadets to be outmaneuvered by others or by the system itself would not foster trust among its participants.

Starting with the Class of 2006, the Army introduced a second pathway for cadets to access their preferred branches in order to address historically low retention rates among junior officers.
This initiative allowed cadets to receive elevated priority for a portion of positions in any branch they selected, provided they committed to an additional three years of \textit{Active Duty Service Obligation} (ADSO) beyond the standard requirement.
At USMA, the standard ADSO is five years. The incentive program became known as the \textit{Branch-of-Choice Active Duty Service Obligation}, or simply the \textit{BRADSO} program.

Given the opposition from many senior Army officers to interfering with a system that allocates positions solely based on performance ratings, this second pathway to branch access was limited, 
as a compromise, to 25\% of the positions in each branch at USMA.\footnote{\label{footnote-USMA}See, for example, the following quote from Appendix E of \cite{Colarusso/Wardynski/Lyle:2010}: 
\begin{quote}
``The branch and post incentives also raised concerns. Devoted supporters of the ROTC and West Point Order of Merit (OML) system for allocating branches and posts objected that low OML cadets could 'buy' their branch or post of choice ahead of higher OML cadets. Since branch and post assignments represent a zero-sum game, the ability of cadets with a lower OML ranking to displace those above them was viewed by some as unfair or as undermining the OML system. However, rather than undermining the legacy system or creating inequities, the branch and post incentives program makes willingness to serve a measure of merit in branching and posting, thus providing taxpayers a fair return on their officer accessions investment."
\end{quote}} 
Priority for the remaining 75\% of the positions at each branch continued to be based on the OML.

Naturally, adopting the BRADSO program required the Army to modify its branching mechanism. The core idea was to extend their original mechanism, SSD-OML, 
which allocates all positions based on a single performance ranking, to accommodate two criteria. The Army's modification was simple.

\paragraph{USMA-2006 Mechanism.}
Cadets rank their branch preferences and specify the branches in which they would invoke BRADSO incentives to obtain preferential access to the final 25\% of positions within those branches.
Based on the submitted profile of cadet strategies, a procedure similar to SSD-OML is used to allocate positions. However, unlike SSD-OML, once the initial 75\% of positions in a given branch are assigned, 
a second criterion is applied to allocate the remaining positions. A cadet who receives a position in a given branch---say, Infantry---is charged an additional three years of ADSO 
only if (i) they are assigned one of the last 25\% of Infantry positions and (ii) they had volunteered for the Infantry BRADSO incentive.

As natural as this modification sounds, it has several failures identified in \cite{switzer:11}. 
Most notably, cadets could receive less-preferred outcomes than some of their peers despite having higher OML rankings, merely because they volunteered for the BRADSO incentive in their assigned branches. 
Formally, the USMA-2006 mechanism fails to satisfy the following axiom.\footnote{For the special case of student placement, this axiom reduces to \textit{no justified envy}. 
Following the terminology in \cite{balinski/sonmez:99}, this axiom is also referred to as \textit{fairness} in \cite{switzer:11} and \cite{sonmez/switzer:13}.}

\begin{definition}[\citealp{switzer:11, sonmez/switzer:13}] \label{def:NPR}
An assignment of branches to cadets along with their ``prices" (i.e., ADSO charges) satisfies \textbf{no priority reversal} if there is no branch $b$ and two distinct cadets $c$, $d$ such that:
\begin{enumerate}
\item Cadet $d$ is assigned a position at branch $b$ at some price $t$.
\item Cadet $c$ strictly prefers the branch-price pair $(b, t)$ to their own assignment.
\item Cadet $c$ has a higher OML ranking than cadet $d$.
\end{enumerate}
\end{definition}

The failure of \textit{no priority reversal} under the USMA-2006 mechanism reflects an inability to properly integrate two allocation criteria into the design. 
For the same reason, the USMA-2006 mechanism also creates incentives to game the system, either by misreporting branch preferences or by concealing their willingness to volunteer for BRADSO incentives.

The root cause of the USMA-2006 mechanism’s failures is twofold. First, the mechanism's strategy space is not rich enough to fully capture cadet preferences. 
For example, if a cadet opts for the BRADSO incentive for their first-choice branch, it is unclear whether they would actually prefer their first-choice branch at the 
higher price over their second-choice branch at the lower, base price---or vice versa. Despite this ambiguity, the USMA-2006 mechanism implicitly assumes that a cadet who volunteers for 
BRADSO for their first-choice branch always prefers a higher-priced position there over a base-price position in their second choice. 
This assumption is embedded in the mechanism’s process, which considers each cadet for all positions in their first-choice branch---including higher-priced options---before moving to base-price positions in their second choice.

Fortunately, addressing this issue is relatively straightforward with a simple adjustment to the strategy space. Instead of asking cadets to submit their preferences over 
branches along with the set of branches for which they volunteer for the BRADSO incentive, cadets could be asked to submit their preferences over branch-price pairs. This minimalist adjustment would resolve the issue.

The second source of failure is more subtle. A cadet who volunteers for the BRADSO incentive for a branch pays the increased price whenever they receive one of the last 25\% of its positions, 
even if they would still have received this position at the base price without volunteering for the BRADSO incentive. From a technical perspective, addressing this second issue is more challenging.

\subsection{Switzer's Reform Proposal for the USMA-2006 Mechanism} \label{sec:Switzerproposedreform}

To address the first root cause of failures in the USMA-2006 mechanism, \cite{switzer:11} modifies the strategy space as discussed above so that cadets submit their preferences for branch--price pairs.

To tackle the more complex second root cause, \cite{switzer:11} introduces a priority ranking of cadet--price pairs for each branch:

\begin{enumerate} 
\item Any pair with an increased price has higher priority than any pair with the base price, regardless of which cadets are involved. 
\item Otherwise, the relative ranking between two pairs with the same price is determined by the OML. 
\end{enumerate}

Under this priority structure, the outcome function of Switzer's proposed mechanism follows a procedure reminiscent of the SSD-OML. 
However, unlike SSD-OML, cadets submit their applications to branches along with their proposed prices. Branches then apply Switzer's priority structure to evaluate these applications. 
Because a cadet may rank the same branch at both the base and the increased price, a branch may reject a cadet tentatively admitted at the base price in favor of another cadet applying at the increased price. 
This introduces an iterative structure similar to the \cite{mcvitie/wilson:70} variant of the deferred acceptance algorithm, in which proposals proceed sequentially rather than simultaneously as in Gale and Shapley’s original version.

It is clear that Switzer's procedure terminates in a finite number of steps, and the resulting mechanism satisfies both \textit{no priority reversal} and \textit{strategy-proofness}, 
effectively addressing the shortcomings of the USMA-2006 mechanism. From a research standpoint, his design is well-behaved. However, from a policy perspective, the proposal may disrupt some delicate balances.

Under Switzer's proposed priority structure, all units within a given branch use the same criteria for allocation, with price always being the primary consideration. 
This design choice was the source of my skepticism regarding its viability for the Army. As discussed in Section \ref{sec:history-BRADSO}, 
the Army’s policy maintained a delicate balance by granting BRADSO volunteers elevated priority for only a fraction of available positions under the USMA-2006 system, not all. 
Although I didn’t know the exact rationale at the time, I strongly believed the limitation wasn’t arbitrary.\footnote{As evidenced by Footnote \ref{footnote-USMA}, this hunch turned out to be correct.} 
For this reason, I viewed Switzer’s design as a potential ``deal-breaker'' and chose to refine it by incorporating both criteria, as the Army had done. 
In my view, while Switzer's proposal successfully addressed some critical shortcomings of the USMA-2006 mechanism, it would likely introduce new ones.

\subsection{Leveraging Matching with Contracts  for a Minimalist Reform} \label{sec:matching-with-contracts}

I found myself wondering why Switzer did not incorporate the Army's dual-criteria priority structure into his proposed mechanism. 
After all, the remaining elements of his proposal could still have been used with this modification, making it more closely aligned with the USMA-2006 mechanism.

After some reflection, I made a number of observations that put Switzer’s proposal into perspective. Incorporating the Army’s dual-criteria priority structure makes the mechanism 
significantly more complex to analyze than the single-criterion version. By contrast, under Switzer’s constructed priority structure, the analysis becomes more manageable.

As illustrated in Figure \ref{fig:DIC-SC-arm}, the key to advancing the ongoing ``discovery--invention cycle''---which expanded through Switzer’s thesis 
and ultimately culminated in a reform of the Army’s branching process---was a series of breakthroughs in basic theory from the mid-2000s to the early 2010s.

The Army's branching process, as studied in \cite{switzer:11}, can be framed as a special case of the celebrated \textit{matching with contracts} model by \cite{hatfield/milgrom:05}, 
a generalization of Gale and Shapley's many-to-one matching model discussed earlier in Section  \ref{sec:two-sided-matching}.
In their generalization, Hatfield and Milgrom allow agents (e.g., students or cadets) and institutions (e.g., colleges or branches) to be matched under multiple arrangements, introducing the concept of a \textit{contractual term} between them. 
Each agent’s preferences are expressed as preferences over contracts---triples that specify the agent, an institution, and the contractual term. 
Each institution, in turn, evaluates its set of contracts using a \textit{choice rule}, a function that selects a set of feasible contracts for the institution from any given set of contracts.

\cite{hatfield/milgrom:05} assume that each institution's choice rule satisfies the following two technical  conditions.\footnote{The assumption of the IRC condition in Definition \ref{def:IRC} is implicit in 
\cite{hatfield/milgrom:05}; see \cite{aygun/sonmez:13} for details.}

\begin{definition}
A choice rule $C(\cdot)$ satisfies \textbf{substitutability} if, whenever a contract is chosen from a set of contracts, it continues to be chosen from every subset obtained by removing other contracts from that set.
\end{definition}

\begin{definition} \label{def:IRC}
A choice rule $C(\cdot)$ satisfies the \textbf{independence of rejected contracts} (IRC) condition if removing an unselected contract from the set of available contracts does not change the set of selected contracts.
\end{definition}

Under these assumptions, \cite{hatfield/milgrom:05} generalized Gale and Shapley's celebrated individual-proposing deferred acceptance algorithm through the following iterative algorithm.  

\paragraph{Cumulative Offer Process.}
At each step, following any given priority ranking of agents---which is immaterial to the final outcome under the substitutability and IRC conditions---the highest-priority agent without a contract ``on hold'' 
proposes their most-preferred contract that has not yet been rejected to the institution associated with that contract. The institution then reviews all offers it has received up to that point, 
including those it previously rejected, retains the best set of contracts based on its choice rule, and rejects the rest. The process ends when no new proposal is made. 

As long as all choice rules satisfy the substitutability and IRC conditions, the algorithm ensures that no previously rejected contract is ever held at a later step, thereby guaranteeing termination in a finite number of steps.

Theorem 4 in \cite{hatfield/milgrom:05} implies that, as long as all choice rules satisfy substitutability and IRC conditions, the \textit{cumulative offer mechanism}---the direct mechanism
that uses the cumulative offer process---satisfies the \textit{no priority reversal} axiom. Moreover, as long as the choice rules also satisfy the following 
additional condition, Theorem 11 in \cite{hatfield/milgrom:05} ensures that the cumulative offer mechanism is also \textit{strategy-proof}.

\begin{definition}
A choice rule $C(\cdot)$ satisfies the \textbf{law of aggregate demand} (LAD) if, whenever the set of available contracts expands, the number of contracts selected weakly increases.
\end{definition}

Critically, not only do the branch choice rules satisfy all three conditions under Switzer’s constructed priority structure, 
but the outcome function of his proposed mechanism also reduces to the \textit{cumulative offer process}. 
Although Switzer was unaware of this connection at the time, his mechanism is a direct application of \cite{hatfield/milgrom:05}.

The analysis becomes more complex under the Army’s dual-criteria priority structure in the USMA-2006 mechanism because the resulting branch choice rules no longer satisfy substitutability, 
placing the application beyond the technical scope of \cite{hatfield/milgrom:05}.

To illustrate, consider a set of contracts that includes both the base-price and increased-price options for a cadet $c$. 
A branch might select the increased-price contract and reject the base-price one, but in a smaller set with less competition it could reverse this decision, instead choosing the base-price contract and rejecting the increased-price one.

In addition to falling outside the scope of \cite{hatfield/milgrom:05}, the Army’s dual-criteria priority structure also lies beyond the reach of Switzer's proposed mechanism, 
since his solution relies on a uniform priority ranking of cadet--price pairs for all positions within a branch.

When I received Switzer’s email in May 2011, I initially thought the Army's branching process was a direct application of \cite{hatfield/milgrom:05}. 
This would have been an exciting development for theory, as it would mark the first real-world application of the matching-with-contracts model, beyond its special case of Gale and Shapley’s many-to-one matching without contractual terms. 
At first glance, the cumulative offer process appeared to work seamlessly with the Army’s dual-criteria priority structure, and the resulting mechanism seemed to address the shortcomings of the USMA-2006 mechanism.

However, to my surprise, I soon realized that the choice rules induced by the Army's dual-criteria priority structure did not satisfy the substitutability condition. 
How could the cumulative offer process still function properly under these circumstances? Something didn’t add up.

Fortunately, recent research in pure theory by \cite{hatfield/kojima:2008, hatfield/kojima:10} provided the much-needed clarification. 
Contrary to a claim in \cite{hatfield/milgrom:05}, \cite{hatfield/kojima:2008} showed that the substitutability condition is not necessary for a well-defined and well-behaved cumulative offer process. 
Moreover, thanks to the important generalization of \cite{hatfield/milgrom:05} developed by \cite{hatfield/kojima:10}, the cumulative offer mechanism continues to satisfy both \textit{no priority reversal} and \textit{strategy-proofness}, 
as long as the choice rules satisfy the following weaker substitutability condition, along with the IRC and LAD conditions.

\begin{definition} 
A choice rule $C(\cdot)$ satisfies the \textbf{unilateral substitutability} condition if, whenever the only available contract of a cadet is chosen from a set of contracts, 
it continues to be chosen from every subset obtained by removing other contracts from that set.
\end{definition}

Although the choice rules induced by the Army’s dual-criteria priority structure fail to satisfy the substitutes condition, they satisfy the milder unilateral substitutability condition as well as IRC and LAD, 
thus ensuring that the cumulative offer process is well-defined and well-behaved for the Army’s branching process.

\paragraph{Multi-Price Cumulative Offer (MPCO) Mechanism.}
Building on this foundation, and adopting the cumulative offer process under the Army’s dual-criteria priority structure, 
the MPCO mechanism modifies USMA-2006 by introducing a modest adjustment to the strategy space and a natural adaptation of the outcome function. 
These changes resolve the shortcomings of USMA-2006---crucially, without altering the Army’s original priority structure.
As a minimalist enhancement of the USMA-2006 mechanism,
\cite{sonmez/switzer:13} proposes the MPCO for the Army.\footnote{Alternatively, 
the MPCO mechanism can be constructed as a descending salary adjustment process in the \cite{kelso/crawford:82} model---a key precursor to \cite{hatfield/milgrom:05}---as shown by \cite{jagadeesan:19}.}

\begin{figure}[!tp]
 \centering
       \includegraphics[scale=1.0]{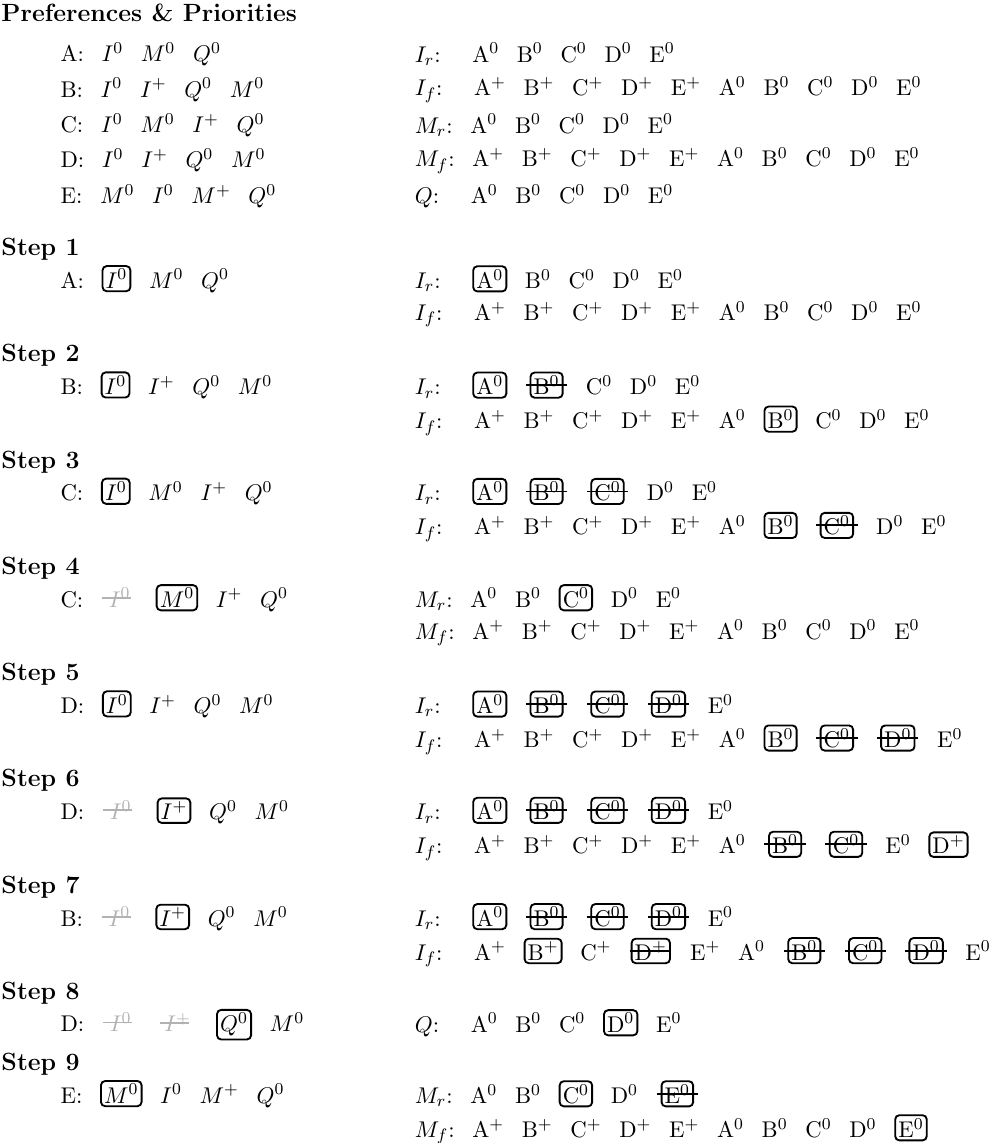}
\caption{Execution of the MPCO mechanism in Example \ref{ex:MPCO}.
At each step, the highest OML-ranked applicant makes a proposal to their highest-ranked un-rejected contract.
Offers are shown in boxes, while earlier proposals that have been rejected are shown in faint gray and struck out with a faint gray line.
For each branch that receives a new proposal, cumulative offers to each slot remain boxed, with rejected ones struck out.} \label{fig:MPCO}
\end{figure}

We next execute the MPCO mechanism and the cumulative offer process using an example, as depicted in Figure~\ref{fig:MPCO}.

\begin{example} \label{ex:MPCO}
There are five cadets---Alp, Banu, Cora, Diya, and Ezra---with earlier names in the alphabet corresponding to higher rankings in the Order of Merit List (OML).
There are three branches: Infantry ($I$), Military Intelligence ($M$), and the Quartermaster Corps ($Q$), with two positions each in Infantry and Military Intelligence, and one in the Quartermaster Corps.

The first positions in Infantry and Military Intelligence---denoted as $I_r$ and $M_r$---and the only position in the Quartermaster Corps
are offered only at the base level of Active Duty Service Obligation (ADSO) (also called \textit{base price}) and the priorities for these positions are determined solely by the OML.
These are called \textit{regular} positions.

The second positions in the Infantry and Military Intelligence---denoted as $I_f$ and $M_f$---are called \textit{flexible ADSO} (or \textit{flexible-price}) positions.
As part of the BRADSO program, these flexible-price positions give priority access to cadets who indicate a willingness to extend their ADSO
by three years and rely on the OML among cadets with the same ADSO willingness.
These flexible-price positions may be awarded with either the base or the extended ADSO.

For the Quartermaster Corps, the choice rule is based solely on the OML, admitting the highest-ranking applicant.
For branches with both regular and flexible-price positions---Infantry and Military Intelligence---the choice rules rely on the slot-specific priorities indicated above, 
filling first the regular position and then the flexible-price position. 

For any cadet and branch, let the superscript $^0$ indicate the contract with the base ADSO, and the superscript $\adsoplus$ indicate the contract with the extended ADSO.
For example, a contract between Alp and Infantry at the base ADSO is written as  $I^0$ in Alp’s preferences (or when he proposes it), and as A$^0$ 
in the priority ordering of an Infantry position (or when Infantry receives it as a proposal).

Using this notation, cadet preferences and the slot-specific priorities of the positions over acceptable contracts are given as follows:

{\small
\[
\begin{array}{llll}
\mbox{A}:  & I^0 - M^0 - Q^0  \qquad \quad \; \;  & I_r: & \mbox{A}^0 -  \mbox{B}^0 -  \mbox{C}^0 -  \mbox{D}^0 - \mbox{E}^0\\
\mbox{B}: & I^0 - I^+ - Q^0 - M^0  & I_f: & \mbox{A}^+ -  \mbox{B}^+ -  \mbox{C}^+ -  \mbox{D}^+ - \mbox{E}^+ - \mbox{A}^0 -  \mbox{B}^0 -  \mbox{C}^0 -  \mbox{D}^0 - \mbox{E}^0\\
\mbox{C}:  & I^0 - M^0 - I^+ - Q^0  \qquad \quad \; \;  & M_r: & \mbox{A}^0 -  \mbox{B}^0 -  \mbox{C}^0 -  \mbox{D}^0 - \mbox{E}^0\\
\mbox{D}: & I^0 - I^+ - Q^0 - M^0  & M_f: & \mbox{A}^+ -  \mbox{B}^+ -  \mbox{C}^+ -  \mbox{D}^+ - \mbox{E}^+ - \mbox{A}^0 -  \mbox{B}^0 -  \mbox{C}^0 -  \mbox{D}^0 - \mbox{E}^0\\
\mbox{E}: & M^0 - I^0 - M^+ - Q^0  \qquad \quad \; \;  & Q: & \mbox{A}^0 -  \mbox{B}^0 -  \mbox{C}^0 -  \mbox{D}^0 - \mbox{E}^0
\end{array}
\]
}

We now process the cumulative offer process, at each step considering the highest-OML cadet without a contract held by any branch.

\textit{Step 1.} Alp proposes to Infantry his base ADSO contract A$^0$. 
Infantry starts its evaluation with its regular position $I_r$ and holds the offer for that position.

\textit{Step 2.} Banu proposes to Infantry her base ADSO contract B$^0$. 
Infantry starts its evaluation with its regular position $I_r$, considers it together with A$^0$, 
and rejects it for this position, continuing to hold A$^0$. 
It then evaluates B$^0$ for its flexible-price position $I_f$ and holds it for that position.

\textit{Step 3.} Cora proposes to Infantry her base ADSO contract C$^0$. 
Infantry starts its evaluation with its regular position $I_r$, considers it together with its earlier cumulative offers A$^0$ and B$^0$, 
and rejects it for this position along with B$^0$, holding A$^0$. 
It then evaluates C$^0$ for its flexible-price position $I_f$ together with B$^0$ and rejects it, continuing to hold B$^0$.

\textit{Step 4.} Cora proposes to Military Intelligence her base ADSO contract C$^0$. 
Military Intelligence starts its evaluation with its regular position $M_r$ and holds the offer for that position.

\textit{Step 5.} Diya proposes to Infantry her base ADSO contract D$^0$. 
Infantry starts its evaluation with its regular position $I_r$, considers it together with its earlier cumulative offers A$^0$, B$^0$, and C$^0$, 
and rejects it for this position along with B$^0$ and C$^0$, holding A$^0$. 
It then evaluates D$^0$ for its flexible-price position $I_f$ together with B$^0$ and C$^0$ 
and rejects it along with C$^0$, continuing to hold B$^0$.

\textit{Step 6.} Having just been rejected by Infantry with her base ADSO contract, 
Diya next proposes to Infantry her extended ADSO contract D$\adsoplus$. 
Ineligible for the regular position $I_r$, Infantry evaluates it for the flexible-price position $I_f$ together with its earlier cumulative offers B$^0$, C$^0$, and D$^0$, 
holding D$\adsoplus$ and rejecting the others, including B$^0$, which it had been holding since Step 2.

\textit{Step 7.} Having been rejected by Infantry with her base ADSO contract after being on hold for several steps, 
Banu next proposes to Infantry her extended ADSO contract B$\adsoplus$. 
Ineligible for the regular position $I_r$, Infantry evaluates it for the flexible-price position $I_f$ together with its earlier cumulative offers B$^0$, C$^0$, D$^0$, and D$\adsoplus$, 
holding B$\adsoplus$ and rejecting the others, including D$\adsoplus$, which it had been holding in the previous step.

\textit{Step 8.} Having been rejected by Infantry with her extended ADSO contract as well, 
Diya next proposes to the Quartermaster Corps her base ADSO contract D$^0$. 
The Quartermaster Corps holds it for its only position.

\textit{Step 9.} Ezra proposes to Military Intelligence his base ADSO contract E$^0$. 
Military Intelligence starts its evaluation with its regular position $M_r$, considers it together with C$^0$, 
and rejects E$^0$, continuing to hold C$^0$. 
It then evaluates E$^0$ for its flexible-price position $M_f$ and, having no offer held there, holds E$^0$.

With no cadet rejected in Step 9, the process ends. The contracts on hold are finalized: Alp receives the regular position at Infantry with the base ADSO,
Banu receives the flexible-price position at Infantry with the extended ADSO, Cora receives the regular position at Military Intelligence with the base ADSO,
Diya receives the position at Quartermaster Corps with the base ADSO, and Ezra receives the flexible-price position at Military Intelligence with the base ADSO.
\end{example}

\subsection{Failed Efforts to Reform the USMA-2006 Mechanism}

Despite several attempts with various officials, our efforts to collaborate with the Army on redesigning their branching mechanism were unsuccessful for many years. 
During this time, my long-time collaborator Parag Pathak began incorporating my research on Army branching into his graduate Market Design course at MIT and independently advocated for it to USMA officials.

In 2019, through Pathak, I finally learned why the Army initially chose not to pursue a reform of the USMA-2006 mechanism. 
I later came to interpret this explanation as the Army’s official justification for its decision. In retrospect, this is unsurprising: 
the Army was unlikely to adopt unsolicited reform proposals from outside researchers without a strong internal rationale and its own process of institutional vetting.

As discussed in Section \ref{sec:history-BRADSO}, the first root cause of the USMA-2006 mechanism's failures is that its strategy space is not sufficiently 
rich to capture all cadet preferences over branch-price pairs. However, this same limitation made any failure of the \textit{no priority reversal} axiom less visible to the Army. 
For example, when a cadet receives their first-choice branch at the increased price but would prefer their second choice at the base price, this information was simply unavailable within the USMA-2006 strategy space. 
Moreover, even if cadets experienced such outcomes, they lacked grounds to object.

By contrast, the second root cause had more tangible consequences. Some cadets unnecessarily paid the increased price for their assigned branch under the USMA-2006 mechanism, 
even though they would have received the same branch at its base price without volunteering for BRADSO incentives. This issue was more apparent, 
as affected cadets could observe that lower OML-ranked cadets received the same branch at the base price. 
However, this challenge could be managed through a simple manual adjustment by waiving the additional ADSO charge for any affected cadet.

Thus, anomalous outcomes under the USMA-2006 mechanism were either obscured by its strategy space or could be corrected manually, and they affected only about 2\% of cadets on average \citep{greenberg/pathak/sonmez:24}. 
Much like the case of Turkey, where the benefits of replacing MCSD with DA were not deemed large enough to justify reform, 
the benefits of replacing the USMA-2006 mechanism with the MPCO mechanism were likewise judged insufficient. 
Regardless of the merits of the alternative, the Army’s mechanism at the time was simply not seen as ``bad enough'' to prompt change.

\subsection{ROTC Mechanism} \label{sec:ROTC}

The U.S. Army relies on two main programs to recruit officers: USMA at West Point and the Reserve Officer Training Corps (ROTC). 
Like USMA, ROTC adopted the BRADSO incentive program in the mid-2000s and changed its branching mechanism. However, ROTC's implementation of BRADSO incentives differed in several ways.

At USMA, the standard Active Duty Service Obligation (ADSO) is five years, whereas for ROTC it is four years for scholarship cadets and three years for non-scholarship cadets. 
Similar to USMA, ROTC's BRADSO program awarded cadets increased priority for a fraction of positions in each branch if they volunteered for an additional three years of ADSO beyond the standard requirement. 
However, while USMA applied BRADSO incentives to 25\% of the positions in each branch and allowed these positions to be awarded at the base price if demand was insufficient,
ROTC applied BRADSO to 50\% of the positions and made them available only at the increased price.

Most critically, ROTC leadership had an additional objective compared to USMA: preventing low-performing cadets from clustering solely in less sought-after branches.

With these specifics in mind, ROTC leadership adopted a mechanism that shares the strategy space of the USMA-2006 mechanism but limits the number of branches cadets may rank. 

\paragraph{ROTC Mechanism.} 
Cadets could rank up to three branches and indicated which, if any, they were willing to pay an increased price for in exchange for elevated priority in half of the positions. 
For the allocation process, ROTC, like USMA, used a procedure similar to SSD-OML but applied three criteria for allocation instead of two. 
Each cadet was evaluated according to their OML standing and placed in the highest-ranked branch for which they qualified, based on the following criteria:

\begin{itemize} 
\item First Criterion: Applied to the first 50\% of positions, where OML standing was the sole factor in allocation.
\item Second Criterion: Applied to the next 15\% of positions, restricted to cadets willing to pay the increased price, with OML determining priority.
\item Third Criterion: Applied to the final 35\% of positions, restricted to cadets in the ``lower half of the OML distribution'' who were willing to pay the increased price, with OML again determining priority.
\end{itemize}

Positions awarded under the first criterion were assigned at the base price, whereas those awarded under the second and third criteria were assigned at the increased price.

Not only does the ROTC mechanism suffer from the shortcomings of the USMA-2006 mechanism, but it also introduces additional issues. 
It fails to satisfy the \textit{no priority reversal} axiom and creates incentives for cadets to game the system by misreporting branch preferences or concealing their willingness for BRADSO incentives. 
The cap of three branches for submitted preferences further exacerbates the incentive for misrepresentation. 
However, the most problematic aspect of the ROTC's BRADSO program is its three-criteria priority system, particularly the third criterion, 
which restricts eligibility for the final 35\% of positions to cadets in the lower half of the OML distribution.

Even within single-branch allocation, this priority system poses significant challenges. Although it aligns with ROTC's OML-based distributional objectives, it inherently produces ``priority reversals,'' 
making these reversals an explicit feature of the system. This visibility, in turn, incentivizes cadets to deliberately lower their OML standing---a phenomenon known as ``tanking.''

In early 2011, the ROTC mechanism piqued my interest, particularly because of its third criterion. Interestingly, the Army had a specific term for the eligibility restriction---the root cause of the 
most visible failures of the mechanism---under this criterion: the \textbf{\textit{Dead Zone}}. The official definition of this phenomenon is provided in the 2011 ROTC Accessions Briefing (\citealp{sonmez:13}, p. 188):

\begin{quote} 
``Dead Zone: The area on the branch bar graph where it is impossible for a cadet to receive a certain branch.'' 
\end{quote}

\begin{figure}[!tp]
    \begin{center}
       \label{fig:ROTC-deadzone-2011}
    \includegraphics[scale=0.47]{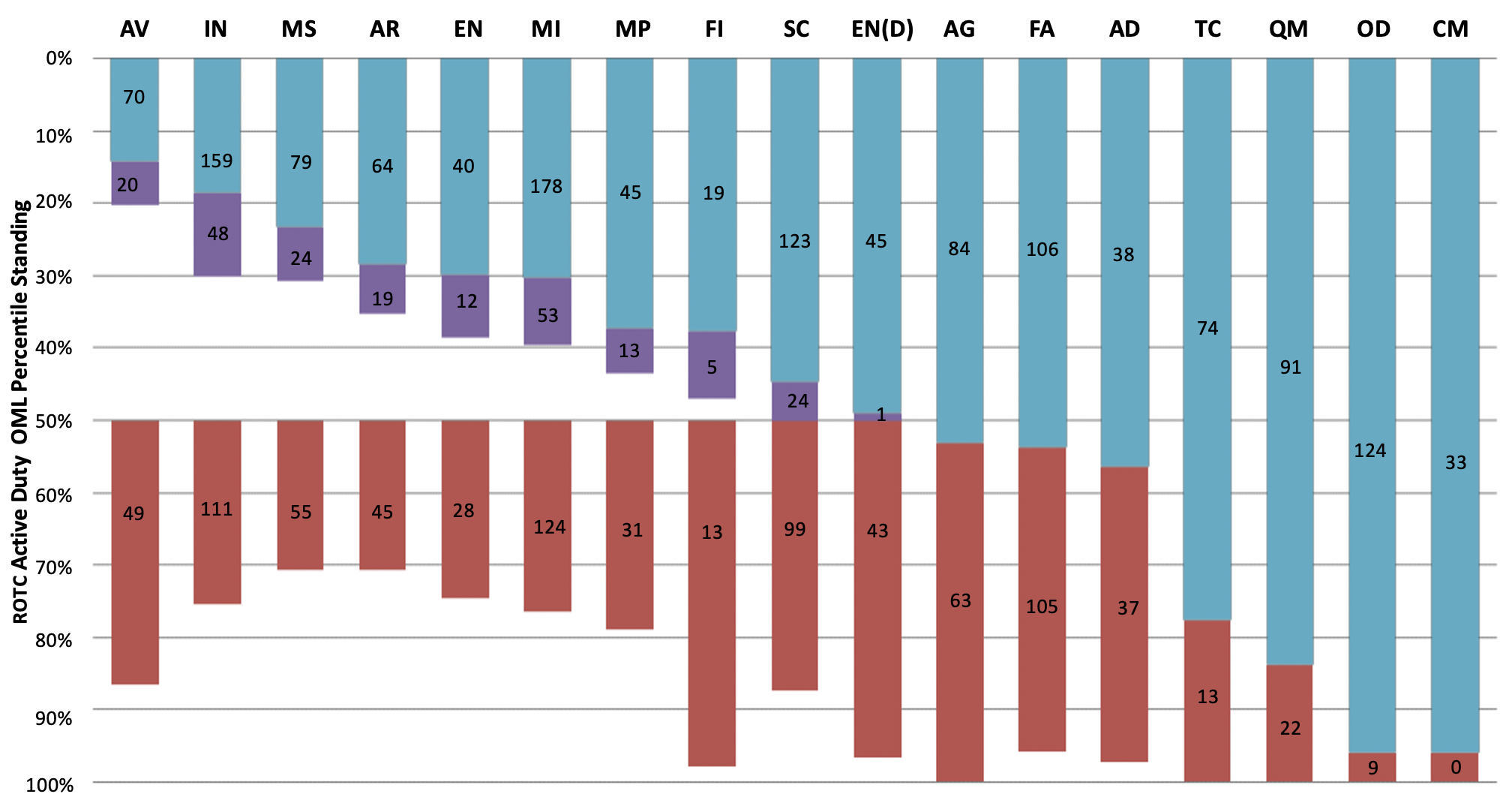}
    \end{center}
    \caption{ROTC branch assignment results for the year 2011. For each branch, 
    the blue region represents the portion of the OML where cadets secured a position within the top 50 percent of positions at the base price. The purple region corresponds to cadets 
    who obtained a position between the 50th and 65th percentiles at the increased price, while the brown region indicates cadets who secured a slot in the last 35 percent, 
    also at the increased price. The numbers in these regions represent the number of cadets in each bracket.
    The blank space between the purple and light brown regions marks the ``dead zone'' for the eight most popular branches: 
    Aviation (AV), Infantry (IN), Medical Service (MS), Armor (AR), Engineering (EN), Military Intelligence (MI), Military Police (MP), and Financial Management (FI). 
    The less sought-after branches, which do not have a dead zone, include Signal Corps (SC), Engineering reserves for those with engineering degrees (EN(D)), 
    Adjutant General’s Corps (AG), Field Artillery (FA), Air Defense Artillery (AD), Transportation Corps (TC), Quartermaster Corps (QM), Ordnance Corps (OD), and Chemical Corps (CM).
    Figure reproduced from Figure 1 in \cite{sonmez:13}, \textit{Journal of Political Economy}, \url{https://doi.org/10.1086/669915}, originally published in grayscale. 
    Reprinted with permission. \textcopyright{} 2013 by The University of Chicago.} \label{ROTC-deadzone-2011} 
\end{figure}

In practice, about half of the branches were susceptible to dead zones. For the most competitive branch, typically Aviation, the dead zone spanned from 20\% to 50\% of the OML distribution. 
Several other combat arms branches entered the dead zone around the 30\% mark of the OML distribution. See Figure \ref{ROTC-deadzone-2011} for the dead zones in 2011.

It soon became evident that the dead zones caused by the third criterion were a much-debated aspect of the ROTC priority system. 
Service Academy Forums were filled with conversations about this problem and the tanking behavior it encouraged.
 A June 2012 discussion, under the heading ``How does ROTC Cadet receive designated branch?'' \citep{ServiceAcademyForums:12}, captures the concern:

\begin{quote} 
``I just went through the accessions process this year. [$\ldots$] DO NOT GET STUCK IN the 30--40\% on the active duty OML. 
Once in those percentiles it becomes mathematically impossible to get a competitive branch (i.e., infantry and aviation).'' 
\end{quote}

A cadet responds:

\begin{quote} 
``OK, that is called the Dead Zone in Branch assignments. Not high enough to get 1st or 2nd Branch choice, but above the 50th percentile line, below which a lot of choice spots are reserved for `bottom halfers'. 
Strange sort of communism within our ranks. OH well, blame Congress.'' 
\end{quote}

Another cadet adds:

\begin{quote} 
``[$\ldots$] These kids in the DEAD ZONE are 20-year-old cadets being faced with a moral dilemma $\ldots$ do your best and kiss your branch choice goodbye, 
or screw up and get your branch choice. I think this is a strange choice to put in front of a young cadet.'' 
\end{quote}

To address the failures of the ROTC mechanism, \cite{sonmez:13} proposed the multi-category cumulative offer (MPCO) mechanism, though in a 
more flexible manner than \cite{sonmez/switzer:13} did for USMA.\footnote{\cite{sonmez:13} refers to the MPCO mechanism as the \textit{Cadet Optimal Stable Mechanism (COSM) under Bid for Your Career (BfYC) Priorities}.}

Since USMA's dual-criteria priority system did not have inherent flaws, \cite{sonmez/switzer:13}, following the principle of minimalist market design, did not interfere with this policy.
Therefore, as in the USMA-2006 mechanism, the proposed reform in \cite{sonmez/switzer:13} involved only two prices: the base price and the increased price.

In contrast, \cite{sonmez:13} intervened in ROTC's three-criteria priority system and proposed a more general version of the MPCO mechanism with multiple prices.
Under this version, priority for the flexible-price positions depended first on the price---with higher prices granting higher priority---and then on the OML. The justification for this more intrusive proposal for ROTC is as follows:

The ``dead zone,'' although highly problematic, served a specific purpose for ROTC: preventing low-performing cadets from clustering exclusively in less sought-after branches.
To achieve this objective, ROTC imposed a quota for higher-achieving cadets---those in the top half of the OML distribution---at each branch, thereby creating the dead zones.

For USMA, addressing the root causes of the failures was relatively straightforward, as the system did not contain deeply flawed elements tied to its key objectives.
This was not the case for ROTC. To remedy certain failures, it was essential to eliminate the dead zones, but this required addressing ROTC's skill-based distributional requirements through alternative means.
I proposed using multiple pricing tiers and potentially increasing the share of flexible-price positions, although other, potentially more effective interventions might also exist.
Ultimately, the effectiveness of my proposal remains an empirical question.

Unlike with the USMA-2006 mechanism, I never discussed my findings on the ROTC mechanism with Army officials.
Interestingly, years later, ROTC also took an interest in the MPCO mechanism, along with USMA.

\subsection{Talent-Based Branching Program and the USMA-2020 Mechanism} \label{sec:USMA-2020}

It took several years and the integration of a second program into USMA's branching system to convince Army officials that the MPCO is indeed the mechanism that best serves the Army's policy objectives.

In 2012, the Army began experimenting with a \textit{Talent-Based Branching} (TBB) program at USMA, introducing heterogeneity into branch priorities.
Under the TBB program, branches rated cadets into one of three tiers: High, Medium, and Low. Until 2019, these ratings were part of a pilot initiative.
For the Class of 2020, the Army decided to integrate the ratings into the branching process, constructing priorities at each branch first by tier and then by OML within the tier.

To avoid undermining the TBB program with BRADSO incentives, cadets who volunteered for BRADSO at a branch received higher priority only within their own tier for that branch.
Since the USMA-2006 mechanism could not accommodate these changes, a new mechanism was designed and implemented in Fall 2019 for the Class of 2020.

\paragraph{USMA-2020 Mechanism.}
Under this new mechanism, which lasted only one year, the Army retained the same strategy space as in previous years: each cadet submitted a ranking of branches and identified any branches for which they 
were willing to pay an additional price in exchange for elevated priority in a subset of positions.
Given a profile of cadet strategies, the Army constructed an adjusted priority order for each branch, incorporating both TBB ratings and willingness to pay the increased price.
The USMA-2020 mechanism then determined branch assignments using Gale and Shapley’s individual-proposing deferred acceptance algorithm, based on the submitted preferences and adjusted priorities.

Regarding the prices---i.e., the ADSO charges---cadets who volunteered for BRADSO incentives at a given branch were charged the increased price, following the reverse-priority order and subject to a cap of 25\% of the positions.
The remaining cadets at the branch were charged the base price.

This was a somewhat rushed design.
In addition to inheriting the two root causes underlying the failures of the USMA-2006 mechanism, the USMA-2020 mechanism faced a third vulnerability:
Even though the number of positions allocated via the BRADSO program was kept at 25\% of the total capacity at each branch, cadets received elevated priority for ``all'' positions in a branch upon volunteering for BRADSO incentives.
While this design choice allowed the heterogeneity in branch priorities to be accommodated using the vanilla version of the deferred acceptance algorithm,\footnote{Indeed, 
the following quote from the \textit{U.S. Army News} story \cite{OConnor:2019} suggests that the design of the USMA-2020 mechanism was inspired by the NRMP's medical match:
``The cadets' branch rankings and the branches' cadet preferences will then determine a cadet's branch using a modified version of the National Resident Matching Program's algorithm,
which won a Nobel Prize for Economics in 2012 and pairs medical school graduates with residency programs.''}
it also resulted in a more problematic version of the no priority reversal failures that could not be manually corrected.

To make matters worse, the same issue also introduced a second type of incentive compatibility failure. Not only could volunteering for BRADSO incentives hurt cadets, 
as it did under the USMA-2006 mechanism, but under the USMA-2020 mechanism, it could also benefit them by granting elevated priority at a branch without requiring them to pay the increased price. 
This resulted in a mechanism that was highly complex to navigate and one with more frequent and costly failures.

Even before our formal analysis in \cite{greenberg/pathak/sonmez:24}, USMA leadership had already recognized issues with the USMA-2020 mechanism. 
A significant concern was the potential erosion of cadets' trust in the Army’s branching process, largely due to a visible lack of incentive compatibility, 
particularly for highly coveted branches like Military Intelligence. Additionally, failures of the \textit{no priority reversal} axiom further undermined cadets' confidence, 
especially when these issues could not be manually corrected ex-post.

In response to these concerns, USMA leadership took proactive steps by conducting a dry run of the USMA-2020 mechanism. 
As emphasized in a September 2019 \textit{U.S. Army News}  article \cite{OConnor:2019},
the goal of this exercise was to inform cadets about potential branch cutoffs, aiming to reduce uncertainties and enhance \textit{transparency} in the branching process.

\begin{quote}
`` `We're going to tell all the cadets, we're going to show all of them, here's when the branch would have gone out, here's the bucket you're in, here's the branch you would have received if this were for real. 
You have six days to go ahead and redo your preferences and look at if you want to BRADSO or not.'  Sunsdahl said. `I think it's good to be transparent. I just don't know what 21-year-olds will do with that information.' ''
\end{quote}

However, as the quote suggests, even with the dry run, USMA leadership recognized the difficulties cadets faced in optimizing their strategies under the USMA-2020 mechanism. 

These challenges marked a turning point in my years-long efforts to reform the Army's branching process. 
In response, the Army initiated a collaboration between Parag Pathak and myself to redesign the branching mechanism, with Lieutenant Colonel Kyle Greenberg leading the reform efforts at USMA. 
Unlike the USMA-2006 mechanism, which did not present enough issues to justify a costly reform, the failures of the USMA-2020 mechanism warranted an overhaul.
If successful at USMA, the redesign would also be considered for ROTC. 

\subsection{Multi-Price Cumulative Offer Mechanism as a Minimalist Reform}

Over the years, the mission of the Army's branching system has evolved. Initially, reflecting the Army's emphasis on hierarchy, 
the system’s primary role was to allocate positions based solely on merit. With the introduction of the BRADSO incentives program in the mid-2000s, 
a second objective emerged: improving officer retention. Later, the TBB program added a third goal---enhancing the match between talent and specialties.

Maintaining trust in the system became another key objective, a task that was relatively straightforward when merit-based allocation was the sole focus 
but increasingly challenging as additional goals were introduced.

In light of these evolving objectives, this section examines why axiomatic methodology is particularly well suited to addressing these challenges 
and why the MPCO mechanism is uniquely compelling for the Army's mission.

Before 2006, the Army's branching system was an application of a simplified version of the student placement model, where all branches uniformly relied on the OML to prioritize cadets. 
The Army used the SSD-OML mechanism, which, as shown in Proposition \ref{prop:SSD}, is the only mechanism that satisfies \textit{Pareto efficiency} and \textit{no justified envy}. 
Additionally, the SSD-OML mechanism satisfies \textit{strategy-proofness}, making it perfectly suited to this era. The \textit{no justified envy} axiom ensured that merit was respected, while the system's \textit{strategy-proofness} fostered trust.

In addition to merit based on OML, the Army introduced a second criterion for allocation with the BRADSO program, which applied to a fraction of positions in each branch.
Later, with the implementation of the TBB program, the Army made the primary allocation criteria branch-specific. This raises a key question:
What is the proper way to extend the \textit{no justified envy} axiom in light of these changes in the Army's broader mission?

Over the years, this axiom---first formalized in \cite{balinski/sonmez:99}---and its generalizations have been central to a range of real-world reforms, 
contributing significantly to several ``discovery--invention cycles'' illustrated in Figure \ref{fig:DIC-SC-arm}.
The Army's branching mechanism was the second of these applications, following school choice.

The essence of the \textit{no justified envy} axiom is to ensure that position assignments respect the ``property rights'' of individuals as defined by institutions.
The original version of this axiom, presented in Definition  \ref{def:NJE}, applies to settings where all positions within an institution are allocated based on a single priority ranking.
While the \textit{no priority reversal} axiom, given in Definition \ref{def:NPR}, is a direct extension of the \textit{no justified envy} principle and central to the Army's mission,
it addresses cadet objections to branch assignments of others only if they are willing to retain their prices. However, this alone does not fully capture the  full scope of the \textit{no justified envy} axiom in the current setting.

To fully realize the essence of \textit{no justified envy}, cadet objections that may alter the price of an assignment must also be considered. 
We next introduce an axiom that addresses the gap between the ``generalized'' \textit{no justified envy} principle and the \textit{no priority reversal} axiom.

\subsubsection{Price-responsiveness Schemes and Respecting Them} 

The Army's mission evolved over time, leading it to experiment with its second allocation criterion.

As discussed in Section \ref{sec:USMA-2020}, the Army reduced the ``strength'' of the BRADSO incentives under the USMA-2020 mechanism to avoid undermining its new TBB program.
When BRADSO incentives were first introduced with the USMA-2006 mechanism, cadets who volunteered for BRADSO at a branch received higher priority for flexible-price positions than their peers who did not, 
regardless of their base priorities. However, under the USMA-2020 mechanism, the Army restricted this advantage to cadets within the same tier. For example, cadets in the Middle tier could not gain 
higher priority over any cadet in the High tier by volunteering for BRADSO.
This adjustment naturally led to a decrease in the number of ADSO years generated by the branching mechanism.

When the Army began collaborating with our team in late 2019, they expressed interest in exploring alternative policies that would better balance the BRADSO and TBB programs. 
Specifically, they were open to considering new approaches for applying the second criterion to allocate flexible-price positions.

In response, building on \cite{sonmez/switzer:13} and \cite{sonmez:13}, we expanded the analysis of the MPCO mechanism in \cite{greenberg/pathak/sonmez:24}, 
arriving at a surprisingly strong result. The minimalist approach yielded a highly refined solution for the Army's branching process. 
This development not only enhanced the practical relevance of the MPCO mechanism but also pushed the boundaries of matching theory.

Given an institution---a branch in our current setting---and its base priority, \cite{greenberg/pathak/sonmez:24} introduced the concept of a \textit{price-responsiveness scheme}. 
This scheme establishes a priority ranking of all individual-price pairs subject to the following conditions:
\begin{enumerate} 
\item Between two pairs with the same price, the individual with the higher base priority also has the higher priority under the price-responsiveness scheme. 
\item Between two pairs with the same individual, the pair with the higher price has the higher priority under the price-responsiveness scheme. 
\end{enumerate} 
In essence, the price-responsiveness scheme preserves the base-priority ranking of individuals unless there is a difference in price, and a higher price only serves to improve an individual's priority.

This concept is analogous to the \textit{marginal rate of transformation} in producer theory, as it captures the trade-off between base priority and price in the allocation of a subset of positions. 
Our formulation allows for a wide range of criteria to allocate an institution's flexible-price positions, including: 
\begin{itemize}
\item The \textit{ultimate} price-responsiveness scheme, which primarily prioritizes individuals based on the price they pay, using the base-priority ranking only as a tie-breaker.
\item A class of \textit{tiered} price-responsiveness schemes, which prioritize individuals based on both their tiers and the prices they pay.
\item A class of \textit{scoring-based} price-responsiveness schemes (in settings where the base priority ranking is determined by a performance score), 
which grants a fixed score advantage to individuals who pay higher prices.
\end{itemize}

It is straightforward to implement the MPCO mechanism with any price-responsiveness scheme by replacing the second allocation criterion with the desired 
scheme when constructing the branch choice rules used in the mechanism.

What, then, is missing from the \textit{no priority reversal} axiom in terms of fully respecting property rights established through two criteria? 
To explore this, let’s consider an example based on the scoring-based price-responsiveness scheme, inspired by the scenario in Section \ref{sec:pursuit}.

\begin{figure}[!tp]
    \begin{center}
       \includegraphics[scale=1.2]{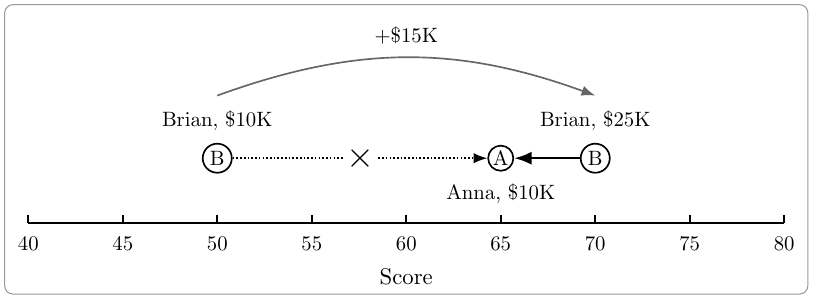}
           \end{center}
\caption{Legitimate claim for a price-reduced version of another individual’s assignment.
The school’s base tuition is \$10K, and seats are allocated based on applicants’ scores on an entry test. For five seats, applicants may obtain a 20-point score boost by paying an additional \$15K, for a total tuition of \$25K.
Suppose that, under the current allocation, fewer than five students pay the higher tuition of \$25K. Anna secures a seat at the base tuition of \$10K with a score of 65, while Brian, with a score of 50, remains unassigned.
Because Brian’s score is lower than Anna’s, he cannot contest her seat at the base tuition (indicated by a crossed-out dotted arrow). However, since not all five flexible-price seats are filled at the higher tuition of \$25K, 
he may contest Anna’s assignment (indicated with a bold arrow) if he is willing to pay the increased tuition.} \label{fig:legitemate-claim-reduced-price}
\end{figure}

\begin{example}  \label{ex:price-responsiveness}
Consider a school with 50 seats and a base tuition of \$10,000. For 45 of the seats, applicant priority is based on a standardized test score out of 100. 
For the remaining 5 seats, priority is determined by an adjusted score, where applicants who pay an increased tuition of \$25,000 gain an extra 20 points added to their test score.

Suppose Anna secures a seat at the base tuition with a score of 65, while Brian remains unmatched. 
The \textit{no priority reversal} axiom only allows Brian to contest Anna’s assignment at the price she paid, in this case, the base tuition. 
If Brian has a score higher than 65, he has a valid claim to Anna’s seat at the same price. Otherwise, he does not.

But what if Brian wants to challenge Anna’s seat by offering to pay the increased tuition instead? The \textit{no priority reversal} axiom is silent on this possibility.

For Brian to have a valid claim to Anna's seat by offering the increased tuition, fewer than 5 seats must have been awarded at the higher price. 
Let’s assume this is the case. Additionally, Brian must have a score higher than 45, since paying the extra tuition only gives him 20 additional points. 
Therefore, if his score is 50, he has a valid claim to Anna’s seat at the increased tuition. However, if his score is 40, he does not, even if fewer than 5 flexible-price seats were awarded at the increased price.
(See Figure \ref{fig:legitemate-claim-reduced-price} for an illustration.)
\end{example}

Example \ref{ex:price-responsiveness} demonstrates that individuals can contest the assignments of others not only at the same prices 
the recipients paid but also at different prices, whether higher or lower. Let’s now formalize these types of challenges within the Army's setting, where two allocation criteria are used.

\begin{definition} \label{def:increased-claim}
Consider an allocation where a cadet $d$ is assigned a position at branch $b$ at price $t$.
Then, cadet $c$ has a \textbf{legitimate claim for a price-elevated version of} cadet $d$'s assignment if there is a higher price $t' > t$ such that:  
\begin{enumerate}
\item Cadet $c$ strictly prefers the branch-price pair $(b, t')$ to their own assignment.
\item The cadet-price pair $(c,t')$ has higher priority under branch $b$'s price responsiveness scheme than the pair $(d,t)$. 
\item Awarding cadet $d$'s seat to cadet $c$ at the increased price $t'$ results in a feasible outcome. 
\end{enumerate}
\end{definition}

\begin{definition} \label{def:reduced-claim}
Consider an allocation where a cadet $d$ is assigned a position at branch $b$ at price $t$.
Then, cadet $c$ has a \textbf{legitimate claim for a price-reduced version of} cadet $d$'s assignment if there is a lower price $t' < t$ such that:  
\begin{enumerate}
\item Cadet $c$ strictly prefers the branch-price pair $(b, t')$ to their own assignment.
\item The cadet-price pair $(c,t')$ has higher priority under branch $b$'s price responsiveness scheme than the pair $(d,t)$. 
\end{enumerate}
\end{definition}

Definition \ref{def:reduced-claim} omits the last condition in Definition \ref{def:increased-claim} because, 
whereas awarding a position at a higher price may be infeasible due to the cap on flexible-price positions, 
awarding a position at a lower price is always feasible.

\begin{definition} \label{def:respecting-PRP}
An allocation \textbf{respects the price responsiveness scheme} if no cadet has a legitimate claim to either a price-reduced or price-elevated version of another cadet's branch-price assignment. 
\end{definition}

The axioms \textit{no priority reversal} and \textit{respecting the price responsiveness scheme} together address all potential objections cadets might have to their peers' branch-price assignments 
within the Army's dual-criteria priority structure. As a result, these two axioms, in conjunction, can be interpreted as a ``generalized'' \textit{no justified envy} axiom in our current setting.

\begin{definition} \label{def:Army-NJE}
An allocation satisfies \textbf{no justified envy} if and only if it satisfies \textit{no priority reversal} and \textit{respects the price responsiveness scheme}. 
\end{definition}

\subsubsection{A Characterization of the MPCO Mechanism} \label{sec:thm-MPCO}

Starting with school choice reforms in Boston, New York City, Chicago, and England in the mid-2000s and early 2010s, the Deferred Acceptance (DA) mechanism
emerged as one of the leading student assignment mechanisms worldwide. Earlier, in Section  \ref{sec:two-sided-matching}, 
Theorem \ref{thm:SP-SOSM-strategy-proofness} established that DA is the unique mechanism that satisfies \textit{individual rationality, non-wastefulness, no justified envy},
 and \textit{strategy-proofness}. Given the prominent role played by the \textit{no justified envy} and \textit{strategy-proofness} axioms in these reforms, 
 this result naturally contributed to the widespread acceptance of DA.

The MPCO mechanism, a generalization of DA for the Army's dual-criteria setting, retains the broad appeal of DA.
In fact, Theorem  \ref{thm:SP-SOSM-strategy-proofness}---which underpinned DA's success---also applies to this setting, demonstrating that the MPCO mechanism is uniquely suited to the Army's branching process.
To present this result, we first generalize the \textit{non-wastefulness} axiom to account for multiple price levels in the current context.

\begin{definition}  \label{def:non-wastefulness-Army}
An outcome satisfies \textbf{non-wastefulness} if no branch $b$ is left with an idle position, unless all cadets prefer their assignments to a position at branch $b$ at the base price.
 \end{definition}
 
 With this adjustment in place, Theorem  \ref{thm:SP-SOSM-strategy-proofness} extends directly:

\begin{theorem}[\citealp{greenberg/pathak/sonmez:24}] \label{thm:MPCO}
The multi-price cumulative offer mechanism is the unique mechanism that satisfies \textit{individual rationality, non-wastefulness, no justified envy}, and \textit{strategy-proofness}. 
\end{theorem}

Even though the MPCO mechanism was initially designed to meet the specific needs of the U.S. Army---where the relevant version was a simplified case involving only two prices---the practical 
relevance of the mechanism and Theorem \ref{thm:MPCO} extends to the most general version of the mechanism in other real-world settings.

A noteworthy application of this framework is a policy implemented across China in major municipalities, including Beijing, Shanghai, and Tianjin, 
from the early 1990s to the mid-2010s, mirroring Example \ref{ex:price-responsiveness}. In this policy, graduating middle school students could boost their exam scores for 
admission to their preferred high schools by opting to pay higher tuition in the centralized public high school admissions process.

This policy, referred to as the \textit{Ze Xiao (ZX)} policy in \cite{wang/zhou:24}, exemplifies a scoring-based price-responsiveness scheme. 
The authors empirically examine this policy in a city that employed an especially elaborate version, incorporating four distinct price levels:

\begin{quote}
``Tuition for public high schools is based on the student’s exam score.
Since scores are the only admission criteria, schools set a cutoff for normally admitted
students. Normal students pay an annual tuition of 1600 yuan (roughly \$260 in 2013),
ensuring affordability. ZX students’ tuition depends on their score: scores within 10
points of the cutoff result in a 3333.3 yuan annual fee; scores 11--20 points above the cutoff
pay 5000 yuan; and scores 21--30 points above the cutoff pay 6000 yuan. No school
can admit a ZX student with a score more than 30 points below its cutoff.'' (p. 1157)
\end{quote}

In a system with a base tuition of 1,600 yuan, this policy corresponds to a scoring-based price-responsiveness scheme, 
offering score boosts of 10, 20, and 30 points, respectively, for increased tuitions of 3,333.3, 5,000, and 6,000 yuan.

\subsection{Policy Impact at USMA and ROTC: Redesigning the U.S. Army's Branching System} \label{sec:USMA-ROTC}

Beginning with the branching of the Class of 2021 cadets to Army branches in Fall 2020, our collaboration with the Army led to the adoption of the MPCO mechanism at USMA, 
incorporating the Army's dual-criteria priority structure \citep{greenberg/pathak/sonmez:24}.
Encouraged by the success of the new system, the Army decided to extend it to ROTC that same year.

Since its inception in 2011, my collaborative research and policy efforts---initially with Switzer and later with Greenberg and Pathak---have systematically followed the principles of minimalist market design.
This intentional and consistent application of minimalist principles, which succeeded in reforming the Army's branching process, further validated the effectiveness of this institutional design paradigm.

\section{Reserve Systems} \label{sec:reserve-systems}

In Sections \ref{sec:schoolchoice} and \ref{sec:Army}, I argue that the failure to satisfy the axioms of \textit{no justified envy} and \textit{strategy-proofness} has been a major catalyst for reforms in both school choice
and the Army's branching system over the past two decades. As a result, many school choice and college admissions systems, as well as the U.S. Army, have abandoned mechanisms that violate these desiderata.
They have instead adopted the DA mechanism or its generalized version, the cumulative offer mechanism, primarily because these mechanisms satisfy them.
As established in Theorems \ref{thm:SP-SOSM-strategy-proofness} and \ref{thm:MPCO}, in many settings these are, in fact, the only mechanisms
that meet these desiderata together with the basic axioms of \textit{individual rationality} and \textit{non-wastefulness}.

In single-institution settings, institutions often determine applicants' priorities---their ``property rights'' over positions---based on criteria such as standardized test results,
demographic characteristics, or the price paid for a position. As discussed in Section \ref{sec:matching-with-contracts}, these criteria can be represented and implemented
through \textit{choice rules} that select a set of collectively feasible contracts from any given set.
In essence, these choice rules represent the property rights of individuals.

Provided that each institution's choice rule satisfies certain technical conditions,
mechanisms like DA or the cumulative offer mechanism can be understood as instruments that extend the implementation of ``property rights'' from single-institution settings to multi-institution settings.
Consequently, exploring the technical conditions that ensure this transition has been an active area of matching theory since the seminal work of \cite{hatfield/milgrom:05}.\footnote{Technical conditions 
such as variants of the substitutability condition and the IIC condition ensure that the algorithms of these mechanisms not only terminate but also generate 
outcomes that satisfy the \textit{no justified envy} axiom, thereby enforcing the underlying property rights in multi-institution settings. 
Additionally, LAC condition ensures that these mechanisms satisfy the \textit{strategy-proofness} axiom, thus enforcing these property rights in an incentive-compatible manner. See \cite{hatfield/kominers/westcamp:21} 
for the mildest known conditions that serve these objectives.}

Up until the early 2010s, despite the theoretical interest in choice rules, our practical design efforts had never interfered with institutional choice rules, unless they were among the root causes of failures in the underlying mechanisms.
The minimalist mindset guiding our holistic approach to both research and policy was a key reason for this stance. 
This remained true until Boston Mayor Thomas Menino's pivotal 2012 State of the City Address, in which he pledged to increase the fraction of students assigned to schools in their neighborhoods.\footnote{One instance 
where a choice rule directly contributed to failures was the ROTC branching mechanism, leading to a proposed mechanism with an alternative choice rule, as discussed in Section \ref{sec:ROTC}.} 
From a pragmatic standpoint, interfering with choice rules would not only deviate from minimalist market design but also risk alienating interest groups adversely affected by our proposals.

This stance was also shaped by an implicit assumption we had held up to that point: that institutions’ choice rules faithfully reflected their intended policies. 
Accordingly, unless technical necessity dictated otherwise, we saw no reason to propose changes that might interfere with those policies. 
Menino’s State of the City Address, however, made clear that this need not always be the case, 
and that designing choice rules to align with an institution’s actual policy objectives may itself constitute an essential exercise in market design \citep{dur/kominers/pathak/sonmez:18}.

\subsection{Subtleties of Reserve Systems} \label{sec:subtleties-reserve}

Consider a common scenario in which an authority allocates multiple identical indivisible goods with unit demand---such as school seats, public jobs, or housing units---based on two criteria. 
The first criterion is a baseline priority ranking determined by factors like standardized test results, application waiting time, or a lottery. 
The second criterion adjusts this ranking by giving preferential treatment to specific groups of applicants, such as underrepresented minorities or individuals with particular qualifications. 
A portion of the available units---the \textit{open} positions---are allocated based solely on the first criterion, while the remaining \textit{reserved} positions are allocated according to the second. 
This process exemplifies a simple two-category version of what is called a \textbf{\textit{reserve system}}.

\begin{figure}[!tp]
 \centering
   \captionsetup{skip=0pt}
       \includegraphics[scale=1.02]{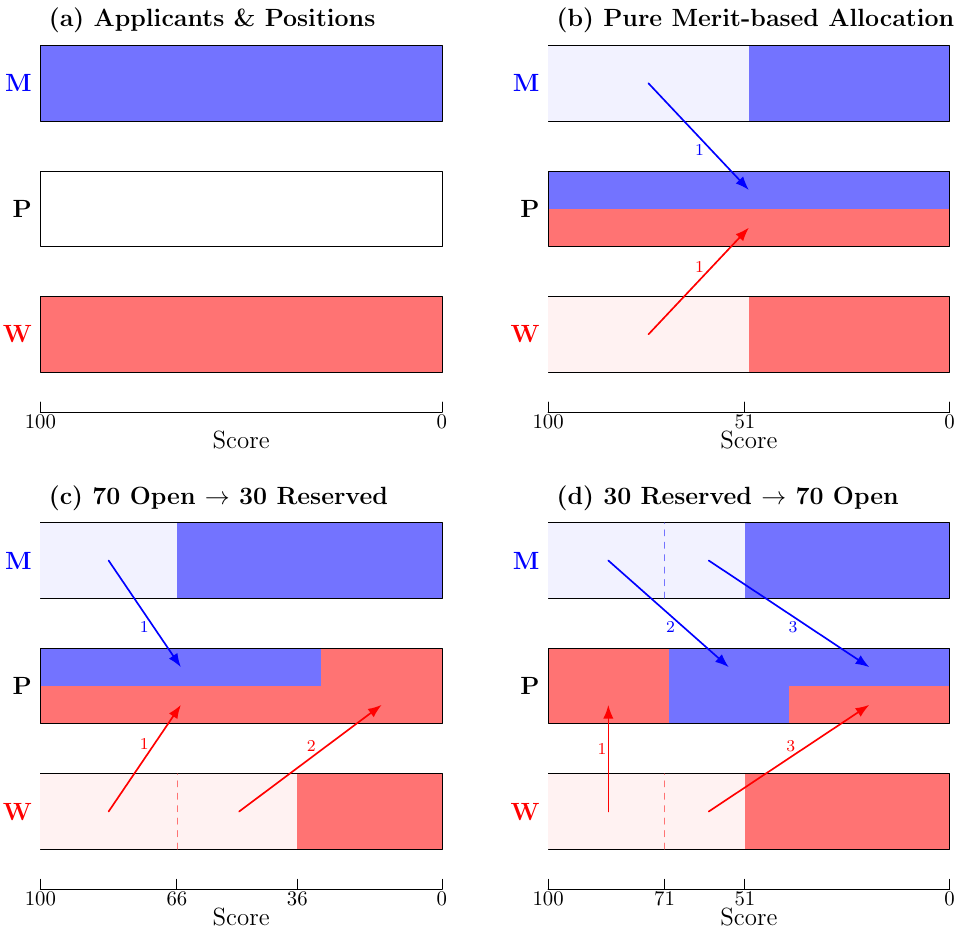}
\caption{Merit-based allocation vs.\ two reserve systems. 
Panel (a) shows a continuum approximation of the setup: the top blue box represents 100 men (M), and the bottom red box represents 100 women (W), each evenly distributed across test scores from 1 to 100 (with higher scores on the left).
The middle empty box represents the 100 available positions (P).
Panel (b) illustrates merit-only allocation: all 100 open positions go to the top scorers (scores $\geq 51$), giving 50 positions to men and 50 to women. 
Panel (c) allocates open positions before reserved ones: the 70 open positions go to the top-scoring individuals, namely 35 men and 35 women (scores $\geq 66$). 
The 30 reserved positions then go to the next highest-scoring women (scores 65--36), giving 65 positions to women and 35 to men in total. 
Panel (d) allocates reserved positions before open ones: the first 30 reserved positions go to the highest-scoring women (scores 100--71). With no women left scoring above 70, 
the next 30 positions, all open, go to men with scores 100--71. The final 40 open positions go to the highest-scoring remaining individuals, split evenly between men and women with scores 70--51, 
giving 50 positions to women and 50 to men.}
\label{fig:reservesystem-example}
\end{figure}

In this setting, since there are two categories of positions with their own priority rankings, the implementation might seem straightforward at first: 
allocate the open positions using the baseline priority ranking and the reserved positions using the modified priority ranking---a procedure that appears simple enough. 
However, as we will show in the next example, depicted in Figure \ref{fig:reservesystem-example},  there is more to this design exercise than meets the eye.

\begin{example} \label{ex:precedence}
A total of 200 applicants---comprising 100 men and 100 women---are competing for 100 identical positions. 
Of these, 70 positions are open to all applicants, while the remaining 30 are reserved specifically for women, providing them with preferential access. 
Selection for both the open and reserved positions is based on scores from a standardized test. 
For both men and women, scores are uniformly distributed between 1 and 100, meaning that for each integer score within this range, there is exactly one man and one woman who achieved that score.

As a baseline scenario, if all 100 positions were open to everyone, the top 100 scorers would be selected, resulting in exactly 50 men and 50 women being hired---specifically, those who scored 51 or higher.
How would the aforementioned positive discrimination policy for women affect this outcome?

Starting with the 70 open positions, suppose we allocate all positions based on the specified criteria. 
Since both groups have the same score distribution, 35 men and 35 women---those with scores of 66 or higher---receive the open positions. 
The remaining 30 reserved positions are awarded to women with scores between 36 and 65. 
Overall, men receive 35 positions, and women receive 65, giving women a considerable advantage under this policy.

This solution, however, is not the only way to implement the stated positive discrimination policy. In particular, the outcome would differ significantly if the process began with the allocation of the 30 reserved positions.
In that case, the 30 reserved positions would all be awarded to women with scores of 71 and higher.
When the 70 open positions are allocated next, women would be at a disadvantage, as the highest-scoring remaining woman has a score of 70.
The 30 men with higher scores than any remaining woman would be awarded the first 30 open positions. The playing field is leveled only at that point, with 40 open positions remaining.
Since the score distribution of the remaining applicants is identical for men and women at this stage, the last 40 open positions would be evenly split between the two groups,
with each receiving 20 positions. 
Consequently, women would receive only 20 open positions compared to men, who receive 50. Overall, both men and women would receive 50 positions each.

The second implementation of the reserve policy is not only less favorable for women than the first method but also fails to provide any benefits compared to a purely merit-based policy without positive discrimination.
\end{example}

The diverging outcomes between the two implementations of the reserve policy in Example \ref{ex:precedence} 
are driven by a crucial aspect that is not specified in the policy description: Is a woman awarded an open position or a reserved one when she qualifies for both?

While not critical for the woman herself, this detail matters for other applicants. If she is awarded an open position, she does not use up a reserved spot, thereby benefiting women with lower scores. 
However, if a reserved position is awarded to a woman who does not need positive discrimination, it effectively converts into an open position.

Not only do the two implementations produce very different outcomes, but they also represent distinct policy objectives. 

\paragraph{Over-and-Above Choice Rule.}
The first implementation, which awards open positions before the reserved ones, ensures that the protected group is not disadvantaged in any way during the allocation of open competition 
positions due to the ``cream skimming'' of higher-scoring members of the target group for the reserved slots. In a manner of speaking, competition remains truly open in this implementation. 
First explored by \cite{dur/kominers/pathak/sonmez:18}, the choice rule that implements this form of reserve policy is known as the \textit{over-and-above} choice rule.

\paragraph{Minimum Guarantee Choice Rule.}
In contrast, the alternative implementation awards reserved positions before the open ones. 
This approach significantly disadvantages the beneficiaries of the reserved positions in the allocation of open positions. 
Aptly called the \textit{minimum guarantee choice rule} and first explored by \cite{hafalir/yenmez/yildirim:13}, 
this policy merely provides a ``minimum guarantee'' to the protected group equal to the number of reserved positions.
If the group would have already received at least as many positions without positive discrimination---as is the case in Example  \ref{ex:precedence}---then 
they do not benefit at all from this policy. What they gain from the reserved positions, they lose from the open ones.

In Figure \ref{fig:reservesystem-example}, Panels (c) and (d) depict the workings of the over-and-above and minimum guarantee choice rules, respectively.

When reserve systems were first studied by \cite{hafalir/yenmez/yildirim:13} in the market design literature, they were presented directly as the
minimum guarantee choice rule,\footnote{\cite{hafalir/yenmez/yildirim:13} refer to the minimum guarantee choice rule policy as \textit{soft reserves}.} without discussion of other possible variants.
The significance of the processing sequence of positions---the \textbf{\textit{order of precedence}}---was first introduced and studied by \cite{kominers/sonmez:13, kominers/sonmez:16}
within a more general framework that incorporates slot-specific priorities into the matching-with-contracts model discussed in Section \ref{sec:matching-with-contracts}.
Building on this use-inspired basic theory---pivotal in several subsequent ``discovery---invention cycles''---\cite{dur/kominers/pathak/sonmez:18} 
examine the simpler setting discussed in this section,
which is widespread in real-world applications, as a special case of that model (see Figure \ref{fig:DIC-SC-arm}).

As illustrated in Example \ref{ex:precedence} and formalized in our next result, 
the over-and-above and minimum guarantee choice rules represent two polar extremes in our current setting.

\begin{theorem}[\citealp{dur/kominers/pathak/sonmez:18}]  \label{thm:precedence}
Suppose there are two types of positions---open and reserved---that are allocated based on a baseline priority order subject to eligibility. 
Let there be a set of applicants who are all eligible for the open positions, 
and a single protected subset of these applicants who are exclusively eligible for the reserved positions. 
Fix the numbers of open and reserved positions, as well as the order of precedence, and consider the choice rule $C(\cdot)$, 
which allocates positions sequentially to the highest-priority eligible individual for each position. Then:
 \begin{enumerate} 
 \item The set of protected applicants who receive a position under the minimum-guarantee choice rule is a subset of the set of protected applicants who receive a position under the choice rule $C(\cdot)$, 
 which in turn is a subset of the set of protected applicants who receive a position under the over-and-above choice rule. 
 \item The set of unprotected applicants who receive a position under the over-and-above choice rule is a subset of the set of unprotected applicants who receive a position under the choice rule $C(\cdot)$, 
 which in turn is a subset of the set of unprotected applicants who receive a position under the minimum-guarantee choice rule. 
 \end{enumerate} 
\end{theorem}

Theorem \ref{thm:precedence} establishes that, while keeping the relative processing sequence of all other positions fixed,
the later a reserved position is processed, the more beneficial it is for its beneficiaries.

At first glance, this result may seem counterintuitive, since processing a reserved position earlier might appear more advantageous to its beneficiaries.
Yet, in reality, allocating a reserved position earlier exposes the beneficiary group to harsher competition for subsequent open positions, potentially reducing their overall benefit.
Panel (d) of Figure \ref{fig:reservesystem-example} illustrates the worst-case scenario---under the minimum guarantee choice rule, where all reserved positions are processed 
before all open ones---in which the beneficiary group loses all advantage relative to pure merit-based allocation.\footnote{In general, however, the loss may be less severe,
with the group retaining part of the benefit when pure merit-based allocation falls short of the minimum guarantee.}

This subtlety in the order of precedence in reserve systems---particularly the distinction between the over-and-above and minimum guarantee policies---is so elusive that it has led to unintended consequences in several real-world applications.
In some cases, decision-makers have completely overlooked this distinction, leaving major loopholes in systems that otherwise meticulously stipulated more straightforward parameters, 
such as the fraction of reserved units.\footnote{See the \textit{The Hindu} op-ed \cite{pathak/sonmez:19} for an example of such a loophole, 
when the \textit{103rd Constitutional Amendment Act, 2019} introduced reservations for economically weaker sections of society.}
Occasionally, politically motivated parties have leveraged these loopholes through reforms with significant distributional consequences.\footnote{See \cite{pathak/rees-jones/sonmez:25} for an example of this, 
when officials of the U.S. Department of Homeland Security reversed the processing sequence of two categories of H-1B visas in response to President Trump’s \textit{Buy American and Hire American Executive Order in 2017}.}
In other instances, as illustrated in our next case study, system operators intended to implement either the over-and-above or 
minimum guarantee policy but ended up implementing the wrong one---either because they were unaware of the distinction or did not know how to operationalize the intended policy, especially in a multi-institution setting.

\subsection{The Demise of Neighborhood Priority in Boston School Choice} \label{sec:demise}

Over the past fifty years, Boston Public Schools (BPS) have undergone several major reforms in their student assignment policies. 
From controversial busing mandates to the adoption of state-of-the-art mechanisms developed in collaboration with economic designers, 
the city has consistently aimed to balance equal educational opportunities, racial diversity, and neighborhood priority.

Building on \cite{dur/kominers/pathak/sonmez:18}, our next application examines Boston's neighborhood priority policy and the unintended consequences of its 
flawed implementation between 2005 and 2013, which ultimately disadvantaged applicants seeking to attend their neighborhood schools.\footnote{The period from 2005 to 2013 provides a 
conservative estimate of the duration of the flawed implementation of Boston's neighborhood priorities. The neighborhood priority policy, intended for use under the DA mechanism during this period, 
was originally adopted in 2000 under the Boston mechanism. The flawed implementation of this policy, discussed in Section \ref{sec:unintended-BPS}, led to the de facto elimination of neighborhood priorities under the 
DA mechanism and would have caused similar unintended consequences under the Boston mechanism had an analogous error occurred, though whether such a flaw existed between 2000 and 2005 remains unclear.}

\subsubsection{Evolution of Neighborhood Priority in Boston} \label{sec:evolution-neighborhood-priority}

In 1974, Judge Arthur Garrity's racial balance plan required more than 17,000 students to be bused across Boston’s public schools to promote desegregation. 
While the plan succeeded in increasing racial diversity, it also faced strong opposition from various community groups, many of whom resented the forced changes to neighborhood assignment.

A significant shift occurred in 1987 when the U.S. Court of Appeals ruled that BPS had achieved ``unitary status,'' meaning schools were as desegregated as the city’s demographics allowed. 
This ruling granted BPS the autonomy to design its own assignment processes, provided they did not lead to resegregation.

In December 1988, BPS introduced a new assignment plan that divided the city’s schools into three zones: East, North, and West. To maintain racial balance, each zone set ideal racial percentages, 
allowing assignments to vary within 10\% of these benchmarks. The plan adopted the Boston mechanism introduced in Section \ref{sec:Boston-mechanism} and 
established priorities for siblings and neighborhood students, with the latter type of priority referred to as \textit{walk zone priority}.

In response to legal challenges to race- and ethnicity-based affirmative action policies across the U.S., the Boston School Committee proactively enacted a major reform in July 1999,
eliminating racial and ethnic classifications in Boston's assignment plan, effective from the 2000--01 school year. Since this reform risked resegregation due to Boston’s demographics,
an additional reform was introduced later that year. Crucial to our current discussion, the new system reduced the scope of walk zone priority from 100\% to 50\% of school seats,
balancing neighborhood priority with broader access for all students. Half of the seats at each school were reserved for students within the walk zone, while the other half were open to students from outside the zone.

In a November 1999 BPS memo, Superintendent Thomas Payzant explained the role of their latest reform:
\begin{quote}
``Fifty percent walk zone preference means that half of the seats at a given school are subject to walk zone preference. The remaining seats are open to students outside the walk zone.''
\end{quote}

In media coverage, the 50--50 seat split was often portrayed as a delicate compromise between advocates of neighborhood assignment and supporters of broader school choice. 
Superintendent Payzant expressed hope that the plan would appease both groups by allowing families to send their children to nearby schools while also providing options for 
families seeking greater choice \citep{dur/kominers/pathak/sonmez:18}.

Since the city relied on a single uniform lottery to break ties across all schools and both types of seats, achieving this compromise required implementing walk zone priorities through the 
\textit{over-and-above} choice rule (Panel (c) in Figure \ref{fig:reservesystem-example}). Otherwise, neighborhood students would face a disadvantage when competing for open seats. 
In the worst-case scenario, as shown in Example \ref{ex:precedence}, implementing the intended walk zone priorities via the \textit{minimum guarantee} choice rule 
would effectively eliminate any positive discrimination for neighborhood students, directly contradicting the spirit of the intended compromise (Panel (d) in Figure \ref{fig:reservesystem-example}).

Following the publication of \cite{abdulkadiroglu/sonmez:03}, 
BPS adopted the DA mechanism in 2005 after two years of community engagement and an evaluation of the Boston mechanism in collaboration with our team of economic designers (see Section \ref{sec:Boston}).
It remains unclear whether BPS properly implemented its walk zone priority with the \textit{over-and-above} choice rule between 2000 and 2005, prior to this reform. 
However, it became evident that BPS failed to do so under the DA mechanism starting in 2005, unintentionally leading to the de facto elimination of its positive discrimination policy for neighborhood applicants.

\subsubsection{Unintended Consequences: De Facto Elimination of Boston's Neighborhood Priority from 2005 to 2013} \label{sec:unintended-BPS}

Mayor Thomas Menino’s January 2012 State of the City Address marked a turning point for student assignments in Boston. 
In his speech, titled ``Finishing the Job on School Assignment,'' Menino underscored the need for stronger, neighborhood-based school communities, 
pointing to the disconnect among families in the same neighborhoods whose children attended different schools, thereby weakening community cohesion.

Menino pledged a new assignment plan focused on placing students closer to home. Together with Superintendent Carol Johnson, he formed a 27-member 
External Advisory Committee (EAC) to collaborate with the community in shaping this new plan.

The 1999 reduction of the scope of walk zone priority from 100\% of school seats to 50\% was intended to strike a balanced compromise (see Section \ref{sec:evolution-neighborhood-priority}). 
In light of Menino’s 2012 address, we examined whether reducing walk zone priority from 100\% of school seats to 50\% had achieved the desired compromise. 
Fortunately, the \textit{strategy-proofness} of the DA mechanism enabled us to explore various counterfactuals:

\begin{itemize} 
\item Maintaining 100\% Walk Zone Priority: What would happen if walk zone priority were reinstated for all seats? 
\item Abandoning Walk Zone Priority Altogether: How would assignments change if walk zone priority were completely removed? 
\end{itemize}

An unexpected outcome emerged from these counterfactuals: the 50--50 walk zone compromise produced results remarkably similar to those in a system with no walk zone priority at all. How could this happen?

To understand this anomaly, we examined how the city implemented its 50--50 walk-zone compromise under the DA mechanism in a multi-institution setting.

Rather than embedding a choice rule that reflected the city’s intended compromise policy directly into the deferred acceptance algorithm, 
the BPS software team implemented the vanilla version by treating each school as two distinct schools, each with half capacity and its own priorities---one with walk-zone priority and the other without. 
Even though sibling priority also applied for both halves, to simplify the discussion, let’s refer to them as the \textit{walk-zone half} and the \textit{open-competition half}.

Since students submit preferences over schools rather than their specific halves, the software team artificially constructed preferences for each student $i$ over the ``expanded'' set of schools:
\begin{enumerate}
\item For any two schools $a$ and $b$, student $i$ prefers both halves of school $a$ to both halves of school $b$ if and only if student $i$ prefers school $a$ to school $b$.
\item For any school $a$, student $i$ prefers the walk-zone half to the open-competition half.
\end{enumerate}

While this construction may have appeared natural and without major consequences, the second condition in fact altered the intended neighborhood priority policy. 
Specifically, it meant that students with sufficient priority for both types of seats---for example, those living in the walk zone with favorable lottery numbers---were systematically considered first for the walk-zone half. 
This led to the implementation of the \textit{minimum guarantee} choice rule for each school, rather than the intended \textit{over-and-above} choice rule \citep{dur/kominers/pathak/sonmez:18}. 
Intuitively, when the open-competition half was considered at a given school, walk-zone applicants systematically had less favorable lottery numbers, 
since those with better numbers had already been considered for seats in the walk-zone half. 
Consequently, the 50--50 compromise produced outcomes very similar to those in a system with no walk-zone priority, significantly shifting the balance to the detriment of neighborhood applicants.

\subsubsection{Phasing Out Neighborhood Priority in Boston: 2013 Reform for Transparency}

Mayor Thomas Menino's 2012 address sparked a year-long, city-wide debate on school choice, resulting in numerous proposals from BPS, the EAC, and community members. 
Over 70 public meetings were held, with input from thousands of parents. A central issue was a puzzling discrepancy: At popular schools, 
neighborhood students were consistently receiving about 50\% of the seats---those reserved for them---but virtually none from the open-competition seats. 
Not only was this highly implausible, but it also undermined the city's goal of balancing the needs of neighborhood students and providing broader access to all students.

As its own minimalist alternatives to the existing three-zone system, BPS introduced five plans in the fall of 2012 aimed at reducing competition 
from non-neighborhood applicants by limiting the number of schools students could rank. 
The proposals included dividing the city into 6, 9, 11, 23 zones, or eliminating school choice entirely in favor of neighborhood-based assignments. 
However, these plans stalled due to disagreements within the EAC, with one camp advocating for increased school choice and the other for neighborhood-based assignments.

Though I didn’t know it at the time, one of the regular participants in the public meetings was Peng Shi, an MIT graduate student in operations research who was auditing my market design course at Boston College. 
Recognizing the gridlock as an opportunity, Shi proposed an entirely new approach during one of the meetings: the ``Home-Based'' system \citep{shi:15}.\footnote{I later learned that Shi 
had been studying the crisis with the aim of informing a reform as early as the spring of 2012.
Before the adoption of the Home-Based system, he developed several alternative proposals, including one---later published in \cite{Ashlagi/Peng:2014}---for improving community cohesion in 
school choice through a correlated-lottery implementation that generalized earlier work by \cite{Liu/Pycia:2016}. Some of these proposals were submitted when the EAC invited suggestions from the wider community in late summer.} 
This system provided each student with a customized list of schools based on their home address, reducing the number of schools 
they could rank while ensuring that highly-rated options were included.\footnote{See \cite{Shi:2022} for an in-depth exploration of Shi's broader methodology, 
which optimizes parameters such as student priority distribution within mechanisms like DA and SC--TTC, parameters that are typically treated as exogenous in economic design.} 
His proposal gained traction, as EAC members were eager to bridge the divide between the committee's opposing camps.

Meanwhile, building on my work on the U.S. Army's branching process (see Section  \ref{sec:Army}), 
Scott Kominers and I began to recognize the importance of the \textit{order of precedence}---the processing sequence of different types of slots---in a matching with contracts
model that incorporates slot-specific priorities \citep{kominers/sonmez:13, kominers/sonmez:16}. 
In collaboration with Umut Dur and Parag Pathak, we further explored the implications of these findings within the specific context of BPS school choice, 
discovering a discord between the Boston's intended walk zone priority policy and its implementation \citep{dur/kominers/pathak/sonmez:18}.

Due to the timing of these discoveries---particularly their implications for the implementation of BPS walk zone priority---we joined the ongoing debate at a relatively late stage, 
a few months after Shi introduced his Home-Based system to the EAC.
After releasing a preliminary version of our research in \cite{dur/kominers/pathak/sonmez:18}, we first engaged with BPS officials, followed by the EAC in January 2013.
In these meetings, we established that BPS's walk zone priority had been rendered ineffective due to the inadvertent use of the minimum guarantee choice rule instead of the over-and-above choice rule,
which better aligned with the intended policy. Consequently, a simple correction of the choice rule offered a minimalist resolution to the crisis, restoring balance without overhauling the entire system.

BPS officials were intrigued by our discovery but had mixed reactions to our proposed minimalist resolution. Had we presented our findings earlier---before Shi's Home-Based system 
gained substantial support within the EAC---a resolution simply correcting the flawed implementation of the walk zone policy might perhaps have been viable.
By the time we identified the issue, however, the EAC had already held dozens of heated meetings, and the Home-Based system had garnered strong backing. 
There was concern that it was too late to revert to the existing three-zone model. In light of this constraint, implementing the 
Home-Based system with the over-and-above choice rule risked disproportionately benefiting neighborhood students.

To address this predicament, BPS officials  were compelled to advocate for a ``compromise'' choice rule---a middle ground 
between the over-and-above and minimum guarantee choice rules---introduced in \cite{dur/kominers/pathak/sonmez:18}.
This compromise involved filling half of the walk zone seats first, 
followed by all open seats, and then the remaining walk zone seats. As implied by Theorem \ref{thm:precedence}, 
while this choice rule increased the number of students assigned to neighborhood schools compared to the current implementation, it did not go as far as the over-and-above choice rule.

Our discovery of the critical role of the \textit{order of precedence}---the sequence in which different types of positions are processed---became a central point in the debate between advocates of 
neighborhood assignment and those supporting increased school choice. Proponents of neighborhood assignment argued that the improper implementation of the 50--50 split had unjustly 
excluded many students from their neighborhood schools and that correcting the order of precedence was essential to honor the School Committee's original policy goals from 1999.

For advocates of broader access, however, even the BPS proposal to transition to the compromise choice rule raised concerns, particularly given the plans to limit school choice menus 
under the Home-Based system. This group opposed any changes to the choice rule as long as the walk zone priority applied to half of the available seats, 
arguing that enough concessions had already been made under the Home-Based system.

In my view, there was also a less benign reason for this faction's insistence on maintaining the minimum guarantee choice rule:
some members of this group appeared tempted to leverage the flawed and misleading implementation of Boston's neighborhood priority policy---the 50--50 split, viewed as a symbol of compromise---to create the
false impression that they were making concessions to neighborhood advocates. While the system emphasized preferential treatment for neighborhood students in half of the seats,
it concealed the advantage enjoyed by non-neighborhood students in the other half.

In February 2013, the EAC made its formal recommendation to the School Committee to adopt the Home-Based system and maintain the existing neighborhood priority policy 
until the issue could be studied further \citep{BSC:February-13}. 

Disappointed with the EAC's recommendation, Pathak and I testified before the School Committee within weeks, arguing that the current implementation of the walk zone priority was not only inconsistent
with the policy approved by the School Committee in November 1999 but also misleading the community regarding the walk zone priority’s role \citep{pathak/sonmez-testimony:13}.
We asserted that correcting this flawed implementation---which falsely conveyed a strong emphasis on walk zone priority in the assignment plan---would enhance \textit{transparency}.
While the city’s policy had, until then, unintentionally misled the public, maintaining it in light of our findings would render this misrepresentation deliberate, a course we argued was not in the public's best interest.

Based on our testimony, Superintendent Johnson chose to embrace \textit{transparency} and, despite the EAC's formal recommendation, 
called for the elimination of the walk zone priority altogether. On March 13, 2013, she announced \citep{johnson:13}:

\begin{quote} ``After viewing the final MIT and BC presentations on the way the walk zone priority actually works, 
it seems to me that it would be unwise to add a second priority to the Home-Based model by allowing the walk zone priority to be carried over. [...]

Leaving the walk zone priority to continue as it currently operates is not a good option. 
We know from research that it does not make a significant difference the way it is applied today: although people may have thought that it did, 
the walk zone priority does not in fact actually help students attend schools closer to home. 
The External Advisory Committee suggested taking this important issue up in two years, 
but I believe we are ready to take this step now. We must ensure the Home-Based system works in an honest and transparent way from the very beginning." 
\end{quote}

Following Johnson's recommendation, the Boston School Committee eliminated the walk zone priority and adopted the Home-Based system, 
implementing the DA mechanism with student-specific school menus \citep{BSC:13}. 
The new plan went into effect for elementary and middle schools in the fall of 2013.

\section{Affirmative Action in India} \label{sec:India}

A common technical challenge faced by U.S. Army officials integrating BRADSO incentives into the branching system and by 
Boston Public Schools (BPS) officials incorporating walk-zone priority into the student assignment plan was designing a mechanism that allocates some positions 
according to one criterion and others according to another. In both cases, officials found it relatively straightforward to define and implement policies based on a single criterion, 
but introducing a second criterion for certain positions proved to be a formidable task.

This challenge extends beyond the Army and BPS. Mechanisms like the Army's branching system and BPS’s school choice plan are technically complex, 
making it unrealistic to expect fully satisfactory designs from officials lacking expertise in formal fields, such as economic design or optimization. 
This underscores the importance of academic market designers expanding their focus beyond traditional economic principles, such as \textit{preference utilitarianism}.

More broadly, the formulation, analysis, and implementation of various normative desiderata represent a promising area where market designers 
have provided---and can continue to provide---valuable guidance in real-world applications. As \cite{Li2017} emphasizes, market design must address ethical considerations, 
since policymakers often need assistance articulating their objectives precisely. Policymakers may be intimately familiar with the details of their environment, 
yet they may struggle to express their ethical requirements in exact terms. Even when they do, they are not typically trained to design mechanisms that adhere to ethical or other principles. 
And yet, they are tasked with designing such systems all the time.

For example, despite lacking expertise in formal analysis, justices in India have been formulating the normative principles that guide the country’s affirmative action policies for decades. 
In many cases, they have even designed the mechanisms intended to implement these principles. This tradition raises the question: What could possibly go wrong with such practices?

The discovery of the erroneous implementation of the walk-zone priority at BPS from 2005 to 2013 (see Section \ref{sec:demise}), 
was the first indication that the subtle concept of \textit{order of precedence}---the processing sequence of different types of positions---may be the source of significant unintended consequences in policy. 
Following this discovery,  I systematically explored whether similar experiences existed worldwide where policymakers, 
lacking clarity on policy distinctions, intended one type of reserve policy but inadvertently implemented another.

One such exploration led to the affirmative action system embedded within the 1950 Constitution of India, which included two familiar policies later codified as concrete reserve policies in the
landmark Supreme Court judgment \textit{Indra Sawhney vs. Union of India} (1992): the \textit{over-and-above} reserve policy and the \textit{minimum guarantee} reserve policy.

Remarkably, unlike the BPS officials, the justices understood the difference between the two policies despite their subtle distinction.
However, applying these policies in India proved far more challenging than implementing the 50/50 walk-zone compromise in Boston,
as it required enacting both affirmative action policies simultaneously. This additional complexity posed a formidable challenge for the judiciary, policymakers,
and system operators in designing and implementing India's affirmative action system.

In this context, much of our focus in the next application is on a case where a flawed affirmative action procedure, mandated by the Supreme Court judgment in 
\textit{Anil Kumar Gupta vs. State of U.P.} (1995), led to numerous lawsuits and disrupted recruitment processes in India for decades. 
Twenty-five years later, the Supreme Court addressed this design flaw in the subsequent judgment \textit{Saurav Yadav vs. State of U.P.} (2020), 
but not before the issues and their remedies were thoroughly analyzed in \cite{sonyen19, sonyen22}.
A brief guide to the key judgments referenced in this section is presented in Table \ref{tab:India-judgments}.


\begin{table}[!tp]
\centering
\small
\begin{threeparttable}
\caption{Key judgments on India’s reservation system and their role in Section~\ref{sec:India}.}
\label{tab:India-judgments}

\setlength{\tabcolsep}{5pt} 
\begin{tabular}{@{}p{0.25\textwidth} p{0.23\textwidth} p{0.38\textwidth} p{0.10\textwidth}@{}}
\toprule
\textbf{Case (Court, Year)} & \textbf{Scope} & \textbf{Role} & \textbf{Section} \\
\midrule
\textit{Indra Sawhney v.\ Union of India} (SC, 1992)
& VR and HR policies (general framework)
& Introduces VR, HR policies, and the ``migration'' concept; leaves joint VR and HR implementation unspecified.
& \ref{sec:IndiaAA}--\ref{sec:horizontal}
\\ \addlinespace[5pt]\midrule

\textit{Anil Kumar Gupta v.\ State of U.P.} (SC, 1995)
& VR and HR policies (joint implementation, identical units)
& Introduces SCI--AKG via the concept of ``adjustments''; restricts Open HR adjustments, leading to NJE failures.
& \ref{sec:horizontal}--\ref{sec:India-EV}
\\ \addlinespace[5pt]\midrule

\textit{Rajesh Kumar Daria v.\ RPSC} (SC, 2007)
& VR and HR policies (joint implementation, identical units)
& Clarifies working of SCI--AKG and the notion of ``adjustments'' under \textit{Anil Kumar Gupta}.
& \ref{sec:horizontal}
\\ \addlinespace[5pt]\midrule

\textit{Saurav Yadav v.\ State of U.P.} (SC, 2020)
& VR and HR policies (joint implementation, identical units)
& Mandates NJE; removes exclusion from Open HR adjustments; endorses 2SMG.
& \ref{sec:India-EV}--\ref{sec:heterogenous}
\\ \addlinespace[5pt]\midrule

\textit{Tamannaben A.\ Desai v.\ Shital A.\ Nishar} (Gujarat HC, 2020)
& VR and HR policies (joint implementation, identical units)
& Introduces procedure equivalent to 2SMG ; mandates in Gujarat.
& \ref{sec:India-EV}
\\ \addlinespace[5pt]\midrule

\textit{Union of India v.\ Ramesh Ram} (SC, 2010)
& VR and HR policies (joint implementation, heterogeneous units)
& Awards vacated Open seats after MRC migration to the general category (govt.~jobs), violating NJE.
& \ref{sec:heterogenous}
\\ \addlinespace[5pt]\midrule

\textit{Tripurari Sharan v.\ Ranjit Kumar Yadav} (SC, 2018)
& VR and HR policies (joint implementation, \ref{sec:heterogenous} units)
& Awards vacated Open seats after MRC migration to the category of the vacating MRC (medical colleges), violating NJE; contradicts \textit{Ramesh Ram}.
& \ref{sec:heterogenous}
\\
\bottomrule
\end{tabular}

\begin{tablenotes}\footnotesize
\item Abbreviations: 
VR: Vertical Reservations; HR: Horizontal Reservations; NJE: No Justified Envy; SCI--AKG: Supreme Court of India--Anil Kumar Gupta; 2SMG: Two-Step Minimum Guarantee; 
MRC: Meritorious Reserved Candidate; SC: Supreme Court; HC: High Court.
\end{tablenotes}

\end{threeparttable}
\end{table}

This is another instance where life imitates science. Similar to the school choice reforms in England and Chicago discussed in Section \ref{sec:Chicago-England},
the amendment introduced in the \textit{Saurav Yadav} (2020) judgment provides external validity for minimalist market design. 
This landmark ruling aligns not only with \cite{sonyen19, sonyen22} regarding the identified issues and their root causes but also with the proposed remedies outlined in these papers.

\subsection{India’s Reservation System and the Quest for Social Justice} \label{sec:IndiaAA}

Affirmative action for various disadvantaged groups is embedded in the 1950 Constitution of India through articles 14--18.\footnote{This segment of the Constitution is referred to as the \textit{Equality Code} in India.} 
In practice, these protective policies are implemented through a \textit{reservation system} that sets aside a certain fraction of positions in public jobs and seats at public schools for various protected groups. 
Naturally, the Constitution itself does not provide any rigorous formula to implement these policies. Nevertheless, in a landmark Supreme Court judgment \textit{Indra Sawhney} (1992), 
the justices formulated these provisions in a rigorous way, at least for the more basic versions of the reservation system.\footnote{Considered the main reference 
on the reservation system in India, this case is popularly known as the \textit{Mandal Commission} case.}

In the first part of this section (up to Section \ref{sec:heterogenous}), following \cite{sonyen22}, we focus on a version of the problem addressed in the Supreme Court judgments \textit{Anil Kumar Gupta} (1995) 
and \textit{Saurav Yadav} (2020). In this setting, a public institution allocates identical positions based on applicants’ merit scores and membership in protected groups---a scenario common in practice across India. 
In Section  \ref{sec:heterogenous}, we extend the analysis to heterogeneous positions, as explored in \cite{sonmez/yenmez:24}, which represents another frequently encountered variant of the problem in the country.

Without affirmative action, the simpler version with identical positions reduces to the most basic version of the student placement problem with a single school. 
The straightforward solution is to admit applicants with the highest merit scores up to capacity. This simple mechanism---a special case of Simple Serial Dictatorship (SSD) 
with a single institution---is adapted in India to accommodate two types of affirmative action policies: a primary policy called \textbf{\textit{vertical reservations}} (VR) and a secondary policy called \textbf{\textit{horizontal reservations}} (HR).

\subsection{Vertical Reservations} \label{sec:vertical-migration-OA}

Under the VR policy, a fraction of positions is exclusively reserved for each of several protected groups.
In contrast to these VR-protected positions, the remaining \textit{open-category} (or \textit{open}) positions are available to all applicants.

Until the \textit{103rd Constitutional (Amendment) Act, 2019}, the target groups for the primary VR policy were restricted to groups whose members had suffered marginalization and discrimination 
based on their hereditary social (i.e., caste or tribal) identities. These groups primarily include Scheduled Castes (SC), Scheduled Tribes (ST), and Other Backward Classes (OBC). 

From a technical point of view, the social identity-based vertical reservation system meant that no individual could be a member of multiple VR-protected groups. 
While VR protections are also provided to individuals in economic disability since 2019 with the enactment of the 103rd Amendment,\footnote{This more recent vertical category is called 
\textit{Economically Weak Segments} (EWS).} members of the earlier and social identity-based VR-protected groups are excluded from its scope,\footnote{This exclusion is highly controversial in India. 
See \cite{sonmez-unver2022} for a case study in minimalist market design that analyzes this exclusion in the 103rd Amendment.} 
thereby still maintaining the \textit{non-overlapping} structure of VR-protected groups in the country.

The defining property of the VR policy is formulated in paragraph 94 of \textit{Indra Sawhney} (1992):

\begin{quote} 
``In this connection it is well to remember that the reservations under Article 16(4) do not operate like a communal reservation. 
It may well happen that some members belonging to, say, Scheduled Castes get selected in the open competition field on the basis of their own merit; 
they will not be counted against the quota reserved for Scheduled Castes; they will be treated as open competition candidates.'' 
\end{quote}

This key mandate from the Supreme Court ensures that VR-protected positions are allocated to members of their target groups who genuinely need positive discrimination, 
rather than to those who could secure an open position based solely on merit.

A subtle but important aspect of the above mandate is its reliance on a basic technical tool commonly referred to as \textit{migration} in Indian legal terminology.\footnote{The concept of 
migration is also known as \textit{mobility} in Indian legal terminology.}

In India, individuals outside any VR-protected category are classified under the \textit{general} category. 
While VR-protected positions are exclusively reserved for members of their respective groups, open-category positions are accessible to all individuals, including those from VR-protected categories.
The justices underscore this distinction in the mandate, explicitly stating that ``the reservations under Article 16(4) do not operate like a communal reservation.''

However, because Indian legal terminology does not distinguish between ``categories of individuals'' and ``categories of positions,'' despite the mismatch between these two concepts, 
the terms ``open'' and ``general'' are used interchangeably in legal discussions. The concept of \textit{migration} likely arises from this ambiguity in legal terminology. 
For example, when a member of the SC receives an open position based on merit, 
they are considered to have ``migrated'' from the SC to the general category. 
Relating this notion to the above mandate from \textit{Indra Sawhney} (1992), individuals who have migrated are those who ``will be treated as open competition candidates.''

Effectively, designers of various reservation systems in the country, despite complex legal requirements, attempt to simplify the system by (1) dividing the set of applicants into separate partitions and (2) using 
SSD to allocate positions within each partition. Critically, however, when a natural partition does not exist, they resort to the notion of migration to artificially create one. 
In my view, reliance on this rudimentary technical tool, together with a second tool discussed in Section \ref{sec:horizontal}---(horizontal) \textit{adjustments}---significantly 
exacerbates the legal and practical challenges of implementing India's reservation system.

The formal analysis of Indian legislation on the reservation system was first presented in \cite{sonyen19, sonyen22} for the basic setting with identical positions and in \cite{sonmez/yenmez:24} 
for the generalized setting with heterogeneous positions. In the basic setting, implementing the VR policy alone, without horizontal reservations, is straightforward using the 
\textit{over-and-above} choice rule---a two-step procedure described in Section \ref{sec:subtleties-reserve} for a single protected group.

When there are multiple VR-protected groups with non-overlapping memberships, as in India, this procedure can be generalized through the following straightforward extension: 
In the first step, open positions are allocated to applicants with the highest merit scores. In the second step, positions reserved for each VR-protected group are allocated to the highest-scoring remaining applicants from each group. 
This process can be thought of as a repeated implementation of SSD---first for the open category and then for each VR-protected category.

\subsection{Horizontal Reservations} \label{sec:horizontal}

The target groups for the secondary HR policy in India are not based on hereditary social identities and include protected groups such as ``Persons with Disabilities'' and ``Women.''
Unlike VR-protected groups, individuals can belong to multiple HR-protected groups.

In Sections \ref{sec:horizontal} through \ref{sec:India-EV}, 
we focus on the simpler case of identical positions and \textit{non-overlapping} HR protections, where each individual belongs to at most one HR-protected group. 
This scenario provides the clearest policy implications, as it aligns with the version of the problem reviewed in \textit{Saurav Yadav} (2020).

In Sections  \ref{sec:overlappingHR} and \ref{sec:heterogenous}, we extend the analysis to \textit{overlapping} HR protections, considering cases where individuals can belong to multiple HR-protected groups. 
We begin with the basic setting of identical positions and later explore the more general scenario involving heterogeneous positions across multiple institutions.

In contrast to the VR policy, which is defined precisely for stand-alone implementation in \textit{Indra Sawhney} (1992), 
the description of the HR policy in paragraph 95 of the same judgment is notably less clear:
\begin{quote}
``Horizontal reservations cut across the vertical reservations that is called inter-locking reservations. To be more precise, suppose 3\% of the vacancies are reserved in favour of physically handicapped persons; this would be a reservation relatable to Clause (1) of Article 16. The persons selected against this quota will be placed in the appropriate category; if he belongs to S.C. category he will be placed in that quota by making necessary adjustments; similarly, if he belongs to open competition (O.C.) category, he will be placed in that category by making necessary adjustments. Even after providing for these horizontal reservations, the percentage of reservations in favour of backward class of citizens remains - and should remain - the same.'' 
\end{quote}

This passage leaves unclear what exactly the ``necessary adjustments'' refer to and how they are to be made. 
The ambiguity surrounding these adjustments and the implementation of the HR policy was later clarified by the Supreme Court in 
\textit{Anil Kumar Gupta} (1995).\footnote{The formulation of the HR policy in \textit{Anil Kumar Gupta} (1995) is further clarified in a subsequent Supreme Court judgment, 
\textit{Rajesh Kumar Daria vs. Rajasthan Public Service Commission (2007)}.} In this judgment, the Supreme Court further elaborated two key characteristics of the HR policy that distinguish it from the VR policy:

\begin{enumerate} 
\item  Positions ``earned'' based on merit (whether from the open category or a VR-protected category) 
still count against the HR-protected positions. Therefore, the HR policy is implemented on a \textit{minimum guarantee} basis. 
\item The HR policy is implemented as \textit{soft reserves}. This means that while members of the protected group receive preferential treatment for HR-protected positions, 
these positions are not exclusively set aside for them. Any HR-protected position that remains after all members of the protected category have received a position can then be awarded to other eligible individuals. 
\end{enumerate}

Possibly to avoid complications arising from overlaps between ``property rights'' gained through these two types of affirmative action policies, 
the justices also recommended implementing the HR policy within each vertical category, including the open category.\footnote{This mode of HR 
policy is referred to as \textit{compartmentalized} horizontal reservations in \textit{Anil Kumar Gupta} (1995).} For example, if there are horizontal reservations
for women, they are provided with preferential access for a fraction of the positions in each category (i.e., open, SC, ST, OBC, etc.) separately.

Within this framework, the first characteristic outlined in \textit{Anil Kumar Gupta} (1995) clarified the legal notion of ``adjustments'' in \textit{Indra Sawhney} (1992) 
in defining horizontal reservations: it simply represented a minimum guarantee within each category.

With this understanding, rather than relying on the rudimentary tool of adjustments, one can employ a computationally more efficient algorithm---the procedure 
used to introduce the minimum guarantee choice rule in Section \ref{sec:subtleties-reserve}---to implement the HR policy within any given category.  

In this algorithm, given a set of applicants eligible for the category, 
the highest-merit beneficiaries of HR-protected groups are awarded HR-protected positions up to capacity in the first step. In the second step, any remaining positions---including unused 
HR-protected positions---are allocated to other individuals with the highest merit scores.

\subsection{SCI--AKG vs. 2SMG: The Supreme Court’s Flawed Choice Rule and Its Minimalist Correction}  \label{sec:SCIAKG-failures}

Building directly on the minimum guarantee choice rule, experts well-versed in formal modeling 
and analysis of reserve systems will find that integrating the HR policy within the VR policy is straightforward.

\paragraph{Two-step Minimum Guarantee (2SMG) Choice Rule.}
With only the VR policy, the reservation system is implemented by repeated application of the SSD for each category, starting with the open category. 
With both VR and HR policies, this procedure can be amended by simply replacing the SSD with the minimum guarantee choice rule for each category. 
Moreover, when VR-protected categories do not overlap---as in India---all of them can be processed simultaneously after the open category, 
thereby justifying the name of the \textit{two-step minimum guarantee} choice rule.

Let's illustrate 2SMG with an example, using a continuum of positions and individuals to simplify its presentation. 
We first describe the setting, which we will also use later to explain the Supreme Court's choice rule in \textit{Anil Kumar Gupta} (1995).

\begin{figure}[!tp]
 \centering
   \captionsetup{skip=0pt} 
       \includegraphics[scale=1.02]{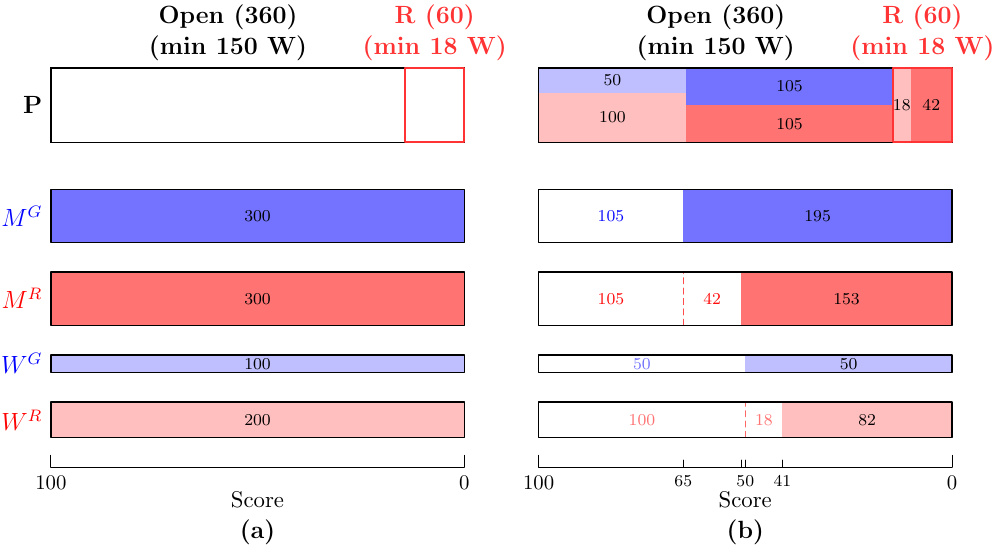}
\caption{Execution of the 2SMG choice rule under its standard formulation. Panel (a) illustrates the continuum economy in Example \ref{ex:VR-HR-implementation}. Building on Example \ref{ex:VR-HR-implementation-2SMG}, 
Panel (b) depicts how the 2SMG choice rule operates for this economy. Abbreviations: P: Positions; W: Women; R: Vertically Reserved; $M^{G}$: General-category men;  $M^{R}$ Reserved-category men;   
$W^{G}$: General-category women;  $W^{R}$: Reserved-category women.}
\label{fig:2SMG-standard}
\end{figure}

\begin{example} \label{ex:VR-HR-implementation}
There is one VR-protected group R and one HR-protected group W (women). 
There is a measure of 360 open positions, with a minimum guarantee of 150 for women. There is a measure of 60 VR-protected positions set aside for 
category R, with a minimum guarantee of 18 for category-R women.
There is a measure of 300 general-category men, 300 category-R men, 100 general-category women, and 200 category-R women, each with scores uniformly distributed between 0 and 100.
Panel (a) of Figure \ref{fig:2SMG-standard} represents this economy.
\end{example}

We now turn to the implementation of the 2SMG choice rule in this setting, depicted in Panel (b) of Figure \ref{fig:2SMG-standard}. 

\begin{subex}\label{ex:VR-HR-implementation-2SMG}
The process starts by allocating the 360 measure of open positions using the minimum guarantee choice rule. 
First, a measure of 150 is allocated to the highest-scoring women from both categories: a measure of 50 general-category women
and a measure of 100 category-R women with scores 50 and higher. 
Next, the remaining 210 measure of open positions is allocated to the highest-scoring remaining individuals. Since there is more than 210 measure of men
with scores higher than 50 (the upper bound of scores for the remaining women), all of these units are awarded to men: 105 measure to general-category men and 105 measure to category-R men with scores 65 and higher. 

The process then continues by allocating the 60 measure of VR-protected positions set aside for category R, again using the minimum guarantee choice rule.
First, a measure of 18 is allocated to the highest-scoring remaining category-R women, namely those with scores between 41 and 50. 
Next, the remaining 42 measure of category-R positions is allocated to the highest-scoring remaining individuals from category R. 
Since there is more than 42 measure of category-R men with scores higher than 41 (the upper bound of scores for the remaining category-R women), these units are awarded to category-R men with scores between 51 and 65.
\end{subex}

 As natural as 2SMG is as a generalization of the over-and-above choice rule to integrate the secondary HR policy with the primary VR policy, 
justices---rarely accustomed to modeling with analytical methods and, understandably, unfamiliar with the minimum guarantee choice rule, 
introduced in the theoretical matching literature nearly two decades after \textit{Anil Kumar Gupta} (1995). 
Instead, they relied on the notion of ``adjustments'' to operationalize the secondary HR policy alongside the primary VR policy.

\paragraph{SCI--AKG Choice Rule.}
The Supreme Court's three-step procedure begins by ignoring the HR policy in the first two steps and focusing solely on the VR policy.
In the first step, open positions are tentatively allocated using SSD. 
In the second step, the remaining VR-protected positions are allocated, again using SSD. 
Together, these two steps form our familiar \textit{over-and-above} choice rule.

In the third step, ``necessary adjustments'' are applied to incorporate the HR policy when needed. For example, if the women’s minimum guarantee is unmet within a specific category, such as SC, 
the lowest-merit SC men tentatively admitted are replaced by the highest-merit unmatched SC women. 
This adjustment helps ensure that the HR policy is accommodated within the SC category.\footnote{\label{footnote:SCI--AKG-sequence}Since adjustments for the open category may require revising 
the allocation of VR-protected positions in Step 2, it is computationally more efficient to delay this step until adjustments for the open category have been finalized.}

Yet even this three-step description does not fully specify a precise choice rule, because the adjustment process in the final step involves a particularly tricky ambiguity. 
Suppose that the minimum guarantee constraints are short by one woman both for the open category and for the SC category. 
If there is an unmatched SC woman, it is unclear whether she should replace a man from the open category or one from SC itself. 

The justices attempted to resolve this difficulty by partitioning the ``adjusters'' into distinct (i.e., non-overlapping) groups. 
In an especially unfortunate mandate in \textit{Anil Kumar Gupta} (1995), they further restricted access for open-category horizontal adjustments to members of the general category only. 
While this mandate conveniently ensured that the set of individuals entitled to horizontal adjustments at each vertical category became uniquely defined for the last step of the procedure, 
it also plagued the resulting choice rule with highly consequential shortcomings. 
Following \cite{sonyen22}, we refer to this procedure, mandated in India between 1995 and 2020, as the \textit{Supreme Court of India--Anil Kumar Gupta} (SCI--AKG) choice rule.

We next illustrate the SCI--AKG choice rule using the economy in Example \ref{ex:VR-HR-implementation}, 
shown in Panel (a) of Figure \ref{fig:2SMG-standard}.\footnote{Strictly speaking, this illustration employs the slight variant mentioned in Footnote \ref{footnote:SCI--AKG-sequence}.}

\begin{figure}[!tp]
 \centering
   \captionsetup{skip=5pt} 
       \includegraphics[scale=1.02]{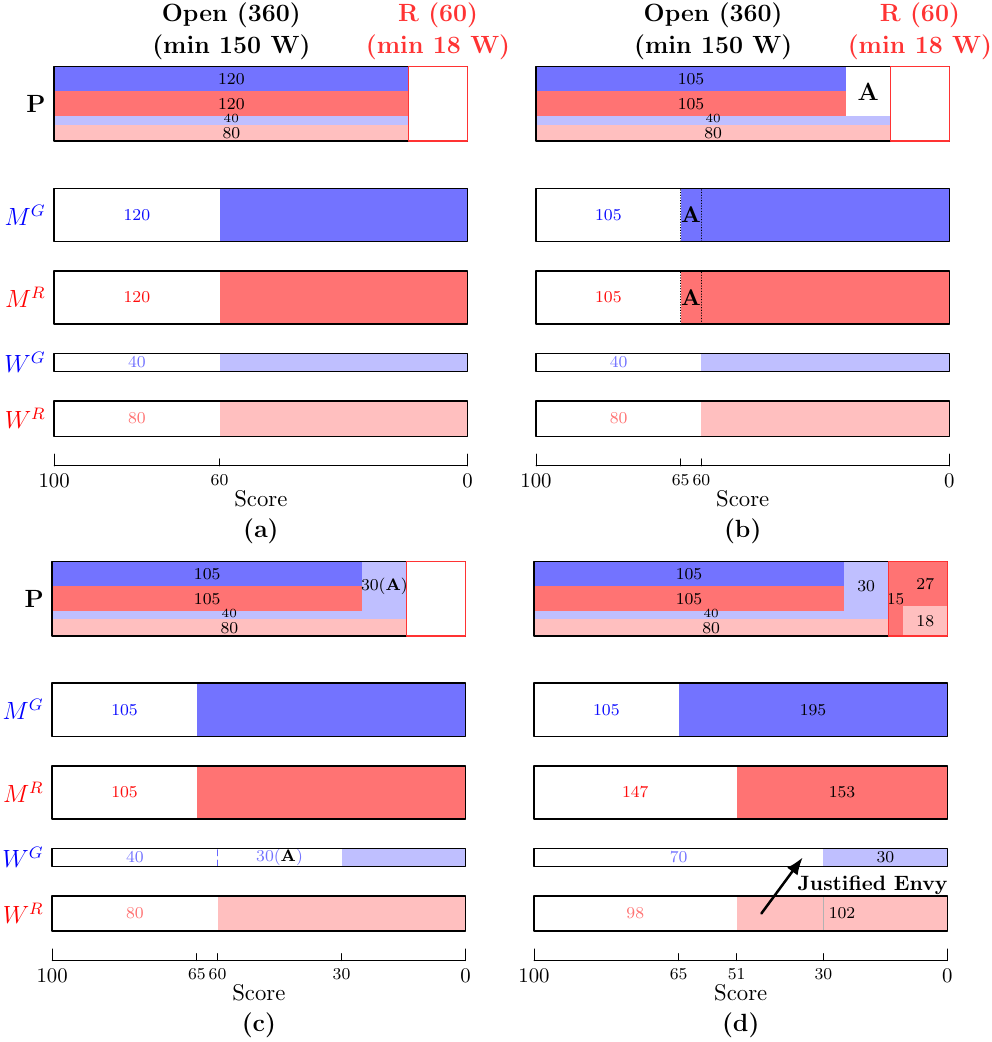}
\caption{Execution of the SCI--AKG choice rule. For the continuum economy in Example \ref{ex:VR-HR-implementation}, given in Panel (a) of Figure \ref{fig:2SMG-standard}, 
Panel (a) here illustrates the allocation of open positions under the VR policy only. Panels (b) and (c) show the subsequent adjustment process (indicated with \textbf{A}) for the open category to accommodate the minimum guarantee 
for women. In Panel (b), a measure of 30 lowest-scoring men (15 from each category) lose their tentative assignments from Panel (a), which in Panel (c) are awarded exclusively to the highest-scoring unmatched women from 
the general category. Panel (d) depicts the allocation of VR-protected positions. Abbreviations: P: Positions; W: Women; R: Vertically Reserved; $M^{G}$: General-category men;  $M^{R}$ Reserved-category men;   
$W^{G}$: General-category women;  $W^{R}$: Reserved-category women.}
\label{fig:SCI--AKG}
\end{figure}

\begin{subex}\label{ex:VR-HR-implementation-SCI--AKG}
The process starts by tentatively allocating the 360 measure of open positions to the highest-scoring individuals across all groups:
a measure of 120 general-category men, a measure of 120 category-R men, a measure of 40 general-category women, and a measure of 80 category-R women with scores of 60 and higher (see Panel (a) of Figure \ref{fig:SCI--AKG}). 
Since women received only a measure of 120 at this stage---below the minimum guarantee of 150---a measure of 30 of the lowest-scoring men tentatively admitted 
must be dropped as part of the horizontal adjustment process for the open category. This results in the removal of a measure of 15 general-category men and a measure of 15 category-R men, 
those with scores between 60 and 65 (see Panel (b) of Figure \ref{fig:SCI--AKG}, where this ``adjustment'' is indicated with \textbf{A}).

To complete the adjustment process for the open category, this measure of 30 positions is awarded to general-category women, since category-R women are deemed ineligible for the
open-category adjustment. Consequently, a measure of 30 general-category women with scores between 30 and 60 receive these adjusted units (see Panel (c) of Figure \ref{fig:SCI--AKG}, 
where this ``adjustment'' is indicated with \textbf{A}).

The process then continues by allocating the 60 measure of VR-protected positions set aside for category R to the highest-scoring members of this category. 
Since no remaining category-R women have scores above 60, the first measure of 15 goes to category-R men with scores between 60 and 65, 
before the process begins awarding positions to category-R women. At this point, the remaining measure of 45 is split between a measure of 27 category-R men and a measure of 18 category-R women 
with scores between 51 and 60. Since the minimum guarantee of a measure of 18 VR-protected units for women is met, no further adjustment is 
necessary, finalizing the process (see Panel (d) of Figure \ref{fig:SCI--AKG}).    

Importantly, since all general-category women with scores of 30 and higher received positions, category-R women with scores between 30 and 51---who remain unassigned---have \textit{justified envy} toward some of these women.    
\end{subex}

Before further discussing the shortcomings of the SCI--AKG choice rule, as illustrated at the end of Example \ref{ex:VR-HR-implementation-SCI--AKG}, 
I am inclined to speculate on what might have led the justices to this unfortunate design, as this will help us identify the root cause of the problem.

In my view, the Supreme Court’s flawed choice rule may have been shaped by the misguided conflation of the terms ``open'' category and ``general'' category, 
as well as by the way the concept of ``migration'' is applied under the VR policy in the Indian legal framework.
Recall that, in the first two steps of the SCI--AKG choice rule, positions are provisionally allocated using the over-and-above choice rule without accounting for the HR policy.
Now, consider an unmatched VR-protected woman at this stage. Why is she deemed ineligible for women’s horizontal adjustments within the ``open'' category?

One possible explanation is as follows:  
Since she has not been provisionally awarded an ``open'' position (see Section \ref{sec:vertical-migration-OA}), she has not yet ``migrated'' to the ``general'' category, according to Indian legal terminology, 
and remains a member of her VR-protected category.
Perhaps, then, because she was not yet considered a member of the ``general'' category at the beginning of the third step, the misleading conflation of ``open'' and ``general'' 
categories might have led the justices to deem her ineligible for an adjustment within the open category.

To understand why SCI--AKG fails, it is instructive to recall a notion from Indian legal terminology that relates this design to 2SMG.

\begin{definition}  \label{def:MRC}
A member of a VR-protected category is considered a \textbf{meritorious reserved candidate} if they qualify for an open category position solely based on merit score, without invoking affirmative action. 
\end{definition}

In the absence of the HR policy, meritorious reserved candidates are precisely those who would be regarded as having ``migrated'' from their VR-protected categories to the general category under the VR policy.

\paragraph{Alternative Formulation for SCI--AKG.}
The justices’ design can be reformulated in terms of meritorious reserved candidates, using repeated applications of the minimum guarantee choice rule.  
In the first step, open positions are allocated using the minimum guarantee choice rule, with eligibility restricted to members of the general category and meritorious reserved candidates. 
In the second step, reserved positions for each VR-protected category are allocated to the remaining eligible applicants, again using the minimum guarantee choice rule.

The only difference between this formulation of SCI--AKG and the 2SMG choice rule lies in the set of applicants competing for open positions in the first step. 
Under the 2SMG choice rule, all applicants are considered. 
By contrast, with the exception of the meritorious reserved candidates, 
the SCI--AKG choice rule restricts eligibility to members of the general category. 
As a result, VR-protected individuals other than meritorious reserved candidates are excluded from accessing HR-protected positions in the open category.

This connection exposes not only the shortcomings of the SCI--AKG choice rule but also their root cause.

Except for the recent EWS category, a member of a VR-protected category is not required to declare membership in their category.\footnote{This contrasts with the 
documentation needed to prove ineligibility for social identity (i.e., caste or tribal) based VR-protected categories, which is a prerequisite for claiming membership in the EWS category. 
This typically involves documentation of forward caste membership in India.} By claiming their VR protections, an individual---unless they are a meritorious reserved candidate---loses their open category 
HR protections under the SCI--AKG choice rule. Thus, this mechanism violates a form of the \textit{incentive compatibility} axiom, first formulated by \cite{aygun/bo:21} in the context of affirmative action at Brazilian colleges.

More evident in practical applications than the failure of incentive compatibility, as we illustrated in Example \ref{ex:VR-HR-implementation-SCI--AKG}, 
the SCI--AKG choice rule also violates a version of the \textit{no justified envy} principle tailored to the current context:

\begin{definition} \label{def:India-NJE}
An allocation of positions across vertical categories (including open) satisfies \textbf{no justified envy} if there exists no category $v$ and two applicants $i$ and $j$, 
both eligible for category $v$, such that: 
\begin{enumerate} 
\item Applicant $j$ is awarded a position in category $v$, 
\item Applicant $i$ receives no position at all, and 
\item Applicant $j$ neither has a merit score as high as that of applicant $i$, 
nor does awarding them a position in category $v$ at the expense of applicant $i$ increase the number of HR-protected positions honored in category $v$. 
\end{enumerate} 
\end{definition}

Consider any category $v$ and two applicants, $i$ and $j$, who are both eligible for this category and satisfy the conditions in Definition \ref{def:India-NJE}.
Awarding a position to applicant $j$ at the expense of applicant $i$ cannot be justified either by meritocracy or by the two affirmative action policies, VR and HR (see Panel (d) of Figure \ref{fig:SCI--AKG}).
Thus, similar to the no justified envy axiom stated in Definition \ref{def:Army-NJE} within the context of the U.S. Army's branching process, Definition \ref{def:India-NJE}
likewise accounts for any objections applicants may raise against an allocation based on multiple assignment criteria.

Central to our application, this crucial social justice axiom fails under the SCI--AKG choice rule precisely because, with the exception of meritorious reserved candidates, 
applicants who declare their VR-protected categories are excluded from competing for open-category HR-protected positions. 
Thus, addressing the root cause of the SCI--AKG choice rule's failures is as simple as removing this exclusion, leading directly to the 2SMG choice rule as a minimalist reform.

For the setting in Example \ref{ex:VR-HR-implementation}, we have already examined the 2SMG choice rule in Example \ref{ex:VR-HR-implementation-2SMG}
using its standard and computationally efficient formulation. 
To highlight its close connection with SCI--AKG, we now re-examine it through an alternative formulation based on horizontal adjustments. 

\begin{figure}[!tp]
 \centering
   \captionsetup{skip=5pt}
       \includegraphics[scale=1.02]{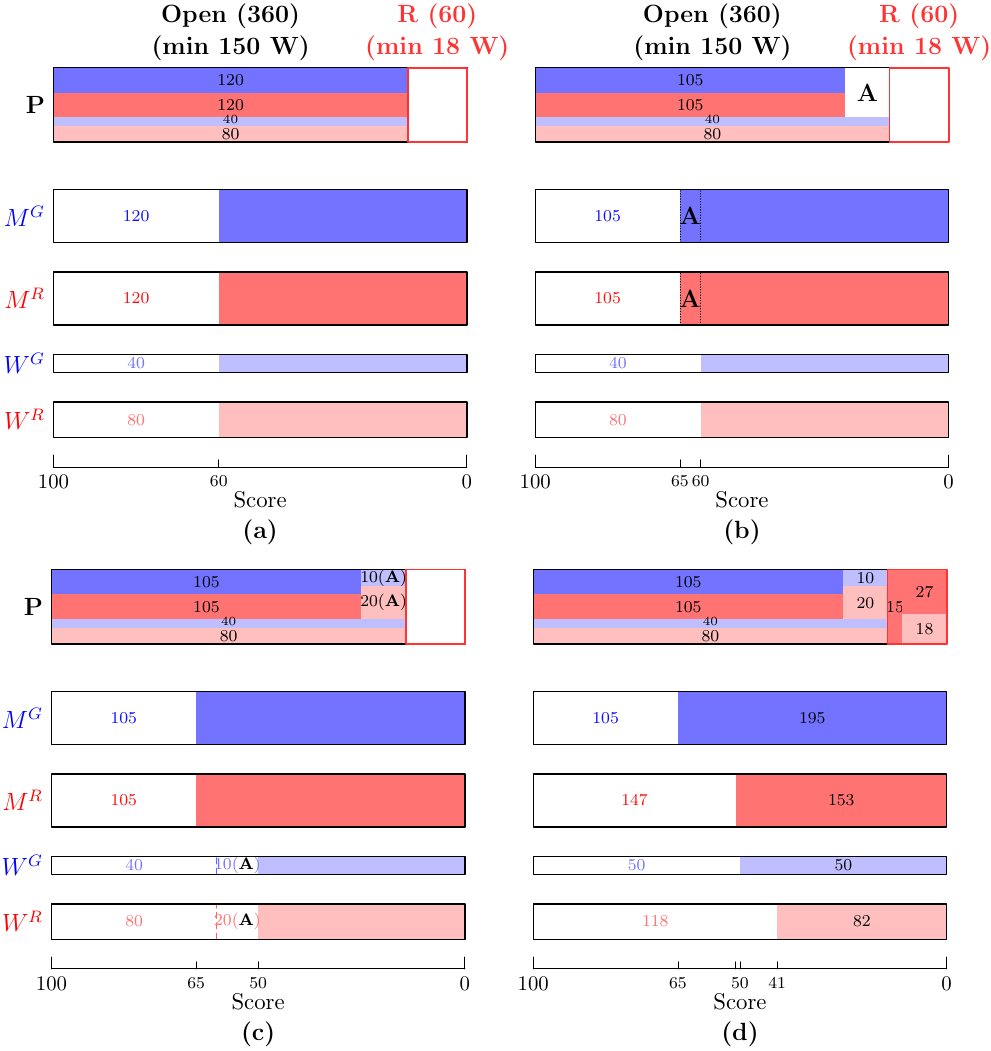}
\caption{Execution of the 2SMG choice rule via the horizontal adjustment process. 
Building on Example \ref{ex:VR-HR-implementation-2SMG-via-adjustments}, 
Panels (a)--(d) trace how the 2SMG choice rule operates with horizontal adjustments to implement the minimum guarantee 
for the continuum economy in Example \ref{ex:VR-HR-implementation}, illustrated in Panel (a) of Figure \ref{fig:2SMG-standard}. 
Compared with Figure \ref{fig:SCI--AKG}, the key difference is in Panel (c), where 
women from VR-protected categories remain eligible for positions created by the horizontal adjustment in Panel (b). 
Positions subject to horizontal adjustment are indicated with \textbf{A} in Panels (b) and (c). Abbreviations: P: Positions; W: Women; R: Vertically Reserved; $M^{G}$: General-category men;  $M^{R}$ Reserved-category men;   
$W^{G}$: General-category women;  $W^{R}$: Reserved-category women.}
\label{fig:2SMG-adjustment}
\end{figure}

\begin{subex}\label{ex:VR-HR-implementation-2SMG-via-adjustments}
As shown in Figure \ref{fig:2SMG-adjustment}, the process is identical to SCI--AKG until the second phase of the open-category horizontal adjustment, 
depicted in Panel (c) (compare Panels (a) and (b) in Figures \ref{fig:SCI--AKG} and \ref{fig:2SMG-adjustment}). At this stage, whereas SCI--AKG awards a measure of 30 adjusted positions to general-category women, 
2SMG awards them to the highest-scoring unmatched women from any category. Consequently, under 2SMG, a measure of 10 general-category women and 20 category-R women with scores 
between 50 and 60 receive these adjusted units (contrast Panel (c) of Figure \ref{fig:2SMG-adjustment} with Panel (c) of Figure \ref{fig:SCI--AKG}, where this “adjustment” is indicated with \textbf{A} in both choice rules).

After this stage, the two processes converge again, allocating the VR-protected positions using the minimum guarantee choice rule for the remaining applicants 
(see Panel (d) in Figures \ref{fig:SCI--AKG} and \ref{fig:2SMG-adjustment}).
\end{subex}

The simplicity of this minimalist remedy likely explains how members of the Indian judiciary eventually identified the correct procedure to implement VR and HR policies concurrently, albeit after a quarter-century delay.

\subsection{External Validity for Minimalist Market Design: Life Imitates Science in a Landmark Judgment} \label{sec:India-EV}

The Indian context offers another compelling case of external validity, both for the specific analysis in \cite{sonyen19, sonyen22} and for minimalist market design as a paradigm, 
where the parallel between research and subsequent judiciary-driven reform is unusually direct. The very failure of the \textit{no justified envy} axiom identified in that research---stemming from the SCI--AKG choice rule’s denial 
of open-category HR protections for VR-protected candidates---materialized in practice, causing persistent disruption for years. 
After decades of litigation, the Supreme Court in \textit{Saurav Yadav} (2020) adopted the same minimalist correction proposed in the research, 
targeting the root cause by removing this denial, and replaced the SCI--AKG choice rule after 25 years.

Before its publication, the initial draft of \cite{sonyen22} was circulated in March 2019 as a Boston College working paper \cite{sonyen19}.
In this work, we established that the Supreme Court-mandated SCI--AKG choice rule had been consistently undermining intended affirmative action measures in real-world practices by 
failing to satisfy the \textit{no justified envy} axiom.\footnote{In \cite{sonyen19}, the SCI--AKG choice rule is referred to as the SCI--VHR choice rule, and the \textit{no justified envy} axiom is referred to as \textit{elimination of justified envy}.} 
We also documented how this failure sparked numerous litigations between 1995 and 2020, causing recurring disruptions in the country's recruitment processes.

This issue became especially problematic in jurisdictions with extensive reliance on HR policies. For example, women in states such as Andhra Pradesh, Bihar, Chhattisgarh, 
Madhya Pradesh, Rajasthan, Sikkim, and Uttarakhand were granted HR protections for 20--35\% of positions through High Court mandates. 
Consequently, numerous High Court judgments in some of these states directly contradicted the Supreme Court’s ruling in \textit{Anil Kumar Gupta} (1995).

An illustrative example is the ruling of the Rajasthan High Court in \textit{Smt. Megha Shetty vs. State of Rajasthan (2013)}. 
In this case, a petitioner from the general category challenged the allocation of open-category HR-protected positions for women to women from the VR-protected category OBC.
 She argued that these women from OBC were not eligible for these positions unless they were selected without invoking the HR policy. 
 Although the OBC women had higher merit scores and the State of Rajasthan had used a more compelling procedure that satisfies the \textit{no justified envy axiom}, 
 the petitioner's case had merit under the SCI--AKG choice rule, which allows for justified envy in such situations.

Nevertheless, dismissing the lawsuit, a three-judge bench of the Rajasthan High Court emphasized:

\begin{quote} 
``The outstanding and important feature to be noticed is that it is not the case of the appellant-petitioner that she has obtained more marks than those 8 OBC (Woman) candidates, 
who have been appointed against the posts meant for General Category (Woman), inasmuch as, while the appellant is at Serial No.184 in the merit list, 
the last OBC (Woman) appointed is at Serial No.125 in the merit list. The controversy raised by the appellant is required to be examined in the context and backdrop of these significant factual aspects.'' 
\end{quote}

This case illustrates that many judges struggled to accept that the Supreme Court-mandated procedure could allow for justified envy, 
as it directly contradicts the spirit of affirmative action.

Motivated by this case and several others that contradicted the ruling in \textit{Anil Kumar Gupta} (1995), 
we proposed the 2SMG choice rule in \cite{sonyen19} as a minimalist reform of the SCI--AKG choice rule.\footnote{As documented in \cite{sonyen19, sonyen22}, 
the challenges arising from the failures of the Supreme Court-mandated SCI--AKG choice rule 
extended far beyond conflicting with high court judgments that upheld choice rules satisfying the \textit{no justified envy} axiom. 
In many instances, state institutions using the SCI--AKG choice rule were challenged in court by VR-protected candidates due to cases of justified envy. 
Some of these cases were ruled in favor of the state, while others favored the plaintiffs. Additionally, there were situations where VR-protected candidates were denied their open-category HR protections 
even though they did not claim their VR protections. Court decisions on these challenges also varied, sometimes favoring the candidates and other times not. 
Moreover, in other cases, VR-protected candidates were denied open-category HR-protected positions even when no positions were reserved for their VR-protected categories.}

Following the circulation of \cite{sonyen19}, the Supreme Court revoked the SCI--AKG choice rule after 25 years in \textit{Saurav Yadav} (2020), thereby providing external validity for minimalist market design.

Over the years, two opposing views implementing the HR policy had prevailed in High Court judgments. 
Under the first view, exemplified by \textit{Smt. Megha Shetty vs. State of Rajasthan (2013)}, the High Courts of Rajasthan, Bombay, Uttarakhand, and Gujarat 
upheld choice rules that satisfy the \textit{no justified envy} axiom. In contrast, under the second view, the High Courts of Allahabad and Madhya Pradesh enforced the SCI--AKG choice rule.

In alignment with the High Courts that adopted the first view, the justices eliminated the exclusion of VR-protected individuals from horizontal adjustments and mandated the 
\textit{no justified envy} axiom in India through their judgment in \textit{Saurav Yadav} (2020). In paragraphs 31--33, the justices ruled:

\begin{quote} 
``31. The second view is thus neither based on any authoritative pronouncement by this Court nor does it lead to a situation where merit is given precedence. 
Subject to any permissible reservations, i.e., either Social (Vertical) or Special (Horizontal), opportunities to public employment and selection of candidates must purely be based on merit.

Any selection which results in candidates getting selected against the Open/General category with less merit than other available candidates will certainly be opposed to principles of equality. 
There can be special dispensation when it comes to candidates being considered against seats or quotas meant for reserved categories, and in theory, 
it is possible that a more meritorious candidate coming from the Open/General category may not get selected. 
But the converse can never be true and will be opposed to the very basic principles which have all the while been accepted by this Court. 
Any view or process of interpretation which will lead to incongruity, as highlighted earlier, must be rejected.

32. The second view will thus not only lead to irrational results where more meritorious candidates may possibly get sidelined, as indicated above, but will, 
of necessity, result in acceptance of a postulate that Open/General seats are reserved for candidates other than those coming from vertical reservation categories. 
Such a view will be completely opposed to the long line of decisions of this Court.

33. We, therefore, do not approve the second view and reject it. The first view, which weighed with the High Courts of Rajasthan, Bombay, Uttarakhand, and Gujarat, is correct and rational.'' 
\end{quote}

Before this judgment, under the SCI--AKG choice rule, HR-protected positions in the open category could not be awarded to members of a 
VR-protected category---unless they were meritorious reserved candidates---even if HR protections were not fully honored in the open category. 
With the removal of this restriction in \textit{Saurav Yadav} (2020), the following axiom also effectively became mandated, although it was not the primary focus of the case.

\begin{definition} \label{def:MAHR}
An allocation of positions across vertical categories (including open) satisfies \textbf{maximal accommodation of HR protections} 
if no individual remains unassigned as long as they can increase the number of HR-protected positions honored in some category. 
\end{definition}

Moreover, as a potential replacement for the revoked SCI--AKG choice rule, the justices in \textit{Saurav Yadav} (2020) 
endorsed the 2SMG choice rule as a procedure that satisfies the \textit{no justified envy} axiom, even though they did not explicitly mandate it.

Central to our purposes, this reform was driven purely by the failure of a key social justice principle---the \textit{no justified envy} axiom---under the SCI--AKG choice rule, 
a rule that had been mandated by the Supreme Court for 25 years. The justices' primary concern was to enforce this essential principle, 
with their advocacy for a specific procedure being a secondary consideration. Therefore, similar to the school choice reforms
for Boston, Chicago, and England (see Section  \ref{sec:schoolchoice}), and the branching reform of the U.S. Army (see Section  \ref{sec:Army}), 
the Supreme Court's choice rule underwent reform as a direct consequence of a significant failure in the existing system based on a vital principle.

There is an additional and subtle aspect of the judgment in \textit{Saurav Yadav} (2020) that is particularly significant, 
even though it may not be widely understood in India. In addition to mandating the \textit{no justified envy} axiom, 
the justices made an important clarification regarding the implementation of VR policy in the presence of HR policy.

In \textit{Indra Sawhney} (1992), it was established that a VR-protected position cannot be awarded to beneficiaries who qualify for an open position based on merit.
However, neither this landmark judgment nor \textit{Anil Kumar Gupta} (1995)---which clarified various operational details of the HR policy---specified who qualifies for an 
open position when HR protections are also considered. This ambiguity was finally resolved in \textit{Saurav Yadav} (2020), where the court legally defined an individual 
deserving of an open position as one who qualifies for an open category position “with or without” invoking HR protections. In paragraph 36, the justices ruled:

\begin{quote}
``Even going by the present illustration, the first female candidate allocated in the vertical column for Scheduled Tribes may have secured higher position than the candidate at Serial No.64. 
In that event said candidate must be shifted from the category of Scheduled Tribes to Open/General category causing a resultant vacancy in the vertical column of Scheduled Tribes. 
Such vacancy must then enure to the benefit of the candidate in the Waiting List for Scheduled Tribes--Female.''
\end{quote}

Due to this crucial clarification, the following axiom is also effectively mandated by \textit{Saurav Yadav} (2020):

\begin{definition}
An allocation of positions across vertical categories (including open) satisfies \textbf{compliance with VR protections} 
if the following three conditions hold for any applicant $i$ who is assigned a VR-protected position: 
\begin{enumerate} 
\item All open-category positions are filled. 
\item For any individual $j$ with a lower merit score than $i$ who is assigned an open-category position, 
replacing $j$ with $i$ would decrease the number of HR-protected positions honored within the open category. 
\item Awarding $i$ an open-category position instead,  with or without replacing another individual, 
does not increase the number of HR-protected positions honored within the open category.\footnote{Prior to \textit{Saurav Yadav} (2020), 
unlike the first two conditions, the necessity of this third condition as a prerequisite for the assignment of VR-protected positions remained unclear.} 
\end{enumerate} 
\end{definition}

Together with the three other axioms also mandated under \textit{Saurav Yadav} (2020)---\textit{no justified envy}, \textit{maximal accommodation of HR protections} and \textit{non-wastefulness}---this clarification 
has a significant consequence in India.

\begin{theorem}[\citealp{sonyen22}] \label{thm:2SMG} 
Consider a setting in which no individual belongs to multiple HR-protected groups.
Then, 2SMG is the unique choice rule that satisfies \textit{non-wastefulness}, \textit{no justified envy}, \textit{maximal accommodation of HR protections}, and \textit{compliance with VR protections}. 
\end{theorem}

Therefore, while the justices have not explicitly mandated the 2SMG choice rule in \textit{Saurav Yadav} (2020), they have, in effect, 
mandated it in the country, as no other choice rule satisfies the explicit mandates of the judgment.\footnote{The 2SMG choice rule is mandated in the state of 
Gujarat under the High Court judgment \textit{Tamannaben Ashokbhai Desai vs. Shital Amrutlal Nishar (2020)}.}

Aligned with the version of the problem reviewed in \textit{Saurav Yadav} (2020), Theorem \ref{thm:2SMG} pertains to the basic version of the problem, where all positions 
are identical and no individual belongs to more than one HR-protected group.\footnote{Strictly speaking, \textit{Saurav Yadav} (2020) 
does not rule out the possibility of individuals belonging to multiple HR-protected groups; it simply does not address this scenario.} 

Next, in Sections \ref{sec:overlappingHR} and \ref{sec:heterogenous}, we explore the implications of this judgment's clarifications on HR policy for more general versions of the problem without these restrictions.

\subsection{Overlapping Horizontal Reservations} \label{sec:overlappingHR}

While \textit{Persons with Disabilities} (PwD) is the only HR-protected group recognized nationwide in India by a Supreme Court judgment,\footnote{HR protections for 
Persons with Disabilities were established in the Supreme Court judgment \textit{Union of India vs. National Federation of the Blind (2013)}.} other groups, such as Women (W), 
receive HR protections in several states through mandates issued by High Courts or state governments. As a result, unlike VR-protected groups, individuals can often belong to multiple HR-protected groups in practical applications.

How does the possibility of overlapping HR protections influence the judgment in \textit{Saurav Yadav} (2020) and our analysis in Sections \ref{sec:horizontal} through \ref{sec:India-EV}?
We address these questions next.

Let's start with a key observation. In India, when institutions announce positions for recruitment or admissions, they categorize these positions based on vertical categories (including open) 
and horizontal protections (or lack thereof). For example, in a setting with four VR-protected categories---SC, ST, OBC, and EWS---and two HR-protected groups---PwD and W---specific numbers of 
positions are designated for pairs such as SC-W or Open--PwD. In practice, this means that when an individual belonging to multiple HR-protected groups is awarded a position, only one of their HR protections is honored. 
Thus, if a woman with disabilities is awarded a position, she fulfills either the HR protection for W or for PwD, but not both.\footnote{This convention is referred to as \textit{one-to-one HR matching} in \cite{sonyen22}.
See \cite{sonmez/yenmez:20} for an analysis of the alternative \textit{one-to-all HR matching} convention, in which an individual fulfills all their HR protections.}

This observation is significant because neither the SCI--AKG nor the minimum guarantee choice rule is well-defined under this convention in cases with overlapping HR protections. 
That is, without additional specifications, both procedures can yield multiple outcomes that align with their basic descriptions.

Let's illustrate this issue with the minimum guarantee choice rule. Under this rule, given a set of eligible applicants, the highest-merit-score beneficiaries of HR-protected groups are awarded 
HR-protected positions up to capacity in the first step. In the second step, the remaining positions---including any unfilled HR-protected positions---are allocated to the highest-scoring individuals who have not yet received a position.

While the first step of this process fully specifies which individuals receive HR-protected positions when each applicant is a beneficiary of at most one HR-protected group, 
it fails to do so when they can be beneficiaries of multiple HR-protected groups. Let's explore why with the following scenario.

\begin{figure}[!tp]
 \centering
       \includegraphics[scale=1.02]{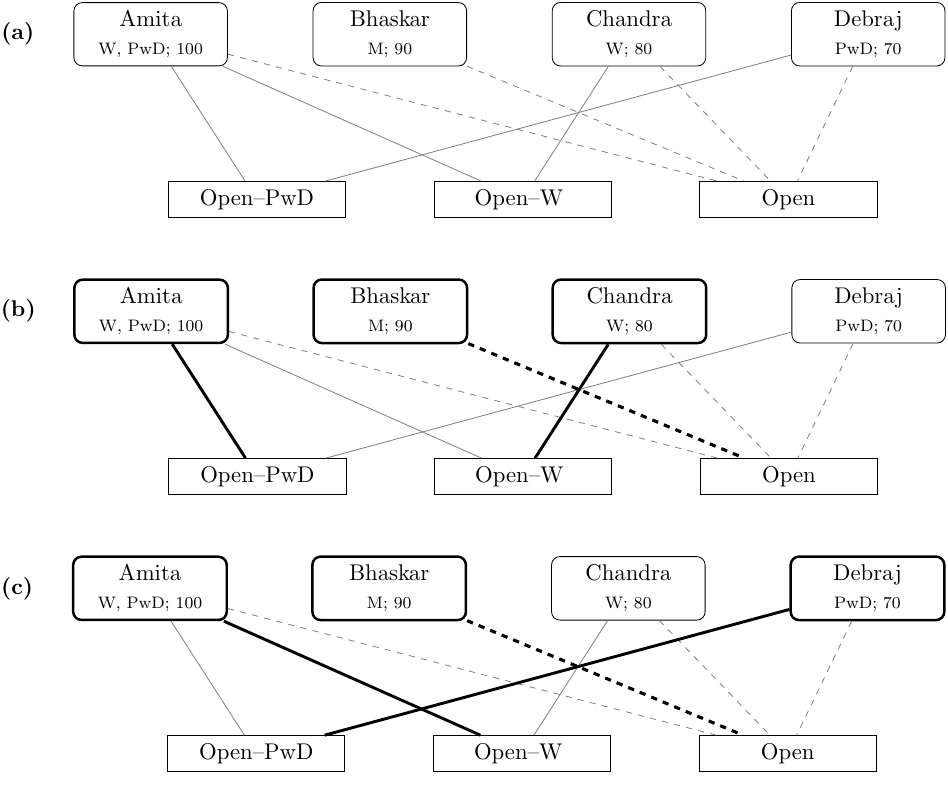}
\caption{Allocation of HR-protected positions under overlapping horizontal reservations.  
For the setting in Example \ref{ex:MG-overlappingHR-NJE}, Panel~(a) shows eligibility in the first phase of the minimum guarantee choice rule for the Open category (indicated by \emph{light unbroken} lines), 
which requires membership in an HR-protected group, and in the second phase (indicated by \emph{light broken} lines), which does not.  
In Panels (b) and (c), assignments are indicated by \textbf{\emph{bold}} versions of these lines. 
Because they each belong to at least one HR-protected group, Amita, Chandra, and Debraj are eligible in the first phase. 
As the highest-scoring applicant, Phase 1 starts with Amita; since she belongs to both HR-protected groups W and PwD, she may be assigned to either of these positions.  
If Amita is assigned the PwD position, then the W position is assigned to the only other woman, Chandra, in the first phase, and the remaining open position is assigned to the highest-scoring remaining applicant, 
Bhaskar, in the second phase (Panel (b)).  
If instead Amita is assigned the W position, then the PwD position is assigned to the only other person with a disability, Debraj, 
in the first phase, and again the remaining open position is assigned to Bhaskar in the second phase (Panel (c)). Abbreviations: W: Women; M: Men; PwD: Persons with Disabilities.}
\label{fig:overlappingHR}
\end{figure}

\begin{example}  \label{ex:MG-overlappingHR-NJE}
There is a single category---say, the open category---with three positions. 
Within the open category, there are two HR-protected groups---PwD and W---each with one HR-protected position.
There are four applicants---Amita, Bhaskar, Chandra, and Debraj---who have the following characteristics:

\begin{itemize} 
\item Amita is a woman with a disability and has a merit score of 100. 
\item Bhaskar is a man without a disability and has a merit score of 90. 
\item Chandra is a woman without a disability and has a merit score of 80. 
\item Debraj is a man with a disability and has a merit score of 70. 
\end{itemize}

Since she has the highest merit score, it is clear that Amita receives one of the HR-protected positions in the first step of the minimum guarantee choice rule. 
However, as a member of both HR-protected groups, she could receive either position, which is not specified in the description of the minimum guarantee choice rule.

If Amita receives the Open--PwD position, then the Open--W position is awarded to Chandra---the only other female applicant---in the first step. 
If, on the other hand, Amita receives the Open--W position, then the Open--PwD position is awarded to Debraj---the only other applicant with a disability---in the first step. 
In either case, Bhaskar receives the third open category position in the second step (see Figure \ref{fig:overlappingHR}). 
\end{example}

While Example  \ref{ex:MG-overlappingHR-NJE}  focuses on the minimum guarantee choice rule, highlighting how the management of HR-protected positions affects the outcome under this rule, 
the issue holds broader significance for any reserve system when protected groups overlap. In both real-life settings and market design literature, a common approach to avoid ambiguities in 
describing reserve systems is to fix the order in which specific positions are awarded by defining an \textit{order of precedence} (see Section \ref{sec:subtleties-reserve}).

Using this approach, the minimum guarantee choice rule can be extended with two possible orders of precedence in Example \ref{ex:MG-overlappingHR-NJE}: 
either (1) the Open--PwD position is awarded first, followed by the Open--W  position, with any unfilled open category positions awarded last; 
or (2) the Open--W  position is awarded first, followed by the Open--PwD position, and any unfilled open category positions awarded last. 
In Example \ref{ex:MG-overlappingHR-NJE}, positions are allocated to Amita, Chandra, and Bhaskar under the first scenario, 
and to Amita, Debraj, and Bhaskar under the second, as depicted in Panels (b) and (c) of Figure \ref{fig:overlappingHR}, respectively.

It could be argued that the first scenario is better than the latter since it denies a position to the lowest merit-score individual, Debraj, rather than to the second-lowest merit-score individual, 
Chandra, who is denied one in the latter. Given that both allocations satisfy the minimum guarantees for both HR-protected groups, why admit a lower-merit-score individual at the expense of an individual with a higher merit score? 
So, in Example \ref{ex:MG-overlappingHR-NJE}, awarding the Open--PwD position before the Open--W  position results in a more ``meritorious'' outcome.

For the same reason, allocating the Open--W  position before the Open--PwD position in this example would also result in a violation of the \textit{no justified envy} axiom, 
as defined in Definition \ref{def:India-NJE}: Debraj would be awarded a position at the expense of Chandra, even though he neither has a higher merit score than she does, 
nor does awarding him a position at her expense increase the number of HR protections that are honored.

The broader issue here, however, is not the specific order of precedence chosen but the rigidity of having a fixed one.
If, for instance, the merit scores of Chandra and Debraj were swapped in Example \ref{ex:MG-overlappingHR-NJE}---with Debraj receiving a merit score of 80 and Chandra 70---it would then be the 
first order of precedence that produces a less meritorious outcome and causes a violation of the \textit{no justified envy} axiom. This observation illustrates that committing to a fixed order of precedence 
may not only sacrifice meritocracy more than necessary to accommodate the HR policy but also lead to a violation of the \textit{no justified envy} axiom.

It is worth emphasizing that, while extending the minimum guarantee choice rule via a fixed order of precedence may lead to a failure of the \textit{no justified envy axiom}, 
it remains unclear whether this constitutes a violation of \textit{Saurav Yadav} (2020)---unlike cases involving non-overlapping HR protections.

Here is why: In the context of the Indian reservation system with non-overlapping HR policies, as well as in earlier settings such as school choice and the U.S. Army's branching process,
whenever there is a violation of \textit{no justified envy}, an individual is able to challenge another’s assignment without involving any third party. 
This makes both the formal definition and practical enforcement of \textit{no justified envy} relatively straightforward.

In the current setting, however, challenging an individual’s assignment may require adjustments to one or more other assignments. 
Specifically, although the challenge is directed at a particular individual, the correction may also involve others. In Example \ref{ex:MG-overlappingHR-NJE}, for instance, 
Chandra’s potential objection to Debraj’s assignment in the first scenario would also entail a change in Amita’s assignment from Open--W to Open--PwD  (see Figure \ref{fig:overlappingHR}). 
While, in my opinion, such a challenge aligns with \textit{Saurav Yadav} (2020),\footnote{As the basis for my opinion, see \textit{Mamta Bisht vs. State of Uttaranchal (2005)} at the Uttarakhand High Court, 
where the plaintiff successfully challenged an individual’s assignment in the open category indirectly. She argued that a third party who had been awarded an SC position should instead be assigned the open category position, 
thereby allowing the SC position to be reassigned to the plaintiff.} as an economist, I cannot assert that this point is clear-cut under reservation jurisprudence.

As a final observation on this point, it is worth noting that for Chandra to challenge Debraj's assignment, she would need to know Amita's disability status.
While the challenge may indeed have merit in principle under \textit{Saurav Yadav} (2020), it may rely on private information, potentially making it more difficult to pursue.

Beyond a possible violation of the \textit{no justified envy} axiom, as seen in our next example, awarding HR-protected positions through a fixed order of precedence can result in another anomaly.

\begin{figure}[!tp]
 \centering
       \includegraphics[scale=1.05]{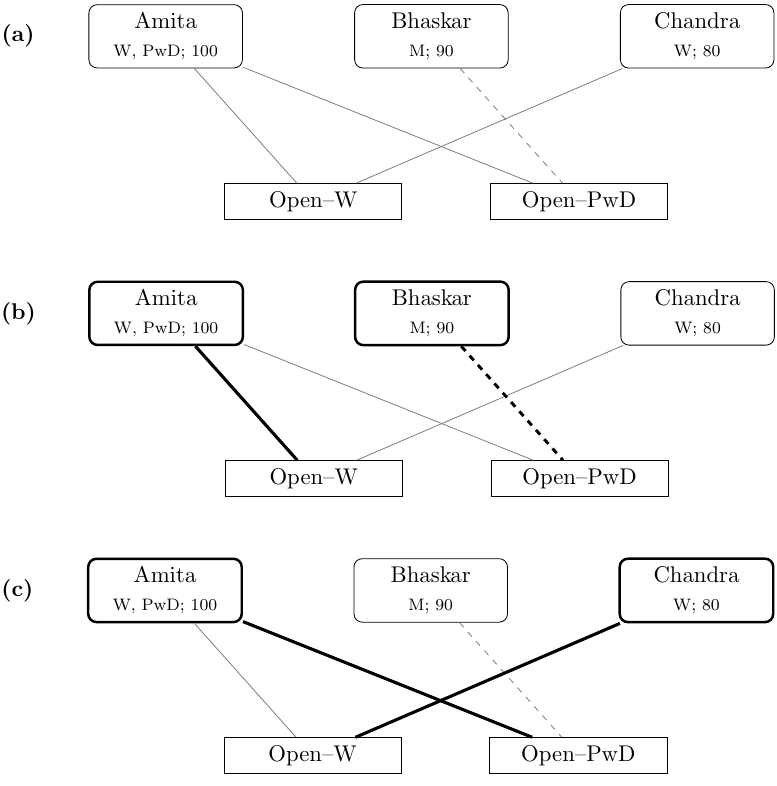}
\caption{Allocation of HR-protected positions under overlapping horizontal reservations with a fixed order of precedence.  
For the setting in Example \ref{ex:MG-overlappingHR-MAHR}, Panel~(a) shows eligibility in the first phase of the minimum guarantee choice rule (indicated by \emph{light unbroken} lines), 
which requires membership in an HR-protected group, and in the second phase (indicated by \emph{light broken} lines), which does not.  
In Panels~(b) and~(c), assignments are indicated by \textbf{\emph{bold}} versions of these lines.  
Panel~(b) presents the order of precedence, where the Open--W position is processed before the Open--PwD position: 
Amita receives the Open--W position and Bhaskar the Open--PwD position, so only the HR-protection for W is honored.  
Panel~(c) presents the counterfactual order of precedence, where the Open--PwD position is processed first: 
Amita is assigned the Open--PwD position and Chandra the Open--W position, thereby honoring both HR protections. Abbreviations: W: Women; M: Men; PwD: Persons with Disabilities.}
\label{fig:overlappingHR-MAHR}
\end{figure}

\begin{example}  \label{ex:MG-overlappingHR-MAHR}
There is a single category---say, the open category---with two positions. Within the open category, there are two HR-protected groups---PwD and W---each with one HR-protected position.
The order of precedence is: Open--W  position first, Open--PwD position second. There are three applicants---Amita, Bhaskar, and Chandra---with the following characteristics:

\begin{itemize} 
\item Amita is a woman with a disability and has a merit score of 100. 
\item Bhaskar is a man without a disability and has a merit score of 90. 
\item Chandra is a woman without a disability and has a merit score of 80. 
\end{itemize}

Amita is awarded the Open--W  position as the highest-merit-score woman. Since she is the only individual with a disability, the Open--PwD position is then assigned to the next highest-merit-score individual, 
Bhaskar, who does not have a disability. Consequently, only one of the HR-protected positions is honored in this outcome. However, if Amita were instead assigned the Open--PwD position, 
Chandra could be awarded the Open--W  position, thereby honoring both HR-protected positions (see Figure \ref{fig:overlappingHR-MAHR}).
\end{example}

Example \ref{ex:MG-overlappingHR-MAHR} shows that generalizations of the minimum guarantee choice rule relying on a fixed order of precedence may violate the \textit{maximal accommodation of HR protections} axiom.

This prompts the question of whether a generalization of the minimum guarantee choice rule exists that satisfies both the \textit{no justified envy} and \textit{maximal accommodation of HR protections} axioms.
In the following, we provide an affirmative answer.

\subsubsection{Smart Reserves and the Two-Step Meritorious Horizontal Choice Rule} \label{sec:meritoriousHR}

We introduce a generalization of the minimum-guarantee choice rule through a \textit{smart reserve} system and 
utilize it to design a choice rule that extends 2SMG to settings with overlapping HR protections.
 
The key innovation is to allocate all HR-protected positions simultaneously by awarding units to the highest merit-score individuals, 
while respecting certain constraints. Specifically, these constraints involve (1) providing minimum guarantees for each HR-protected group and (2) maximizing the total number of HR-protected positions that are honored.

Let's illustrate this procedure using Example \ref{ex:MG-overlappingHR-NJE}, depicted in Figure \ref{fig:overlappingHR}. 
We will consider individuals one at a time based on their merit scores, from highest to lowest. Under the minimum guarantee policy, HR-protected positions must be awarded before the remaining positions in the category.

The highest merit-score individual, Amita, belongs to both HR-protected groups, so she must receive an HR-protected position, which could be from either group.
Instead of immediately fixing her assignment, we leave it flexible, determining which position she receives based on the allocation that enables higher-ranked individuals to secure positions later.

The second-highest merit-score individual, Bhaskar, is not a member of any HR-protected group; he will have to wait, although he is certain to receive an open category position, 
as there is one not tied to any HR-protected group. 

Next is Chandra, who belongs to only one HR-protected group: W. 
Here Amita’s assignment becomes decisive: for Chandra to receive an HR-protected position, namely the Open--W position, 
Amita must be assigned the Open--PwD position (contrast Panel (b) with Panel (c) in Figure \ref{fig:overlappingHR}). 

With the HR-protected positions now settled, Bhaskar---the highest merit-score individual who has not yet received a position---receives the remaining open position.

Critically, if the scenario were slightly different and Debraj---a man with a disability---had a higher merit score than Chandra, he would receive the Open--PwD position, 
resulting in Amita being assigned the Open--W position instead.
Leveraging the flexibility in Amita's assignment to allocate positions to higher-merit applicants is the ``smart'' aspect of this procedure.

\paragraph{Meritorious Horizontal Choice Rule.}
This idea extends naturally regardless of the specifics of the HR-protected groups or their overlapping memberships. 
First, HR-protected positions are allocated in a ``greedy'' manner to the highest merit-score individuals, maximizing the total number of these positions that are honored.\footnote{Technically, this requires solving a 
\textit{maximum matching problem} in a bipartite graph. In an unweighted bipartite graph, the objective is to find a matching with maximum cardinality. 
The \textit{Hopcroft-Karp-Karzanov} algorithm \citep{Hopcroft/Karp:73, Karzanov:1973} solves this problem efficiently, and even faster algorithms exist for specific classes of bipartite graphs.} 
Next, the remaining positions in the category are allocated to the highest merit-score eligible individuals who have not yet received a position. 
Introduced in \cite{sonmez/yenmez:20}, we refer to this choice rule as the \textit{meritorious horizontal choice rule}.\footnote{The terminology \textit{meritorious horizontal choice rule} was first introduced in \cite{sonyen22}. 
Originally, \cite{sonmez/yenmez:20} referred to this choice rule as the \textit{horizontal envelope choice rule}.}

The following result explains why this choice rule is the appropriate extension of the minimum guarantee choice rule in cases where HR-protected groups have overlapping memberships.

\begin{theorem}[\citealp{sonyen22}]   \label{thm:MeritoriousHorizontal} 
Consider a setting with a single category and multiple HR-protected groups with overlapping memberships. 
Then, the meritorious horizontal choice rule is the unique choice rule that satisfies \textit{non-wastefulness}, \textit{no justified envy}, and \textit{maximal accommodation of HR protections}. 
\end{theorem}


Having clarified the proper generalization of the minimum guarantee choice rule for the version of the problem with overlapping HR protections, a natural extension of the 2SMG mechanism emerges. 

\paragraph{Two-step Meritorious Horizontal (2SMH) Choice Rule.}
In the first step, open-category positions are allocated using the meritorious horizontal choice rule. 
 In the second step, 
VR-protected positions are allocated, category by category, to eligible individuals who did not receive an open position in the first step, again using the meritorious horizontal choice rule.

The next result establishes that Theorem \ref{thm:2SMG} naturally extends to the version of the problem involving identical positions and overlapping HR protections.
In this setting, the 2SMH choice rule takes on the central role previously held by the 2SMG choice rule in the simpler version with identical positions and non-overlapping HR protections.

\begin{theorem}[\citealp{sonyen22}] \label{thm:2SMH} 
Consider a setting in which individuals can belong to multiple HR-protected groups.
Then, 2SMH is the unique choice rule that satisfies \textit{non-wastefulness}, \textit{no justified envy}, \textit{maximal accommodation of HR protections}, and \textit{compliance with VR protections}. 
\end{theorem}

By Theorem \ref{thm:2SMH}, and subject to the caveats discussed earlier in this section---namely, 
when objections to assignments may involve third parties---there exists a unique choice rule that fulfills the mandates of \textit{Saurav Yadav} (2020) 
even when the HR-protected groups are overlapping.

\subsection{Allocation of Heterogenous Positions} \label{sec:heterogenous}

Up to this point, our focus in Section \ref{sec:India} has been on settings with identical positions. Another common variant in India involves the centralized allocation of 
heterogeneous positions across multiple institutions. This variant is most  prevalent in the allocation of prestigious government jobs---such as those in the Indian Administrative Service, 
Indian Police Service, and Indian Forest Service---and in the assignment of college seats in medical and engineering schools. 
As in the simpler case with identical positions, VR and HR policies are also enforced here, codified in \textit{Indra Sawhney} (1992). 
However, unlike in the case of identical positions, the Supreme Court refrained from formulating and enforcing an explicit procedure for implementing these policies in the context of heterogeneous positions.

Consequently, government agencies and higher education institutions have had to design and implement their own mechanisms to manage this more complex variant. 
Given the 25 years of challenges discussed earlier even for the simpler case with identical positions, it is perhaps unsurprising that these designs and their implementation have also faced 
ongoing litigation and disruptions in recruitment and admissions processes.

Similar to several other applications we have explored---such as Turkish college admissions and the U.S. Army’s branching processes---system operators in India have commonly relied on 
mechanisms that use the Simple Serial Dictatorship (SSD) as their core engine. To implement the VR policy, and paralleling the case of identical positions, the practice of allocating open-category positions before 
VR-protected ones and applying the concept of migration continues in the allocation of heterogeneous positions.

In a typical mechanism, open-category positions across ``all'' institutions are tentatively allocated using an SSD. The highest-merit individual is provisionally assigned their top choice, 
the second-highest-merit individual receives their top remaining choice among the open-category positions, and so on. As in the allocation of identical positions, VR-protected individuals 
who receive an open-category position are granted the status of meritorious reserved candidates (see Definition  \ref{def:MRC}). 
Once open-category positions are provisionally assigned, VR-protected positions are then tentatively allocated to eligible individuals in a similar manner, again using SSD in a second phase. 
Adjustments for HR policy can be made at any stage.

At this point, this methodology encounters a technical challenge. Since a meritorious reserved candidate may receive two distinct tentative assignments between the two phases, 
they ``migrate'' to the institution-category pair associated with their preferred position. This migration leaves positions vacated by meritorious reserved candidates that must subsequently be reassigned. 
Ironically, while the Supreme Court initially refrained from mandating a specific mechanism for the allocation of heterogeneous positions, it has, following several litigations,
issued explicit directives on how these vacated positions must be reallocated.

In formulating these directives based on the mechanics of this SSD-based methodology, several key Supreme Court judgments have made fundamental errors.
Two judgments, in particular, have set influential precedents. In \textit{Union of India vs. Ramesh Ram (2010)}, positions vacated by meritorious reserved candidates 
were exclusively awarded to general category candidates for government jobs.
By contrast, in \textit{Tripurari Sharan vs. Ranjit Kumar Yadav (2018)}, these positions were exclusively awarded to VR-protected individuals from the category of the 
vacating meritorious reserved candidates for medical college seats.
Not only do these mandates contradict each other, but both are implausible, as the highest-scoring individuals deserving the vacated positions could be from any category, 
whether unmatched or tentatively holding less-preferred positions.

The creation of such ``artificial'' property rights for specific categories of individuals based on flawed mechanics not only represents a misguided attempt to resolve the underlying issues with the SSD-based mechanism 
but also directly contradicts the axiom of \textit{no justified envy}.
Thus, the root cause of these legal challenges lies in the excessive reliance on the concept of ``migration,'' 
further exacerbated by the practice of awarding ``artificial'' property rights to arbitrary groups based on the tentative assignments of meritorious reserved candidates. 
This reliance on migration is an ill-equipped tool for addressing applications with heterogeneous positions and represents a methodological failure at the core of the legal inconsistencies 
and litigations in India.\footnote{A simple search on Indian Kanoon, a free search engine for Indian law, reveals that as of November 2024, 
there have been 959 cases related to the ``migration of meritorious reserved candidates'' at the Supreme Court or various high courts.}

To address these challenges, it is important to recognize that accommodating the principles outlined in \textit{Indra Sawhney} (1992) or their refinements in \textit{Saurav Yadav} (2020) 
cannot be achieved solely by relying on the concept of migration---unless an arbitrary number of rounds of migrations are allowed through an iterative procedure such as the 
individual-proposing deferred acceptance algorithm. The necessity of iterative procedures is analogous to the need for such methods to find \textit{stable} outcomes in two-sided matching markets.

\subsubsection{A Minimalist Resolution: 2SMH--DA Mechanism}

Even though \textit{Saurav Yadav} (2020) pertains to the version of the problem with identical positions, it also clarifies ambiguities in \textit{Indra Sawhney} (1992) 
concerning VR and HR policies, including for the more general version with heterogeneous positions. This observation suggests a minimalist resolution for this more general problem with 
heterogeneous positions and overlapping HR-protected groups: using the individual-proposing deferred acceptance algorithm (see Section \ref{sec:two-sided-matching}), 
where each institution determines its set of best applicants with the 2SMH choice rule. Let's refer to this preference revelation mechanism as \textbf{\textit{2SMH--DA}}.

In essence, this problem can be thought of as the more basic version of the problem involving identical positions, embedded within the school choice problem (see Section \ref{sec:schoolchoice})---particularly the 
Turkish student placement variant, where the axiom of \textit{no justified envy} plays an indispensable role.
In this sense, 2SMH--DA, which simply integrates the respective solutions for these two applications, 
emerges as a very natural solution to India’s affirmative action problem with heterogeneous positions. 

To further illustrate why this mechanism is uniquely appealing for this version of the problem, 
we extend the following two axioms from \textit{Saurav Yadav} (2020) into this more general setting.

\begin{definition} \label{def:India-NJE-heterogenous}
An allocation of positions across vertical categories (including open) at multiple institutions satisfies \textbf{no justified envy} if there exists no institution-category pair $(s,v)$ 
and two applicants $i$ and $j$, both eligible for category $v$ at institution $s$, such that: 
\begin{enumerate} 
\item Applicant $j$ is awarded a position at institution $s$ in category $v$, 
\item Applicant $i$ strictly prefers a position at institution $s$ to their current assignment, and 
\item Applicant $j$ neither has a merit score as high as that of applicant $i$, nor would awarding them a position at institution $s$ in category $v$ at the expense of applicant $i$ 
increase the number of HR-protected positions honored at institution $s$ in category $v$.
\end{enumerate}
\end{definition}

\begin{definition} \label{def:MAHR-heterogenous}
An allocation of positions across vertical categories (including open) at multiple institutions satisfies \textbf{maximal accommodation of HR protections} if,
for any individual $i$, institution $s$, and vertical category $v$, whenever assigning individual $i$ a position in category $v$ at institution $s$ (instead of their current assignment, 
possibly replacing another individual who was assigned a position in category $v$ at institution $s$) increases the number of HR-protected positions honored in that category at institution $s$, 
it must be the case that individual $i$ strictly prefers their current assignment over institution $s$.
\end{definition}

Our next result establishes that, although multiple mechanisms satisfy the mandates of \textit{Saurav Yadav} (2020) for the more general version of the 
problem in India with heterogeneous positions, 2SMH--DA remains the only plausible option among them.

\begin{theorem}[\citealp{sonmez/yenmez:24}] \label{thm:2SMH-heterogenous} 
Consider a setting with heterogeneous positions and overlapping HR-protected groups. Then:
\begin{enumerate}
\item 2SMH--DA satisfies \textit{individual rationality}, \textit{non-wastefulness}, \textit{no justified envy}, 
\textit{maximal accommodation of HR protections}, and \textit{compliance with VR protections}, and \textit{Pareto dominates} any other mechanism that also satisfies these axioms.
\item 2SMH--DA is the unique mechanism that satisfies \textit{individual rationality}, 
\textit{non-wastefulness}, \textit{no justified envy}, \textit{maximal accommodation of HR protections}, \textit{compliance with VR protections}, and \textit{strategy-proofness}. 
\end{enumerate}
\end{theorem}

\section{Medical Resource Rationing During COVID-19} \label{sec:pandemic} 

The COVID-19 pandemic revealed another vital resource allocation problem, where minimalist market design proved highly effective: 
emergency rationing of scarce medical resources \citep{pathak/sonmez/unver/yenmez:24}.
Thanks to the central focus this approach  places on stakeholder objectives, over the course of a year, we were 
able to communicate our ideas  with dozens of healthcare officials when these ideas were most useful. The same focus  
also helped our team to convince many of these officials or experts in medical ethics and public healthcare
to advocate for and adopt the \textit{reserve systems} we designed for equitable allocation of vaccines and therapies
on real time, as the need arise  \citep{persad/peek/emanuel:2020, Schmidt:2020, Schmidt-et-al:2020, Sonmezetal-categorized:2021, Schmidt-NatureMedicine:2021, 
rubin-et-al:21, pathak/sonmez/unver:2021, pathak/persad/sonmez/unver:22, white-et-al:22, McCreary:23}. 

\subsection{The Toughest Triage}

At the onset of the COVID-19 pandemic the world faced a challenge unlike any other for decades. 
In a perspective piece  in the \textit{New England Journal of Medicine}, 
\cite{Truog:2020} reflected upon some of these challenges the world was beginning to face. 
\begin{quote}
``Although rationing is not unprecedented, never before has the American public been faced with the prospect of 
having to ration medical goods and services on this scale.'' (p. 1973)
\end{quote}
In the same issue of the \textit{New England Journal of Medicine}, 
\cite{ezekiel:2020} outlined the desirable ethical principles for allocation of scarce medical resources in crisis situations, and 
urged healthcare officials worldwide to develop the guidelines and procedures to implement them:
\begin{quote}
``The need to balance multiple ethical values for various interventions and in different circumstances is 
likely to lead to differing judgments about how much weight to give each value in particular cases. 
This highlights the need for fair and consistent allocation procedures that include the affected parties: 
clinicians, patients, public officials, and others. 
These procedures must be transparent to ensure public trust in their fairness.'' (p. 2054)
\end{quote}

At the time \cite{ezekiel:2020} and \cite{Truog:2020}  were first published online in March 2020, the rationing of 
ventilators and intensive care units (ICUs) was already underway in Northern Italy \citep{Rosenbaum:2020}. 

Upon exploring with Parag Pathak and Utku \"{U}nver  how these ethical principles were operationalized for the allocation of ventilators and ICUs, 
we made an important discovery----one that made it clear it was time to put our expertise to good use \citep{pathak/sonmez/unver:2020}. 
Little did we know that, for nearly a whole year from that point on, the three of us would end up setting aside all inessential activities and devoting all our time and energy to contributing to the global fight against COVID-19.

\subsection{Priority Point System: Prevalence, Challenges, and Constraints} 

As their recommended allocation rule for crisis rationing of ventilators and ICUs, state guidelines in the U.S. almost uniformly 
suggest some form of a \textbf{\textit{priority point system}}. In these guidelines, a patient’s point score is most often determined by a measure of mortality risk called the 
\textit{Sequential Organ Failure Assessment} (SOFA) score.\footnote{This procedure can be thought as a simple serial 
dictatorship  induced by the SOFA score.} 
I refer to this specific priority point system as the \textit{SOFA system}.\footnote{SOFA scoring system is first introduced in \cite{SOFA:1996} on behalf
of the Working Group on Sepsis-Related Problems of the European Society of Intensive Care Medicine. 
Under this system,  each of six organ groups lungs, liver, brain, 
kidneys, blood clotting and blood pressure is assigned a score of 1 to 4,  with higher scores for more severely failed organs.
The total across the six organ groups determine the SOFA score of the patient, which then determines her priority for 
a ventilator or an ICU. In recent years, however, many expressed concerns that the SOFA system 
discriminates against certain groups. See, for example, \cite{SOFA-disparities:2021} for one such study.} 

While a single utilitarian ethical principle, \textit{saving most lives}, determined who received a ventilator or an ICU bed in crisis situations under most state guidelines in the U.S., 
we made the following observation in two of the most detailed guidelines from New York State and Minnesota: During their deliberations in preparing these guidelines, 
members of both task forces sought to accommodate a second ethical principle known as the principle of \textit{instrumental value}. 
Under this principle, individuals who can help save additional lives are given special consideration in the allocation of vital resources.

As seen in the quotes from these guidelines below, however, members of neither task force could determine how to operationalize the \textit{instrumental value} 
principle alongside their primary principle of \textit{saving most lives} without risking outcomes they deemed unacceptable.

\begin{quote}
``[...] it is possible that they [essential personnel] would use most, if not all, 
of the short supply of ventilators; other groups systematically would be deprived access.''\\
\mbox{}\hfill \cite{minnesota:10},  MN Dept. of Health\\ 

``[...] may mean that only health care workers obtain access to ventilators in certain communities. 
This approach may leave no ventilators for community members, including children; this  alternative was unacceptable to the Task Force.''\\
\mbox{}\hfill  \cite{nys:15}, NY Dept. of Health
\end{quote} 

Task forces responsible for developing healthcare crisis management guidelines are typically composed of experts in medicine and medical ethics, 
as well as healthcare officials. Much like officials in our earlier market design applications---such as those in public school systems, the military, transplant surgery, 
and the judiciary in India---members of these task forces often lack formal expertise in designing allocation mechanisms. 
Moreover, the literature on medical ethics and public healthcare has traditionally relied on various priority systems to allocate scarce medical resources, even when balancing multiple ethical values.

For example, at the onset of COVID-19, the leading allocation system that accommodated multiple ethical principles for the allocation of ventilators was a priority point system by \cite{white:2009}. 
Through a modification of the SOFA system, their system incorporated three ethical principles: it utilized a coarsening of patient SOFA scores to accommodate the principle of \textit{saving most lives} 
and added points to address two additional principles, \textit{saving the most years of life} and the \textit{life-cycle} principle.\footnote{The \textit{life-cycle} principle asserts that younger patients, 
who have had the least opportunity to live through various stages of life, should be given higher priority for the scarce medical resource.}

In their efforts to incorporate the \textit{instrumental value} principle in addition to the \textit{saving most lives} principle, members of the New York State and Minnesota task forces experimented with a similar methodology, 
following standard practices in their fields. They considered awarding extra points to essential personnel in addition to their SOFA system scores. 
However, since this approach could potentially allocate all units to essential personnel in situations of extreme scarcity, 
they not only abandoned this procedure but also discarded the \textit{instrumental value} principle altogether.

Essentially, the task forces in New York State and Minnesota attempted to design a new ventilator rationing system through a minimalist refinement of the SOFA system, 
but they failed to develop an acceptable solution. However, another minimalist refinement of the SOFA system---a \textbf{\textit{reserve system}} similar to those explored in Sections 
\ref{sec:reserve-systems} and \ref{sec:India}---could have provided a potentially acceptable alternative.

\subsection{Pandemic Reserve System: A Minimalist Resolution}

Instead of adjusting the scoring rule within the SOFA system, New York State and Minnesota task forces could have prioritized a portion of the ventilators using patient SOFA scores and employed a separate priority ranking for the remaining units. 
To accommodate the \textit{instrumental value} principle, this secondary ranking would assign heightened priority to essential personnel. 
This approach is similar to the dual-criteria reserve system discussed in Section  \ref{sec:subtleties-reserve}.

Reserve systems have been prevalent in real-world settings as a prominent tool for reaching compromises between groups with competing objectives.
Such settings include ensuring equitable representation of minorities and women in legislative assemblies \citep{Htun_2004, Dahlerup:2007, krook:2009, HUGHES_2011};
implementing affirmative action in the allocation of government positions and public school seats \citep{Correaetal-Chile:19, dur/pathak/sonmez:20, aygun/bo:21, sonyen22};
balancing school seat priorities between neighborhood populations and the broader community \citep{dur/kominers/pathak/sonmez:18};
and allocating immigration visas in the U.S. \citep{pathak/rees-jones/sonmez:25}.
However, given the diversity of allocation criteria for medical resources across jurisdictions, the urgent need to allocate scarce medical resources 
globally during the pandemic called for examining a more general class of reserve systems \citep{pathak/sonmez/unver/yenmez:24}.

Until the onset of the COVID-19 pandemic in early 2020, virtually all reserve systems studied in the literature shared a common feature: 
they relied on a baseline priority ranking of individuals, typically determined through a performance metric, 
such as the results of a centralized test. All portions of the scarce resource were allocated based on this baseline ranking, but for some portions, preferential treatment was also given to certain groups of individuals.

While the task forces in New York State and Minnesota could have adopted a similar approach by allocating a fraction of ventilators with heightened priority for essential personnel, 
a more flexible reserve system was generally needed. 
This flexibility was crucial for implementing the ethical principles outlined in \cite{ezekiel:2020}. Our new application of market design in public healthcare called for
a reserve system in which the priority rankings for various portions of the scarce resource could be determined by criteria that were possibly, but not necessarily, unrelated---without relying on a fixed baseline ranking.

Collaborating with Bumin Yenmez and working around the clock to make a timely contribution to the global fight against COVID-19, 
Pathak, \"{U}nver, and I completed and circulated the first version of our paper in April 2020 \citep{pathak/sonmez/unver/yenmez:20a}, 
just two weeks after beginning this critical resource allocation project. The final version was later published in \cite{pathak/sonmez/unver/yenmez:24}.

Formally, in a reserve system, units of the scarce resource are divided into multiple segments called \textit{categories}, and units in each category are allocated 
based on a category-specific priority ranking. It is essentially a \textbf{\textit{categorized priority system}}, a term we used with our collaborators from 
medical ethics and emergency healthcare in a follow-up paper \cite{Sonmezetal-categorized:2021} when we first introduced the idea to members of the broader medical community. 

While the concept of a reserve system is intuitive, as we have previously discussed in depth, it contains a subtle aspect that can cause unintended distributional consequences if 
not properly understood by system operators: specifically, what happens when an individual has sufficiently high priority for the scarce resource across multiple categories?
(See Example \ref{ex:precedence} and Figure \ref{fig:reservesystem-example}.)

We observed a prime example of this issue in Section \ref{sec:demise}, where the flawed implementation of walk zone priorities at Boston Public Schools from 2005 to 2013 led to unintended outcomes.
Similarly, in Section \ref{sec:India}, we examine this problem in the context of affirmative action policies and court rulings in India. 
This subtlety has also affected the implementation of H-1B visa assignments in the U.S. \citep{pathak/rees-jones/sonmez:25}. 
To prevent confusion among healthcare officials implementing our proposed reserve system for allocating vital medical resources during the COVID-19 pandemic, 
we focused our formal analysis in \cite{pathak/sonmez/unver/yenmez:20a, pathak/sonmez/unver/yenmez:24} on addressing this subtle aspect of the problem.

As a side note, while \cite{pathak/sonmez/unver/yenmez:20a} marked the first time we used minimalist market design in real-time to offer a resolution to an important crisis, 
it was not the last. When the contested \textit{103rd Constitutional (Amendment) Act, 2019} was deliberated in the hearings of a recent 
Supreme Court judgment \textit{Janhit Abhiyan vs. Union of India (2022)} in November 2022, in \cite{sonmez-unver2022}, 
we offered a real-time resolution before a divisive decision was declared by the court in a split 3-2 vote, with strong protests from the two dissenting justices.\footnote{Indeed, about a week before the split decision, 
in \cite{deshpande/sonmez:2022}, I collaborated with Ashwini Deshpande, a prominent expert on the Indian reservation system, to publish an op-ed in \textit{The Hindu}. 
Together, we cautioned Indian society on the subtleties of the case.} 
Though aspired market design projects typically differ from commissioned ones in the flexibility of their timeframe, 
these experiences highlight the promise of minimalist market design in providing timely resolutions to important crises when they are most needed by society.

\subsection{Teaming Up with Three Leaders in Bioethics}

Timing is often everything in market design. A mere two days after \cite{pathak/sonmez/unver/yenmez:20a} was circulated, the following headline from a lead story in the \textit{Boston Globe} caught our attention \citep{Kowalczyk:20}:

\begin{quote}
``Who gets a ventilator? New gut-wrenching state guidelines issued on rationing equipment

Preference given to medical personnel, people who are healthy, younger"
\end{quote}

Not only did this story directly relate to the paper we just circulated, but it also highlighted, in its headline, one of our main motivations regarding the priority for medical personnel! Naturally, we were intrigued. The story started as follows:
\begin{quote}
``Massachusetts health officials issued guidelines Tuesday to help hospitals make gut-wrenching decisions 
about how to ration ventilators, should they become overwhelmed with coronavirus patients and run out of critical treatments.

The guidance, which is not mandatory, asks hospitals to assign patients a score that gives preference 
to healthier patients who have a greater chance of surviving their illness, and living longer overall. 
It gives additional preference to medical personnel who are vital to treating others, 
and to women further along in pregnancy. In the event of tie scores, younger patients are given priority.

`There is a great sense of urgency,' said Dr. Robert Truog, director of the Center for Bioethics at Harvard Medical School 
and a pediatric intensivist who was part of the group that helped develop the guidelines. 
`We realize this all needs to be in place soon. It’s very important to have current guidelines 
that provide very concrete advice to hospitals about how to allocate these resources.' ''
\end{quote}

According to this account, unlike the task forces in New York State and Minnesota, a similar committee in Massachusetts developed a system that 
``gives additional preference to medical personnel who are vital to treating others.'' However, upon reading the \textit{Original Guidance} issued by the 
\textit{Crisis Standards of Care Advisory Working Group} \citep{massachusetts:2020}, we observed that no such preference was given to medical personnel under the recommended system. 
Indeed, the principle of \textit{instrumental value} played no role whatsoever in the recommended allocation rule.

The system recommended in the \textit{Original Guidance}  was basically a simplified version of the priority point system of \cite{white:2009}, with only two ethical principles instead of three. 
Naturally, we wondered how the \textit{Boston Globe} could possibly make such a major error in its coverage. Subsequently, we observed that parts of the \textit{Original Guidance} 
used ambiguous language, mentioning possible preference given to medical personnel who are vital to treating others, even though the 
final recommendation did not reflect this consideration.\footnote{The ambiguous language in the \textit{Original Guidance} 
was later removed in a \textit{Revised Guidance} issued on April 20, 2020, possibly due to an inquiry by our team about this inconsistency to Robert Truog, 
who was a member of the working group mentioned in the \textit{Boston Globe} story.}

While this issue in the \textit{Original Guidance} serves as a ``red herring'' in this key document, it still provided us with a very important clue. 
Why would an otherwise very clear document involve a confusing discussion of the \textit{instrumental value} principle unless members of the working group found 
themselves in a dilemma similar to those faced by the task forces in New York State and Minnesota?

Based on this hunch, later that evening we sent an email to Robert Truog, who was mentioned in the \textit{Boston Globe} story as a member of the working group. 
Sharing our ideas with him was especially important---not only was he the Director of the Center for Bioethics at Harvard Medical School, 
but he was also the first author of \cite{Truog:2020}, one of the two articles that initially sparked our interest in the topic. 
In our email, we attached our paper \cite{pathak/sonmez/unver/yenmez:20a}, mentioned that we were inspired by the \textit{NEJM} symposium that included his paper, 
and expressed our strong interest in exchanging ideas with him in a Zoom meeting soon.

Truog responded positively to our meeting request the following day and even suggested including two other leaders in medical ethics, Govind Persad and Douglas White.

Naturally, we were thrilled with his prompt response. Persad was a renowned expert in legal and ethical aspects of healthcare and the second author of \cite{ezekiel:2020}, 
the main reference for our entire exercise. White was a Professor of Critical Care Medicine at the University of Pittsburgh and directed the University of Pittsburgh Program on 
Ethics and Decision Making in Critical Illness. He was also the first author of \cite{white:2009}, the primary reference on multiple ethical principle priority point systems for ventilator rationing.

Our meeting was one of the most successful and consequential in my career. From that point on, Persad, Truog, and White embraced the reserve system for allocating scarce medical resources during the pandemic. 
Together, we advocated for the reserve system to the broader bioethics and medical communities in \cite{Sonmezetal-categorized:2021}. 
At the suggestion of our new partners and to emphasize its minimal deviation from the priority system these communities were accustomed to, we referred to it as a \textit{categorized priority system} in this paper.

It was our minimalist approach that secured this ``dream'' meeting. We had correctly identified a desideratum---the \textit{instrumental value principle}---which 
was not reflected in the working group's recommended system for the crisis allocation of ventilators in Massachusetts. 
By suggesting a minor adjustment to their system, we addressed this oversight and provided a viable solution to the task force's ongoing challenge.

Although our solution was not adopted for the crisis allocation of ventilators, over the next year it was implemented for other 
scarce medical resources---such as antiviral therapies, monoclonal antibodies, and vaccines---thanks to our collaboration with our new partners 
(see Table \ref{table:Covid-Table} for notable reserve systems adopted during the COVID-19 pandemic).


\begin{table}[!tp]
  \centering
    \caption{Notable reserve systems adopted during the COVID-19 pandemic.}
    \label{table:Covid-Table}

    \includegraphics[scale=0.5]{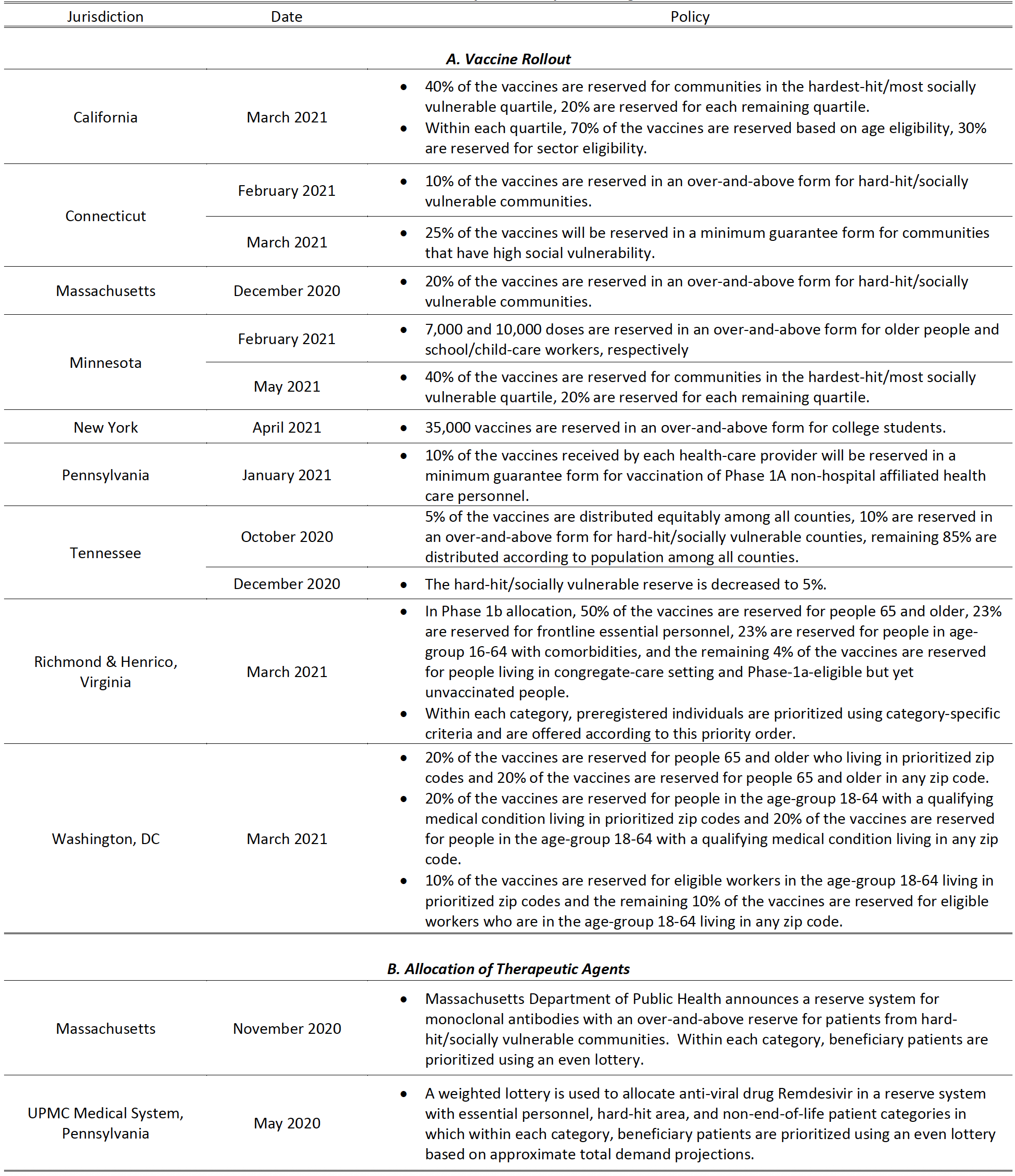}

    \begin{tablenotes}\footnotesize
      \item[] Note: Reproduced from \cite{pathak/sonmez/unver/yenmez:24}, \textit{Management Science}, licensed under \href{https://creativecommons.org/licenses/by/4.0/}{CC BY 4.0}.
    \end{tablenotes}
\end{table}


\subsubsection{Policy Impact on Allocation of COVID-19 Therapies in Pennsylvania} \label{sec:Pennsylvania}

Only a month after our initial collaboration with our new partners, we witnessed the first tangible outcome of our joint efforts. 
By May 2020, public attention had shifted from triage rationing of ventilators and ICU beds to the equitable allocation of scarce therapies. 
On May 1st, 2020, Gilead’s \textit{Remdesivir} became the first antiviral medicine to receive Emergency Use Authorization (EUA) from the U.S. Food and Drug Administration (FDA) for the treatment of COVID-19. 
Due to overwhelming demand for this medication, the University of Pittsburgh Medical Center (UPMC) 
developed and adopted a weighted lottery system to implement a transparent and fair approach to allocate scarce medications for treating COVID-19 patients.

Under the leadership of Douglas White, the UPMC system was designed by our team as a special case of a reserve system 
and was implemented in Western Pennsylvania during the shortage period of Remdesivir \citep{white-et-al:22}. The software for the system was also developed and provided by our core team of three economic designers.

The UPMC system was later endorsed by the Commonwealth of Pennsylvania for the allocation of scarce COVID-19 therapies \citep{Pennsylvania:2020}.

\subsection{Policy Impact During the COVID-19 Vaccine Rollout} \label{sec:vaccine}

Over the next several months, we engaged in various outreach activities to introduce the reserve system to bioethics and emergency medicine communities. 
Through these efforts, we introduced the reserve system to several groups and began collaborating with Harald Schmidt, a bioethicist at the University of Pennsylvania, 
who expressed particular interest in utilizing it to mitigate disparities in healthcare access \citep{Schmidt-et-al:2020}.

\subsubsection{Debates in the U.S. on Priorities for the COVID-19 Vaccine}

By August 2020, phase 3 clinical trials for two of the most promising COVID-19 vaccines were already underway, and much of our focus had shifted to the upcoming vaccine rollout. Since the beginning of the pandemic, there had been vigorous debates on equitable vaccine allocation. Initially, these debates were purely focused on the structure of priority tiers under a presumed priority system.

For example, in June 2020, Melinda Gates made the following public statement  \citep{melindagates:2020}:
\begin{quote}
``We know there are 60 million healthcare workers around the world who are keeping everybody safe. 
They deserve to get this vaccine first. From there, you want to do tiering in various countries to make sure your most vulnerable populations get it. 
In our country, that would be Blacks and Native Americans, people with underlying health conditions and the elderly.''
\end{quote}

However, not everyone agreed with Melinda Gates. An opposing view was expressed in a commentary from the Cato Institute \citep{olson-cato:2020}:
\begin{quote}
``A federal advisory committee recommending priorities for the eventual distribution of a COVID-19 vaccine has floated a very bad idea: 
according priority to some beneficiaries over others because of their race. If implemented, the regime would very likely be struck down by courts as unconstitutional.''
\end{quote}

More generally, rather than focusing on who receives priority, the biggest challenge under any priority system was the notion that some groups receive absolute priority over others \citep{goldhill-quartz:2020}:
\begin{quote}
``There's no easy solution. Healthcare experts at ACIP and the National Academy have been 
debating the question among themselves, and no decision will be uncontroversial. 
But when there aren't enough vaccines for all, someone must inevitably come first.''
\end{quote}

Due to its role as a practical tool to reach a compromise between various groups with opposing views, our proposed reserve system started to gain traction for the upcoming vaccine rollout in this environment.

\subsubsection{NASEM Framework for Equitable Vaccine Allocation}

In July 2020, the Centers for Disease Control and Prevention (CDC) and the National Institutes of Health (NIH) commissioned the National Academies of Sciences, 
Engineering, and Medicine (NASEM) to formulate recommendations on the equitable allocation of a COVID-19 vaccine. 
NASEM promptly appointed a committee of distinguished experts, and in September 2020, it released a discussion draft of its Preliminary Framework for the Equitable Allocation of COVID-19 Vaccine for public commentary 
\citep{NASEM-preliminary:20}. Despite proposing a four-tier priority system in its discussion draft, NASEM also included the following statement:
\begin{quote}
``Equity is a crosscutting consideration: In each population group, vaccine access should be prioritized for geographic areas
identified through CDC's Social Vulnerability Index.'' (p. 57)
\end{quote}

At this point, two of our collaborators, Govind Persad and Harald Schmidt, played key roles in bringing the reserve system to the attention of the NASEM committee.

In response to the NASEM discussion draft, \textit{JAMA} published the viewpoint ``Fairly Prioritizing Groups for Access to COVID-19 Vaccines" 
by \cite{persad/peek/emanuel:2020}, explicitly endorsing our proposed reserve system in their conclusion:
\begin{quote}
``Dividing the initial vaccine allotment into priority access categories and
using medical criteria to prioritize within each category is a promising
approach. For instance, half of the initial allotment might be prioritized
for frontline health workers, a quarter for people working or living in
high-risk settings, and the remainder for others. Within each category,
preference could be given to people with high-risk medical conditions.
Such a categorized approach would be preferable to the tiered ordering
previously used for influenza vaccines, because it ensures that multiple
priority groups will have initial access to vaccines." (p. 1602)
\end{quote}

Not only was our collaborator Govind Persad the first author of this viewpoint, but Ezekiel Emanuel---the first author of the main reference \cite{ezekiel:2020} in these debates---was its senior author. 
Thus, our proposed reserve system received a highly visible endorsement for the upcoming vaccine rollout at a critical moment.

In response to NASEM's discussion draft of the Preliminary Framework for the Equitable Allocation of COVID-19 Vaccine, our collaborator Harald Schmidt submitted a written commentary. 
Upon receiving an invitation from the committee, he also provided oral remarks, where he inquired about the recommended mechanism to prioritize members of communities identified through the CDC's Social Vulnerability Index.

Anticipating this need and in collaboration with Schmidt, we circulated an NBER working paper \cite{pathak-schmidt-etal:2020} weeks earlier, 
illustrating how a tiered priority system can be easily modified into a reserve system as a minimalist refinement by incorporating equity through an index of social vulnerability. 
Schmidt brought this exercise in minimalist market design to the committee's attention as a possible mechanism to embed equity into their framework.

In October 2020, NASEM announced its Framework for Equitable Allocation of COVID-19 Vaccine \citep{NASEM:2020}. 
Following the recommendation in \cite{persad/peek/emanuel:2020} and using the exact formulation from \cite{pathak-schmidt-etal:2020}, 
the NASEM framework formally recommended a 10 percent reserve for people from hard-hit areas:

\begin{quote}
``The committee does not propose an approach in which, within each phase, all
vaccine is first  given to people in high SVI areas. Rather the committee proposes
that the SVI be used in two ways. First as previously noted, a reserved 10 percent
portion of the total federal allocation of COVID-19 vaccine may be reserved to
target areas with a high SVI (defined as the top 25 percent of the SVI distribution
within the state)."\\ 
\mbox{}\hfill \citealp{NASEM:2020}, p. 133
\end{quote}

As evident from the above quote, NASEM clearly emphasized the distinction of their recommendation from a tiered priority system.

\subsubsection{Vaccine Rollouts: U.S. Jurisdictions Adopting Reserve Systems}

Within days of the NASEM recommendation, Tennessee became the first state to adopt a reserve system for its vaccine rollout in October 2020.

Although the FDA granted emergency use authorization for the Pfizer--BioNTech and Moderna vaccines in early December, much to our disappointment, 
Tennessee remained the only state to have adopted the reserve system by that time. Consequently, in early December 2020, in collaboration with Ariadne Labs from the Harvard Chan School of Public Health 
and the Department of Medical Ethics and Health Policy from the University of Pennsylvania, we co-hosted an online symposium titled ``Vaccine Allocation and Social Justice'' \citep{ariadne:2020}. 
One of our aims was to demonstrate to policymakers how easily equity can be incorporated into the vaccine rollout through a reserve system.

The symposium not only helped us better understand the needs, challenges, and perspectives of various jurisdictions but also directly contributed to two important developments.

First, it provided our co-organizers from Ariadne Labs with a natural opportunity to bring the reserve system to the attention of the committee responsible for vaccine rollout in Massachusetts. 
Similarly, it gave our team the chance to introduce the reserve system to California's Surgeon General, Dr. Nadine Burke Harris.

In several group meetings, we presented the reserve system to Harris and her team, advocated for its adoption in California as a means to build equity into their upcoming vaccine rollout, 
and coached members of her team on the subtleties of the system.

In December 2020, Massachusetts became the second state to adopt a reserve system for its vaccine rollout, implementing a 20\% over-and-above reserve for hard-hit communities. 
In March 2021, California adopted a particularly ambitious reserve system, incorporating reserve categories for both educators and hard-hit populations. 

During the initial vaccine rollout from December 2020 to May 2021, at least 12 states and several major cities---including New York City, Chicago, and Washington, D.C.---adopted our 
proposed reserve system at various phases \citep{Schmidt-NatureMedicine:2021, pathak/persad/sonmez/unver:22}.

While equity and social justice considerations were the driving force behind the acceptance of the reserve system in most jurisdictions, they were not the only reasons. 
The Phase 1b COVID-19 vaccine rollout in Richmond and Henrico, Virginia, deployed on March 8, 2021, serves as an illustrative example \citep{Richmond-1b:2021}.\medskip

\paragraph{Richmond and Henrico Reserve System for COVID-19 Vaccine Rollout.}
In their Phase 1b allocation, the Richmond and Henrico health districts gave the following specification of categories, 
their respective shares of COVID-19 vaccines, and the factors shaping category-specific priorities:

\begin{enumerate}
\item \textbf{Phase 1a \& Congregate Care} (4\% of units)
\item \textbf{Adults age 65+} (50\% of units)
\begin{itemize}
\item Age (older residents have higher priority) 
\item Race and ethnicity (Black, Hispanic/Latinx, and American Indian or Alaska Native residents have higher priority)
\item Burden of disease in the area where a person lives
\item Social Vulnerability Score (SVI) of the area where a person lives
\end{itemize}
\item  \textbf{Frontline Essential Workers} (23\% of units)
\begin{itemize}
\item Age (older residents have higher priority) 
\item Race and ethnicity (Black, Hispanic/Latinx, and American Indian or Alaska Native residents have higher priority)
\item Burden of disease in the area where a person lives
\item Social Vulnerability Score (SVI) of the area where a person lives
\end{itemize}
\item  \textbf{People ages 16--64 with Comorbidities} (23\% of units)
\begin{itemize}
\item Age (older residents have higher priority) 
\item Race and ethnicity (Black, Hispanic/Latinx, and American Indian or Alaska Native residents have higher priority)
\item Socioeconomic status (residents who are under and uninsured have higher priority)
\end{itemize}
\end{enumerate}

Notably, the Richmond and Henrico reserve system utilized the full generality of the model presented in \cite{pathak/sonmez/unver/yenmez:20a, pathak/sonmez/unver/yenmez:24}. 
Specifically, while categories 2 and 3 used the same baseline priority ranking (subject to eligibility), category 4 employed a related but different priority ranking.

\subsection{Policy Impact on Allocation of mAb Therapies in Massachusetts} \label{sec:mAb}

When we first communicated with Robert Truog in April 2020 as a member of the Crisis Standards 
of Care Advisory Working Group in Massachusetts, our initial focus was on equitable allocation of ventilators.  
By May 2020, public attention shifted to allocation of therapies against COVID-19,
which started to receive Emergency Use Authorization (EUA) from the U.S. Food and Drug Administration (FDA).  

During that period, we met with Dr. Emily Rubin from Massachusetts General Hospital, another member of the Working Group in Massachusetts. 
At the time, she considered our reserve system for the allocation of the antiviral Remdesivir at the Mass General Brigham healthcare system, 
but in this chaotic period when healthcare personnel were extremely overwhelmed, this possibility did not materialize.

Six months later, in November 2020, the FDA authorized EUA for three monoclonal antibodies (mAb) for the treatment of COVID-19. 
This is when Rubin reconnected with us and inquired whether we could support the Working Group as they prepared the Guidance for the Allocation of COVID-19 Monoclonal Antibody Therapeutics.
Our group supported the Working Group with a reserve system design tailored to the specifications for Massachusetts policies and provided them with an Excel spreadsheet implementation of the system. 
Later in November, Massachusetts officially recommended the following reserve system in its Guidance.\medskip 

\paragraph{Massachusetts Reserve System for Allocation of mAb Therapeutics.}
In the Massachusetts Reserve System for the Allocation of mAb Therapeutics, the state specified the categories, 
their shares of infusions, the factors influencing category-specific priorities, and the processing sequence of categories:

\begin{enumerate} 
\item \textbf{Open Category} (80\% of Infusions)
\begin{itemize}
\item \textit{Eligibility}: All patients. 
\item \textit{Priorities}: 
\begin{itemize}
\item[*] \textit{Tier 1}: Age $\geq$ 65 or Body Mass Index (BMI)$\geq$ 35
\item[*] \textit{Tier 2}: Other patients who satisfy the EUA criteria of the FDA. 
\item[*] Within each tier, priority ranking is determined with a uniform lottery. 
\end{itemize}
\end{itemize}

\item \textbf{Vulnerable Category} (20\% of Infusions)
\begin{itemize}
\item \textit{Eligibility}: Patients who live in a census tract in the top half of the Social Vulnerability Index (SVI) or in a town or city with an incidence rate in the top quartile.
\item \textit{Priorities}: 
\begin{itemize}
\item[*] \textit{Tier 1}: Age $\geq$ 65 or Body Mass Index (BMI)$\geq$ 35
\item[*] \textit{Tier 2}: Other patients who satisfy the EUA criteria of the FDA. 
\item[*] Within each tier, priority ranking is determined with a uniform lottery. 
\end{itemize}
\end{itemize}
\item[$\blacktriangleright$] \textbf{Processing Sequence of Categories}:  Open category first, followed by the vulnerable category (Over-and-Above reserve).
\end{enumerate}

Subsequently, the Massachusetts Reserve System for the Allocation of mAb Therapeutics was implemented by the 
Mass General Brigham healthcare system. \cite{rubin-et-al:21} report the successful deployment of this system for allocating scarce 
COVID-19 therapies in Massachusetts and how it increased access to socially vulnerable patients.

The following quotes from the editorial \cite{Makhoul/Drolet:21}, written in response to \cite{rubin-et-al:21}, 
not only highlight the success of the reserve system in allocating various scarce medical resources during the COVID-19 pandemic but also illustrate the promise of minimalist market design in interdisciplinary efforts.

\begin{quote}
``This work contributes to a growing body of evidence that reserve systems offer a pragmatic framework for
equitably allocating scarce resources. [...]

As reserve systems become more prevalent, it is important to acknowledge and understand the
psychological effects on participants. Not only do reserve systems enable policymakers to allocate
resources equitably, but they also signal to participants that expert judgment has been used to design a system
for maximal societal benefit. Participants eligible for prioritized categories (eg, patients from high-SVI zip
codes) may feel more adequately safeguarded. [...]

Health equity must not only be demonstrated objectively but must also be perceived by participants in the system. [...]

Despite challenges related to administering a time-sensitive,
novel therapeutic during a pandemic,  \cite{rubin-et-al:21} demonstrate that a reserve system can be used
effectively on an individual patient level to prioritize access
for certain groups.'' (p. 2005-06)
\end{quote}

\subsection{Post-Pandemic Policy Impact: Crisis Care Guidance in Oregon} \label{sec:Oregon}

In the wake of the pandemic, numerous states established advisory committees to evaluate and overhaul their crisis standards of care, designed for adoption during public health emergencies or overwhelming disasters. 
Oregon is among these states.

In May 2023, the Oregon Health Authority invited public input on a preliminary framework developed by the Oregon Resource Allocation Advisory Committee \citep{ORAAC:23}. 
Within this document, the committee delineated six criteria under consideration for crisis care triage and explored their potential utilization either independently or as part of a multi-criteria approach.

As their preferred methodology for implementing multiple criteria, the committee endorsed the reserve system \citep{ORAAC:23}:

\begin{quote}
``In a major health emergency situation, ease of implementation is a feature that needs to be taken seriously. Models such as those described below can all be 
implemented readily via a methodological approach known as Categorized Priority System (sometimes also called a Reserve System). 
During the COVID-19 pandemic, systems that combined factors such as survivability, level of disadvantage and essential worker status were successfully developed for
purposes including allocating vaccines, tests and treatments. Custom-made, free- of-charge software has been developed to facilitate implementation.'' (p. 20)
\end{quote}

\section{Liver Exchange} \label{sec:LE}


Throughout my career, I have consistently sought projects with high practical value, and thanks to the broader persuasion strategy embedded into minimalist market design, 
the majority of my academic pursuits have borne fruit by influencing real-life policies and reshaping institutions.
This is particularly true for ideas I found especially valuable.

There is, however, one notable exception---an idea that could easily double the efficacy of kidney exchange but, as of this writing, has failed to fulfill its practical promise. 
This exception is \textit{incentivized kidney exchange} (IKE). Interestingly, of all the ideas I relentlessly pursued in policy spheres, 
though not exactly ``radical'' (cf. \citealp{posner/weyl:18}), it also happens to be one of the few ideas that is not minimalist!

As discussed in detail earlier in Section \ref{sec:IKE}, the core idea of IKE centers on mitigating welfare losses that occur when kidney patients receive living donations 
from blood-type-compatible---but not blood-type-identical---donors (for example, a blood-type A patient with a blood-type O donor). 
By participating in kidney exchange instead, these patients can help others receive transplants while still securing one for themselves.
IKE introduces an ``institutional'' incentive for such patients to participate in kidney exchange by granting them priority on the deceased-donor (DD) list should a future failure of their transplant kidney occur. 
 Its non-minimalist nature stems from the need to integrate two separate ``institutions,'' namely, the allocation mechanisms for DD kidneys and living-donor kidneys, 
 which are typically managed independently in the U.S. and elsewhere.

However, as the welfare gains from IKE are explicitly tied to an ``institutional'' incentive mechanism, realizing them poses greater challenges.
If only we could discover a ``biological'' incentive mechanism inherent to the transplantation procedure, perhaps we would have better luck transforming this idea into practice.

In the mid-2010s, we finally discovered one such ``biological'' incentive mechanism we had been seeking---though for another organ: the liver. 
Little did I know that this discovery would not only pave the way for a series of subsequent breakthroughs but also, in an initiative uniquely close to my heart, 
culminate in a program---the \textbf{\textit{Banu Bedestenci S\"{o}nmez Liver Paired Exchange System}}---dedicated to my beloved wife, 
whom we tragically lost not long after our discovery \citep{BBS--LPE:25}.

\subsection{Living-Donor Liver Transplantation and Liver Paired Exchange} \label{sec:LDLT}

In principle, donor exchange can be implemented for any organ where living-donor transplantation is possible, and the liver is one such organ. 
Despite potential donors having only one liver (unlike the kidney), it is often possible to donate a portion of it---typically either the right lobe, the left lobe, or a part of the left lobe known as Segments 2-3 (see Figure \ref{fig:liver-anatomy}).

\begin{figure}[!tp]
    \begin{center}
       \label{fig:liver-anatomy}
    \includegraphics[scale=0.6]{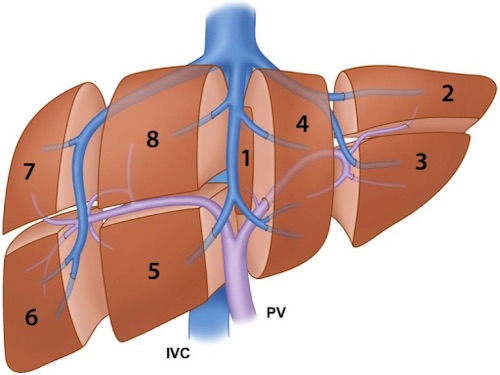}
    \end{center}
\caption{Liver anatomy. Segments 1-4 form the left lobe and Segments 5-8 form the right lobe. In some cases, Segments 2--3 of the left lobe can be transplanted into pediatric patients.
Abbreviation:  IVC: inferior vena cava;  PV: portal vein. Reproduced from \cite{Orcutt:2016}, \textit{Frontiers in Surgery},
licensed under \href{https://creativecommons.org/licenses/by/4.0/}{CC BY 4.0}.} \label{fig:liver-anatomy}
\end{figure}

However, based on three main reasons, \"{U}nver and I did not give much thought to liver exchange in the mid-2000s when we were heavily engaged in kidney exchange:

\begin{enumerate}
\item \textbf{Prevalence of Living-Donor Transplantations:} In contrast to kidney, living-donor transplantation for the liver was considerably less common in the U.S. and globally during the early 2000s.
\item \textbf{Donor Mortality and Morbidity Rates:} Given the complexity of liver dissection, the rates of donor mortality and morbidity were significantly higher for liver transplantation compared to kidney transplantation.
\item \textbf{Precedence in Transplantation Literature:} Unlike kidney exchange, there were no reported cases of liver exchange in the transplantation literature.
\end{enumerate}

In the early 2000s, living-donor kidney transplants in the U.S. surpassed 6,000 annually. By contrast, after peaking at 524 in 2001, 
the number of \textit{living-donor liver transplants} (LDLTs) declined sharply in the following years \citep{Ratnera:2010}.\footnote{After falling to an all-time low of 219 cases in 2009, 
the annual number of LDLTs gradually rose over the next 25 years, reaching a record high of 658 in 2023 \citep{OPTN:25}.}

\begin{quote}
``The death of a living organ donor can tarnish the reputation of transplantation in general, and live donor transplantation, more specifically. [...]
Prior to that well-publicized live donor's death, live donor liver transplantation was increasing in frequency nationwide.
Five hundred twenty-four live donor liver transplants were performed in 2001. Subsequent to the donor death, there was a 31\% decline in live donor liver transplantation in 2002,
to 363 live donor liver transplants performed. By 2009, live donor liver transplantation in the United States had decreased to 219 cases annually." (p. 2578)
\end{quote}

A high-profile liver donor fatality in 2002 not only fueled this decline but also triggered heightened scrutiny of LDLT \citep{josefson:2002, Ratnera:2010}.

In addition to the elevated risks faced by liver donors and relatively few LDLTs, the lack of documented liver exchanges in the literature before 2010 further diminished the appeal of this otherwise intriguing potential application. 
To put it plainly, given our focus on building upon and improving existing transplantation practices---a feature of minimalist market design---liver exchange did not emerge as a compelling prospect.
 
While much of our attention was on increasing the efficiency of kidney exchange, news of liver exchange cases from two Asian countries piqued our interest.

In January 2010, the journal \textit{Liver Transplantation} published two papers detailing cases of liver exchanges in South Korea and Hong Kong \citep{Hwang:2010, Chan:2010}.
The study by \cite{Hwang:2010} provided an in-depth analysis of sixteen transplants conducted at \textit{ASAN Medical Center} (Seoul, Korea) through eight 2-way liver exchanges over the six-year period between July 2003 and June 2009.
These transplants constituted approximately 1\% of the LDLTs at this center, recognized as the largest-volume center worldwide. 
On the other hand, \cite{Chan:2010} reported a single 2-way liver exchange performed on an emergency basis on January 13, 2009, at \textit{The University of Hong Kong, Queen Mary Hospital}. 

The successful implementation of liver exchange in these two centers served as compelling proof of concept, encouraging us to pay closer attention to this potential application of market design. 
Yet, in the early 2010s, the annual number of LDLTs in the U.S. remained consistently in the 200s. 
Indeed, not long after the liver exchanges from South Korea and Hong Kong were reported, the death of another donor made headlines in the U.S. \citep{Cohen:2012}.

Even though we were not fully convinced that the conditions were ripe to pursue a project in liver exchange, these developments still compelled us to learn about the risks of \textit{partial hepatectomy}, 
the surgery in which part of the liver is removed from a donor. It was then that we made a key discovery----a ``biological'' incentive mechanism that could, at least in theory, motivate compatible pairs to participate in liver exchange. 
Building on this idea, as well as another concept driven by a liver transplant technique called ``dual-lobe liver transplant,'' we extensively researched liver exchange throughout much of the mid-2010s 
as part of a team that included Haluk Ergin, Utku \"{U}nver, and me \citep{ergin/sonmez/unver:17, ergin/sonmez/unver:20}.

\subsection{A Biological Incentive Mechanism for Liver Exchange} \label{sec:biological}

To minimize the risk of death during or immediately after liver transplantation, an adult patient must receive a liver graft comprising at least 40\% of the size of their dysfunctional liver. 
Receiving a graft smaller than this limit significantly elevates the risk of patient death due to the \textit{small-for-size syndrome}. 
Given that the liver's weight is approximately 2\% of body weight, meeting this criterion necessitates a \textit{graft-to-recipient weight ratio} (GRWR) of at least 0.8\%. 
In principle, eligible donors can contribute either their larger right lobe (typically 60--75\% of their liver) or the left lobe (typically 25--40\% of their liver) (see Figure \ref{fig:liver-anatomy}). 
However, these constraints imply that the left lobe of a donor is often too small for an adult patient. Consequently, in most cases, 
eligible donors can only donate their right lobe to adult patients.\footnote{For donor safety, the remnant liver after hepatectomy needs to be at least 30\% of the original liver volume. 
This means right lobes larger than 70\% of the liver cannot be safely transplanted.}

Another crucial statistic reveals that, in much of the world, donor mortality and morbidity rates are much higher for right-lobe hepatectomy than 
for left-lobe hepatectomy \citep{Lee:2010, mishra/etal:18}.\footnote{\cite{Lee:2010} reports mortality rates of 0.4--0.5\% for right hepatectomy and 0.1\% for left hepatectomy.
\cite{mishra/etal:18} reports morbidity rates of 28\% for right hepatectomy and 7.5\% for left hepatectomy.} 
Consequently, while the left lobe is frequently deemed unsuitable for transplantation, it represents a much safer option for the donor. 
This observation forms the basis for a ``biological" incentive that may compel blood-type-compatible patient--donor pairs to participate in liver exchange.

Consider a scenario involving a blood type A patient with a petite blood type O donor. The patient can only receive the larger right lobe of the donor. 
In a typical liver exchange pool, there are often numerous blood type O patients who are incompatible with their corresponding blood type A donors.\footnote{In matching theory, 
we refer to such pairs as ``under-demanded'' \citep{roth/sonmez/unver:05, roth/sonmez/unver:07}.} Frequently, some of these patients---pediatric or slender individuals---require smaller liver grafts. 
While the first patient is compatible with their donor, unlike in kidney transplantation, a donor exchange between the two pairs becomes a visibly appealing option. 
This is because the first donor can undergo a much less risky left-lobe hepatectomy through liver exchange, as opposed to the considerably higher-risk direct donation to their own patient.

Thus, the need for size compatibility in liver transplantation, together with the differing risks of right- versus left-lobe hepatectomy, creates a strong ``biological'' incentive for many compatible pairs. 
Building on this insight, our work in \cite{ergin/sonmez/unver:20} developed an efficient and incentive-compatible liver exchange mechanism centered on 2-way exchanges---the only type attempted before 2022, 
since no center had yet undertaken the more complex 3-way or larger exchanges.

By utilizing this mechanism and leveraging aggregate statistics from South Korea, even with only 2-way exchanges, 
we demonstrated not only the potential to increase the number of LDLTs by over 30\% but also the ability to raise the proportion of less risky left-lobe transplants. 
All that remained was to find a center willing to pursue this idea. In 2019, we identified such a center in our homeland, Turkey.

\subsection{Living-Donor Liver Transplantation at Malatya, Turkey}

Shortly after the world’s first LDLT was performed in the U.S. in 1989 \citep{Strong:1990}, Turkey followed with its first case in 1990 \citep{Haberal:2014}. 
Like many Asian and Muslim-majority countries---including South Korea, Japan, India, and Saudi Arabia---Turkey faces cultural challenges with deceased donor organ transplantation. 
The introduction of LDLT provided a compelling alternative; however, until 2005, the annual number of LDLTs in Turkey remained below 100.

In 2005, the first LDLT was performed at the medical school of  \.{I}n\"{o}n\"{u}  University in Malatya, Turkey, under the leadership of Dr. Sezai Y{\i}lmaz. 
By 2013, the annual LDLT count in Turkey neared 1,000, with over a quarter carried out by the Malatya team. In less than a decade, this team had become one of the highest-volume centers globally and the largest in Europe. 
This achievement is especially remarkable, given that it emerged from a public university in Eastern Turkey, an economically and socially less developed region compared to the rest of the country.

The transplantation team and facilities at  \.{I}n\"{o}n\"{u}  University grew so significantly that, in 2018, they performed three LDLTs simultaneously. 
Despite their substantial accomplishments and the extensive experience accumulated by the team, they remained relatively unknown throughout much of Turkey, let alone the world.

In 2019, a distinguished surgeon and the director of a prominent Ivy League university hospital in the U.S. visited the Malatya team. 
After learning about their 2018 experience with three simultaneous LDLTs, he suggested that \.{I}n\"{o}n\"{u} University promote this remarkable feat by applying for a \textit{Guinness World Record}. 
Once this idea resonated with the university leadership, Y{\i}lmaz informed him that, by then, his Malatya team had developed the capacity to perform five simultaneous liver transplants.

In a bid for a Guinness World Record, a team of more than 100 medical personnel led by Y{\i}lmaz performed five simultaneous liver transplants in June 2019. 
The operations lasted more than thirteen hours and were witnessed and registered by a public notary. The achievement naturally became headlines in Turkish media. 

However, not everyone shared the enthusiasm for achieving a world record through five simultaneous liver transplants.
Following formal complaints submitted to the Health Ministry by a world-renowned Turkish transplant surgeon---the founding father of transplantation in Turkey---seeking disciplinary action against the group,
 a number of reporters also heavily criticized the Malatya team. While government officials remained supportive of the team, 
 Y{\i}lmaz recognized that the Guinness World Record attempt risked overshadowing the accomplishment itself and withdrew the application.

I first learned about the leadership of the Malatya team in liver transplantation in Turkey sometime in early 2019. In May 2019, a month before their publicity crisis unfolded, 
I emailed Y{\i}lmaz to propose a collaboration to establish a single-center liver exchange program at their institute. One of his juniors responded positively on his behalf, 
expressing the group's willingness to collaborate on an initiative to benefit their patient population. Excited by their response, I followed up with a more detailed message, elaborating on the collaboration I envisioned.

To my disappointment, no one in the Malatya group replied to my second message for two months. 
Unaware of the developments surrounding the Guinness World Record application and the publicity crisis that followed a month after my initial correspondence, 
I decided in early July 2019---while visiting my parents in Turkey---to follow up on my earlier email. I wrote to Y{\i}lmaz that I was in the country and would be delighted to visit his team in 
Malatya if an in-person meeting would be more helpful. He responded promptly, apologizing for the delay due to his heavy schedule, sharing his phone number, 
and expressing that they would be very happy to host me. I called him right away and arranged a visit with his team a few days later.

Once I secured a meeting with them, I conducted some internet searches about the Malatya team. It was then that I learned about their 
Guinness World Record application and all the bad publicity that came with it. Amazed that they were capable of performing five simultaneous liver transplants, 
I couldn't help but ask myself the following question: Couldn't these simultaneous transplants be interpreted as a test of their logistical capacity for the potential liver exchange program we had been discussing?

With these thoughts in mind, I arrived in Malatya in mid-July 2019.

\subsection{How a Personal Tragedy Influenced My Policy Efforts in Turkey} \label{sec:Banu}

My initiative in Malatya wasn't my first policy effort in Turkey. Ever since my failed attempt to reform the centralized Turkish college admissions mechanism in 1997
(see Section \ref{subsec:failed-Turkish-policyeffort}), I have made several other attempts. 
\"{U}nver and I made quite a few of our initial discoveries in kidney exchange while working at Ko\c{c} University in the early 2000s, 
fueling our enthusiasm to advance these innovations in our homeland.

The aspiration to save lives through these ideas worldwide was a primary motivation for our decision to leave Ko\c{c} University in 2004 and move to the U.S. 
This significant life choice was made possible by the considerable sacrifice of my wife, Banu, whom I had married only two years earlier. Prior to our relocation, 
Banu was a successful dentist in \.{I}stanbul and had never contemplated leaving Turkey. By choosing to move abroad, she transformed my professional aspirations into our joint dream, 
and became a homemaker in a foreign country, leaving behind her career and social life to support my efforts to improve the world through research.

While my policy aspirations for kidney exchange and school choice came true shortly after I started my new position at Boston College, our family, which had recently grown with the birth of my son Alp Derin, 
was devastated by a health crisis. In 2006, Banu was diagnosed with an aggressive form of breast cancer. For about a decade, until we lost our beloved Banu in 2016, I slowed my research and policy efforts, focusing much on my family.

During this period, I conducted extensive research on breast cancer and, drawing on insights from Banu's decade-long treatments, I even attempted to initiate a project on cancer clinical trial design. 
Unfortunately, this project did not materialize. While Banu spent her last few months with us in hospice at home between April and August 2016, partly as a coping mechanism, 
I occasionally dedicated time to our liver exchange project. The potential of this project to save lives was the sole motivation for me to engage in any research activity during such a challenging time. 
In that period, the most arduous in my life, an unexpected development unfolded.

Through a series of initiatives by  \"{O}m\"{u}r Budak, the Boston Consul General of Turkey, \"{U}nver and I were able to convince the Turkish Health Minister, Recep Akda\u{g}, 
to establish a national living-donor organ exchange system in Turkey for both kidney and liver. 
As a tribute to Banu's personal sacrifices that contributed to the global success of kidney exchange over the years, the system 
would be named the \textit{Banu Bedestenci S\"{o}nmez Living-Donor Organ Exchange System}.  Just as we made this breakthrough, we lost our beloved Banu on August 25th, 2016...

For about a year, we interacted with bureaucrats at the Turkish Health Ministry to arrange various details for the planned system. 
By the end of June 2017, the final form of a protocol between the Health Ministry and our team of two economic designers was ready. 
However, just as we prepared to sign the protocol, Akda\u{g}'s tenure as Health Minister expired in July 2017. 
We never heard from the Health Ministry bureaucrats after this date, and the protocol was never signed.

After this heavy setback, we gave up on the idea of a nationwide system in Turkey  and began exploring the possibility of pursuing this initiative at a single transplant center. 
It took us another two years to find the right one: The Liver Transplantation Institute at \.{I}n\"{o}n\"{u} University in Malatya.

\subsection{Forging a Partnership with a Renowned Liver Transplantation Team}

When I flew to Malatya in July 2019, I didn't know what to expect. Despite numerous successful policy initiatives in the U.S. over the years, 
my interactions in Turkey consistently left me empty-handed, often for no apparent reason and after considerable effort. Our latest encounter with the Health Ministry, in particular, had dealt a significant blow to our spirits.

At the same time, considering the remarkable professional and logistical capacity of the \.{I}n\"{o}n\"{u} University liver transplantation team to conduct five simultaneous transplants, 
and the publicity crisis they faced while trying to highlight this capacity, I felt that this time could be very different.

I saw an opportunity for a partnership that could be exceptionally valuable for both sides. By collaborating with us, the Malatya team could hit two birds with one stone: 
they could become a global leader in liver exchange by increasing their number of LDLTs by an unprecedented 20--40\%, and simultaneously clarify to the 
Turkish public the compelling justification for building the capacity to perform multiple simultaneous transplantations. In many ways, the conditions resembled those 
we faced right before our first interaction with Delmonico (see Section \ref{sec:NEPKE}), although, given their ongoing publicity crisis, the stakes were higher for the Malatya team.

In Malatya, I encountered a very friendly group eager to learn about the liver exchange system we envisioned. After a productive discussion with their leader, Sezai Y{\i}lmaz, 
I delivered a comprehensive presentation to the entire team. During my visit, I shared with them the (never signed) protocol drafted by the Health Ministry two years earlier to establish a national system in honor of my beloved wife, Banu.
Y{\i}lmaz proposed that they could readily adapt this protocol for a single-center system and promptly arranged a meeting with Prof. Dr. Ahmet K{\i}z{\i}lay, the President of  \.{I}n\"{o}n\"{u} University. 
With enthusiastic support from the president, we reached an agreement to establish and jointly manage the \textbf{\textit{Banu Bedestenci S\"{o}nmez Liver Paired Exchange (BBS--LPE) System}}. 

I returned to my parents' house in Ankara full of hope. Two months later, in September 2019, we formalized our partnership by signing a protocol. 
I was overjoyed---not only because I could finally make a tangible contribution to my homeland, but also because I could honor the memory of the person dearest to me in a profoundly meaningful way.

Initiating the operation of this system, however, took nearly three additional years. The team at \.{I}n\"{o}n\"{u} University wanted to ensure that both the Health Ministry and the Higher Education Council supported our joint initiative. 
Securing formal approval took several months. Just as these bureaucratic requirements were completed in early 2020, a global crisis hampered its implementation: COVID-19. 
We could only commence the operation of the system in June 2022, almost three years after reaching an agreement. However, we did start, and began with a major breakthrough.

 \subsection{Banu Bedestenci S\"{o}nmez Liver Paired Exchange System}   \label{sec:BBS--LPE}

In June 2022, we received the initial dataset from Malatya and began running the system.\footnote{Before the deployment of the computerized system in June 2022, 
the Malatya team conducted 2-way exchanges on an emergency basis on two occasions: first in October 2021 and second in January 2022.} 
Just days before my first visit to Turkey in three years, during the first week of July 2022, we surprised the Malatya team with an unexpected outcome that no one had anticipated, especially not at such an early stage.

\subsubsection{Breaking New Ground: The First-Ever 4-Way Liver Exchange}

After several entries into the liver exchange pool, the number of patient--donor pairs reached 12.\footnote{Although I use the phrase ``patient--donor pairs,'' 
it is not uncommon for patients to have multiple willing donors. Therefore, the intended meaning is patients with one or more donors.} 
Upon receiving a new set of data, I couldn’t believe my eyes. As always, I reviewed it before \"{U}nver ran the software he’d coded for the system. 
Immediately, I spotted a 4-way exchange alongside a separate 2-way exchange. Remarkably, it was already possible to match half of the patients in the pool!
By then, even a 3-way liver exchange was unprecedented.\footnote{As it turns out, with the support of Alex Chan---now an Assistant Professor at Harvard Business School, 
and a graduate student at Stanford Business School at the time---a world-first 3-way liver exchange was carried out in Pakistan a few months earlier on March 17, 2022. 
We became aware of this development in December 2022, when its report appeared online in \textit{JAMA Surgery} \citep{Salman:23}.}

Once \"{U}nver ran the software, he confirmed that the matching I had identified was one of several ``maximum-size'' matchings. 
Although the largest possible exchange was a 5-way---an exchange the Malatya team was capable of implementing---performing a 4-way exchange together with a 2-way exchange would actually save one additional life. 
In fact, maximizing the number of transplants required exactly this combination of a 4-way and a 2-way exchange. 
Relying only on 2-way and 3-way exchanges would result in one fewer patient being transplanted, and restricting to 2-way exchanges alone would mean two fewer patients. 
(See Figure \ref{fig:FirstLPE-July2022} for feasible transplants and liver exchanges during the first week of July 2022.)
We immediately surprised Y{\i}lmaz with the good news.

\begin{figure}[!tp]
    \begin{center}
       \label{fig:FirstLPE-July2022}
    \includegraphics[scale=0.7]{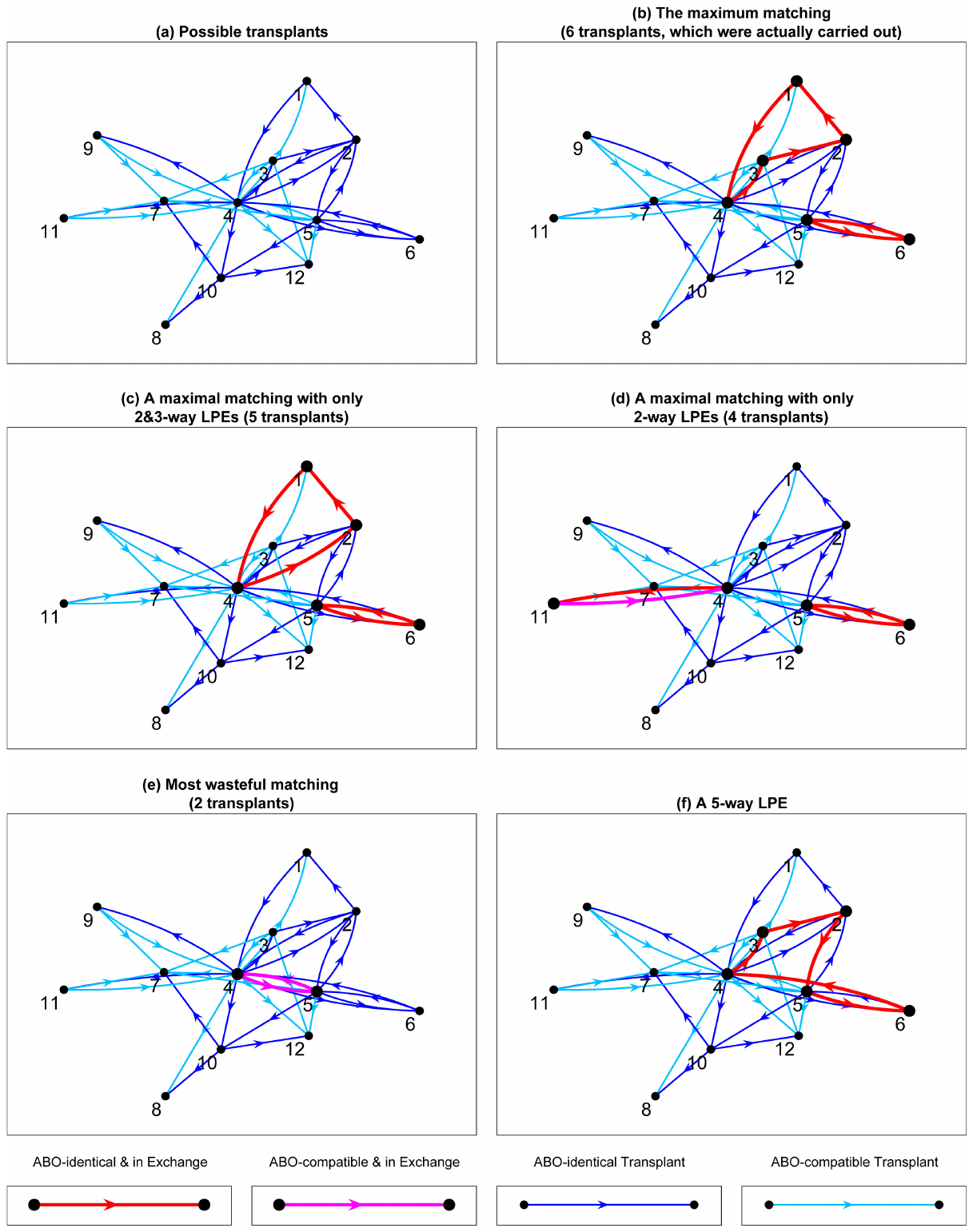}
    \end{center}
\caption{Possible transplants and liver exchanges under the Banu Bedestenci Sönmez Liver Paired Exchange System at the Liver Transplantation Institute, İnönü University, Malatya, Turkey, during the first week of July 2022. 
Each node in the graphs indicates a patient--donor pair. 
Abbreviation: LPE: Liver paired exchange. Reproduced from \citet{Malatya:2023}, \textit{American Journal of Transplantation},
licensed under \href{http://creativecommons.org/licenses/by-nc-nd/4.0/}{CC BY-NC-ND 4.0}.} \label{fig:FirstLPE-July2022}
\end{figure}

Even though it was quite late in Turkey---around 9 PM---when we shared the news, Y{\i}lmaz and much of his team were still at the hospital. 
Although we didn’t realize it at the time, it later became clear that this level of commitment is the norm for the Malatya team. 
In a public hospital located in one of the most isolated and often neglected parts of the country, 
Y{\i}lmaz built not only a highly skilled team but also an exceptionally dedicated one.

Y{\i}lmaz promptly assembled his team, and by midnight, they had selected the exchanges they planned to carry out within a few days: the 4-way and 2-way exchanges that made the maximal size matching.
The next morning, after informing the patients and their donors and confirming their continued interest in participating in the system, 
the Malatya team scheduled the 4-way exchange for July 5, 2022, and the 2-way exchange for July 7, 2022.

The Malatya team successfully conducted a world-first 4-way liver exchange on its scheduled day, just one day before I flew to Turkey. 
As soon as I reached Ankara the next day, I dropped off my son, Alp Derin, at my parents' house and immediately flew to Malatya.
On July 7th, 2022, I personally witnessed the operations for the week's second liver exchange---this time a 2-way exchange. 
I also visited the patients and donors recovering from their surgeries for the 4-way exchange conducted two days earlier.

\subsubsection{Heightened Importance of Size Incompatibility in Liver Exchange} \label{sec:leftlobe}

Among the eight participants in the 4-way exchange, there was one donor I was especially eager to meet: a 38-year-old Afghan refugee with 
blood type O who was ``size incompatible'' with his 12-year-old son, who had blood type A. Since the father-son duo were blood type-compatible but not identical, 
they were precisely the type of pair we aimed to include in the liver exchange (see  Section \ref{sec:biological}). 
Critically, however, the reason this donor could not donate to his son was one we had not yet considered. 

In our earlier liver exchange research,
we assumed that such pairs join the liver exchange when the left lobe of the donor is too small for their co-registered patient \citep{ergin/sonmez/unver:20}. 
This assumption formed the basis of our ``biological'' incentive mechanism embedded within liver exchange. 
However, the current situation was entirely different: the right lobe of the father was too large for the young recipient, and his smaller left lobe was unsuitable for donation due to multiple narrow hepatic arteries.
Until this point, our focus had been exclusively on size incompatibility arising from small grafts. This experience revealed that feasible grafts could also be too large. 
This realization meant there were other, perhaps more relevant, forms of ``biological'' incentive mechanisms inherent to the problem. 
Consequently, the presence of highly sought-after blood type-compatible but non-identical pairs in liver exchange would likely be more common than we had previously anticipated.

The BBS--LPE System effectively transformed individual misfortunes into collective opportunities.
The size incompatibility between the Afghan refugee and his pediatric patient, typically a source of despair, became a beacon of hope for others.
In addition to this pediatric patient, three adult patients also received liver transplants: one faced blood-type incompatibility with their registered donor, while the other two had donors with insufficiently sized right lobes.
Notably, among the four pairs, only one encountered blood-type incompatibility, while the remaining three faced size incompatibility.

When we asked Y{\i}lmaz how often prospective donors are unable to donate their left lobes, we received another surprising answer: it is extremely common. 
According to their transplantation norms, no more than 30\% of donors who could safely donate a lobe were eligible to donate their left lobes!
This information revealed another fundamental aspect of liver exchange. Donors who could safely donate their left lobes had very high value for the system because they are the only ones who can donate to pediatric patients.

This breakthrough had significant implications for our recently launched liver exchange program. It became increasingly clear that size incompatibility 
would play an even more substantial role than we had initially anticipated \citep{Malatya:2024}. Moreover, this new possibility amplified the importance of the Malatya team's capacity to perform larger-size exchanges.

\subsubsection{The Pilot Phase in the First Year}
    
For the year leading up to August 2023, we were disappointed that the system generated only a few additional exchanges, with numbers much lower than we had hoped. 
On the bright side, these included a 3-way exchange in April 2023, marking the third such exchange globally. This achievement followed the world’s first 3-way exchange conducted in Pakistan in March 2022 \citep{Salman:23} 
and a second 3-way exchange in India in December 2022, which was performed after our world-first 4-way exchange \citep{Soin:2023}. 
Including the two 2-way exchanges performed before the computerized system was deployed in June 2022, the Malatya team had conducted a total of 15 liver exchange transplants by August 2023.

While the contribution of liver exchange to the number of LDLTs performed by the Malatya team was greater than that of any other liver exchange 
system worldwide up to that time, it still fell short of our expectations. For example, the 10 liver exchange transplants performed by the Malatya team in 2022 represented 4.3\% of all LDLTs they carried out that year, 
whereas we had been hoping for at least 20\%. Several factors account for this relatively slow start.

Established within the constraints of a public university hospital with limited resources and a burdened staff, the responsibility of 
explaining the system to prospective patients and their donors fell predominantly on Y{\i}lmaz during the early phases of the system's implementation. 
Moreover, the system remained undisclosed to the public for over a year, even after their groundbreaking 4-way exchange, primarily due to the publicity 
crisis stemming from their misunderstood Guinness World Record attempt in 2019. The Malatya team hesitated to publicize their achievement unless it could be presented in a scholarly context.

However, the most significant setback in the system's first year was a devastating magnitude-7.8 earthquake---a once-in-a-century disaster---that struck the region in February 2023. 
The Malatya team faced immense challenges in performing LDLTs during the four months following the catastrophe, 
which razed cities and claimed over 50,000 lives, according to official figures. Although much of the hospital's infrastructure withstood the earthquake, Malatya itself was severely impacted. 
This disaster marked the most formidable crisis ever encountered by the Liver Transplantation Institute, posing a serious threat to its leadership in liver transplantation in Turkey and raising questions about the institute's future trajectory.

Y{\i}lmaz must be one of the most perceptive and hardest-working individuals I have ever known. 
Despite being devastated by the situation, he navigated through this challenging period by collaborating with us to document our experience with the BBS--LPE System for scholarly publication, 
with a specific focus on our groundbreaking 4-way exchange. 

Fortunately, the editorial process in medicine moves much faster than in economics. 
The timely online publication of \cite{Malatya:2023} in July 2023 not only provided an opportunity to introduce the BBS--LPE System to the general 
Turkish public but also reversed the unfortunate dynamics that had gripped the entire Malatya team.

\subsubsection{From Pilot to Global Leadership: First Multi-Way Exchanges and Unprecedented Volume}

Following the publication of \cite{Malatya:2023}, İn\"{o}n\"{u} University leadership hosted a media event at the end of July 2023 to highlight the achievements of the BBS--LPE System. 
Participants from both the 4-way and 3-way exchanges attended, sharing their experiences and praise with the wider public. 
Emerging from a region still recovering from a once-in-a-century catastrophe, this success quickly captured national attention. 
Within weeks, patient participation surged, marking the conclusion of the BBS--LPE System’s pilot phase and the beginning of a period of rapid expansion.

Over the next 27 months, the system facilitated 315 additional liver exchange transplants---averaging more than 11 per month---bringing the cumulative total to 330 through October 2025.
This rate exceeded that of the most active liver-exchange programs worldwide by an order of magnitude.
Before the BBS--LPE System’s introduction, only one center---\textit{Medanta} in India---had performed as many as ten liver-exchange transplants in any calendar year.

\begin{figure}[!tp]
    \begin{center}
       \includegraphics[scale=0.75]{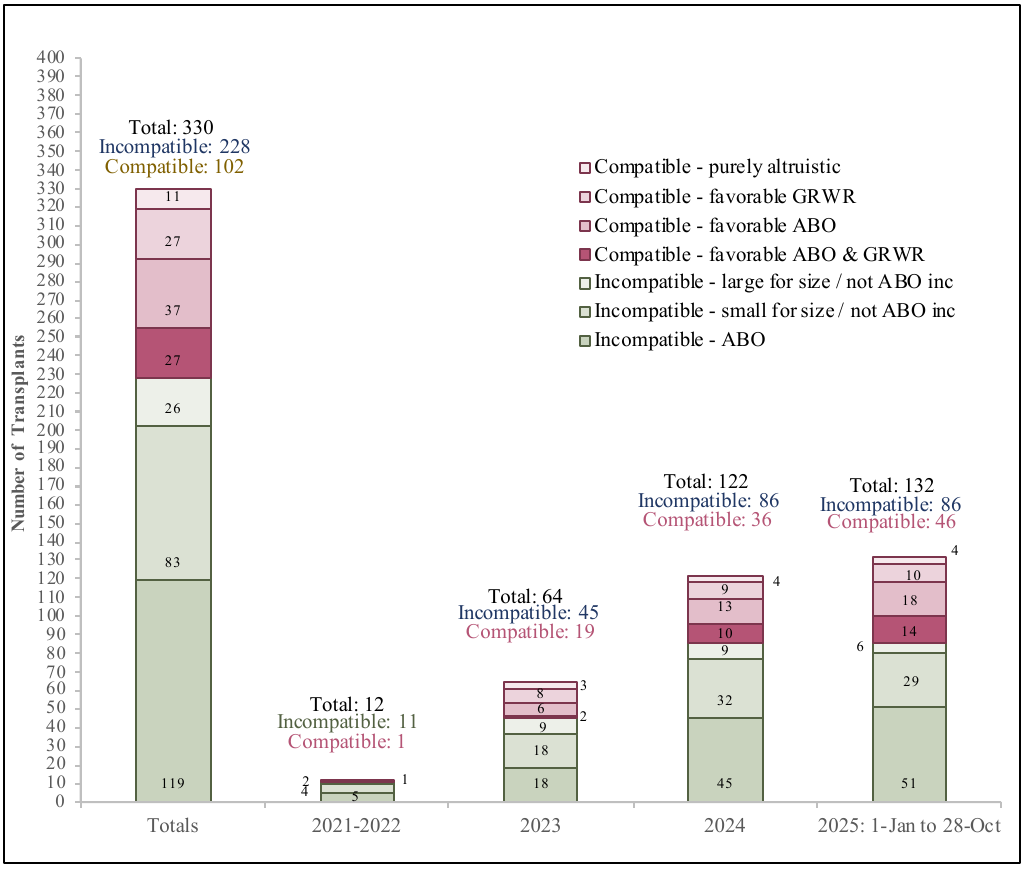}
           \end{center}
\caption[BBS--LPE System: Exchange Decomposition by Donor Compatibility]{BBS--LPE system: decomposition of liver exchanges by patient compatibility with their donors, as of October 2025. 
Acronyms: LPE: liver paired exchange; GRWR: graft-to-recipient weight ratio; ABO inc: ABO incompatible.}  \label{fig:BBS--LPE-October2025-type} 
\end{figure}

Despite the devastating earthquake that disrupted life across the region for months, the BBS--LPE System enabled 64 liver-exchange transplants in 2023, accounting for 27.7\% of the 231 LDLTs performed by the Malatya team that year.
Building on this momentum, it facilitated 122 transplants in 2024---an unprecedented 45.2\% of the 270 LDLTs performed at the Institute.
Further scaling up the program in 2025, the team had already performed 132 transplants by the end of October, with numbers continuing to rise.
When its earlier exchanges are included (two transplants in 2021 and ten in 2022), the team’s cumulative total reached 330 liver-exchange transplants through October 2025 (see Figure \ref{fig:BBS--LPE-October2025-type}).

\begin{figure}[!tp]
    \begin{center}
    \includegraphics[scale=0.75]{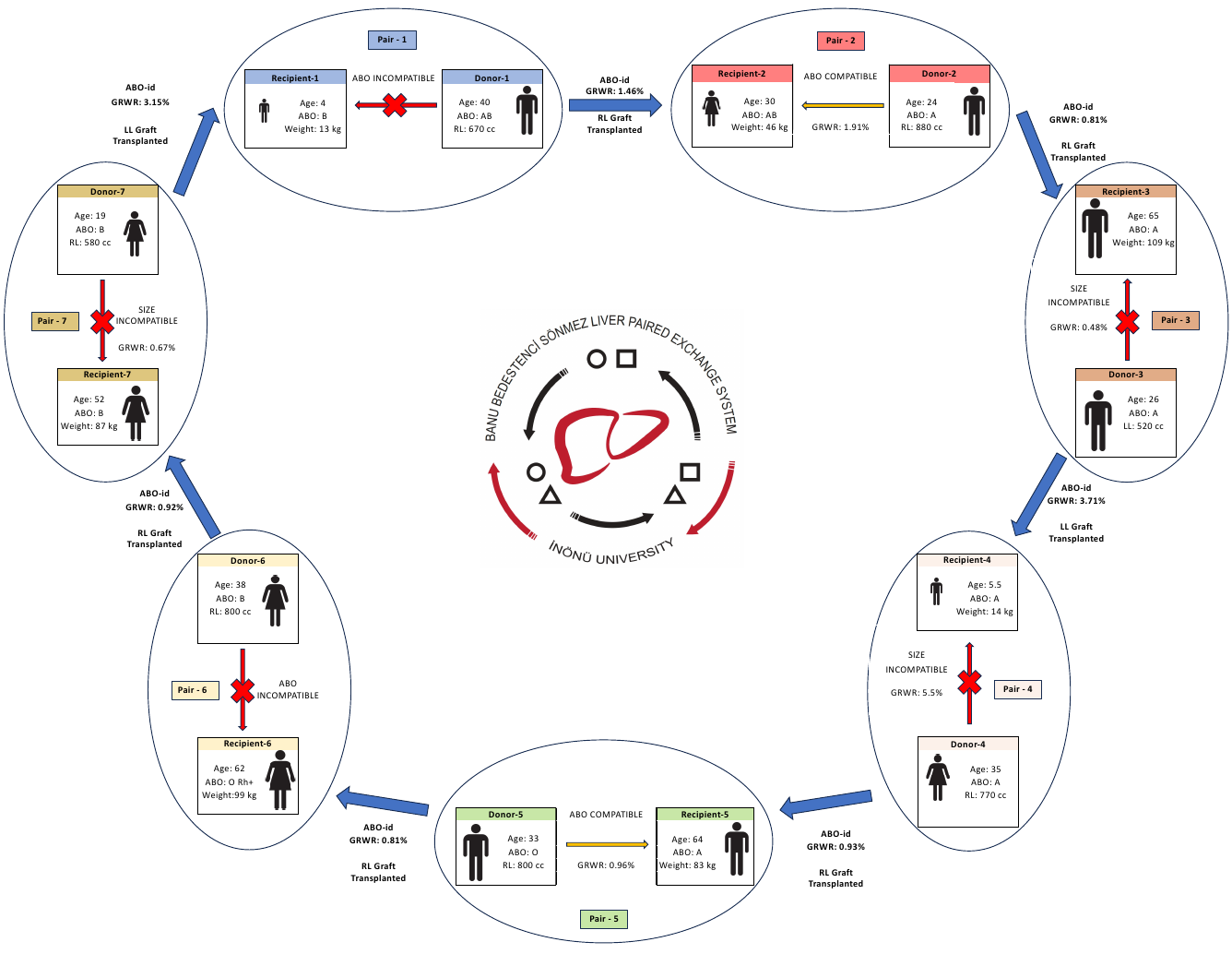}
    \end{center}
    \caption[World’s First 7-Way Liver Exchange Performed by the BBS--LPE System (Malatya, July 2024)]{Illustration of the world's first 7-way liver exchange, performed by the BBS--LPE System at the Liver Transplantation Institute, 
    İn\"{o}n\"{u} University, Malatya, Turkey, on July 2, 2024. The exchange involved two ABO-incompatible pairs, two adult pairs with undersized grafts (GRWR $<$ 0.8\%), one pediatric pair with an oversized graft (GRWR $>$ 4.0\%), and two compatible pairs who received more favorable ABO-identical grafts through exchange. Abbreviations: RL: Right lobe, ABO-id: ABO identical, ABO-c: ABO compatible but not identical, GRWR: Graft-to-recipient weight ratio.}
    \label{fig:7-way}
\end{figure}

For perspective, fewer than 250 liver exchange transplants were reported worldwide between 2002 and 2022, before the launch of the BBS--LPE System. 
Ushering in a series of unprecedented milestones, in just three years, the Malatya team surpassed that two-decade global total, establishing its leadership in transplant volume and complexity:

\begin{itemize}
\item  \textbf{World’s First Multi-Way Exchanges:} Building on its world-first 4-way liver exchange in July 2022, they performed the first 5-way, 
6-way, and 7-way liver exchanges in October 2023, January 2024, and July 2024, respectively (Figure \ref{fig:7-way}).
\item \textbf{Record-Breaking Cumulative Volume:} By March 2024, they became the first to surpass 100 cumulative liver exchange transplants, in January 2025 the first to surpass 200, and in September 2025 the first to surpass 300.
\item \textbf{Unmatched Annual Performance:} In October 2024, they became the first to exceed 100 liver exchange transplants in a single calendar year---a milestone repeated in September 2025.
\end{itemize}

\begin{figure}[!tp]
  \centering
  \makebox[\textwidth][l]{
       \includegraphics[scale=0.70]{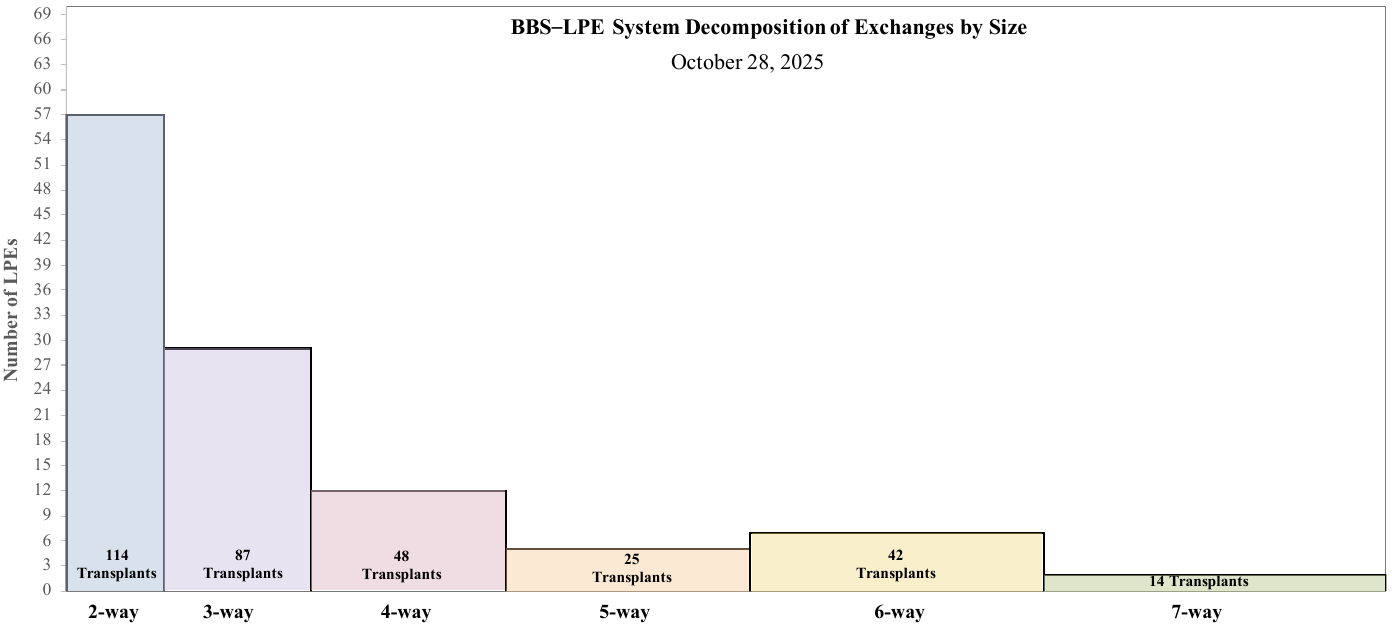}
  }
\caption[BBS–LPE System: Exchange Decomposition by Cycle Size]{BBS–LPE system: 
decomposition of liver exchanges by the number of patient–donor pairs per exchange, as of October 2025. 
For each $k$-way exchange, bar width is proportional to the number of patients in each exchange, 
and bar area is proportional to the total number of patients who received transplants via $k$-way exchanges. 
Acronym: LPE: Liver paired exchange.}
 \label{fig:BBS--LPE-October2025-size} 
\end{figure}

As of October 2025, the BBS--LPE System accounts for both 7-way exchanges ever performed worldwide, all seven 6-way exchanges, all five 5-way exchanges, 
all thirteen 4-way exchanges, and twenty-nine 3-way exchanges---representing more than 80\% of all 3-way liver exchange transplants ever reported globally (see Figure~\ref{fig:BBS--LPE-October2025-size}).

\subsubsection{Comparison with Other Prominent Liver Exchange Programs}  

Prior to the introduction of the BBS--LPE System, India was the most active country globally for liver exchange. 
According to \cite{Kute:2025}, 259 liver exchange transplants were performed in the country between January 2007 and March 2025 across nine centers. 
This corresponds to 1.8\% of the 14,341 LDLTs conducted in India during this period.

Almost half of these transplants were performed at Delhi-based \textit{Medanta}---the global leader in liver exchange prior to BBS--LPE---which carried out 123 liver exchange transplants, 
including three 3-way and fifty-seven 2-way exchanges. These accounted for 2.6\% of the 4,799 liver transplants performed at the center as of March 2025 \citep{Soin:2023, Kute:2025}. 
Another Delhi-based center, \textit{Max Saket Hospital}, performed 76 liver exchange transplants in thirty-eight 2-way exchanges as of March 2025, 
constituting 1.6\% of the 4,773 total liver transplants at that center \citep{Agarwal:2020, Kute:2025}.\footnote{Although \cite{Kute:2025} does not disclose center names, 
\cite{Soin:2023} for \textit{Medanta} and \cite{Agarwal:2020} for \textit{Max Saket Hospital}---further verified through personal communication with 
Dr. Dhiraj Agarwal regarding the latter---confirm that these two centers are the highest- and second-highest-volume centers in India for kidney exchange.}

Beyond India, another historically important program is at \textit{ASAN Medical Center} in South Korea, a pioneer in kidney exchange. 
Liver exchange activity at the center and across the country has slowed in recent years with the expansion of \textit{ABO-incompatible liver transplantation}. 
Although the \textit{ASAN} team performed the world’s first liver exchange in 2002, only 26 liver exchange transplants were conducted between January 2003 and December 2011, 
corresponding to 1.2\% of the 2,182 LDLTs performed at the center during that period \citep{Jung:2014}. 
Nationwide, just 52 liver exchange transplants were reported in South Korea from 2002 to 2018, representing only 0.4\% of the country’s LDLTs \citep{Kim:2022}.

Notably, all liver exchanges at \textit{Max Saket Hospital}, and the vast majority at \textit{ASAN Medical Center}, were limited to 2-way exchanges between blood-type incompatible pairs: 
one pair with a blood-type A patient and B donor, and another with a blood-type B patient and A donor. This restriction is likely the most significant factor explaining the relatively small number of exchanges at these centers.

By contrast, in the U.S., liver exchange programs remain in their early stages. As of October 2025, a total of 196 liver exchange transplants had been reported nationwide, 
with a peak of 49 in 2023 \citep{OPTN:25}. Most activity has been concentrated at two centers. The program at the \textit{University of Texas Health San Antonio} reported 23 liver exchange transplants from 2019 through October 2023, 
conducted through four 2-way exchanges, six 2-way non-directed donor chains, and one 3-way non-directed donor chain \citep{SanAntonio-Liver:2024}. 
Similarly, the program at the \textit{University of Pittsburgh Medical Center} reported 20 liver exchange transplants between January 2019 and July 2021, 
carried out through three 2-way exchanges and seven 2-way non-directed donor chains \citep{UPMC-Liver:2022}.

In Turkey, liver exchange predates our establishment of the BBS--LPE system at \.{I}n\"{o}\"{u} University's Liver Transplantation Institute. 
\cite{Dinckan:2025} reports that 18 liver exchange transplants were performed in nine 2-way exchanges at \textit{\.{I}stinye University Hospital Multiorgan Transplant Center} 
between January 2015 and December 2022. This represents one of the earliest large-scale liver exchange initiatives outside India and South Korea.
Table \ref{tab:global-LE} summarizes the major liver exchange programs worldwide that have reported 10 or more cumulative transplants.


\begin{table}[!tp]
\footnotesize
\centering
\begin{threeparttable}
\caption{Major liver exchange programs with 10 or more cumulative transplants. For each country, the first row reports the national total, followed by rows for individual programs.}
\label{tab:global-LE}
\setlength{\tabcolsep}{4pt}
\renewcommand{\arraystretch}{1.06}
\begin{tabular}{@{}p{0.16\textwidth} p{0.47\textwidth} p{0.07\textwidth} p{0.25\textwidth}@{}}
\toprule
\addlinespace[4pt] 
\textbf{Country} & \textbf{Program / Operator} & \textbf{Start} & \textbf{Documented volume} \\ 
\addlinespace[4pt] 
\midrule
\addlinespace[2pt] 

Turkey
& Multiple single-center programs
& 2015
&  $\geq$348  \; (by 10/2025) \\
\addlinespace[2pt] 
\cline{2-4}
\addlinespace[3pt] 
& \.{I}n\"{o}n\"{u} University, Liver Transplantation Institute
& 2022
& 330  \;  (by 10/2025) \\
\addlinespace[3pt] 
& \.{I}stinye University Hospital
& 2015
&  18 \; (01/2015--12/2022)\\
\midrule

India 
& Multiple single-center programs
& 2007
&   $\geq$259  \; (by 03/2025) \\
\cline{2-4}
\addlinespace[3pt] 
& Medanta
& 2009
& 129  \;  (by 03/2025) \\
\addlinespace[3pt] 
& Max Saket Hospital
& 2012
&  78 \; (2012--03/2025)\\
\midrule

United States 
& Multiple single-center programs
& 2014
&  196  \; (by 10/2025) \\
\cline{2-4}
\addlinespace[3pt] 
&  University of Texas Health San Antonio
& 2019
& 23  \;  (01/2019-10/2023) \\
\addlinespace[3pt] 
& University of Pittsburgh Medical Center
& 2019
&  20 \; (01/2019--07/2021)\\

\midrule
South Korea
& Multiple single-center programs
& 2002
&   $\geq$52  \; (2002--2018) \\
\cline{2-4}
\addlinespace[3pt] 
& ASAN Medical Center
& 2003
& 26  \;  (01/2003--12/2011) \\
\addlinespace[1pt] 
\bottomrule
\end{tabular}
\end{threeparttable}
\end{table}

\subsection{Malatya's Journey to Global Leadership: Implementing Best Practices} 

Our two decades of experience in kidney exchange have revealed three key factors that boost the efficacy of living-donor exchange programs. 
The first is the ability to incorporate altruistic non-directed donors (NDDs) into the system and amplify their impact through NDD chains. 
This approach enabled the \textit{National Kidney Registry} to become the leading kidney exchange platform in the U.S. (see Section \ref{sec:NSF}) and helped the 
U.K. and Canada establish two of the most successful national kidney exchange systems in the world (see Section \ref{sec:KE-2008}). 
Unfortunately, because altruistic donation is illegal in Turkey, this was not an option for the Malatya team.

The other two crucial factors are the capacity to conduct larger exchanges and, more importantly, the inclusion of compatible pairs in the system 
(see Figures \ref{fig:BBS--LPE-October2025-type} and \ref{fig:BBS--LPE-October2025-size}). 
These are the areas in which the Malatya team truly excelled.

When our collaboration began in 2019, the Malatya team had already demonstrated the ability to perform five simultaneous liver transplants. 
Although the Liver Transplantation Institute at \.{I}n\"{o}n\"{u} University was equipped to handle this capacity, they had the personnel to support up to seven. 
Following the world’s first 4-way exchange in July 2022, the team added two operating rooms to the existing ten at the Institute---all located on a single floor---enabling 6-way exchanges within its premises. 
As of September 2025, they had utilized their expanded capacity for 7-way exchanges twice, using additional operating rooms within the larger hospital complex.

However, while highly significant, this impressive logistical capacity was not the single most important driver of their extraordinary performance after the program’s pilot phase.
Rather, it was the integration of ``biological'' incentive mechanisms---encouraging the inclusion of compatible pairs---that ultimately played the leading role in their success.

Following the highly successful media event at the end of July 2023, Y{\i}lmaz inquired about additional policies that could enhance the efficacy of the BBS--LPE System. 
Reminding him that our 4-way exchange in July 2022 had been made possible by a blood-type O donor paired with a blood-type A patient, 
we asked what other strategies might encourage such pairs to participate in liver exchange. He explained that there are several reasons why blood-type compatible pairs may benefit from participating in liver exchange.

We already discussed one such reason in Section  \ref{sec:leftlobe}. 
Since donors can rarely donate their smaller left lobes, liver exchange becomes a compelling possibility whenever their patient needs a smaller graft. 
Another reason is the opposite scenario, which we explored in \cite{ergin/sonmez/unver:20}: 
donors who are blood-type compatible with their patients may have grafts that are too small for their patients. We had two such pairs in our 4-way exchange.

These scenarios have been utilized by the BBS--LPE System since its deployment. However, three additional factors could further enhance its efficacy.

Firstly, liver transplants using blood-type identical grafts are generally better for patients than those using blood-type compatible but non-identical grafts \citep{Pittsburgh:1986, Koukoutsis-etal:2007}.
 Although the patient may not have antibodies against the transplanted liver, the liver itself can temporarily carry antibodies against the patient’s body. 
 This phenomenon, known as \textit{passenger lymphocyte syndrome}, often leads to \textit{hemolytic anemia} in recipients of blood-type non-identical grafts. 
 While this condition typically resolves on its own, it can occasionally result in severe complications, including death. In fact, in medical practice, transplanting a blood-type compatible but 
 non-identical graft is referred to as \textit{minor ABO incompatibility} \citep{deBruijn-etal:2017}. This creates an additional ``biological'' incentive to enhance the efficacy of the BBS--LPE System.

Secondly, the size of the graft plays a crucial role in the success of a liver transplant. While a graft-to-recipient weight ratio (GTWR) between 0.8\% and 2\% is considered acceptable for adult transplantation, 
the ideal range is 1\% to 1.2\% \citep{Malatya:2024}. The Malatya team was particularly cautious about cases where the GTWR exceeded 1.6\%, as larger grafts can pose surgical challenges. 
This concern creates another natural ``biological'' incentive to encourage compatible pairs to participate in liver exchange.

Finally, for adult female patients, grafts from their children---especially if they are male---have been shown to have inferior long-term graft and patient survival, likely due to sensitization during pregnancy \citep{Dagan-etal:2020}. 
This creates yet another natural ``biological'' incentive to encourage pairs consisting of a mother patient with her compatible offspring donor to participate in liver exchange.

In addition to integrating these ``biological'' incentive mechanisms into the BBS--LPE System, the Malatya team has also allowed for ``altruistic'' participation from compatible pairs, 
even when there is no clear benefit for the pair, provided that the patient receives an equally suitable graft without delay.

The adoption of these policies since August 2023 has been the primary driver behind the exceptional outcomes of the BBS--LPE program following its pilot phase. 
Our hope is that these practices will evolve into established standards in liver exchange, enhancing its global efficacy by orders of magnitude.

\section{Contributions, Insights and Directions on Behavioral Market Design} \label{sec:behavioral}

While I have primarily relied on theory as the formal apparatus for my pursuits in minimalist market design, many applications have revealed intriguing insights, questions, 
and puzzles in behavioral economics. In this section, I explore some of the most significant ones.

\subsection{Leveling the Playing Field in School Choice} \label{sec:leveling-the-playingfield}

After my initial, unsuccessful effort to reform the Turkish college admissions system in 1997---my first experience with policy engagement---I came to understand that achieving 
policy impact as an external critic was unlikely, no matter the merits of my proposals, unless the flaws in the existing system were  glaring. In the late 1990s, I set my sights on the Boston mechanism as a prime example of 
such a flawed system where I might have better luck. 

To build a compelling case against this mechanism in favor of the DA or SC--TTC mechanisms, 
Yan Chen and I conducted a laboratory experiment involving all three \citep{chen/sonmez:06}.\footnote{Building on \cite{chen/sonmez:06}, a substantial body of literature has employed lab 
experiments to study preference manipulation in school choice mechanisms. For a comprehensive review of this literature and other behavioral studies in education market design, 
see \cite{rees-jones/shorrer:23}. For a broader review of behavioral studies in market design, see \cite{Chen/Cramton/List/Ockenfels:21}.} 
In our experimental setting, both DA and SC--TTC outperformed the Boston mechanism.

A key perk of adopting a \textit{strategy-proof} mechanism is that it enables the central planner to offer families straightforward advice, 
reassuring them that truthfully revealing their preferences is a safe strategy. This reassurance remains valuable, even if participants choose not to follow it or remain skeptical.

In our experiment, we deliberately withheld this key information, not disclosing the \textit{strategy-proofness} of DA and SC--TTC to participants. 
Our goal was to explore whether these mechanisms could still outperform Boston without any external guidance. 
Both mechanisms did, with DA performing best. However, a substantial number of participants still submitted non-truthful preferences under both DA and SC--TTC---albeit less frequently than under the Boston mechanism.

Encouraged by these experimental results, we grew confident that replacing the Boston mechanism with either DA or SC--TTC could yield even greater benefits in real-world settings, especially when families are properly informed and guided.

In a related theoretical analysis, Haluk Ergin and I performed a complete-information equilibrium analysis of the Boston mechanism, obtaining the following result.

\begin{theorem}[\citealp{ergin/sonmez:06}] \label{thm:ergin-sonmez-06}
Fix a school choice problem and consider the preference revelation game induced by the Boston mechanism. 
The set of Nash equilibrium outcomes of this game is equal to the set of outcomes that satisfy \textit{individual rationality, non-wastefulness,} and \textit{no justified envy}.\footnote{Equivalently, 
invoking Proposition \ref{prop:NJE-stability}, the set of Nash equilibrium outcomes of this game is equal to the set of \textit{stable} outcomes of the induced many-to-one matching market.} 
\end{theorem}

Since the outcome of the DA mechanism---the student-optimal stable matching---satisfies all three axioms and \textit{Pareto dominates} any other outcome that also satisfies them, 
we have the following immediate corollary, which lends additional support in favor of DA over the Boston mechanism:

\begin{corollary}[\citealp{ergin/sonmez:06}] \label{cor:ergin-sonmez-06}
Fix a school choice problem. The outcome of the DA mechanism under truthful dominant strategies either \textit{Pareto dominates} 
or is equal to any given Nash equilibrium outcome of the preference revelation game induced by the Boston mechanism.\footnote{\label{fn:Ergin-Sonmez:2006}\cite{ergin/sonmez:06} show that the complete information assumption 
is key for both Theorem \ref{thm:ergin-sonmez-06} and Corollary \ref{cor:ergin-sonmez-06}. 
The characterization result in Theorem \ref{thm:ergin-sonmez-06} not only fails to extend to an incomplete information environment, 
but in such an environment a student may also
prefer a Bayesian Nash equilibrium outcome of the Boston mechanism to the outcome of the DA mechanism under truthful dominant strategies.}

\end{corollary}

During our first meeting with Boston Public Schools (BPS) officials in October 2003, 
the experimental evidence from \cite{chen/sonmez:06} and the theoretical results from \cite{ergin/sonmez:06} were instrumental in helping to persuade 
BPS officials that it might be in the city's best interest to consider replacing the Boston mechanism with either the DA or SC--TTC mechanism (see Section  \ref{sec:Boston}).

Even though the reform process took close to two years for due diligence, it was clear from the beginning that both BPS leadership and the members of the Task Force formed to evaluate the city's assignment process  
were supportive of a reform that would increase the \textit{transparency} of the student assignment mechanism. 
While the decision between DA and SC--TTC remained uncertain for some time, it was clear that there was little interest in retaining the Boston mechanism. Everything was progressing just as I had hoped.

Just as it seemed that there was near-unanimous support for discarding the Boston mechanism in favor of a \textit{strategy-proof} one, 
at the June 8, 2005 public hearing on Boston's student assignment process, the leader of a parents group---the West Zone Parent Group (WZPG)---made the following statement in their community testimony:

\begin{quote} 
``There are obviously issues with the current system. If you get a low lottery number and don’t strategize or don’t do it well, then you are penalized. But this can be easily fixed. 
When you go to register after you show you are a resident, you go to table B and the person looks at your choices and lets you know if you are choosing a risky strategy or how to re-order it.

Don’t change the algorithm, but give us more resources so that parents can make an informed choice.'' 
\end{quote}

We later learned that WZPG was a well-informed group of approximately 180 members who met regularly before admissions periods to discuss school choice for elementary schools, 
devising elaborate strategies to optimize their children's assignments. For example, the meeting minutes from October 27, 2003, state (\citealp{pathak/sonmez:08}, p. 1636):

\begin{quote} 
``One school choice strategy is to find a school you like that is undersubscribed and put it as a top choice, OR, find a school that you like that is popular and put it as a first choice 
and find a school that is less popular for a `safe' second choice.'' 
\end{quote}

WZPG was the first group in nearly two years to express interest in retaining the Boston mechanism. 
I wasn’t surprised that well-informed and strategic families---or even organized parent groups---were able to leverage the manipulability of Boston's system to 
their advantage at the expense of other families. In effect, they were gaining priority over others through their informed and strategic participation.

What did surprise me, however, was their willingness to publicly defend this advantage. Here is a statement from a member's testimony during the November 4, 2004 
public hearing of the Boston School Committee \citep{abdulkadiroglu/che/yasuda:2011}.

\begin{quote}
``I’m troubled that you’re considering a system that takes away the little power that parents have to prioritize... what you call this strategizing as if strategizing is a dirty word...''
\end{quote}

Despite WZPG’s opposition, the broader sentiment remained unchanged. In fact, it reinforced Payzant's earlier remark that ``a strategy-proof mechanism levels the playing field.'' 
Consequently, the Boston School Committee voted in July 2005 to replace the Boston mechanism with the DA mechanism (see Section \ref{sec:Boston}).

Subsequent analysis of school choice data from BPS under the Boston mechanism confirmed the widespread presence of both highly strategic and relatively naive preference submissions \citep{aprs:06}. 
For example, among the latter group, 19\% of families listed two consistently oversubscribed (i.e., popular) schools as their top two choices, leading to 27\% of these children being unassigned.

The prevalence of naive preference submissions in the Boston data suggested that the complete information equilibrium analysis in \cite{ergin/sonmez:06}, 
which assumed all families acted strategically, might not have been a realistic benchmark for Boston’s school choice. Motivated by this realization, in \cite{pathak/sonmez:08}, 
we introduced a simple generalization by partitioning families into ``strategic'' and ``sincere,'' assuming that sincere families always report their preferences truthfully.

To describe the main result from \cite{pathak/sonmez:08}, consider the following construction. 
Given a school choice problem, student preferences remain unchanged, but each school’s priorities are modified based on the preferences of sincere students:

\begin{enumerate} 
\item Students are grouped into tiers, with higher tiers corresponding to higher priority. Priority within a tier is determined by the school’s original priority. 
\item All strategic students are placed in the highest-priority tier. 
\item For sincere students, their tier is determined by how highly they rank the school: students who rank the school as their first choice are 
placed in the highest-priority tier along with the strategic students. In general, the higher a sincere student ranks a school, the higher their tier. 
\end{enumerate}

This construction results in a unique priority profile, which we refer to as the \textit{adjusted} priority profile.

In this model, which accounts for both strategic and sincere students, we have the following generalization of Theorem  \ref{thm:ergin-sonmez-06}:

\begin{theorem}[\citealp{pathak/sonmez:08}] \label{thm:pathak-sonmez-08}
Partitioning the students into strategic and sincere groups, fix a school choice problem and  
consider the preference revelation game induced by the Boston mechanism. 
The set of Nash equilibrium outcomes of this game is equal to the set of outcomes that satisfy \textit{individual rationality, non-wastefulness,} 
and \textit{no justified envy} for the modified school choice problem with the adjusted priority profile. 
\end{theorem}

When all students are strategic, the adjusted priority profile is the same as the original priority profile, and therefore Theorem \ref{thm:pathak-sonmez-08}
reduces to Theorem \ref{thm:ergin-sonmez-06}.
When some students are sincere, under complete information equilibrium, each strategic student effectively gains priority at each school over any sincere student, 
except those who rank the school as their first choice under their truthful preferences. In this sense, the Boston mechanism rewards both being well-informed and strategic, 
which may explain the testimony from the leader of the WZPG. This idea is further captured in the following result:

\begin{theorem}[\citealp{pathak/sonmez:08}] \label{thm2:pathak-sonmez-08}
Partitioning the students into strategic and sincere groups, fix a school choice problem and  
consider the preference revelation game induced by the Boston mechanism. 
There exists a Nash equilibrium outcome in this game that is weakly preferred by each strategic student to the outcome of the DA mechanism 
in truthful dominant strategies.\footnote{With the BPS data used in \cite{pathak/sonmez:08}, 
the Nash equilibrium outcome of the preference revelation game induced by the Boston mechanism is virtually unique under all partitions of students into strategic and sincere.}
\end{theorem}

\textit{Transparency}, along with informational and strategic equity---``leveling the playing field''---for families who could not or did not strategize effectively, emerged as key objectives for 
BPS leadership during the school choice reform process from 2003 to 2005. Meanwhile, a group of families---the WZPG---sought to preserve their informational and strategic advantage over others. 
Since this objective was not publicly articulated as such, it does not align with my definition of ``legitimate.'' To gain insight into the possible motives of this group, we relied on a simple behavioral model.

We now turn to a sharply contrasting case study from Seattle, where the school assignment process remained opaque for nearly a decade, 
possibly to serve objectives of decision-makers that, by my definition, do not qualify as ``legitimate.''

\subsection{Opting for Obscurity: The Curious Case of Seattle School Choice} \label{sec:Seattle}

In early 2007, Chris MacGregor, a Seattle parent and software engineer, began exploring the city's school choice mechanism for his son, who was starting kindergarten that year. 
He soon faced an unexpected challenge: conflicting advice from high-level officials, such as school principals, on how to rank schools on the selection form optimally \citep{MacGregor:08}:

\begin{quote} 
``Some folks will tell you that you should never list a `long shot' (a school you don’t think you’ll get into) as your first choice---they call it `throwing away your first choice,' 
implying there is special magic about that first slot on the form. (It turns out this is not true.) Others (not many) will tell you it doesn’t matter.''
\end{quote}

While many officials seemed to suggest that the city was using the Boston mechanism, a few hinted that it actually employed a \textit{strategy-proof} mechanism.

At the time MacGregor received this contradictory advice, I was also under the strong impression that Seattle was using the Boston mechanism. 
In fact, during our earlier analysis of this mechanism's limitations in the late 1990s and early 2000s, we came across an article from the \textit{Seattle Press}, 
published in consecutive years, cautioning parents against listing schools where they had low priority as their top choices \citep{Mas:98}.

\begin{quote} 
``This is why you have such an excellent chance of getting into your reference school if you make it your top choice. 
Choosing another neighborhood’s reference school, however, puts a lot of kids in line ahead of yours. That reduces your chances of getting in, particularly if the school has small classes.'' 
\end{quote}

Uncertain about which mechanism was in use, MacGregor negotiated with Seattle Public Schools for three weeks and eventually secured permission to review the source code of the school choice system. 
His analysis confirmed that the mechanism in place was DA, validating the assertions of the few officials who had correctly stated that strategizing would not improve outcomes under the current system.

Up until then, all of us in the market design community believed that no jurisdiction worldwide had used the DA mechanism for student assignment prior to the publication of \cite{abdulkadiroglu/sonmez:03}. 
Thanks largely to MacGregor's efforts, we discovered that Seattle had transitioned from the Boston mechanism to the DA mechanism in 1999, 
through what was known as the \textbf{\textit{Barnhart--Waldman (BW) amendment}}. 
This amendment---named after two school board members familiar with the U.S. entry-level medical job market’s use of DA---led to the change.
However, the switch was poorly communicated, if at all, resulting in widespread confusion among both the families participating in the system and the officials responsible for explaining it.

In \cite{pathak/sonmez:13}, we identified several indicators that the BW amendment was not well understood, even though a \textit{strategy-proof} mechanism 
should have enabled the school district to advise families that submitting truthful preferences was a safe strategy. Confusion about the BW amendment 
even surfaced during a 2001 deposition of the school board president in a court challenge to Seattle’s school choice plan.\footnote{Page 58 of the 
U.S. Court of Appeals for the Ninth Circuit, No. 01--35450, \textit{Parents Involved in Community Schools vs. Seattle School District No.1}, 2001.}

\begin{quote} 
``Q: Can you explain for me what the Barnhart/Waldman Amendment is and how it works?

A: If I could I’d be the first. [...] 
Before we adopted that amendment, all the first choices were processed in one batch and assignments made. If you did not get your first choice, 
it is my understanding that all the students who did not get the first choice fell to the bottom of the batch processing line, and then they would process the second choices, et cetera. 
Barnhart/Waldman says that after all the first choices are processed, in the next batch, if you don’t get your first choice, you don’t fall to the bottom of the list but you are then processed, 
your second choice, with all the other second choices together. The result is that instead of a high degree of certainty placed---or of value placed on first choice,
people can list authentically their first, second and third choices and have a higher degree of getting their second and third choice if they do not get their first choice. Now, was that clear as mud?'' 
\end{quote}

Even in court deposition, the school board president seems unaware that, under the DA mechanism, when students are considered for their second-choice schools, 
they are evaluated not only alongside other students who ranked that school as their second choice, but also with students from the first batch who ranked it as their first choice.

Seattle's poorly understood and inadequately publicized reform---switching from the Boston mechanism to the DA mechanism---serves as a cautionary tale about the importance of \textit{transparency}. 
Simply using a \textit{strategy-proof} mechanism does not mean much unless it is communicated effectively to all participants. 

This also raises another issue. In Section  \ref{sec:leveling-the-playingfield}, we have seen that even with full \textit{transparency}, some participants in Boston utilized the available information better than others. 
In Seattle, judging from the inconsistent advice provided by officials, such as school principals, it is likely that while some families understood that truthful preference revelation was in their best interest, 
others did not. These families surely gained an even greater advantage than members of the WZPG in Boston compared to less informed families. 
In general, inconsistent communication of various aspects of an institution can even be exploited by decision-makers to benefit some groups at the expense of others.

As troubling as Seattle’s lack of \textit{transparency} was for nearly a decade, what unfolded after MacGregor’s discovery was even more surprising---dare I say, eye-opening.

In June 2009, the Seattle School Board proposed a new student assignment plan to transition from the DA mechanism back to the Boston mechanism. 
The district publicly justified this reform as part of its ``New Student Assignment Plan,'' explicitly stating that 
``one of the goals of the new assignment plan is to reduce transportation costs'' \citep{SPS:09}.

This proposal nevertheless faced significant opposition from families. 
As \cite{Walkup:09} documents, many of the concerns echoed those raised in \cite{abdulkadiroglu/sonmez:03} regarding the limitations of the Boston mechanism.

First, moving to a manipulable mechanism would encourage more families to engage in strategic behavior when ranking schools. 
Since the algorithm would no longer ignore strategic choices, many would feel compelled to game the system. 
This, in turn, would prevent the district from obtaining accurate information about genuine family preferences, and the most sought-after programs might appear less popular if families believe they stand little chance of admission.

Second, access to historical data would become critical, giving an advantage to those who had it. Information on past choices and seat availability would become a valuable resource, 
and if the district did not make this data publicly accessible, families with access to it would have a substantial edge in the choice process.

Third, concerns arose about the persistence of misinformation, particularly among families accustomed to the DA mechanism. In the early years of the new plan, 
many might continue to follow the previously correct advice to rank schools based on their true preferences. Therefore, clear, timely, and ongoing communication about the new system would be crucial.

Opposition was not limited to technical critiques. Contemporaneous public commentary captured the depth of frustration with the reform. 
For example, a widely circulated blog post titled ``Ten+ Reasons Why the Seattle Public Schools Superintendent Should Be Fired With Cause'' 
listed the new assignment plan itself as a reason for dismissal, citing its effects on overcrowding, under-enrollment, and re-segregation \citep{SeattleEd:11}.\footnote{Indeed, 
the superintendent was dismissed shortly thereafter. In March 2011, the school board voted to remove her following revelations of financial improprieties, 
with board members noting that key audit findings had not been disclosed to them \citep{Samuels:11}.}

Despite the opposition, Seattle adopted the Boston mechanism beginning in the 2011--12 school year.

Perplexed by this reversal---the only instance we were aware of at the time in which a jurisdiction had replaced a school choice mechanism with a more manipulable one---we corresponded with 
several committee members involved in the decision. 
In \cite{pathak/sonmez:13}, we reported the only plausible explanation we encountered: that the BW amendment (i.e., the DA mechanism) 
would increase transportation costs by encouraging student mobility, as parents could freely express their true preferences without risking a seat at their neighborhood schools. 
This interpretation is fully consistent with the district’s own stated rationale in 2009, which explicitly identified reducing transportation costs as a goal of the new assignment plan \citep{SPS:09}. 
If this was indeed the case, it likely became a concern only after the \textit{strategy-proofness} of the city's DA mechanism was brought to public attention, thanks to MacGregor's efforts.

If reducing transportation costs was indeed a factor in Seattle’s reversal to the Boston mechanism, it would exemplify an objective that does not fully meet my definition of ``legitimate.'' 
Rather than directly addressing school choice, decision-makers may have sought instead to limit the mobility it introduced indirectly by reverting to the Boston mechanism. 
This rationale would also be consistent with the decision to keep the \textit{strategy-proofness} of the school choice mechanism opaque for nearly a decade---from 1999 to 2008.

These two episodes from Seattle’s school choice policies---first, concealing the details of its mechanism, including its \textit{strategy-proofness}, between 1999 and 2008, 
and then switching to the Boston mechanism shortly after a parent exposed this \textit{strategy-proofness}, despite families' concerns about its manipulability---raise several important questions.

Why was such a significant reform not uniformly communicated to all families? Was this an oversight, or were there other factors at play---perhaps a ``hidden'' objective of reducing mobility and transportation costs? 
How much advantage did the uneven communication give to families who learned that the mechanism in place was \textit{strategy-proof}? 
Could this information have been strategically provided to some families and withheld from others? 
Why did the city revert to the Boston mechanism in the second phase, after a parent discovered and publicized the \textit{strategy-proofness} of the DA mechanism?

Beyond Seattle’s experience, could the Boston mechanism be a preferred mechanism for decision-makers who favor neighborhood-based assignment policies but avoid directly advocating for them due to political costs?

Although this is a plausible hypothesis, our next case study from Taiwan suggests there may be even more effective ways to advance 
politically sensitive objectives---such as increasing neighborhood-based assignments---by carefully selecting and framing school choice mechanisms.

\subsection{Nudging School Choices: Taiwan’s Deduction Mechanism} \label{sec:Taiwan}

In 2014, Taiwan enacted the \textit{Senior-High School Education Act}, centralizing admissions to public high schools within each district and assigning students based on their scores 
from a Comprehensive Assessment Examination (CAE). Under this Act, each district uses a direct mechanism known as the \textbf{\textit{Taiwan Deduction (TD)}} mechanism---a hybrid of the 
Boston and DA mechanisms, first studied in \cite{dur/pathak/song/sonmez:22}. A key parameter of the TD mechanism is the \textbf{\textit{deduction rule}}, 
a policy variable that specifies a score ``penalty'' applied to students' lower-ranked schools in their applications. This deduction rule, an important policy lever, is determined by each district.

Formally, the deduction rule is a weakly increasing sequence of integers starting at zero, specifying how many points are subtracted from a student's CAE score for each school in their preference ranking.
Specifically, for a given school ranked $k^{\footnotesize\mbox{th}}$ in a student's preferences, their effective score for that school is calculated by reducing their CAE score by the $k^{\footnotesize\mbox{th}}$ 
element of the deduction sequence. For example, with a deduction rule of $0, 1, 2, 3, 4, \ldots$, a student loses 1 point for their second-ranked school, 2 points for their third-ranked school, and so on.
Given a profile of student preferences, the outcome of the TD mechanism is determined using the \textit{individual-proposing deferred acceptance mechanism}, 
but with student priorities based on their effective scores rather than their baseline CAE scores.

The DA mechanism is a special case of the TD mechanism, where the deductions remain uniformly zero throughout the sequence. In contrast, the Boston mechanism---also a special case---represents the 
opposite extreme, with a deduction rule in which the difference between any two consecutive elements exceeds the entire grading scale of the CAE. 
For example, if the score range is 0 to 100, a deduction rule such as $0, 101, 202, 303, 404, \ldots$ ensures that a student's effective score is determined primarily by the order in which 
they rank schools in their preferences. Consequently, schools consistently give higher priority to students who rank them higher, with CAE scores serving only as a tie-breaker.

In practice, the deductions implemented in Taiwan have been relatively modest, especially when compared to the more drastic deductions implied by the Boston mechanism.
For example, in 2014, Jibei---the largest district with 139 schools---allowed students to rank up to 30 schools and, within a score range of 90 points, used the following deduction rule:
\[ 0,1,2,3,4,5,6,7,8,9,10,10,10,10,10,10,10,10,10,10,12,12,12,12,12,12,12,12,12,12
\]
That same year, the largest deduction in any major district occurred in Zhongtou, where a 30-point reduction within a 100-point score range was applied to choices ranked between 31 and 50.

Despite these relatively modest deductions, it is evident that---partly due to the mechanism's explicit penalties for lower-ranked choices---TD mechanisms 
other than DA are all susceptible to preference manipulation. Thus, of all TD mechanisms, only DA satisfies the axiom of \textit{strategy-proofness}. 
Similarly, none of the TD mechanisms, except for DA, satisfy the axiom of \textit{no justified envy}.\footnote{It’s important to note that the school priority rankings induced 
by the effective scores under the TD mechanism are artifacts of the mechanism, not inherent to the problem. Therefore, \textit{no justified envy} is defined based on the original student scores from the CAE.}

Given experiences with mechanisms that fail these axioms worldwide, it is hardly surprising that many families were frustrated with the TD mechanism due to its failure to satisfy the axioms of 
\textit{no justified envy} or \textit{strategy-proofness}. For instance, an article in \textit{The China Post} \citep{Wei:2014} reported a parent’s outrage over students being penalized with point 
deductions for incorrectly filling out their preferences, which resulted in high-achieving students being assigned to the same schools as lower-performing peers.

What stands out, however, in Taiwan’s case is the unprecedented scale and intensity of the protests it sparked across the country, lasting several years. A key source of public discontent during these 
protests was the role of deductions within the TD mechanism. As noted in a 2014 \textit{Taipei Times} article \citep{I-chia:2014}, hundreds of parents and teachers protested, 
arguing that the deduction mechanism unfairly penalized children for setting high goals, with points deducted each time they were rejected by a school on their list.

Protesters voiced concerns that the point deduction system undermined their ability to choose the best educational options for their children. 
Some parents felt the admission process was akin to gambling with their children's futures due to inadequate guidance on filling out school preferences. 
Teachers were also frustrated, with one reporting that six of her ten highest-achieving students failed to gain admission to any school.

These protests led to widespread demands for abolishing the point deduction mechanism and redesigning the entrance process. 
Some even called for accountability and punishment of officials responsible for the flawed system.

In response, Education Minister Chiang Wei-ling and K--12 Education Administration Director General Wu Ching-shan issued formal apologies in July 2014 for the confusion surrounding 
Taiwan’s new high school admissions system \citep{CNA:2014}. 

However, despite the protests, Taiwan continued to use score deductions within the TD mechanism, albeit with softer penalties starting in 2015. For example, Jibei, the largest district, introduced the following deduction rule:
\[ 0,0,0,0,0,1,1,1,1,1,2,2,2,2,2,3,3,3,3,3,4,4,4,4,4,4,4,4,4,4
\]
Under this five-school \textit{grouping system}, no points were deducted from the top five choices, 1 point from choices 6 to 10, 2 points from choices 11 to 15, 
3 points from choices 16 to 20, and 4 points from choices 21 to 30.

Despite the introduction of a more lenient TD mechanism in 2015, protests persisted. The National Education Parents' Alliance, blaming the deduction system for leaving more than 
9,000 of the 230,000 applicants nationwide without school placements, staged demonstrations in front of the Ministry of Education \citep{Yang:2015}. Protesters argued that the five-school 
grouping system was unacceptable and demanded a switch to a ten-school grouping as the bare minimum they could tolerate. They lamented that the system turned the admissions process into a gamble for mid-tier students, 
causing significant anxiety among both students and parents due to deductions tied to preference rankings. Furthermore, they questioned whether the system’s true objective was to financially benefit private schools \citep{Yuan-ling:2015}.

Jibei officials dismissed the parents' demands through a secret ballot, further aggravating the situation, and continued with the five-school grouping.

Taiwan’s experience with school choice provides valuable insights and raises critical questions that could benefit from behavioral economics perspectives.

One pressing question is why officials insisted on retaining the deduction system, even after formally apologizing for it and significantly reducing the penalties.
This persistence, despite potential political costs, may be tied to the concept of a \textit{nudge}, as defined by \cite{Thaler/Sunstein:08}:

\begin{quote} 
``A nudge, as we will use the term, is any aspect of the choice architecture that alters people’s behavior in a predictable way without forbidding any options or significantly changing their economic incentives.'' 
\end{quote}

The deductions in the TD mechanism, in general, do not function purely as a nudge, since they directly affect student options and incentives.
However, suppose a central planner were to describe the Boston mechanism using its TD mechanism formulation with deductions.
If students respond to this formulation with more conservative strategies than they do with the standard formulation, then the deductions, in this case,
would function purely as a nudge.

In this sense, beyond their direct effect on choices and incentives,
the TD mechanism’s deductions also act as a nudge, encouraging families to adopt more conservative school rankings.\footnote{While nudges are typically 
viewed as tools to guide individuals toward beneficial decisions---i.e., paternalistic nudges---in this case, they represent a form of nudging ``for bad,'' as discussed in \cite{Glaeser:06, Thaler:2015, Schmidt/Engelen:2020}.}

But why would officials nudge students toward more conservative strategies? Do they have hidden objectives? Some Jibei parents speculated that the goal was to benefit private schools financially. 
However, I find a more plausible explanation in the argument proposed by one of our Boston College graduate students, Charles Po, a native of Taiwan. He suggests that the hidden objective was 
more likely aimed at discouraging mobility and promoting neighborhood assignments, similar to the Seattle case. If this is true, is Jibei’s modest deduction system based on five-school groupings 
sufficient to achieve this goal, while an even milder deduction system based on ten-school groupings---reflecting parent demand---would fall short?

Normally, one would expect such modest penalties not to significantly change outcomes unless they greatly alter students' preference reporting patterns. 
Given that officials were willing to maintain the modest deductions with the five-school grouping despite the political cost, did they anticipate a considerable shift in preference reporting compared to truthful preference revelation?

Another related question is why the TD mechanism provoked much fiercer opposition, even more so than the disproportionately more punitive Boston mechanism. 
For a possible answer, we can draw insight from \textit{prospect theory} \citep{kahneman/tversky:79}, which suggests that people respond more strongly to 
losses than to gains relative to a reference point. In this context, the reference point is the baseline ranking of students based on the CAE. 
Since deductions apply to all students, depending on how highly a student ranks a school, their relative priorities may improve compared to some students and deteriorate compared to others. 
However, families exhibiting \textit{loss aversion} reacted more strongly to the loss of priority for lower-ranked choices, relative to this reference point, than to gains for higher-ranked choices---especially 
when the loss was explicitly presented as a penalty in the mechanism description. Faced with the need to adjust their preferences under pressure from the system, families felt strong resentment.

The Taiwanese experience shows that when families feel the system punishes students for truthfully reporting their preferences and undermines their hard-earned achievements, it can spark widespread public backlash.

On a broader level, how does the particular formulation of a mechanism affect participant behavior? I would conjecture that if one formulation includes a built-in nudge, 
as in the case of the TD mechanism, and another does not, there may be a significant effect. For example, consider the Boston mechanism historically used in many school districts in the U.S. 
and other countries such as China. As we discussed, although its failures are more severe than those of the TD mechanism adopted in Taiwan, it never generated protests of the magnitude seen in Taiwan. 
Would that still be the case if, instead of its traditional formulation, it were communicated in the form of a TD mechanism with very heavy deduction points?

Along similar lines, consider the following hypothetical laboratory experiment: In a school choice setting similar to that of \cite{chen/sonmez:06}, the Boston mechanism is simulated. 
Unlike in \cite{chen/sonmez:06}, however, participants are divided into two randomized groups. One group receives the standard description of the mechanism, while the other receives a description under the TD formulation. 
Would the preference reporting differ between the two groups? I would expect that subjects in the TD arm would use much more conservative strategies, leading to lower average payoffs for them.

If the subjects' preference reporting differs between the two groups, could certain participants be nudged to submit more conservative strategies simply by providing them with an alternative formulation of the existing mechanism?

\subsection{Framing Institutions: How Descriptions Shape Their Adoption} \label{sec:framing}

Taiwan’s experience with its school choice mechanism illustrates how the framing of an institution can significantly influence both participant behavior and its acceptance by central planners, 
revealing the potential role of behavioral economics while exposing the limitations of traditional \textit{mechanism design} in some practical contexts. In this abstract framework, 
a mechanism consists of a message space for each agent and an outcome function that maps message profiles to outcomes. It assumes that neither the process by which this mapping is 
derived nor how it is communicated affects participant behavior or the mechanism’s appeal to central planners. 
However, Taiwan’s case suggests that these assumptions may not always hold in real-world settings.\footnote{This observation aligns with the recent work of 
\cite{Gonczarowski/Heffetz/Thomas:23, Gonczarowski/Heffetz/Ishai/Thomas:24} and \cite{Katuscak/Kittsteiner:24} on mechanism descriptions that make \textit{strategy-proofness} easier to observe. 
Given a direct mechanism $\phi$, an individual $i$, and the reported types (preferences, in our context) of all other individuals, the menu for $i$ under $\phi$ is 
constructed by applying the mechanism across all possible types for $i$. \cite{Hammond:79} shows that a direct mechanism is \textit{strategy-proof} if and only if each agent always receives 
one of their best outcomes from their menu. Focusing on \textit{strategy-proof} mechanisms, \cite{Gonczarowski/Heffetz/Thomas:23, Gonczarowski/Heffetz/Ishai/Thomas:24} 
and \cite{Katuscak/Kittsteiner:24} suggest that the \textit{strategy-proofness} of a mechanism can be made evident to participants through its menu descriptions, 
with experimental evidence indicating that truthful preference revelation increases under such descriptions. Their work highlights that, irrespective of the mechanism’s specific formulation, 
how it is communicated can significantly influence participant behavior.}

As explored in Section  \ref{sec:Taiwan}, the explicit penalties for lower-ranked choices in the TD mechanism activated strong moral concerns among participants. 
This aligns with growing evidence from behavioral economics, which shows that people respond to the language in which actions are framed, particularly when it engages moral considerations \citep{capraro/halpern/perc:24}. 
The TD mechanism's formulation, with its emphasis on deductions, provoked unprecedented public protests, highlighting how framing can impact perceptions and reactions.

Throughout this monograph, we have observed several instances where one formulation of an institution can trigger moral concerns, while another does not. 
For example, drawing from the Taiwanese experience, the reaction to the Boston mechanism might have been much stronger globally if it had been presented through its TD formulation, 
where the priority ``penalty'' is more pronounced, compared to its traditional version, where this aspect is more subtle.

Another example concerns the minimum guarantee choice rule, as formulated by \cite{hafalir/yenmez/yildirim:13} (see  Section  \ref{sec:subtleties-reserve}), 
versus its alternative formulation based on the ``adjustments'' method used in India to implement horizontal reservations (see  Section \ref{sec:horizontal}). 
Suppose there is a minimum guarantee of 30 positions for women out of 100 positions. In the first formulation, the reserved positions are first allocated to the highest-merit-score women, 
and the remaining positions are then allocated to the highest-merit-score individuals who have not yet received a position. In the second formulation, 
all positions are tentatively allocated to the highest-merit-score individuals in the main phase, regardless of gender, and if the minimum guarantee of 30 is not met, 
necessary adjustments are made by replacing the lowest-merit-score men with the highest-merit-score unassigned women.

Consider the first formulation. As illustrated in Example \ref{ex:precedence} (and depicted in Figure \ref{fig:reservesystem-example}), 
women are disadvantaged in the second batch of positions because their highest-merit-score members have already received assignments, while the highest-merit-score men remain.
However, this disadvantage may not be obvious to the layperson.\footnote{Indeed, 
indicative of a conservative estimate of this oversight, an online experiment by \cite{pathak/rees-jones/sonmez:23} reveals that about 40\% of subjects believe the order in which positions are allocated is irrelevant.} 
In contrast, in the second formulation, it is clear that the benefits for women under this affirmative action policy are conditional, only activating if women do not receive 30 positions without affirmative action. 
Thus, the relative modesty of this policy is more apparent in the second formulation. Therefore, a decision-maker seeking to adopt a more potent policy might be more inclined to 
choose the minimum guarantee choice rule when presented with its original formulation, perhaps overlooking its modesty, than if they were presented with the adjustment-based formulation.

As another example, consider the house allocation model with existing tenants, discussed in Section   \ref{sec:minimalist-YRMH--IGYT}, and the equivalence presented in Theorem  \ref{thm:technocratic}. 
Depending on which of the two formulations of the same mechanism is presented to decision-makers, different reactions may be anticipated. 
One formulation clearly highlights the preferential treatment of newcomers over existing tenants, which may trigger moral concerns, 
while the other obscures this aspect through a core-based technocratic description, potentially avoiding such concerns.

Similarly, as we discussed in Section  \ref{sec:Boston}, when Boston Public Schools (BPS) was evaluating two \textit{strategy-proof} mechanisms---DA and SC--TTC---as potential replacements for the Boston mechanism, 
the Student Assignment Task Force formally recommended SC--TTC. However, BPS leadership opted for DA due to various concerns related to the implicit ``trading of priorities'' carried out under SC--TTC, 
a feature prominently highlighted in \cite{abdulkadiroglu/sonmez:03}. Some of the concerns were that the trading shifted focus to priorities rather than the reasons they are awarded, 
created the possible impression that the mechanism could be manipulated (even though it could not), and led to a lack of \textit{transparency} due to ``behind-the-scenes mechanized trading."

This is not the only case where the notion of ``trade'' triggered moral concerns in our policy interactions. 
When we launched the ``New England Program for Kidney Exchange'' with Dr. Francis Delmonico and his colleagues in the transplantation community in 2004, 
they initially objected to its name because of the word ``exchange.'' They were concerned that our initiative could be associated with a monetized market. 
Though the concept of ``donor exchange'' is originally due to \cite{Rapaport:1986}---the transplant surgeon who first proposed it---over time, 
many in the transplantation community adopted the term ``kidney paired donation" as less loaded terminology.

To summarize, beyond the framing of institutions---whether a mechanism, a choice rule, a set of regulations, or even an ad hoc system---their features, 
and even their names, can shape broader attitudes. How these institutions are presented and communicated can evoke moral concerns or resistance, ultimately influencing their acceptance and effectiveness.

\section{Minimalist Design Across Diverse Domains} \label{sec:additionalapplications}

Sections \ref{sec:HA} through \ref{sec:LE} delineate the progression and refinement of minimalist market design through applications in several matching markets, 
showing how theory and practice have interacted in ``discovery---invention cycles.''
Those sections reflect my own expertise, relying on theory as the primary analytic tool.

Yet the pillars of minimalist market design---respecting institutional missions, avoiding nonessential changes, and embedding a persuasion strategy---are not confined to matching markets or to theory-driven work.
They can also guide promising interventions in other domains and with other methods.

This section presents seven additional minimalist designs---proposed or implemented---developed by other researchers and practitioners that illustrate the framework’s versatility. 
The domains of these reforms range widely: a government auction, the economics job market, online reputation systems, financial exchanges, peer-to-peer platforms, 
permissionless blockchain consensus, and an industrial process in cement production. 
Some were commissioned, others initiated independently; some are centralized, others decentralized. 
Across these diverse settings, designers adopted the same minimalist strategy: identify the root cause of a failure---whether in rules, information channels, incentive alignment, or material inputs---and craft a targeted fix.

Taken together with the cases in Sections \ref{sec:HA} through \ref{sec:LE}, these interventions demonstrate that small, surgical adjustments can yield significant improvements in efficiency, equity, 
incentive compatibility, robustness, and environmental sustainability.

\subsection{Mitigating Tacit Collusion in FCC Spectrum Auctions} \label{sec:FCC}

The FCC’s spectrum auctions---a landmark application that marked the  beginning of market design---have undergone a number of minimalist reforms aimed at mitigating tacit collusion.

Since 1994, the FCC has used \textit{simultaneous ascending auctions} (SAA) to sell spectrum licenses---rights to use particular radio frequencies in specified geographic markets \citep{milgrom:2000, Milgrom:2004, Klemperer:2004}. 
Bidding proceeds in rounds: in each, bidders may raise the price on any license by at least a set increment. The auction ends when a round concludes with no new bids on any license, 
at which point the standing high bidders win and pay their final bid amounts.

To ensure transparency, the FCC initially disclosed after each round all bids, bidder identities, and changes in eligibility. 
Bidders could submit any amount above the minimum or withdraw bids without limit, enabling them to relinquish licenses when complementary ones became unattainable, 
thereby mitigating the ``exposure problem'' and increasing efficiency.\footnote{If the final price of a license ended up lower than a withdrawn bid, 
the withdrawing bidder was charged a penalty equal to the difference between the withdrawn bid and the final price.}

While these design choices enhanced transparency and efficiency, they also created avenues for tacit collusion. 
Three features proved especially problematic: (i) unrestricted manual bid entry, which allowed coded signaling; (ii) unlimited bid withdrawals, which could be used to punish rivals; 
and (iii) full disclosure of bidder identities after each round, which enabled retaliation.

These vulnerabilities were not merely theoretical; they manifested in actual bidder behavior.
For example, analyses of Auction 4 (Broadband PCS A \& B Block Auction) and Auction 11 (Broadband PCS D, E \& F Block Auction) document widespread coded signaling via the trailing digits of bids \citep{Weber:1997, Cramton/Schwartz:2002}.
Bidders encoded messages such as desired license numbers in the last three or four digits, used telephone-keypad encodings to spell company names, adopted firm-specific signatures,
inserted actual market numbers or recognizable phone numbers, and coordinated market divisions or threatened retaliation.
Some bidders cross-referenced contested markets with “punishment” markets, and when identities were public, retaliatory bidding often required no special digits at all.

Auction 11 marked a turning point, as the scale of coded trailing digits, bid withdrawals, and retaliatory bidding made it impossible for the FCC to overlook these vulnerabilities \citep{Cramton/Schwartz:2002}.
The gravity of the problem was further underscored in November 1998, when the Department of Justice filed civil antitrust suits against three companies, alleging bid coordination through coded bids.
All three cases quickly resulted in consent decrees.\footnote{\emph{United States v. Omnipoint Corp.}, No. 1:98CV02750 (D.D.C. Nov. 10, 1998); 
\emph{United States v. Mercury PCS II, L.L.C.}, No. 1:98CV02751 (D.D.C. Nov. 10, 1998); \emph{United States v. 21st Century Bidding Corp.}, No. 1:98CV02752 (D.D.C. Nov. 10, 1998).}

In response---and consistent with the minimalist ethos---the FCC targeted two root causes of tacit collusion while preserving the core SAA design, beginning with Auction 16 (800 MHz SMR Auction):
\begin{enumerate}
\item \textit{Click-box bidding.} The FCC replaced manual entry with menu-based ``click-box'' bids. 
Bidders were required to choose from discrete options---typically one to nine FCC-specified increments above the standing high bid---thereby eliminating the channel for coded signaling.
\item \textit{Limits on withdrawals.} Each bidder was permitted to withdraw bids in at most two \textit{rounds} over the course of the auction, 
directly constraining the retaliatory strategies observed in earlier auctions.
\end{enumerate}

For more than a decade, the FCC continued full information disclosure, publicly posting after each round all bids on each license, bidder identities, and eligibility changes. 
While this policy advanced transparency---a central FCC objective---it also facilitated retaliation \citep{mcmillan:94, Cramton:1997, Cramton/Schwartz:2002, klemperer:02, Marshall/Marx:2009}.

Beginning with Auction 73 (700 MHz Band Auction), the FCC introduced a further minimalist reform, 
this time targeting the third root cause of tacit collusion while preserving the core SAA design \citep{Bajari/Yeo:2009}:
\begin{itemize}
\item[3.] \textit{Anonymous bidding.} The FCC replaced full disclosure with anonymous bidding: after each round, 
only the standing high bid for each license was released, while bidder identities, competing bids, and eligibility information were withheld until the auction concluded. 
This eliminated a key channel for retaliation while maintaining transparency about prices.
\end{itemize}

These reforms exemplify minimalist market design: targeted rule changes that eliminated several avenues for collusion while preserving the core institution. 
Open, multi-round bidding---and its informational benefits---remained intact. In short, the FCC corrected key flaws, substantially curbing tacit coordination without undermining the SAA framework.

\subsection{Signaling for Interviews in the AEA Job Market} \label{sec:AEA}

The economics job market for new Ph.D.s---a large, decentralized matching process coordinated through the annual 
\textit{Allied Social Science Associations} (ASSA) meetings---faced growing coordination frictions by the mid-2000s  \citep{Coles-etal:2010}.
With thousands of applications per candidate cohort and hundreds per employer, screening became congested, and many employers refrained from interviewing highly qualified 
candidates they feared were beyond their reach. As a result, promising matches were often overlooked because interest could not be credibly conveyed.

In response to these market frictions---and aligned with the minimalist ethos---the \textit{American Economic Association} (AEA) convened an Ad Hoc Committee on the Job Market in 2005, 
chaired by Alvin Roth and including John Cawley, Peter Coles, Phillip Levine, Muriel Niederle, and John Siegfried. Beginning with the 2006--07 job market, 
the AEA adopted a minimalist reform recommended by the committee: a \textit{signaling mechanism}.\footnote{\cite{Avery/Levin:2010} and \cite{Coles/Kushnir/Niederle:2013} 
provide theoretical analyses of the potential gains from signaling in the context of matching markets.}

This reform directly targeted the root cause of a coordination failure---employers’ inability to distinguish routine applications from genuine interest---while preserving the market’s decentralized structure.
Under the signaling mechanism, each candidate was permitted to send up to two official signals of special interest to employers. 
The strict cap created scarcity and credibility: a signal conveyed real opportunity cost, while its absence carried no negative inference. 
Crucially, the broader process remained untouched.\footnote{Strictly speaking, through a second minimalist reform, the committee also introduced an aftermarket---the ``scramble''---between unmatched candidates 
and employers with unfilled positions \citep{Coles-etal:2010}.} Candidates continued to apply widely, departments retained full discretion in interviews and offers, and the market itself remained decentralized.

Early evidence showed that the reform was both effective and non-disruptive \citep{Coles-etal:2010}. More than half the candidate pool sent signals in each of the first three years. 
Employers took them seriously: candidates who signaled a department were significantly more likely to secure an interview there than similar candidates who did not. 
The mechanism alleviated the ``out-of-reach candidate'' problem by allowing applicants to credibly convey special interest, reduced redundant interviews, and leveled the playing field for candidates from less visible programs.

This intervention is an example of minimalist market design. The AEA’s job market committee introduced a surgical fix: a simple, voluntary channel for credible communication 
that improved coordination while preserving the core institution---while at the same time reducing missed matches and congestion. In short, the AEA fixed only what was broken, retaining the existing system while alleviating a key inefficiency.

\subsection{Rebate-for-Feedback Mechanism on Taobao} \label{sec:Taobao}

Reputation systems are central to trust in online marketplaces, yet they continue to suffer from well-documented shortcomings, 
including the underprovision of feedback and ratings inflation \citep{Resnick/Zeckhauser:2002, Bolton/Katok/Ockenfels:2004, Cabral/Hortacsu:2010}. 
For example, on platforms like \textit{eBay} only about half of transactions generate any review, and the vast majority of those are positive, 
leaving the reputation system largely uninformative. Buyers often refrain from leaving negative reviews for fear of retaliation, while sellers struggle with a “cold-start” barrier when introducing new products without prior feedback.

To address these issues, \cite{Li:2010} and \cite{Li/Xiao:2014} proposed a minimalist fix: a \textit{Rebate-for-Feedback} (RFF) mechanism.
Sellers could voluntarily offer rebates to buyers who left reviews. Importantly, the rebate was tied to the act of leaving ``informative'' feedback, not to whether the review was positive.
This design tackles both underprovision of feedback and ratings inflation: buyers have an incentive to report candidly, and sellers who expect satisfied customers are willing to pay for feedback as a credible signal of product quality.
In equilibrium, the mechanism allows good sellers to separate themselves from bad ones while leaving the decentralized marketplace structure intact.

Informed by \cite{Li:2010, Li/Xiao:2014} and with the direct guidance of Linfang Li---an ``outsider'' critic of its feedback system---\textit{Taobao}, 
Alibaba’s online marketplace and the world’s largest of its kind, put RFF into practice by launching an official program in March 2012 \citep{Li/Tadelis/Zhou:2020}.\footnote{In Footnote 7, 
\cite{Li/Tadelis/Zhou:2020} report that ``Li suggested the RFF mechanism to Alibaba Research toward the end of 2011, and several months later, Taobao launched the RFF mechanism.''}

Under this system, sellers could attach a rebate (partial refund or store coupon) to any product listing, awarded to buyers who returned to leave a review after purchase.
Taobao mediated the process: once a review was submitted and classified as ``informative,'' the platform automatically transferred the rebate from seller to buyer.
This guaranteed that buyers were rewarded even for negative evaluations, preserving the credibility of the signal.

This minimalist tweak aligned incentives without altering the decentralized nature of the marketplace.
Sellers gained a credible tool to build reputation, buyers were rewarded for candid reviews, and the feedback system became more detailed and informative---all without platform micromanagement.

Evidence from \cite{Li/Tadelis/Zhou:2020} shows that Taobao’s RFF mechanism worked as predicted.
Sellers targeted rebates toward ``cold-start'' items lacking prior feedback.
Buyers responded strongly: products with RFF saw sales rise by about 36\%.
Reviews also became longer and more detailed, without biasing ratings toward the positive.

Possibly contributing to an outsider critic’s success in convincing a major company---\textit{Taobao}---to reform its reputation system, the RFF mechanism exemplifies minimalist market design.
By targeting a specific incentive misalignment, it solved the problem with a simple, voluntary mechanism while preserving the core marketplace structure.
Sellers gained a credible tool to signal quality, buyers were rewarded for candid reviews, and the reputation system became more informative and trustworthy.

\subsection{Frequent Batch Auctions for High-Frequency Trading} \label{sec:FBA}

Modern electronic exchanges use the \textit{continuous limit order book} (CLOB) to match buy and sell orders in real time.
In this system, orders are ranked by price and then by arrival time, so even a millisecond advantage can secure priority. In the 2000s and 2010s, this feature sparked a \textit{high-frequency trading} (HFT) ``arms race,'' 
as firms invested heavily in progressively faster technology to capture these razor-thin advantages \citep{Budish/Cramton/Shim:2014, Budish/Cramton/Shim:2015}. 

Evidence shows that continuous trading at millisecond scales creates predictable arbitrage opportunities that the fastest traders systematically exploit.
For example, when news moves the price of one asset, related markets do not adjust instantly; in that brief gap, algorithms can profitably trade against stale quotes before market-makers can react.
More competition among HFT firms did not eliminate these profits---it only raised the speed threshold for capturing them, fueling a costly arms race with little social value.
In short, the continuous-time market design itself was enabling a never-ending speed race at the expense of market liquidity and fairness \citep{Budish/Cramton/Shim:2015}.  

To address this flaw, \cite{Budish/Cramton/Shim:2015} proposed a minimalist reform: replacing continuous trading with \textit{frequent batch auctions}.
The idea is simple: time becomes discrete rather than continuous. Instead of executing orders one by one, the exchange collects all orders over a short interval (e.g., 100 milliseconds to 1 second) 
and clears them in a uniform-price double auction, which sets the single market-clearing price that balances supply and demand. 
Within the same interval, arriving a few milliseconds earlier provides no advantage. Speed competition disappears, giving way to competition on price and liquidity provision.

In theory, shifting to frequent batch auctions would thus render latency-arbitrage strategies ineffective and remove the incentive for socially wasteful investment in millisecond-optimized trading technology.
Traders could quote with less fear of getting ``sniped,'' and overall market efficiency would improve, all without slowing trade beyond a fraction of a second. 
Crucially, the open, competitive nature of the marketplace is preserved: buyers and sellers still meet in a continuous flow of trading opportunities, with only the market’s timing mechanism altered.

The \cite{Budish/Cramton/Shim:2015} proposal of frequent batch auctions is a clear example of minimalist market design.
It targets a specific failure---the incentive to race for speed---while preserving the core institution of continuous trading.
Traders would still meet through a transparent order book and price discovery process, but updates would arrive in small periodic bursts rather than a perfectly continuous stream.

Although major exchanges have not yet adopted frequent batch auctions,\footnote{The proposal entered the regulatory conversation in 2016, 
when Eric Budish submitted a comment letter to the U.S. Securities and Exchange Commission (SEC) (File No. 10-222), 
urging regulators to consider frequent batch auctions as a way to improve fairness and efficiency by eliminating the incentive for costly speed races \citep{Budish:SEC2016}.} related ideas have been tested in practice.
For example, the \textit{U.S. Investors Exchange} (IEX) introduced a 350-microsecond “speed bump” to blunt fast sniping on its venue \citep{Hu:2019}.
Likewise, \textit{Turquoise}---a \textit{London Stock Exchange} platform---runs random midpoint call auctions every few seconds to reduce HFT advantages \citep{Rosov:2014}.
These incremental steps echo the spirit of the \cite{Budish/Cramton/Shim:2015} proposal, reflecting a broader recognition of the problems created by speed-based competition.

\subsection{Anti-Discrimination Measures on Airbnb and Uber} \label{sec:Airbnb-Uber}

\textit{Airbnb} and \textit{Uber} each faced evidence of racial bias in their marketplaces. In both cases, reforms followed the minimalist ethos: targeted interface modifications to suppress discriminatory channels while leaving the broader market institutions intact.  

On Airbnb, discrimination first came to light through a randomized field experiment. 
In 2016, a Harvard Business School working paper---later published in \cite{Edelman/Luca/Svirsky:2017}---showed that booking requests from African American--sounding names were about 
16\% less likely to be accepted than identical requests with white-sounding names. 
Even before its formal release, the findings were first publicized in December 2015 across several major media and university outlets, 
putting Airbnb under intense public pressure.  

Importantly, as ``outsider'' critics, \cite{Edelman/Luca/Svirsky:2017} not only documented discrimination but also offered concrete minimalist design recommendations to reduce opportunities for bias without altering the core marketplace. 
They argued that platforms control which information is transmitted between parties and can therefore suppress identity cues that facilitate discrimination. 
Specifically, they proposed that Airbnb (i) \textit{conceal guest names} at the booking stage, just as it already withheld phone numbers and email addresses; 
and (ii) \textit{expand the ``Instant Book'' option}, under which hosts accept reservations without screening guests in advance.  

Airbnb’s policy response closely followed these suggestions. In September 2016, the platform introduced a ``Community Commitment'' 
requiring all users to pledge nondiscrimination, paired with anti-bias training and an expansion of ``Instant Book'' \citep{Kamen:2016}.  
Two years later, in October 2018, Airbnb announced that guest profile photos would no longer be displayed until after booking confirmation \citep{Airbnb:2018}. 
Help Center documentation clarifies that, at the request stage, hosts now see only a guest’s first name, review history, and verification status, 
with full names and photos revealed only once a booking is confirmed \citep{AirbnbHelp:2025b, AirbnbHelp:2025a}.  
These changes directly reflected the \cite{Edelman/Luca/Svirsky:2017} recommendations to suppress discriminatory identity cues while preserving the 
request--accept workflow.\footnote{Subsequent research reinforced this reformed design. 
\cite{Cui/Li/Zhang:2020} showed that even a single positive review on a guest profile was sufficient to close the racial acceptance gap, 
highlighting the value of directing host attention toward reputation rather than names or photos. 
Airbnb’s reforms thus aligned closely with both the original policy recommendations and subsequent evidence: minimalist interface changes that left the market institution intact while reducing discrimination.}  

Uber faced a parallel challenge. In a randomized field experiment, \cite{Ge/Knittel/MacKenzie/Zoepf:2020} showed that drivers were more likely to cancel rides for passengers with African American--sounding names, leading to longer wait times.  
This revealed a structural vulnerability in Uber’s marketplace: although rider photos and full names were initially withheld, name-based cues were disclosed after acceptance, enabling cancellations.  
As ``outsider'' critics, \cite{Ge/Knittel/MacKenzie/Zoepf:2020} not only documented discrimination but also offered a minimalist policy recommendation: 
platforms can reduce bias by controlling what information is revealed to drivers before pickup.  

Motivated by these findings, and in collaboration with Uber’s research team, \cite{Knittel-etal:2024} conducted a randomized controlled experiment of minimalist interface interventions.  
Two tweaks were tested: (i) enlarging the font of the passenger’s star rating on the request screen to make reliability signals more salient; and (ii) revealing the rider’s name earlier, rather than only after acceptance.  
The results showed that emphasizing ratings eliminated the racial cancellation gap, whereas revealing names earlier had no effect.  
Uber thus addressed bias with nothing more than a font-size change---while leaving its core matching and pricing systems intact.\footnote{The interventions were implemented directly 
by Uber’s research team in a live environment but, as of September 2025, it is not publicly documented whether the larger font-size display of rider ratings has been retained on the platform.}  

Both cases exemplify minimalist market design: Airbnb reduced discrimination by suppressing identity cues, while Uber did so by amplifying reputation signals. 
In each case, modest, surgical interventions improved fairness while leaving the broader institutions intact.

\subsection{Stubborn Nakamoto for Permissionless Consensus} \label{sec:StubbornNakamoto}

\textit{Nakamoto consensus}, first implemented in Bitcoin, is the protocol that maintains a permissionless, decentralized ledger without trusted authorities or barriers to entry \citep{nakamoto:2008}.
 Its purpose is to provide two properties essential to a payment system:
\emph{consistency}, meaning that all honest nodes agree on a single history of transactions, which in turn ensures that payments cannot be reversed or double-spent; and
\emph{liveness}, meaning that whenever a client submits a transaction, the transaction is eventually incorporated into the log. 

The design works as follows: transactions are broadcast to a public pool of pending records, known as the \textit{mempool}---the collection of valid but unconfirmed transactions awaiting inclusion in a block.
Specialized participants called \textit{miners} gather transactions from this pool into blocks. To add a block, they must solve a \textit{proof-of-work puzzle}, 
a cryptographic challenge that requires significant computational effort and secures the system against tampering. Each block references the one before it, forming a chain.
Meanwhile, \textit{nodes}---computers running the blockchain software, which may operate as miners, validators, or simply record-keepers---verify that blocks and transactions are valid and collectively maintain the ledger.
By convention, they adopt the \textit{longest valid chain} (the one with the most accumulated proof-of-work) as the authoritative record.
Rewards flow to miners who successfully extend this chain, aligning incentives with the continued growth and security of the ledger.\footnote{\cite{Budish:2025} provides a formal economic analysis of 
Nakamoto consensus---examining incentives, rewards, and attack risks---while also offering an accessible introduction for readers unfamiliar with computer science notions.}

Occasionally two miners successfully extend the chain at almost the same time, producing a temporary ``fork'' in which different parts of the network see different continuations of the chain.
Initially these forks are equal in length, but once one branch is extended the protocol instructs all nodes to treat it as the valid history.
Within Nakamoto consensus, nodes are often classified by behavior: \textit{honest nodes} follow the protocol exactly, while \textit{adversarial nodes} may deviate in pursuit of strategic advantage or to exploit the system.
Since honest miners and validators alike adopt the longest valid chain by definition, the protocol resolves routine forks automatically and without coordination.\footnote{Transactions contained only in the discarded branch are not lost. 
If they remain valid, they typically return to the mempool and may be included in later blocks on the winning chain.}

But not all forks are innocuous. The very mechanism that smooths over harmless ones also exposes a deeper vulnerability.
If adversaries command a majority of the network’s computational power---a so-called 51\% attack---they can consistently outpace the honest chain, reverse transactions, and undermine the system’s consistency. 
As \cite{Budish:2025} emphasizes, this possibility is the ``Achilles heel of all forms of permissionless consensus,'' since the protocol offers no internal defense once an adversary gains majority control.\footnote{Several proof-of-work 
cryptocurrencies with relatively low hash rates have experienced successful 51\% attacks. Documented cases include Ethereum Classic in 2019 and 2020, Bitcoin Gold in 2018 and 2020, and Vertcoin in 2018, 
where attackers were able to reorganize the blockchain and execute large double-spending attacks against exchanges. 
As the MIT Digital Currency Initiative notes, such attacks are feasible on smaller networks because renting sufficient hash power on the open market is relatively inexpensive \citep{MITDCI:51percent}. 
By contrast, Bitcoin itself has not suffered a 51\% attack, reflecting the far higher cost of amassing majority control over its mining power, 
though mining pool concentration has occasionally raised concerns---for example, when Ghash.io briefly exceeded 50\% of Bitcoin’s total hash rate in 2014.}
Even short of that threshold, nodes cannot tell whether a longer chain reflects ordinary network variation or the deliberate withholding of blocks by an attacker---whether to reverse transactions, 
to orphan honest blocks through \textit{selfish mining} strategies,\footnote{\cite{Eyal/Sirer:2018} introduced the concept of selfish mining, showing that even miners with less than 
majority power can strategically withhold and release blocks to orphan honest work and gain disproportionate rewards. This illustrates that block withholding can be profitable even without double-spending, 
underscoring Nakamoto consensus’s broader vulnerability.} or to censor targeted transactions by discarding any block that contains them.
Because block finality is only probabilistic, an attacker can rent hash power, mine an alternative history in secret, and later release it to erase past transactions.

The incentive structure compounds the problem. Miners are rewarded for producing valid-looking blocks whether they extend an honest chain or an adversarial one. 
Each miner thus has reason to adopt whichever chain appears longer, even if this helps an attacker, while the costs of attacks---price declines, wasted effort, and lost trust---are borne collectively. 
This is a coordination failure: all honest miners would be better off rejecting adversarial forks, but the protocol gives them no way to coordinate on that response.

These vulnerabilities stem from two root causes: \textit{indistinguishability} and \textit{ambiguous finality}, since nodes adopt the longest chain with no internal test for adversarial forks; 
and \textit{undifferentiated incentives}, since all valid-looking work is rewarded equally, leaving no mechanism to deter attacks. 

This key insight underlies a minimalist reform proposal by \cite{Leshno/Pass/Shi:2025}, who introduce \textit{Stubborn Nakamoto}, 
an amendment of the original protocol that accepts a small cost in liveness---short, rule-based pauses during attacks---in order to strengthen consistency and eliminate the profitability of double-spends. 
Rather than promising progress at all times, the protocol behaves like Nakamoto’s original under normal conditions but activates 
temporary halts as a circuit breaker precisely when uninterrupted responsiveness cannot be guaranteed.

To operationalize this idea, Stubborn Nakamoto targets only these root causes through two precise adjustments to finalization, while leaving proof-of-work, anonymity, open entry, 
and the blockchain structure intact. Under ordinary conditions, it behaves just like the original protocol; the modifications activate only when conflict arises:
\begin{enumerate}
\item \emph{Explicit, staged finality.} Blocks become “certified” once they are k-deep in the chain, meaning they have been extended by several subsequent blocks 
(often described as being ``buried under'' others). Before treating such a block as final, nodes pause briefly to check for conflicts. 
If no conflicting block appears during this pause, the block is finalized permanently and cannot be reversed.

\item \emph{Strategic waiting and stubborn rejection.} The waiting period turns competing branches into a detectable signal. If a longer conflicting chain appears after the pause, 
it must be adversarial.\footnote{The pause is chosen to exceed the maximum time needed for honest blocks to propagate. 
If two honest miners generate blocks at the same height, all honest nodes will see both during the waiting window. 
Thus, a conflicting chain that only appears after the pause must have been withheld deliberately and is therefore treated as adversarial.} 
Under the protocol’s rules, all honest nodes then reject it stubbornly, even if longer, rather than blindly following it.
\end{enumerate}

These small changes internalize what in the original protocol required messy, ex post ``social forks.'' 
In Nakamoto’s design, recovering from a major attack depended on the community coordinating outside the protocol to reject the attacker’s chain.
 Stubborn Nakamoto instead builds this coordination directly into the rules: if one honest node rejects an attack, every honest node does so as well, automatically and together. 
 The effect is a fundamental shift in failure modes. 
 
 Standard Nakamoto risked \textit{consistency failure}, where transactions thought final could later be undone. 
 Stubborn Nakamoto trades this risk for rare, temporary delays in finalization, akin to circuit breakers in financial markets. 
 Liveness is preserved---transactions are eventually confirmed---but finalization halts until the threat passes, ensuring that no false history is ever accepted. 
 Restarting requires only a modest ``recovery oracle,'' the state of one honest node, so no new governance layer is needed.

The incentive effects are potentially profound. The protocol is designed to prevent adversarial branches from being adopted, making it much harder for attackers to profit by renting hash power or bribing miners. 
Under these rules, standard double-spend strategies become far less attractive, since withheld chains will not be accepted once revealed. 
This improvement is achieved without altering issuance, difficulty adjustment, or entry rules. Although no protocol can eliminate every possible disruption, 
the reform substantially raises the cost of attacks and shifts incentives back toward honest participation.

Viewed through the lens of minimalist market design, Stubborn Nakamoto preserves the original mission by keeping participation open and competition free, 
targets only the root causes of failure---ambiguous finality and incentive misalignment---while leaving the rest of the institution untouched, and, finally, embeds persuasion in the protocol itself, 
since the labels and waiting period create common knowledge that renders the honest response focal and automatic. 
In short, the reform fixes only what is broken: finality becomes explicit and conflict-responsive, honest nodes act as one against attacks, and the openness and competition that give Nakamoto consensus its value remain intact. 
This is the essence of minimalist reform---small, surgical adjustments that yield major improvements while conserving the institution’s core mission.

\subsection{Decarbonizing Cement through Mineral Substitution} \label{sec:cement}

\textit{Portland cement}, the world’s most widely used cement, is one of the largest industrial sources of carbon dioxide (CO$_2$), 
responsible for roughly 7--8\% of global emissions \citep{Barbhuiya/Kanavaris/Das/Idrees:2024}.
These emissions arise primarily from the industry’s reliance on \textit{limestone} (calcium carbonate, CaCO$_3$), the principal raw material in its production.
When limestone is heated, it decomposes into \textit{lime} (CaO) and CO$_2$: in Portland cement production, lime is essential, 
while CO$_2$ is released inevitably as a byproduct.\footnote{\textit{Clinker} is an intermediate material produced in rotary kilns by heating lime (CaO) from limestone together with other minerals. 
Portland cement is then made by grinding clinker and blending it with small amounts of other materials.}
This single reaction accounts for roughly 50--60\% of the sector’s total carbon footprint \citep{Benhelal/Shamsaei/Rashid:2021, WEF:2024}.\footnote{The remainder 
comes mainly from burning coal or natural gas to achieve kiln temperatures of up to \SI{1500}{\celsius}, with smaller contributions from grinding and transport.}

Because these emissions result from \textit{carbonate} decomposition,\footnote{A \textit{carbonate} is a compound containing the chemical group CO$_3$.} they remain even if cement plants use only carbon-free fuels.
A central challenge for climate policy, therefore, lies in the reliance on limestone: heating CaCO$_3$ inevitably releases carbon in the form of CO$_2$. Process emissions alone thus account for more than half of cement’s footprint.

A potential minimalist solution targets this root cause: rather than reinventing the entire cement-making process, 
it replaces the carbon-bearing input with a carbon-free source of lime.
In this spirit, to produce lime without releasing CO$_2$, 
\cite{Atakan:2024} and \cite{Faucher-etal:2025} independently propose substituting limestone with minerals from the \textit{calcium sulfate} family, 
such as \textit{gypsum} (CaSO$_4\cdot$2H$_2$O) or \textit{anhydrite} (CaSO$_4$).
Unlike limestone, these minerals contain no carbon; when heated, they yield CaO---\textit{lime}, 
essential for cement production---and \textit{sulfur oxides} (typically SO$_2$ or SO$_3$), which can be captured and converted into \textit{sulfuric acid} (H$_2$SO$_4$).
The fix is surgical: substitute a carbon-bearing mineral with a carbon-free one that still yields lime.

While the emissions-reduction rationale for using calcium sulfate instead of limestone is novel in \cite{Atakan:2024} and \cite{Faucher-etal:2025}, the industrial concept itself is not.
The \textit{M\"{u}ller--K\"{u}hne} process, developed in Germany in the 1910s, substituted anhydrite for limestone in cement production and was utilized for decades at commercial scale in 
Germany, France, Austria, South Africa, Poland, and the UK \citep{Ribas:1991, Faucher-etal:2025}.
Interestingly, its primary motivation was to produce sulfuric acid, with cement production serving only as a secondary output.
The process was abandoned in the 1970s, when cheap sulfur from oil refining undercut the value of acid output, eroding its economic advantage \citep{Faucher-etal:2025}.
This history demonstrates that existing high-temperature processing technology, supply chains, and product quality can all be maintained using this alternative input.

With the rising burden of CO$_2$ emissions---economic, environmental, and regulatory---since the 1970s, \cite{Atakan:2024} and \cite{Faucher-etal:2025} argue that the cost--benefit balance has shifted.
Today, unlike in that era, the central consideration is not the economics of acid production but the feasibility of supplying low-carbon cement at scale.
They show that replacing limestone with gypsum can cut process emissions by more than half while also producing sulfuric acid.
Gypsum is abundant and already widely used in cement plants, making the process largely compatible with existing infrastructure.
The main challenge lies in scaling supply and securing market uptake.

This reform exemplifies the minimalist design ethos applied to an \textit{industrial process}: it preserves the fundamental structure of existing cement production while altering only the raw material inputs.
No new heating technology, no novel chemistry, no disruption to supply chains---yet it achieves substantial reductions in process emissions.
The intervention addresses a major root cause while leaving the broader production system intact.

\section{Epilogue} \label{sec:epilogue}

Economic designers employ a range of objectives and analytical frameworks for their policy-oriented projects, typically selecting the framework based on the problem at hand. In this monograph, 
I introduce minimalist market design as a new framework that complements mainstream approaches, adding to the arsenal of practically oriented economic designers.
My main thesis is that by adopting minimalist market design---even as an outsider critic---an economic designer can build a persuasive case for change grounded in practical, 
use-inspired basic research, thereby increasing the chances of adoption and implementation by decision-makers.

In Sections \ref{sec:HA} through \ref{sec:LE}, I present a series of applications and case studies executed through minimalist market design, 
several of which have played significant roles in the success of the field. For instance, minimalist market design has enabled two of the most prominent applications in market design: 
school choice and kidney exchange. It has facilitated numerous policy reforms and has even provided real-time benefits during the COVID-19 pandemic, 
highlighting the framework's potential to be deployed swiftly to address a range of societal challenges. Moreover, it has received external validity from 
England and Chicago in the context of school choice, and from India in Supreme Court judgments on affirmative action. 
These examples illustrate its potential as a reliable framework for uncovering intended institutions that failed to materialize, often due to a lack of formal expertise among their designers.

These successes have been achieved by a small group of economic designers. Now imagine the possibilities if the approach were systematically embraced across the field.

\begin{figure}[!htbp]
    \begin{center}
      \vspace{-10mm}
    \includegraphics[scale=0.82]{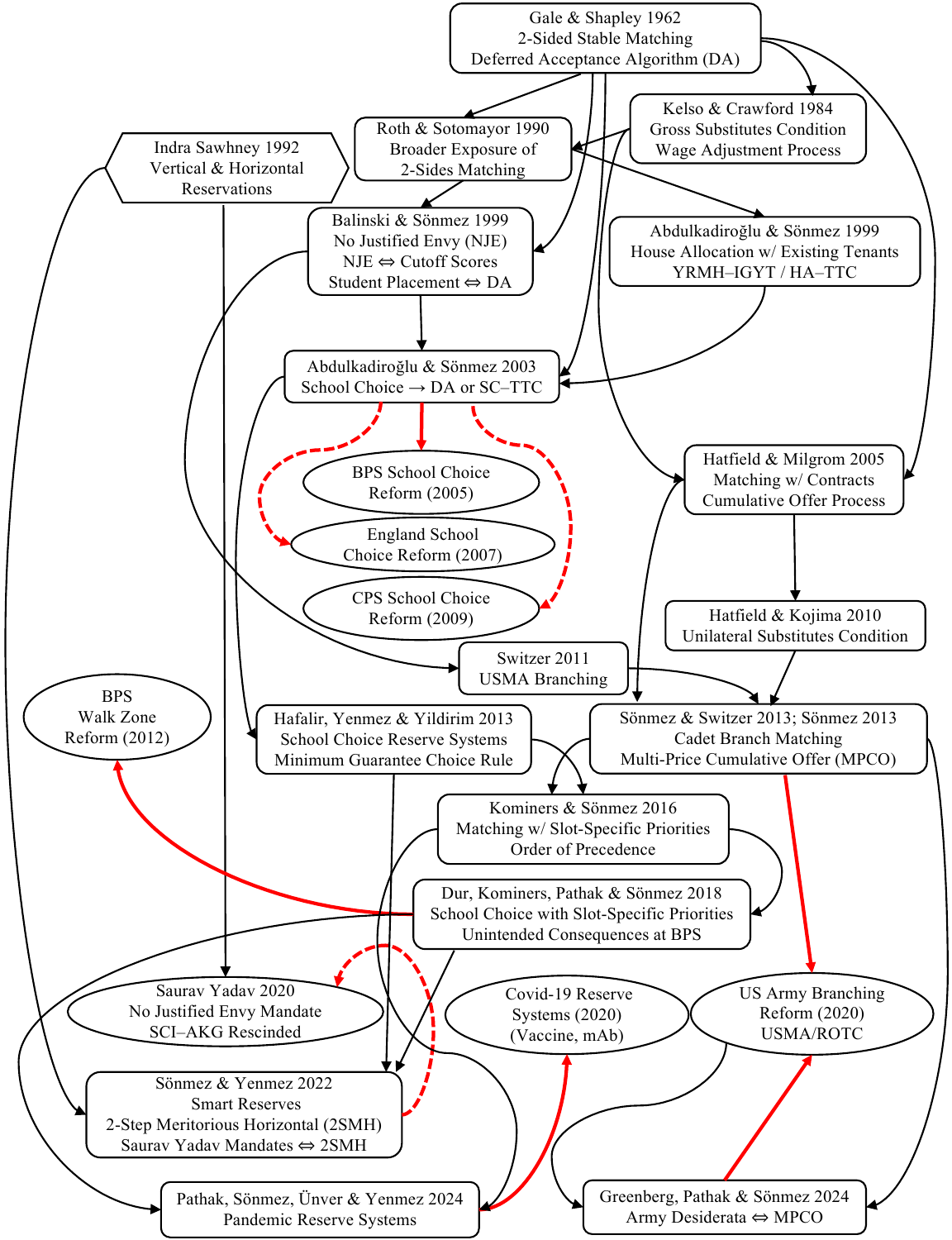}
    \end{center}
      \captionsetup{skip=-3pt}
\caption{Discovery---Invention Cycles linking innovations and policy events in Sections \ref{sec:schoolchoice}, \ref{sec:Army}, \ref{sec:reserve-systems}, \ref{sec:India} and \ref{sec:pandemic}. 
Red arrows indicate direct policy impact; dashed red arrows indicate external validity.
Abbreviations: YRMH--IGYT: You Request My House, I Get Your Turn; HA--TTC: House Allocation-Top Trading Cycles; SC--TTC: School Choice-Top Trading Cycles; 
BPS/CPS: Boston and Chicago Public Schools; mAb: monoclonal antibody; USMA: United States Military Academy; ROTC: Reserve Officers’ Training Corps.}\label{fig:DIC-SC-arm} 
\end{figure}

\begin{figure}[!htbp]
    \begin{center}
    \includegraphics[scale=0.85]{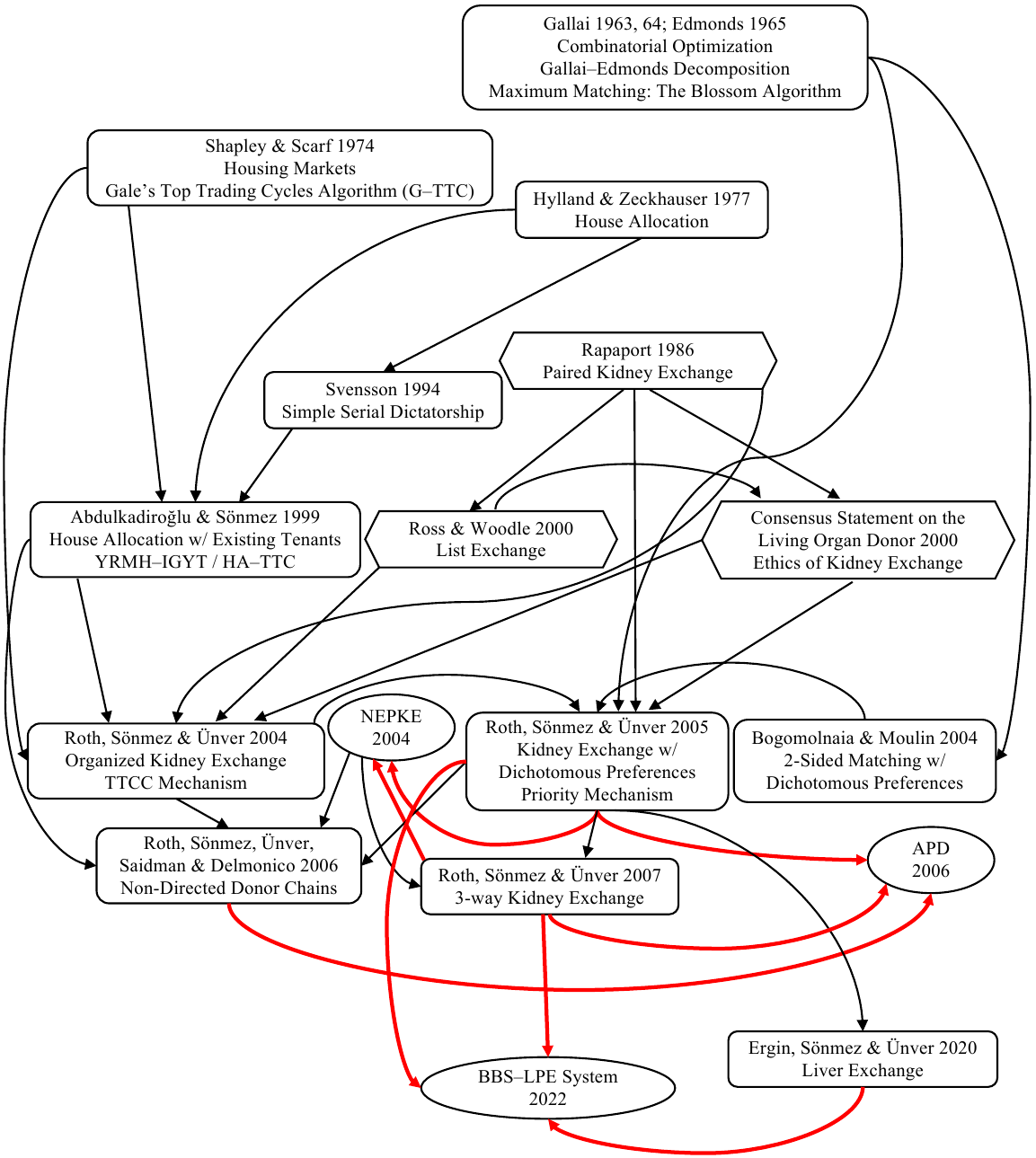}
    \end{center}
    \caption{Discovery---Invention Cycles linking innovations and policy events in Sections \ref{sec:HA}, \ref{sec:KE} and \ref{sec:LE}. 
Red arrows indicate direct policy.
Abbreviations: YRMH--IGYT: You Request My House, I Get Your Turn; HA--TTC: House Allocation-Top Trading Cycles; TTCC: Top Trading Cycles and Chains; 
NEPKE: New England Program for Kidney Exchange; APD: Alliance for Paired Donation; BBS--LPE: Banu Bedestenci S\"{o}nmez--Liver Paired Exchange.} \label{fig:DIC-KE-arm} 
\end{figure}

In detailing the applications in Sections \ref{sec:HA} through \ref{sec:LE}, I present a comprehensive timeline of the intellectual evolution of the innovations in these 
sections---research discoveries and practical inventions---highlighting how they continually feed into one another through ``discovery---invention cycles.'' 
Figures \ref{fig:DIC-SC-arm} and \ref{fig:DIC-KE-arm}  illustrate how the flow between theory and application is reciprocal---at times, discoveries inspire practical designs, while at other times,
practical challenges motivate new theoretical advances. Throughout, I emphasize the intuitions and connections linking the key inventions.

This structure serves two purposes. 
First, while the monograph is not intended as a technical book on matching markets, it can serve as a resource for those interested in a semi-formal exposition of several key applications. 
Second, the success and evolution of these applications lend strong support to the merits of a holistic approach to basic and practice-oriented research, 
as advocated by \cite{Stokes:1997} through use-inspired basic research in \textit{Pasteur's Quadrant} and by \cite{Narayanamurti/Odumosu:2016} through \textit{Discovery--Invention Cycles}. 
In particular, minimalist market design emerges as a natural framework to conduct such research.

Building on three pillars---respecting institutional missions, avoiding nonessential changes, and embedding a persuasion strategy---this monograph presents an idealized version of minimalist market design, 
with the exception of Sections \ref{sec:behavioral} and \ref{sec:additionalapplications}.
To implement these pillars, I rely on theory as the primary formal tool. Influenced in part by my training under William Thomson, 
I employ the axiomatic methodology in many applications to define institutional missions, diagnose shortcomings, and propose solutions.

At the same time, theory in general---and the axiomatic methodology in particular---is only one way to pursue minimalist market design. 
Other approaches, such as behavioral, computational, and empirical methods, can also be fruitfully employed to address institutional questions and advance the three pillars. 
Accordingly, while this monograph emphasizes a theoretical path, many of its insights may prove useful to policy-oriented economists who adopt different analytical tools.

In this spirit, Section \ref{sec:behavioral} examines how behavioral economics can inform both research and policy within the minimalist market design framework, 
while Section \ref{sec:additionalapplications} highlights minimalist applications developed by other researchers using a variety of methods.

I conclude the monograph by addressing two potential criticisms of minimalist market design: whether its conservatism limits its promise, 
and whether its lack of technical guidance generally limits its relevance to specific areas such as matching markets.

\subsection{Minimalist Market Design as a Conservative Framework} \label{sec:conservatism}

Admittedly, the first two pillars make minimalist market design a conservative framework, and critics might argue that this conservatism could impede bold interventions. 
I recognize that if an institution's mission has significant flaws, or if the existing system contains numerous questionable elements, minimalist market design may not be the ideal approach for analysis or reform. 
However, when the institutional mission is at least partially clear and the system is reasonably sound and time-tested, this approach helps avoid many of the critiques often directed at 
mainstream economics and market design---particularly those concerning ``economics imperialism.''

Gary Becker, often seen as the father of economics imperialism, argues that the mainstream economic approach can provide a unified framework for understanding all human behavior \citep{Becker:76}. 
Advocates of mainstream economics' dominance over other social sciences, such as Edward Lazear, suggest that economics is a genuine science, 
employing methodologies that produce testable implications verified through rigorous statistical techniques \citep{Lazear:2000}. 
They argue that economics stands apart by modeling rational individuals who engage in utility-maximizing behavior, focusing on equilibrium outcomes and efficiency---qualities they believe justify 
economics' expansion into intellectual areas previously considered outside its domain.

Unsurprisingly, this perspective often attracts skepticism from other disciplines. Critics point to the underlying assumptions of neoclassical economics, 
accusing economists of being dismissive of alternative ways of understanding social phenomena \citep{Rothschild:2008}. 
They argue that this insularity has reinforced disciplinary boundaries and made economic theories seem disconnected from reality.

Market design has also received its fair share of these critiques. \cite{santos:11} asserts:

\begin{quote} 
``Choice architecture and design economics promote a particular version of economics imperialism that goes beyond the mere export of its concepts to 
territories traditionally occupied by disciplines other than economics. They actually aim at inculcating economic calculus in human deliberation 
and introducing market-like forms of social interaction where they have been absent.'' (p. 721)
\end{quote}

Compared to proponents of economics imperialism, minimalist market design offers a far less intrusive approach when applying economic formalism and 
methodologies to non-economic fields and practical applications.\footnote{For a similar perspective against economic imperialism, see also \cite{Rodrik:2021}, 
who emphasizes that while economists' approaches can complement those of other scholarly disciplines, they cannot substitute the often qualitative methods employed in these fields. } 
It adopts an interdisciplinary stance without insisting on neoclassical principles.\footnote{Importantly, 
minimalist market design does not reject the main principles of neoclassical economics; it simply does not \emph{assume} them.} By adhering to its first two pillars, 
minimalist market design largely avoids the criticisms raised by Santos and other critics of economics imperialism, which may explain its success in interdisciplinary collaborations. 
While often operating outside the boundaries of traditional economics---extending into areas such as education, constitutional law, and medical sciences---it does not promote profit-seeking or strategic behavior. 
Instead, as explored throughout this monograph, the minimalist approach seeks to eliminate the need for such behavior.

Some critics, drawing on the ``performativity thesis'' of \cite{Callon:98}, argue that economic designers create a performative framework in real-world implementations 
\citep{mirowski/khah:2007, santos:11, breslau:2013, hitzig2018, hitzig2020}. They contend that this process shapes participants and institutions to align with theoretical models, 
effectively making reality conform to economists' constructs.\footnote{Two additional concerns related to the performativity thesis are the \textit{normative gap} 
and \textit{algorithmic bias}. In applications where social justice issues are important, \cite{hitzig2020} highlights a concern that market designers' formulation of social goals and 
methods can diverge from policymakers' objectives, leading to what she calls the normative gap. This occurs when the goals implemented by economic engineers do not fully align with the intentions of policymakers. 
Meanwhile, algorithmic bias arises as markets increasingly rely on rules and algorithms to achieve objectives, with some researchers warning that these designs might unfairly advantage 
certain groups \citep{hitzig2018, vanBasshuysen:2022}. When executed properly, minimalist market design can address both issues. 
In cases where there is a unique mechanism satisfying stakeholders' goals, neither a normative gap nor algorithmic bias arises. 
When multiple solutions are possible, best practices involve maintaining \textit{informed neutrality} among reasonable normative positions \citep{Li2017} and 
providing policymakers with a representative list of minimalist interventions. This approach ensures that the design aligns with the institution’s mission and stakeholders' objectives, 
thereby minimizing both the normative gap and algorithmic bias.}

As \cite{santos:11} contends:

\begin{quote} 
``To put it another way, the efficacy of design economics ultimately hinges on determining the extent to which economists are able to implement 
their models in the real world and make reality conform to their theoretical constructs [...]''  (p. 719)
\end{quote}

While minimalist market design is undeniably performative in its aim to influence policy, its performativity fundamentally differs from the form criticized by these scholars. 
Rather than imposing theoretical constructs onto reality, minimalist market design seeks to align institutional practices with stakeholders' genuine objectives, correcting flaws based on the institution's own mission. 
Consequently, this approach can be seen as \textit{virtuous performativity}. As \cite{vanbasshuysen:2023} suggests, the legitimacy of practical institution design can be 
evaluated based on the concept of ``felicity''---the adequacy of the design in achieving its intended purpose:

\begin{quote} 
``Because the purpose of such design processes is to create or change a target in desirable ways, 
the designed mechanism can be evaluated with respect to its `felicity'---according to whether it is adequate for this purpose.'' (p. 1462)
\end{quote}

In this light, minimalist market design emphasizes designing mechanisms that are adequate and effective in fulfilling the institution's mission as defined by its stakeholders. 
External validity, such as those from Chicago and England in school choice (see Section \ref{sec:Chicago-England}) and from India in Supreme Court judgments on affirmative action (see Section \ref{sec:India-EV}), 
provide strong evidence that designs guided by minimalist market design fare well in terms of their felicity. 
These cases show that minimalist market design respects the institution's original mission and enhances its effectiveness without imposing external theoretical constructs.

By concentrating on stakeholders' objectives and the institution's intrinsic goals, minimalist market design fosters collaboration across disciplines and 
effectively addresses concerns about economic imperialism and performativity. 
It offers a practical pathway to reform that respects existing structures while providing solutions to real-world problems.

\subsection{Minimalist Market Design as a Conceptual Paradigm}

Minimalist market design is a conceptual framework for institution design, rather than a technical one that provides specific formal techniques for execution.

Unlike frameworks such as \textit{Myersonian mechanism design} \citep{myerson:81}, which focus on developing technical methods for implementation, 
minimalist market design does not prescribe explicit technical methods. Instead, it leaves it to the researcher to find or develop the appropriate tools to carry out the program. 

There are, however, two notable exceptions. First, applications in matching markets represent an exception, as the framework evolved alongside my research in this area 
through a series of discovery--invention cycles detailed in this monograph (see Figures \ref{fig:DIC-SC-arm} and \ref{fig:DIC-KE-arm}). Second, applications where the axiomatic methodology serves as a suitable formal approach 
may also qualify as exceptions.\footnote{See \cite{budish:12} for a contrast between the Myersonian and axiomatic approaches as technical frameworks for market design.}

While the lack of a uniform technical methodology may pose difficulties for young policy-oriented researchers, 
I do not view this absence as a weakness of minimalist market design. On the contrary, for a methodology aimed at holistic and interdisciplinary efforts in research and policy, I believe this is a necessity.

As the saying goes, ``if all you have is a hammer, everything looks like a nail.'' When researchers limit their perspectives to a specific field and rely on rigid methodologies, 
it inevitably influences how problems are modeled, often leading to significant divergence between the real-world issue and the researcher's conceptual framework. 
For example, a game theorist may focus primarily on private information, an operations researcher on optimization, or a computer scientist on computational complexity. 
Approaching real-world policymaking this way can risk creating a performative framework, where participants and institutions are shaped to fit the researcher's model---effectively 
making reality conform to the construct, as discussed in Section \ref{sec:conservatism}.

For fruitful interdisciplinary collaborations and the potential to generate sound policy advice, an economic designer is best served by keeping an 
open mind about various perspectives---especially in the home area of the intended application---and formal methods.


\bibliographystyle{econ-aea}

\bibliography{Minimalist-Market-Design-October25-ArXiv-v8}

\newpage

\section*{Supplemental Materials}
\setcounter{figure}{0}
\renewcommand{\thefigure}{A\arabic{figure}}

\begin{figure}[!htbp]
    \begin{center}
       \label{fig:OSYM}
    \includegraphics[scale=0.73]{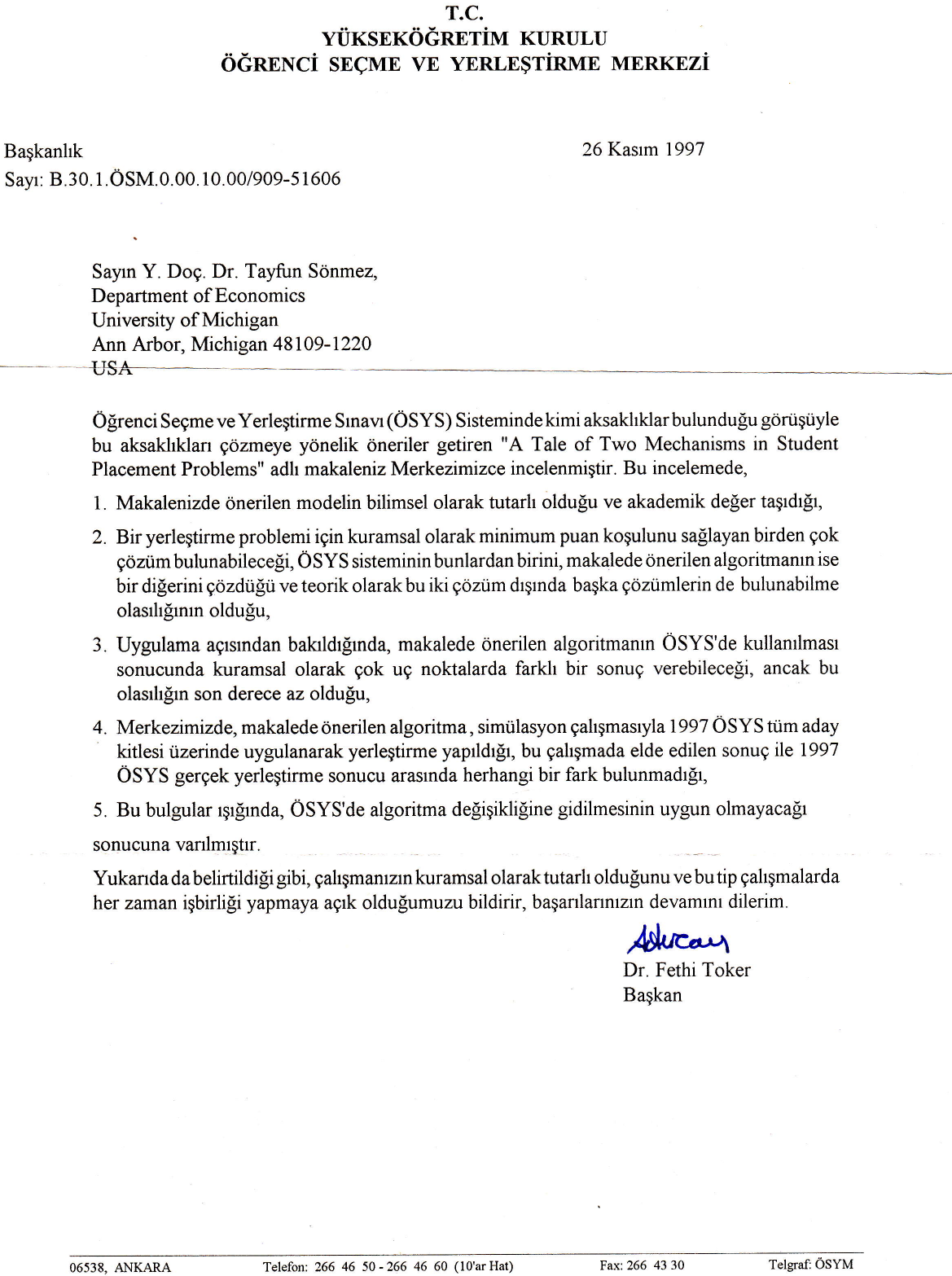}
    \end{center}
    \caption{Response letter from leadership at \"{O}SYM, declining my proposal to reform the mechanism 
    that assigns high school graduates to colleges in Turkey.} \label{OSYM-letter} 
\end{figure}

\begin{figure}
    \begin{center}
       \label{fig:OSYM-English}
    \includegraphics[scale=0.75]{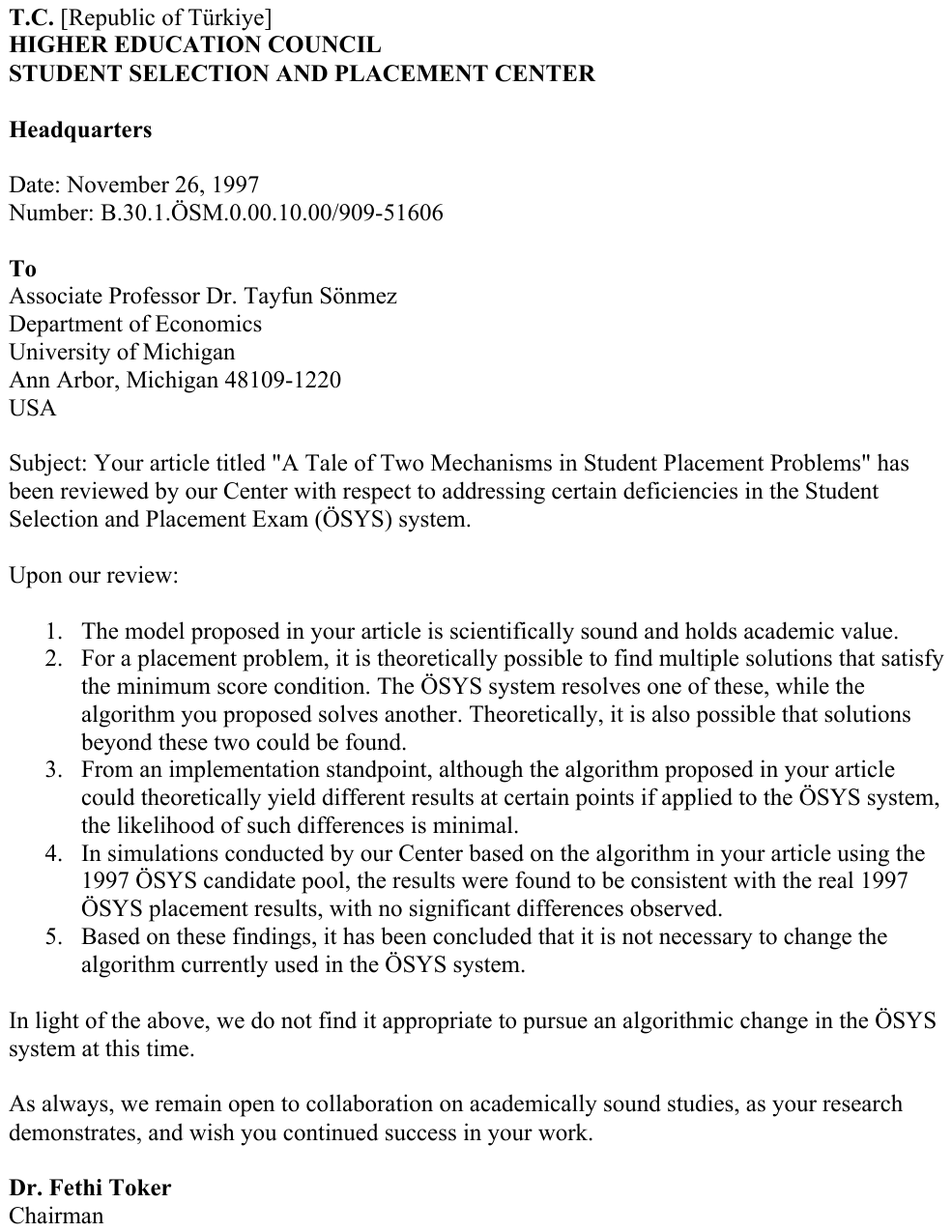}
    \end{center}
    \caption{The English translation of the response letter shown in Figure \ref{OSYM-letter}, from the leadership at \"{O}SYM, the public office in charge of assigning 
    high school graduates to Turkish colleges. The English translation is provided by ChatGBT 40.} \label{OSYM-letter-English} 
\end{figure}

\end{document}